# A Lie Theoretic Framework for Controlling Open Quantum Systems

Corey Patrick O'Meara

TECHNISCHE UNIVERSITÄT MÜNCHEN

Technische Universität München
Department Chemie
Fachgebiet Organische Chemie

# A Lie Theoretic Framework for Controlling Open Quantum Systems

Corey Patrick O'Meara

Vollständiger Abdruck der von der Fakultät für Chemie der Technischen Universität München zur Erlangung des akademischen Grades eines

**Doktors der Naturwissenschaften**

genehmigten Dissertation.

| | | |
|---|---|---|
| Vorsitzender: | | Univ-Prof. Dr. B. Reif |
| Prüfer der Dissertation: | 1. | Univ-Prof. Dr. S. J. Glaser |
| | 2. | Univ-Prof. Dr. K. Hüper,<br>Julius-Maximilians-Universität Würzburg |

# Abstract

This thesis focuses on the Lie theoretical foundations of controlled open quantum systems. We describe Markovian open quantum system evolutions by Lie semigroups, whose corresponding infinitesimal generators are part of a special type of convex cone - a Lie wedge. The Lie wedge associated to a given control system therefore consists of all generators of the quantum dynamical semigroup which are physically realisable as a result of the interplay between the coherent, and incoherent processes the quantum system is subject to. For $n$-qubit open quantum systems, we provide a parametrisation of the largest physically relevant Lie algebra, i.e. the system algebra, which these Lie wedges are contained in - the Lindblad-Kossakowski Lie algebra. This parametrisation provides several useful benefits.

Firstly, it allows us to construct, for the first time, explicit forms of these system Lie wedges and their respective system Lie algebras. The comparison between these two provides new information which one cannot get from only considering the usual system Lie algebra - thereby providing evidence that the Lie wedge is the true fingerprint of an open quantum system.

Secondly, we analyse which control scenarios yield Lie wedges that are closed under BCH-multiplication and therefore generate Markovian semigroups of time-independent quantum channels. Lie wedges of this form are called Lie semialgebras and we completely solve this open problem by proving Lie wedges only specialise to this form when the coherent controls have *no effect* on *both* the inherent drift Hamiltonian and incoherent part of the dynamics.

Thirdly, the parametrisation of the Lindblad-Kossakowski Lie algebra points to an intuitive separation between unital and non-unital dissipative dynamics. Namely, the non-unital component of the open system dynamics is described by an affine transformation, a part of which can be interpreted as a *translation.* These translation operators are then exploited to construct purely dissipative fixed-point engineering schemes to obtain either pure or mixed states as a system's unique fixed point. Precisely, we introduce the novel concept of relating the target state's symmetries to translation operators, which then provides a geometric interpretation of the fixed point analysis. This yields a unified procedure for determining many possible purely dissipative dynamics which drive any initial state to any desired final state which has its non-zero eigenvalues being non-degenerate. The generalisation to obtain *any* target state as the unique fixed point is briefly remarked upon as an extension of this work.

# Zusammenfassung

Die vorliegende Arbeit befasst sich mit den Lie-theoretischen Grundlagen kontrollierter offener Quantensysteme. Wir beschreiben die Entwicklung Markovscher offener Quantensysteme durch Lie-Halbgruppen, deren zugehörige infinitesimale Generatoren Teil eines konvexen Kegels - nämlich des Lie-Keils - sind. Der Lie-Keil für ein bestimmtes Kontrollsystem besteht daher aus allen Generatoren der quantendynamischen Halbgruppe, die physikalisch erzeugt werden können, indem das Quantensystem einem Zusammenspiel zwischen kohärenten und inkohärenten Prozessen ausgesetzt wird. Für ein offenes Quantensystem mit $n$ Qubits geben wir eine Parametrisierung der größtmöglichen relevanten Lie-Algebra, d.h. der Systemsalgebra, in der diese Lie-Keile enthalten sind, an: dies ist die Lindblad-Kossakowski Lie-Algebra. Diese Parametrisierung bietet mehrere nützliche Vorteile.

Erstens erlaubt sie uns erstmalig, explizite Formen dieser System-Lie-Keile und ihrer entsprechenden System-Lie-Algebren zu konstruieren. Der Vergleich zwischen diesen beiden liefert neue Erkenntnisse, die nicht allein durch die Betrachtung der üblichen System-Lie-Algebra gewonnen werden können. Dies weist darauf hin, dass der Lie-Keil der wahre Fingerabdruck eines offenen Quantensystems ist.

Zweitens analysieren wir, für welche Kontrollszenarien Lie-Keile unter BCH-Multiplikation abgeschlossen sind und deswegen Markovsche Halbgruppen zeitunabhängiger Quantenkanäle erzeugen. Lie-Keile dieser Form werden Lie-Halbalgebren genannt. Wir lösen dieses offene Problem durch den Beweis, dass Lie-Keile nur dann diese spezielle Form annehmen, wenn die kohärenten Kontrollen *keinen Effekt* auf *sowohl* den inhärenten Drift-Hamiltonian *als auch* den inkohärenten Teil der Dynamik haben.

Drittens deutet die Parametrisierung auf eine intuitive Trennung zwischen unitaler und nicht-unitaler dissipativer Dynamik hin. Die nicht-unitale Komponente des offenen Quantensystems wird durch eine affine Transformation beschrieben, deren einer Teil als Translation interpretiert werden kann. Diese Translationsoperatoren werden dann genutzt, um rein dissipative Fixpunkt-Konstruktionsschemata zu erstellen, um entweder reine oder gemischte Zustände als den eindeutigen Fixpunkt eines System zu bestimmen. Konkret führen wir als neues Konzept ein, die Symmetrien des Zielzustandes mit Translationsoperatoren in Verbindung zu setzen, was uns dann eine geometrische Interpretation der Fixpunktanalyse erlaubt. Dies liefert ein vereinheitlichtes Verfahren zur Bestimmung vielfältiger rein dissipativer Dynamiken, die einen beliebigen Ausgangszustand in jeden gewünschten Endzustand lenken, dessen von Null verschiedene Eigenwerte nicht-degeneriert sind. Als Erweiterung dieser Arbeit wird kurz besprochen, wie dieses Verfahren so verallgemeinert werden kann, dass *jeder beliebige* Endzustand as eindeutiger Fixpunkt erreicht werden kann.

# Declaration

I hereby declare that parts of this thesis are already published or plan to be submitted to scientific journals:

- C. O'Meara, G. Dirr and T. Schulte-Herbrüggen, *Illustrating the Geometry of Coherently Controlled Unital Open Quantum Systems*, IEEE Trans. Autom. Contr. **57**, 2050–2054, (2012)
- C. O'Meara, G. Dirr and T. Schulte-Herbrüggen, *Systems Theory for Markovian Quantum Dynamics: A Lie-Semigroup Approach*, in preparation

# Eidesstattliche Versicherung

Ich versichere, dass ich die von mir vorgelegte Dissertation selbständig angefertigt, die benutzten Quellen und Hilfsmittel vollständig angegeben und die Stellen der Arbeit, die anderen Werken im Wortlaut oder dem Sinn nach entnommen sind, in jedem Einzelfall als Entlehnung kenntlich gemacht habe; dass diese Dissertation noch keiner anderen Fakultät oder Universität zur Prüfung vorgelegen hat; dass sie — abgesehen von unten angegebenen Teilpublikationen — noch nicht veröffentlicht worden ist sowie, dass ich eine solche Veröffentlichung vor Abschluss des Promotionsverfahrens nicht vornehmen werde. Die Bestimmungen dieser Promotionsordnung sind mir bekannt. Die von mir vorgelegte Dissertation ist von Herrn Prof. Dr. S. J. Glaser betreut worden.

**Publikationsliste**

- C. O'Meara, R. Pereira, *Self-Dual Maps and Symmetric Bistochastic Matrices*, Lin. Multilin. Alg. **61**, 23–34, (2013)
- C. O'Meara, G. Dirr and T. Schulte-Herbrüggen, *Illustrating the Geometry of Coherently Controlled Unital Open Quantum Systems*, IEEE Trans. Autom. Contr. **57**, 2050–2054, (2012)

# Acknowledgements

I would first like to thank Dr. Thomas Schulte-Herbrüggen for giving me this amazing opportunity to live and study here in Munich. His constant support and imaginative ideas over the years really helped develop this work farther than I had hoped. The constant pushing finally did pay off! I'd also like to thank my unofficial second advisor, Dr. Gunther Dirr, for being ultra-patient and supportive over the years working with me. I've learned so much from your clean and precise math style, I hope I take your rigorousness with me to other places in life. Special thanks to my "doctor father" Dr. Steffen Glaser for always being there when I needed his guidance (or signature!). A huge thank you to Dr. Knut Hüper for agreeing to be a second referee and to represent Würzburg on such short term notice as well as Dr. Bernd Reif for agreeing to Chair my examination.

A special thanks to Dr. Robert Zeier and Dr. Ariane Garon for providing feedback on many issues and giving sound, technical advice. To the rest of the Glaser Group and friends: I couldn't have done this without the collective of everyone encouraging me both here and in Canada. Thanks to Dr. Uwe Sander for helping me with German over the years and giving colourful commentary on the state of academia. Special mention goes to Dr. Rob Fisher and Phil Nehammer for putting up with my shenanigans and listening to me talk about work every week for 3.5 years now.

To the Elite Network of Bavaria (ENB) and the Quantum Computing, Control and Communication (QCCC) Ph.D. program for providing me with the opportunity to work and live here. Many thanks to Simone Lieser and Martha Fill for making my transition easier and for helping me with paperwork along the way.

To Julia, thanks for being understanding throughout these wild years and even crazier last few months!

And finally to my family: Heather, Shawn, Kirk, Grandma Dee, Grandpa Jim and Grandad - thanks for everything big and small each of you have done for me - it made the distance seem much shorter.

# Contents

# Introduction

Typical quantum systems are inevitably bombarded by neighbouring systems which alter their current state and drive them into another. Usually, the interaction between the environment and a quantum system which is encoding some information results in a loss of information or coherence. Nonetheless, this environmental noise has recently been shown to provide some remarkable *advantages* in many areas of quantum computation. Concomitantly to some general foundations mostly on Markovian dynamics [10, 8, 55, 56, 49, 50, 16, 54, 45, 9, 48, 51, 46, 47, 52], relaxation may actually be exploited as an additional resource, in particular in dissipative quantum computing [32, 14, 53, 33], fixed-point engineering [10, 8, 9], memory design and simulation [7, 44] as well as noise-switching [11].

From a control theorists perspective, one can ask certain fundamental questions such as controllability, accessibility, reachability, and stabilizibility of a quantum system subject to the open environment - certainly, an overarching question is: to what extent can we manipulate the quantum system to do as we desire while outside influences interfere with our controls? An overarching framework of systems theory comprising coherent and incoherent dynamics is therefore desirable to answer such questions. Here we set out for a unified picture in terms of Lie theory. Following the well-established use of Lie groups for characterizing closed quantum systems (see, e.g., Jurdjevic [28], Dirr and Helmke [15] or d'Alessandro [12]), we take the next step towards the open system picture by applying Lie semigroup theory [23, 38, 24, 36], to describe open Markovian quantum systems. First steps in this direction were made for controllability of single qubit systems [2] and beyond [35]. Furthermore, Markovian quantum maps can be *defined* as being infinitesimally divisible [55, 56] or exponentially generated [16]. In fact, one can further show that (in the connected component), the Lie and the Markov properties of quantum maps are one-to-one in the sense that every Markovian quantum map has a representation as a Lie semigroup, while non-Markovian maps do not [16].

Having this property allows many powerful Lie theoretic tools which were already developed in the mathematics literature (see, e.g., [23]) to be used for solving problems related to Markovian semigroups of quantum maps. For example, the set of Lindblad generators which generate the corresponding one parameter semigroups of quantum channels form a closed convex cone called a Lie wedge. Clearly then, this Lie wedge is contained in some smallest vector space - which turns out to be the systems dynamic Lie algebra. This system Lie algebra is usually considered to be the main algebraic structure that describes the open (or closed) system - however, in this work we show that the Lie wedge associated to a given system yields geometric and algebraic properties of the control system which arn't visible by analysing the system Lie algebra alone.

This thesis is structured as follows:

In Chapter 1 we provide the reader with the necessary tools and background on Lie semigroups, quantum control theory, Markovian quantum maps and their generators - and in particular, the overarching connection between all of them.

In Chapter 2 we set out to provide a useful parametrisation of the so-called Lindblad-Kossakowski (LK) Lie algebra. This is the largest physically allowed system algebra which can be generated by any quantum control system. For the case of $n$-qubit systems, we explicitly parameterise the different components in terms of a Pauli basis. This parametrisation is then inherited by the overall LK-wedge (the largest physically possible) which generates all coherently controlled Markovian quantum maps as well as by all the individual (sub)wedges of Markovian control systems. Another benefit of the parametrisation is that it allows one to interpret certain geometric time evolutions that the Lindblad generator induces. Namely, we show that part of the dissipative dynamics for non-unital systems can be expressed as a translation-type operator (an affine shift) which drives elements of the set of states along directions of the state space.

In Chapter 3 we provide a systematic way of engineering unique fixed points of the Markovian quantum system using purely dissipative noise. We use the translational part of the (non-unital) system algebra to our advantage and relate it to the symmetries of the target fixed point. This turns out to be part of a unified symmetry picture which we show provides solutions for obtaining either pure or mixed state fixed points uniquely. In particular, we obtain new solutions for obtaining GHZ states, W States, Stabilizer States (or for example the toric code subspace) as well as certain Dicke States as unique fixed points (or subspaces) of a system. We finally show that using this symmetry picture and allowing for full Hamiltonian control, one can arrive at any target state - which has its non-zero eigenvalues being non-degenerate - by a multitude of ways. Thereby opening the door to future work in optimising the choice of these different solutions.

Chapter 4 then considers Hamiltonian drift control at large. It is devoted to characterizing a plethora of concrete examples of unitarily controlled unital and non-unital Markovian systems in terms of their explicit Lie wedges. The structures emerge from one-qubit and two-qubit examples and are discussed in view of extensions to general $n$-qubit systems. Several illustrative tables are provided where a collection of information is given such that the dimension of each systems system algebra and the relative dimension of the systems Lie wedge can be compared. We use these notions to argue that a systems Lie wedge can truly be regarded as a fingerprint of the open quantum system since its structure is more unique to the system setup than the systems Lie algebra itself.

In Chapter 5 we focus on systems that simplify to solutions of time-independent Lindblad master equations. They are characterized by Lie wedges that are closed under Baker-Campbell-Hausdorf (BCH) multiplication and thus specialise to the form of Lie semialgebras. As shown earlier in [16], evolutions generated by those Lie semialgebras can be realized as *time-independent* Markovian maps. In turn, this is a necessary (but not sufficient) precondition for experimental implementation without switching controls. We show that Lie wedges only specialize to this form of Lie wedge when the coherent controls have *no effect* on *both* the inherent drift Hamiltonian and incoherent part of the dynamics.

# Chapter 1

# Background

We start out by recalling some basic notions and notations of Lie subsemigroups [23] and their application for characterising reachable sets of quantum control systems modelled by controlled Lindblad-Kossakowski master equations [16].

## 1.1 Introduction to Lie Semigroups

Let $\mathbf{G}$ be a linear Lie group, i.e. a path-connected subgroup of the general linear group $GL(\mathbf{H})$, where $\mathbf{H}$ is a finite dimensional real or complex (Hilbert) space, and let $\mathfrak{g}$ be its corresponding matrix Lie algebra. Thus $\mathfrak{g}$ is a Lie subalgebra of $\mathfrak{gl}(\mathbf{H})$. For $\mathbf{H} = \mathbb{R}^n$ or $\mathbf{H} = \mathbb{C}^n$ we write as usual $GL(n, \mathbb{R})$ and $GL(n, \mathbb{C})$, respectively, instead of $GL(\mathbf{H})$ as well as $\mathfrak{gl}(n, \mathbb{R})$ and $\mathfrak{gl}(n, \mathbb{C})$, respectively, instead of $\mathfrak{gl}(\mathbf{H})$.

Then a subset $\mathbf{S} \subset \mathbf{G}$ which is closed under the group operation in the sense $\mathbf{S} \cdot \mathbf{S} \subseteq \mathbf{S}$ and which contains the identity $\mathbb{1}$ is called *subsemigroup* of $\mathbf{G}$. The largest subgroup within $\mathbf{S}$ is written $E(\mathbf{S}) := \mathbf{S} \cap \mathbf{S}^{-1}$. Moreover, a closed convex cone $\mathfrak{w} \subset \mathfrak{g}$ is called a wedge with the largest linear subspace of $\mathfrak{w}$, $E(\mathfrak{w}) := \mathfrak{w} \cap (-\mathfrak{w})$, denoting the *edge of the wedge* $\mathfrak{w}$. Now, $\mathfrak{w} \subseteq \mathfrak{g}$ forms a *Lie wedge* of $\mathfrak{g}$ if it is invariant under the adjoint action of the subgroup generated by the edge $E(\mathfrak{w})$, i.e. if it satisfies

$$e^{\mathrm{ad}_A}(\mathfrak{w}) = e^A \, \mathfrak{w} \, e^{-A} = \mathfrak{w} \tag{1.1}$$

for all $A \in E(\mathfrak{w})$. Clearly, the edge of a Lie wedge always forms a Lie subalgebra of $\mathfrak{g}$.

For any closed subsemigroup $\mathbf{S}$ of $\mathbf{G}$ consider its tangent cone $L(\mathbf{S})$ at the identity $\mathbb{1}$ given by

$$L(\mathbf{S}) := \{A \in \mathfrak{g} \mid \exp(tA) \in \mathbf{S} \;\text{ for all } t \geq 0\} \,. \tag{1.2}$$

Then $L(\mathbf{S})$ is a Lie wedge of $\mathfrak{g}$ satisfying $E\big(L(\mathbf{S})\big) = L\big(E(\mathbf{S})\big)$. Yet, the 'local-to-global' correspondence between Lie wedges and closed connected subsemigroups is much more subtle than the correspondence between Lie (sub)algebras and Lie (sub)groups: for instance, several connected subsemigroups may share the same Lie wedge $\mathfrak{w}$ in the sense that $L(\mathbf{S}) = L(\mathbf{S}')$ for $\mathbf{S} \neq \mathbf{S}'$, or conversely there may be Lie wedges $\mathfrak{w}$ which do not correspond to any subsemigroup, i.e. $\mathfrak{w} = L(\mathbf{S})$ fails for all subsemigroups $\mathbf{S} \subset \mathbf{G}$.

Hence one introduces the important notion of a *Lie subsemigroup* $\mathbf{S}$ characterised by the equality

$$\mathbf{S} = \overline{\langle \exp L(\mathbf{S}) \rangle_S} \,, \tag{1.3}$$

where the closure is taken in $\mathbf{G}$ and $\langle \exp L(\mathbf{S}) \rangle_S$ denotes the subsemigroup generated by $\exp L(\mathbf{S})$, i.e. $\langle \exp L(\mathbf{S}) \rangle_S := \{e^{A_1} \cdots e^{A_n} \mid A_1, \ldots, A_n \in L(\mathbf{S}),\, n \in \mathbb{N}\}$. Thus, precisely this type of subsemigroup can be completely 'reconstructed' by its Lie wedge. Moreover, a Lie wedge $\mathfrak{w}$ is said to be *global* in $\mathbf{G}$, if there is a Lie subsemigroup $\mathbf{S} \subset \mathbf{G}$ such that

$$L(\mathbf{S}) = \mathfrak{w} \ . \tag{1.4}$$

Thus, one has the identity $\mathbf{S} = \overline{\langle \exp \mathfrak{w} \rangle}_S$.

Whenever a Lie wedge $\mathfrak{w} \subset \mathfrak{g}$ specialises to be compatible with the Baker-Campbell-Hausdorff (BCH) multiplication

$$A \star B := A + B + \tfrac{1}{2}[A, B] + \cdots = \log(e^A e^B) \quad \forall A, B \in \mathfrak{w} \tag{1.5}$$

defined via the BCH series, it is termed *Lie semialgebra.* For this to be the case, there has to be an open BCH neighbourhood $\mathcal{B} \subset \mathfrak{g}$ of the origin such that $(\mathfrak{w} \cap \mathcal{B}) \star (\mathfrak{w} \cap \mathcal{B}) \subseteq \mathfrak{w}$. An equivalent useful definition for being a Lie semialgebra is the tangential condition

$$[A, T_A\mathfrak{w}] \subseteq T_A\mathfrak{w} \quad \text{for all } A \in \mathfrak{w} \ , \tag{1.6}$$

with $T_A\mathfrak{w}$ denoting the tangent space of $\mathfrak{w}$ at $A$. For more detail see Chapter 5.

Only in Lie semialgebras the exponential map of a zero-neighbourhood relative to $L(\mathbf{S})$ yields a $\mathbb{1}$-neighbourhood relative to $\mathbf{S}$. In contrast, as soon as $\mathfrak{w}$ is merely a Lie wedge without the stronger structure of a Lie semialgebra, there are elements in $\mathbf{S}$ that are arbitrary close to the identity without belonging to any one-parameter semigroup completely contained in $\mathbf{S}$. Hence the conceptual importance of Lie semialgebras lies in the fact that at least locally around the identity $\mathbb{1}$ the image of $\mathfrak{w}$ under the exponential map yields $\mathbf{S}$ without taking further products, cf. Eqn. (1.3). For details, a variety of illustrative examples, and a lucid overview of the entire subject, see [23] and [24].

For later applications of the above concepts to controlled open quantum system and the computation of their associated Lie wedges (see Theorem 2) the following corollary which results from the so-called Globality Theorem in [23] will be of vital importance.

**Corollary 1.1.1** ([23])**.** *Let $\boldsymbol{G}$ be a (linear) Lie group with Lie algebra $\mathfrak{g}$ and let $\mathfrak{w}' \subset \mathfrak{w}$ be two nested Lie wedges in $\mathfrak{g}$. Then $\mathfrak{w}'$ is global in $\boldsymbol{G}$ if the following conditions are satisfied:*

*(a) $\mathfrak{w}$ is global in $\boldsymbol{G}$;*

*(b) the edge of $\mathfrak{w}'$ is given by $E(\mathfrak{w}') = E(\mathfrak{w}) \cap \mathfrak{w}'$;*

*(c) the edge of $\mathfrak{w}'$ is the Lie algebra of a closed Lie subgroup of $\boldsymbol{G}$.*

In other words, if the edge of the wedge $\mathfrak{w}'$ follows the intersection $E(\mathfrak{w}') = E(\mathfrak{w}) \cap \mathfrak{w}'$ and $\mathfrak{w}$ is global, then $\mathfrak{w}'$ is also a global Lie wedge, provided $\exp E(\mathfrak{w}')$ generates a closed subgroup.

Thus we set the frame to analyse the time evolution of Markovian (i.e. memory-less) open quantum systems in the differential geometric picture of Lie wedges. The first connection between Lie wedges and time-independent Markovian quantum maps was detailed in [16].

## 1.2 Markovian Quantum Dynamics in Terms of Lie Semigroups

Markovian quantum dynamics follows the Lindblad-Kossakowski master equation

$$\dot{X}(t) = -\mathcal{L}\, X(t) \tag{1.7}$$

where usually $X(t)$ is identified with the density operator $\rho(t)$, (i.e. $\rho(t) \geq 0$, and $\operatorname{tr}\rho(t) = 1$). Here and henceforth, $(\cdot)^\dagger$ denotes the adjoint (complex-conjugate transpose). For ensuring complete positivity, $\mathcal{L}$ has to be of Lindblad form [37], i.e.

$$\mathcal{L}(\rho) = \mathrm{i}\,\mathrm{ad}_H(\rho) + \Gamma(\rho)\ , \tag{1.8}$$

with $\mathrm{ad}_{H_j}(\rho) := [H_j, \rho]$ and

$$\Gamma(\rho) := \tfrac{1}{2}\sum_k V_k^\dagger V_k \rho + \rho V_k^\dagger V_k - 2 V_k \rho V_k^\dagger\ . \tag{1.9}$$

Since we restrict to finite dimensional systems in the following, the *Hamiltonian H* is represented by a Hermitian $N \times N$ matrix while the *Lindblad terms* $\{V_k\}$ may be arbitrary $N \times N$ matrices. The equation of motion given by Eqn. (1.7) acts on the vector space of all $N \times N$ Hermitian operators, $\mathfrak{her}(N)$ leaving the set of all density operators $\mathfrak{pos}_1(N) := \{\rho \in \mathfrak{her}(N)\,|\, \rho \geq 0, \operatorname{tr}\rho = 1\}$ invariant.

In [16] we showed that the set of all Lindblad generators $-\mathcal{L}$ allowing a representation as in Eqn. (1.8) has an interpretation as a Lie wedge. To see this, consider the group lift of Eqn. (1.7), where now $X(t)$ denotes an element in the general linear group $GL(\mathfrak{her}(N))$. Moreover, define the set of all completely positive (CP), trace-preserving (TP) invertible linear operators acting on $\mathfrak{her}(N)$ as $\mathbf{T}^{\mathsf{CPTP}}$, i.e.

$$\mathbf{T}^{\mathsf{CPTP}} := \{T \in GL(\mathfrak{her}(N))\,|\, T \text{ is CP and TP}\}$$

and let $\mathbf{T}_{\mathbf{0}}^{\mathsf{CPTP}}$ denote the connected component of the identity. Then, $\mathbf{T}^{\mathsf{CPTP}}$ is exactly the set of so-called invertible *quantum maps*. A quantum map $T$ is said to be *time independent Markovian* if it is a solution of Eqn. (1.7), or more precisely, if $T = \mathrm{e}^{-t\mathcal{L}}$ for some fixed Lindblad generator $\mathcal{L}$ and some $t \geq 0$. More generally, $T$ is *time dependent Markovian* if it is a solution of Eqn. (1.7), where now $\mathcal{L} = \mathcal{L}(t)$ may vary in time (see also [55, 56, 43]). Let us denote the set of all *time independent Markovian* and *time dependent Markovian* quantum maps by **TIM** and **TDM**, respectively.

Trivially, one has **TIM** $\subset$ **TDM**. Note that in the literature, sometimes just the time independent Markovian maps **TIM** are briefly called 'Markovian'. This leads to confusion when it comes to 'non-Markovian' quantum maps: In accordance with the divisibility criteria in [55, 56], which can be taken over to Lie semigroups [16] and a recent review including broader discussions of terminology [43], here and henceforth, we call a quantum map *non-Markovian* if it is *not* time dependent Markovian (**TDM**) and thus (by **TIM** $\subset$ **TDM**) not time independent Markovian (**TIM**) either. In turn, in this terminology *Markovian quantum maps* comprise both, time dependent and time independent Markovian maps. Thus all Markovian quantum maps arise as solutions of time dependent or time independent Lindblad-Kossakowski equations of motions, while the non-Markovian quantum maps can only be represented by Kraus maps and not by solutions of Eqn. (1.7). These stipulations are made precise in the following fundamental result:

**Theorem 1** ([16])**.** *The setting of Lindblad [37] and Kossakowski [18] and the divisibility characterisations by Wolf and Cirac [55] can be embraced by the following formulation in terms of Lie semigroups:*

*(a) The* Lie wedge *of* $\boldsymbol{T}_{\mathbf{0}}^{\mathsf{CPTP}}$ *denoted by* $L(\boldsymbol{T}_{\mathbf{0}}^{\mathsf{CPTP}})$ *is given by the* set of all Lindblad generators*, i.e. by the set of all operators of the form*

$$-\mathcal{L} := -\big(\mathrm{i}\,\mathrm{ad}_H + \Gamma\big) \tag{1.10}$$

*with $H \in \mathfrak{her}(N)$ and $\Gamma$ as in Eqn. (1.9). It is global and generates a Lie semigroup that exactly coincides with the closure of all time dependent Markovian quantum maps in the sense*

$$\overline{\mathbf{TDM}} = \overline{\langle \exp L(\boldsymbol{T}_{\mathbf{0}}^{\mathsf{CPTP}}) \rangle}_S \,. \tag{1.11}$$

*Thus $\overline{\mathbf{TDM}}$ excludes all non-Markovian maps in $\boldsymbol{T}_{\mathbf{0}}^{\mathsf{CPTP}}$, as the non-Markovian maps are exactly those that prevent $\boldsymbol{T}_{\mathbf{0}}^{\mathsf{CPTP}}$ from being a Lie subsemigroup.*

*(b) The set of time independent Markovian quantum maps* **TIM** *is by definition a collection of one-parameter subsemigroups, i.e.*

$$\mathbf{TIM} = \exp L(\boldsymbol{T}_{\mathbf{0}}^{\mathsf{CPTP}}) \,. \tag{1.12}$$

*Neither* **TIM** *takes the form of a semigroup nor $L(\boldsymbol{T}_{\mathbf{0}}^{\mathsf{CPTP}})$ the form of a Lie semialgebra. However,* **TIM** *comprises (at least near the identity) all Lie subsemigroups of $\boldsymbol{T}_{\mathbf{0}}^{\mathsf{CPTP}}$ the Lie wedges of which specialize to Lie semialgebras.*

*(c) The results can be connected as*

$$\overline{\mathbf{TDM}} = \overline{\langle \exp L(\boldsymbol{T}_{\mathbf{0}}^{\mathsf{CPTP}}) \rangle}_S = \overline{\langle \mathbf{TIM} \rangle}_S \,, \tag{1.13}$$

*where all closures are taken relative to $GL(\mathfrak{her}(N))$.*

*Proof.* The above statement is largely a rearrangement of results already proven in previous works. Part (a) and (c) follow immediately from Theorems 3.2 - 3.4 as well as Corollaries 3.1 and 3.2 in [16]. While Eqn. (1.12) of part (b) follows by definition, Corollary 3.3 in [16] implies that neither $L(\mathbf{T}_{\mathbf{0}}^{\mathsf{CPTP}})$ constitutes a Lie semialgebra nor **TIM** a subsemigroup, see also Theorem 2.2 in [16]. Finally, the remarkable fact that **TIM** includes (at least locally) all Lie subsemigroups with Lie wedges taking the form of semialgebras follows from Theorem 2.2 in [16]. □

To summarize in simplified terms: Even close to identity (i.e. in the connected component $\mathbf{T}_{\mathbf{0}}^{\mathsf{CPTP}}$) the completely positive, trace-preserving maps $\mathbf{T}^{\mathsf{CPTP}}$ do *not* form a Lie subsemigroup of $GL(\mathfrak{her}(N))$. Moreover (in the connected component $\mathbf{T}_{\mathbf{0}}^{\mathsf{CPTP}}$), one finds two important division lines: (i) the border between (time dependent) Markovian maps and non-Markovian maps (i.e. neither time dependent Markovian nor time independent Markovian) is drawn by the Lie-semigroup property, while (ii) the demarcation between time dependent and time independent Markovian maps results from the fact that $L(\mathbf{T}_{\mathbf{0}}^{\mathsf{CPTP}})$ does not specialize to a Lie semialgebras.

A similar result as Theorem 1 holds for the closed subgroup $\mathbf{T}_u^{\mathsf{CPTP}}$ of all *unital* invertible quantum maps and it connected $\mathbb{1}$-component $\mathbf{T}_{u,\mathbf{0}}^{\mathsf{CPTP}}$. Since the corresponding Lie wedges are of the utmost importance for the further presentation of our results we henceforth denote them by

$$\mathfrak{w}^{LK} := L(\mathbf{T}_{\mathbf{0}}^{\mathsf{CPTP}}) \qquad \text{and} \qquad \mathfrak{w}_0^{LK} := L(\mathbf{T}_{u,\mathbf{0}}^{\mathsf{CPTP}}) \,. \tag{1.14}$$

We note that the notation between the distinction of unital and non-unital wedges will hold throughout this thesis. Namely, any Lie wedge related to a unital system ($\Sigma$) will be denoted $\mathfrak{w}_0$, whereas one which is related to a non-unital system will lack the subscript.

Furthermore, we can define the so-called Lindblad-Kossakowski Lie algebra as follows. It is the smallest Lie algebra which contains the Lindblad-Kossakowski Lie wedge i.e.

$$\mathfrak{g}^{LK} := \langle \mathfrak{w}^{LK} \rangle_{\mathrm{Lie}} \,, \quad \text{and} \quad \mathfrak{g}_0^{LK} := \langle \mathfrak{w}_0^{LK} \rangle_{\mathrm{Lie}} \,, \tag{1.15}$$

for non-unital and unital systems, respectively. Providing a parametrisation of the basis elements of these Lie algebras will be the focus of Section 1.3 since we can then explicitly describe the structure of Lie wedges which are contained within them.

Next, let us recall some implications of the above semigroup theory to coherently controlled open quantum systems, i.e. to quantum system of the following Lindblad-Kossakowski form

$$(\Sigma) \qquad \dot{X}(t) = -\mathcal{L}_{u(t)}\, X(t), \qquad X(0) \in GL(\mathfrak{her}(N)) \tag{1.16}$$

where now $\mathcal{L}_{u(t)}$ depends on a possibly time-dependent control $u(t) \in \mathbb{R}^m$. More precisely, $\mathcal{L}_{u(t)}$ is given by

$$\mathcal{L}_{u(t)}(\rho) = \mathrm{i}\,\mathrm{ad}_{H_{u(t)}}(\rho) + \Gamma(\rho)\ , \tag{1.17}$$

with the operator $\Gamma$ as given by Eqn. (1.9) and

$$H_{u(t)} := H_d + \sum_{K=1}^{m} u_k(t) H_k \tag{1.18}$$

Here, the term 'coherently controlled' accounts for the fact that the controls affect only the Hamiltonian part of Eqn. (1.17). By the above discussion, we have established that that solutions of Eqn. (1.16) are Markovian quantum maps, and thus we can define the *system semigroup* $\mathbf{P}_\Sigma$ and *system group* $\mathbf{G}_\Sigma$ associated to $(\Sigma)$ as

$$\mathbf{P}_\Sigma = \langle T_u(t) = \exp(-t\mathcal{L}_u) \,|\, t \geq 0, u \in \mathbb{R}^m \rangle_S\ , \quad \text{and} \quad \mathbf{G}_\Sigma = \langle \mathbf{P}_\Sigma \rangle_G\ , \tag{1.19}$$

respectively, and hence $\mathbf{G}_\Sigma$ is just the Lie group generated by system semigroup. In this group lifted scenario, the *reachable set* of $(\Sigma)$ is defined as the set of all maps $X(T)$ for $T \geq 0$ that can be reached from the unity $X(0) = \mathbb{1}$ under the dynamics of $(\Sigma)$, i.e.

$$\mathrm{Reach}_\Sigma(\mathbb{1}) := \bigcup_{T \geq 0} \mathrm{Reach}\,(\mathbb{1}, T)\ . \tag{1.20}$$

Since $\mathrm{Reach}\,(\mathbb{1}, T_1) \cdot \mathrm{Reach}\,(\mathbb{1}, T_2) = \mathrm{Reach}\,(\mathbb{1}, T_1 + T_2)$, its clear that $\mathrm{Reach}_\Sigma(\mathbb{1})$ is a subsemigroup of $GL(\mathfrak{her}(N))$. The following well-known result [36, 16] allows us to associate to each coherently controlled open quantum system $(\Sigma)$ a unique Lie wedge $\mathfrak{w}_\Sigma$.

**Theorem 2** ([36, 16]). *Let $(\Sigma)$ be given as in Eqn. (1.16), $\boldsymbol{P}_\Sigma$ and $\boldsymbol{G}_\Sigma$ be the system semigroup and system group given by Eqn. (1.19) and assume that $\boldsymbol{G}_\Sigma$ is a closed subgroup of $GL(\mathfrak{her}(N))$. Then for the closures taken relative to $GL(\mathfrak{her}(N))$ we have that*

*(a)* $\overline{\boldsymbol{P}}_\Sigma = \overline{\mathrm{Reach}_\Sigma(\mathbb{1})}$

*(b)* $\overline{\boldsymbol{P}}_\Sigma \subset \overline{\mathbf{TDM}}$ *is a Lie subsemigroup of $GL(\mathfrak{her}(N))$.*

*(c) The interior of $\overline{\boldsymbol{P}}_\Sigma$ and the interior of* $\mathrm{Reach}_\Sigma(\mathbb{1})$ *coincide.*

*(d) The Lie wedge $\mathfrak{w}_\Sigma := L(\overline{\boldsymbol{P}}_\Sigma) \subset \mathfrak{w}^{LK}$ is the smallest Lie wedge of $\mathfrak{gl}(\mathfrak{her}(N))$ which is global and covers all evolution directions of the form $\mathcal{L}_u = \mathrm{i}\,\mathrm{ad}_{H_u} + \Gamma$, $u \in \mathbb{R}^m$*

Due to the property that $\mathfrak{w}_\Sigma$ is also the largest subset of $\mathfrak{gl}(\mathfrak{her}(N))$ to which the evolution directions can be extended without enlarging the closure of the reachable set, it is often called the *Lie saturate* of $(\Sigma)$ in control theory cf. [30, 29, 36]. Furthermore, we will refer to $\mathfrak{w}_\Sigma$ (later dropping the index $\Sigma$) as the *Lie wedge associated to* the control system ($\Sigma$), and they will be the focus of Section 4.1. There, we will provide an explicit form of these Lie wedges for any given control system and then provide several illustrative examples.

**Remark 1.** 1. *A main result of this thesis is the parametrisation of the Lindblad-Kossakowski Lie algebra $\mathfrak{g}^{LK}$ (and its unital subalgebra $\mathfrak{g}_0^{LK}$) cf. Eqn.* (1.15)*. This Lie algebra is the largest possible system algebra any open quantum system may have. In Section 1.3 we provide an explicit representation of $\mathfrak{g}^{LK}$ (and hence $\mathfrak{g}_0^{LK}$), and show that they are equivalent to Eqns.* (1.45) *and* (1.46)*, respectively. A simple criterion which guarantees that the system group $\boldsymbol{G}_\Sigma$ is a closed subgroup of $GL(\mathfrak{her}(N))$ is the accessibility of* ($\Sigma$) *which is equivalent in the non-unital case to the fact that the system algebra $\mathfrak{g}_\Sigma$ coincides with $\mathfrak{g}^{LK}$ and in the unital case with $\mathfrak{g}_0^{LK}$, where $\mathfrak{g}^{LK}$*

2. *If* ($\Sigma$) *does not meet the above closedness assumption of $\boldsymbol{G}_\Sigma$ one can restate Theorem 2 relatively to $\boldsymbol{G}_\Sigma$, i.e. all closures have to be taken with respect to $\boldsymbol{G}_\Sigma$.*

## 1.3 Lindblad-Kossakowski Operators: Representations and Properties

Given a Lindblad-Kossakowski operator $\mathcal{L} : \mathfrak{her}(N) \to \mathfrak{her}(N)$ as in Eqn. (1.7), i.e. $\mathcal{L}$ is the infinitesimal generator of a completely positive semigroup of linear operators $T(t) := \mathrm{e}^{-t\mathcal{L}}$, $t \geq 0$. As already mentioned, it is well-known by the seminal work of Lindblad [37] that $\mathcal{L}$ can be represented in the following forms:

$$\mathcal{L} := \mathrm{i}\,\mathrm{ad}_\mathrm{H} + \Gamma \qquad \text{with} \qquad \Gamma(\rho) := \tfrac{1}{2}\sum_{k=1}^{m}\left(V_k^\dagger V_k\rho + \rho V_k^\dagger V_k - 2V_k\rho V_k^\dagger\right), \tag{1.21}$$

where $H$ is Hermitian and $V_k \in \mathfrak{gl}(N, \mathbb{C})$. Equivalently, Kossakowski, Gorini and Sudarshan [18] have shown that $\mathcal{L}$ allows the representation

$$\mathcal{L} := \mathrm{i}\,\mathrm{ad}_{\mathrm{H}'} + \Gamma' \qquad \text{with} \qquad \Gamma'(\rho) := -\tfrac{1}{2}\sum_{j,k}^{N^2-1} a_{jk}\Big([B_j, \rho B_k^\dagger] + [B_j\rho, B_k^\dagger]\Big), \tag{1.22}$$

where $H'$ is Hermitian, $A := (a_{jk})_{j,k=1,\dots N^2-1}$ is a positive semi-definite $(N-1)\times(N-1)$ matrix, and $B_1, \dots, B_{N^2-1}$ is an orthonormal basis of $\mathfrak{sl}(N, \mathbb{C})$. Here and henceforth, the matrix $A$ is called *GKS-matrix* of $\mathcal{L}$ (relative to $B_1, \dots, B_{N^2-1}$), where GKS stands for Gorini-Kossakowski-Sudarshan [18].

**Remark 2.** *Note that the $V_k$ terms given by Eqn.* (1.21) *are not at all unique even if phase factors are disregarded and renumbering is admitted. In contrast, the GKS-matrix $A$ of Eqn.* (1.22) *is uniquely determined once an orthogonal basis $B_1, \dots, B_{N^2-1}$ is fixed. This follows from a somewhat tedious calculation, see e.g. [34], Lemma 2.4 and Prop. 2.24.*

Another straightforward calculation yields that GKS-matrices with respect to different orthogonal basis sets differ only by a unitary conjugation and thus Eqn. (1.21) can be obtained from Eqn. (1.22) via a diagonalising transformation of $A$. Conversely, expanding Eqn. (1.21) in an orthonormal basis $B_0 = \mathbb{1}_N, B_1, \dots, B_{N^2-1}$ of $\mathfrak{gl}(N, \mathbb{C})$ readily yields Eqn. (1.22), cf. [34].

The subsequent largely known results clarify some further uniqueness aspect of the above representations. First, let us introduce the following terminology: a Lindblad-Kossakowski operator $\mathcal{L}$ is called *purely dissipative* if $\mathcal{L}$ is orthogonal to $\mathrm{ad}_{\mathfrak{su}(N^2)} := \{\mathrm{i}\,\mathrm{ad_H} \mid H \in \mathfrak{her}_0(N)\}$, i.e. if

$$\langle \mathcal{L}, \mathrm{i}\,\mathrm{ad_H}\rangle = 0 \tag{1.23}$$

for all $H \in \mathfrak{her}_0(N)$, where, $\langle\cdot,\cdot\rangle$ denotes the Hilbert-Schmidt scalar product on $\mathfrak{gl}\big(\mathfrak{her}(N)\big)$. That is

$$\langle \Phi, \Psi\rangle := \mathrm{Tr}(\Phi^*\Psi) := \sum_{k=0}^{N^2-1} \mathrm{tr}\left(\Phi^*(B_k)\Psi(B_k)\right), \tag{1.24}$$

where $B_0, \dots, B_{N^2-1}$ is any orthonormal basis of $\mathfrak{her}(N)$. Likewise, $\Gamma$ in Eqn. (1.21) is called *purely dissipative* if $\Gamma$ is orthogonal to $\mathrm{ad}_{\mathfrak{su}(\mathrm{N})}$. Note that the expression $\mathrm{Tr}(\Phi^*\Psi)$ in Eqn. (1.24) boils down to the ordinary trace of matrices once a matrix representation of the linear maps $\Phi$ and $\Psi$ is fixed. A simple sufficient characterisation for $\Gamma$ being purely dissipative is given by the following result, the proof of which is shifted to Appendix A.

**Lemma 1.3.1.** *If $V_1, \dots, V_m \in \mathfrak{sl}(N, \mathbb{C})$, i.e. if $V_1, \dots, V_m$ are traceless, the operator $\Gamma$ given by Eqn.* (1.21) *is purely dissipative.*

Next, we associate to each representation of $\Gamma$ as in Eqn. (1.21) a $\mathbb{R}$-linear map $\kappa : \mathbb{C}^m \to \mathfrak{her}(N)$ defined by

$$\kappa(\alpha) := \tfrac{\mathrm{i}}{2} \sum_{k=1}^{m} \left(\alpha_k V_k^\dagger - \overline{\alpha}_k V_k\right). \tag{1.25}$$

The map $\kappa$ allows us to characterise how the shifting $V_k \to V_k' = V_k + \alpha_k \mathbb{1}_N$ effects the representation given by Eqn. (1.21) and to obtain a necessary and sufficient condition for $\Gamma$ being purely dissipative.

**Lemma 1.3.2.** *Let $V_k' = V_k + \alpha_k \mathbb{1}_N$ and let*

$$\Gamma(\rho) = \tfrac{1}{2} \sum_{k=1}^{m} \left(V_k^\dagger V_k \rho + \rho V_k^\dagger V_k - 2 V_k \rho V_k^\dagger\right) \tag{1.26}$$

*and*

$$\Gamma'(\rho) = \tfrac{1}{2} \sum_{k=1}^{m} \left(V_k'^\dagger V_k' \rho + \rho V_k'^\dagger V_k' - 2 V_k' \rho V_k'^\dagger\right). \tag{1.27}$$

*Then one has the identity*

$$\Gamma - \Gamma' = \mathrm{i}\,\mathrm{ad}_{\mathrm{H}_0}, \qquad \textit{with} \qquad H_0 = \kappa(\alpha), \tag{1.28}$$

*where $\alpha := (\alpha_1, \dots, \alpha_m)$.*

**Proposition 1.3.1.** *The operator $\Gamma$ given by Eqn.* (1.21) *is purely dissipative if and only if $\alpha := (\mathrm{tr}\, V_1, \dots, \mathrm{tr}\, V_m)$ is in the kernel of $\kappa$.*

*Proof.* "$\Longrightarrow$": Let $\Gamma$ be purely dissipative and define $V_k' := V_k - \mathrm{tr}(V_k)\mathbb{1}_N$. Then, by Lemma 1.3.2, we obtain the equality

$$\Gamma = \mathrm{i}\,\mathrm{ad}_{\mathrm{H}_0} + \Gamma' \qquad \text{with} \qquad H_0 = \kappa(\alpha) \qquad \text{and} \qquad \alpha := (\mathrm{tr}\, V_1, \dots, \mathrm{tr}\, V_m). \tag{1.29}$$

By Lemma 1.3.1, $\Gamma'$ is purely dissipative, too, and therefore $\mathrm{ad}_{\mathrm{H}_0}$ has to vanish. Since $H_0 = \kappa(\alpha)$ is by construction traceless, $\mathrm{ad}_{\mathrm{H}_0} = 0$ implies $H_0 = 0$. i.e. $\alpha$ is in the kernel of $\kappa$.

"$\Longleftarrow$": Assume that $\alpha$ is in the kernel of $\kappa_\Gamma$. Then Eqn. (1.29) reduces to $\Gamma = \Gamma'$, where $\Gamma'$ is purely dissipative by Lemma 1.3.1. Hence, $\Gamma$ is purely dissipative. □

The above considerations suggest to decompose any Lindblad-Kossakowski operator into a "Hamiltonian" and a "purely dissipative" part. The following theorem clarifies the uniqueness to this decomposition.

**Theorem 3.** *Let* $\mathcal{L} := \mathrm{i}\ \mathrm{ad}_{\mathrm{H}} + \Gamma$ *be a Lindblad-Kossakowski operator given by Eqn.* (1.21)*. Then there exists a unique decomposition of the form*

$$\mathcal{L} := \mathrm{i}\ \mathrm{ad}_{\mathrm{H}_0} + \Gamma_0 \tag{1.30}$$

*with* $H_0 \in \mathfrak{her}_0(N)$ *and* $\Gamma_0$ *purely dissipative. Moreover, if the non-vanishing eigenvalues* $\sqrt{\lambda_1} > \cdots > \sqrt{\lambda_{m_0}}$ *of a GKS-matrix of* $\mathcal{L}$ *(and thus of all GKS-matrices of* $\mathcal{L}$*) are distinct then there exist orthonormal* $C_1, \dots, C_{m_0} \in \mathfrak{sl}(N, \mathbb{C})$*, which are unique up to phase factors, such that*

$$\Gamma_0(\rho) = \tfrac{1}{2} \sum_{k=1}^{m_0} \gamma_k \big( C_k{}^\dagger C_k \rho + \rho C_k{}^\dagger C_k - 2 C_k \rho C_k{}^\dagger \big) \tag{1.31}$$

*Proof.* Existence and Uniqueness (part 1): Let $\mathcal{L} := \mathrm{i}\ \mathrm{ad}_{\mathrm{H}} + \Gamma$. As in the proof of Proposition 1.3.1 one has the representation $\Gamma = \mathrm{i}\,\mathrm{ad}_{\mathrm{H}_0} + \Gamma'$, where $\Gamma'$ is purely dissipative. Hence

$$\mathcal{L} = \mathrm{i}\ \mathrm{ad}_{\mathrm{H}} + \Gamma = \mathrm{i}\ \mathrm{ad}_{\mathrm{H}} + \mathrm{i}\,\mathrm{ad}_{\mathrm{H}_0} + \Gamma' = \mathrm{i}\,\mathrm{ad}_{H + \mathrm{H}_0} + \Gamma' \tag{1.32}$$

and therefore $H_0' := H + H'$ and $\Gamma_0 := \Gamma'$ prove existence of Eqn. (1.30). Moreover, the two components $\mathrm{i}\ \mathrm{ad}_{\mathrm{H}_0'}$ and $\Gamma_0$ are uniquely determined as Eqn. (1.30) constitutes an orthogonal decomposition of $\mathcal{L}$. Thus $H_0'$ is also unique since the map $H \mapsto \mathrm{ad}_{\mathrm{H}}$ restricted to $\mathfrak{her}_0(N)$ is one-to-one.

Existence and Uniqueness (part 2): Now, let $B_1, \dots, B_{N^2-1}$ be an orthogonal basis of $\mathfrak{sl}(N, \mathbb{C})$ and let $A \geq 0$ be the corresponding GKS-matrix. Assume that $A$ has the following non-vanishing distinct eigenvalues $\sqrt{\lambda_1} > \cdots > \sqrt{\lambda_{m_0}}$. Then it is well-known that $A$ can be decomposed as

$$A = \sum_{k=1}^{m_0} \sqrt{\lambda_k} \alpha_k \alpha_k^\dagger \,, \tag{1.33}$$

where $\alpha_k \in \mathbb{C}^{N^2-1}$ are orthonormal eigenvectors which are unique up to phase factors. Then the substitution of Eqn. (1.33) into Eqn. (1.22) yields Eqn. (1.31) with $C_k := \sum_{l=1}^{N^2-1} \alpha_{kl} B_l$ and $k = 1, \dots, m_0$. This settles the existence of Eqn. (1.31). Uniqueness follows readily from the fact any two GKS-matrices are unitarily conjugate. More precisely, if $D_1, \dots, D_{n_0}$ and $\sqrt{\mu_1} > \cdots > \sqrt{\mu_{n_0}}$ give rise to another representation of the form Eqn. (1.31). Then, one has

$$\begin{pmatrix} \sqrt{\lambda_1} & 0 & & & \\ 0 & \ddots & \ddots & & \\ & \ddots & \sqrt{\lambda_{m_0}} & 0 & \\ & & 0 & 0 & \ddots \\ & & & \ddots & \ddots \end{pmatrix} = U \begin{pmatrix} \sqrt{\mu_1} & 0 & & & \\ 0 & \ddots & \ddots & & \\ & \ddots & \sqrt{\mu_{n_0}} & 0 & \\ & & 0 & 0 & \ddots \\ & & & \ddots & \ddots \end{pmatrix} U^\dagger \tag{1.34}$$

where $U$ is the unitary matrix which describes the change of basis from $C_1, \dots, C_{N^2-1}$ to $D_1, \dots, D_{N^2-1}$, i.e.

$$C_k = \sum_{l=1}^{N^2-1} u_{kl} D_l \tag{1.35}$$

Here, $C_1, \dots, C_{N^2-1}$ and $D_1, \dots, D_{N^2-1}$ denote arbitrary orthonormal extensions of $C_1, \dots, C_{m_0}$ and $D_1, \dots, D_{N^2-1}$, respectively. Equation (1.34) implies immediately $m_0 = n_0$ and $\lambda_k = \mu_k$ for all $k$. Moreover, if all $\lambda_k$ are distinct, $U$ has to be of the block form

$$U = \begin{pmatrix} U_1 & 0 \\ 0 & U_2 \end{pmatrix} \qquad \text{with} \qquad U_1 = \begin{pmatrix} e^{i\varphi_1} & & \\ & \ddots & \\ & & e^{i\varphi_{m_0}} \end{pmatrix} \qquad \text{and} \qquad U_2 \in U(N^2-1-m_0) \tag{1.36}$$

and thus $C_1, \dots, C_{N^2-1}$ and $D_1, \dots, D_{N^2-1}$ differ only by a phase factor. □

**Remark 3.** *Note again that in general the Lindbald terms $V_k$ of the representation given by Eqn.* (1.21) *are by no means unique unless orthogonality is required. This is comparable to the fact that a positive matrix can be decomposed in may different ways into rank-*1 *projectors if the projectors are not mutually orthogonal.*

Based on Theorem 3, the operators $i\,\mathrm{ad}_{H_0}$ and $\Gamma_0$ in Eqn. (1.30) are called *the Hamiltonian part* and *the (purely) dissipative part* of $\mathcal{L}$, respectively. In particular, the above proof has shown (via diagonalisation of the GKS matrix $A$) that Eqn. (1.22) already constitutes the unique decomposition into Hamiltonian and (purely) dissipative part whenever all $B_k$ are traceless. Moreover, one has the following trivial consequence.

**Corollary 1.3.1.** *Let $\mathcal{L} := i\,\mathrm{ad}_{H_0} + \Gamma_0$ be the unique decomposition of $\mathcal{L}$ given by Eqn.* (1.30) *and let $\Gamma_0$ be purely dissipative. Then one has the equivalence*

$$\mathcal{L} \text{ purely dissipative} \quad \Longleftrightarrow \quad H_0 = 0 \tag{1.37}$$

Furthermore, a Lindblad-Kossakowski operator $\mathcal{L}$ is said to exhibit *no intrinsic Hamiltonian dynamics* if there exists a representation of the form $\mathcal{L} = \Gamma$.

**Lemma 1.3.3.** *Let $\Gamma$ be of the general Lindblad-Kossakowski form as in Eqn.* (1.9)*. Then decomposing each Lindblad operator $V_k \in \mathfrak{gl}(N, \mathbb{C})$ as $V_k = C_k + iD_k$ with $C_k, D_k \in \mathfrak{her}(N)$ gives*

$$\Gamma = \tfrac{1}{2} \sum_{k=1} \Big( \big(\mathrm{ad}^2_{C_k} + \mathrm{ad}^2_{D_k}\big) + i\,\big(\mathrm{ad}_{C_k} \circ \mathrm{ad}^+_{D_k} - \mathrm{ad}_{D_k} \circ \mathrm{ad}^+_{C_k}\big) \Big)\,, \tag{1.38}$$

*where $\mathrm{ad}^+_{C_k}$ and $\mathrm{ad}^+_{D_k}$ are anti-commutator super-operators and moreover, if $\{C_k, D_k\}_+ = 0$, then*

$$\Gamma = \tfrac{1}{2} \sum_{k} \Big( \big(\mathrm{ad}^2_{C_k} + \mathrm{ad}^2_{D_k}\big) + 2i\,\big(\mathrm{ad}_{C_k} \circ \mathrm{ad}^+_{D_k}\big) \Big)\,. \tag{1.39}$$

*Proof.* Equations (1.38) and (1.39) follow by straightforward algebra and that fact that $\mathrm{ad}_{C_k} \circ \mathrm{ad}^+_{D_k} = -\mathrm{ad}_{D_k} \circ \mathrm{ad}^+_{C_k}$ if $C_k D_k = -D_k C_k$. □

Define the Lie algebra isomorphisms

$$\mathrm{voc} : \mathfrak{gl}\big(\mathfrak{her}(N)\big) \to \mathfrak{gl}(N^2, \mathbb{R}) \tag{1.40}$$

and

$$\widehat{(\cdot)} : \mathfrak{gl}\big(\mathbb{C}^{N\times N}\big) \to \mathfrak{gl}(N^2, \mathbb{C}) \tag{1.41}$$

as follows: Let $\mathcal{B} := \mathcal{B}_0 \cup \{\mathbb{1}_N\}$, where $\mathcal{B}_0$ is any (orthogonal) basis of $\mathfrak{her}_0(N)$, and let $\mathcal{E} := \{E_{kl} \mid 1 \leq k, l \leq N\}$ be the standard basis[1] of $\mathfrak{gl}(N^2, \mathbb{C})$. Then,

$$\Phi \mapsto \mathrm{voc}(\Phi) := [\Phi]_{\mathcal{B}} \tag{1.42}$$

and

$$\Phi \mapsto \hat{\Phi} := [\Phi]_{\mathcal{E}} \tag{1.43}$$

where $[\Phi]_{\mathcal{B}}$ and $[\Phi]_{\mathcal{E}}$ denote the matrix representation of $\Phi$ with respect to $\mathcal{B}$ and $\mathcal{E}$. For simplicity, we always presume that $\mathcal{B}$ is ordered such that $\mathbb{1}_N$ corresponds to the last basis vector and that $\mathcal{E}$ carries the standard lexicographic order. Clearly, then $\hat{\Phi}$ is given by the usual Kronecker product formalism, e.g. for the adjoint operator $\mathrm{ad}_A : \mathfrak{gl}(N^2, \mathbb{C}) \to \mathfrak{gl}(N^2, \mathbb{C})$, $B \mapsto \mathrm{ad}_A(B) := [A, B]$ one has

$$\widehat{\mathrm{ad}}_A = (\mathbb{1}_N \otimes A - A^\top \otimes \mathbb{1}_N)\,. \tag{1.44}$$

In the following, we refer to Eqns. (1.42) and (1.43) as *vector of coherence* and *superoperator representation*, respectively. We also have the following result which describes the inclusions of four special Lie algebras in operator representation.

**Lemma 1.3.4.** *Let $\mathfrak{g}^{LK}$ and $\mathfrak{g}_0^{LK}$ denote the Lindblad-Kossakowski algebra and its unital subalgebra. Moreover, let $\mathfrak{g}^E$ and $\mathfrak{g}_0^E$ denote the following subsets of $\mathfrak{gl}\big(\mathfrak{her}(N)\big)$:*

$$\mathfrak{g}^E := \big\{\Phi \in \mathfrak{gl}\big(\mathfrak{her}(N)\big) \;\big|\; \mathrm{Im}\,\Phi \subset \mathfrak{her}_0(N)\big\} \tag{1.45}$$

*and*

$$\mathfrak{g}_0^E := \big\{\Phi \in \mathfrak{gl}\big(\mathfrak{her}(N)\big) \;\big|\; \mathrm{Im}\,\Phi \subset \mathfrak{her}_0(N),\ \mathbb{1}_N \in \ker\Phi\big\}\,. \tag{1.46}$$

*Then $\mathfrak{g}^E$ and $\mathfrak{g}_0^E$ are* real *(Lie) subalgebras satisfying the inclusion relation:*

$$\begin{array}{ccc} \mathfrak{g}_0^{LK} & \xrightarrow{\text{inc}} & \mathfrak{g}_0^E \\ \big\downarrow{\scriptstyle \text{inc}} & & \big\downarrow{\scriptstyle \text{inc}} \\ \mathfrak{g}^{LK} & \xrightarrow{\text{inc}} & \mathfrak{g}^E \end{array}$$

For an in depth analysis of these Lie algebras, the dimensions of their respective Cartan-decompositions and their natural embeddings into larger Lie algebras see Appendix A. Notably, the extended version of Lemma 1.3.4 in Appendix A provides explicit relations between the representations of the above Lie algebras in operator, superoperator and coherence vector representations.

One of the central points presented in this extended Lemma 1.3.4 is that we have the isomorphism

$$\mathfrak{g}^E \overset{\text{iso}}{=} \mathfrak{gl}(N^2 - 1, \mathbb{R}) \oplus_s \mathbb{R}^{N^2-1}\,, \tag{1.47}$$

where $\oplus_s$ is given by the *semidirect sum.* Clearly, elements of the form $(0, b)$ which belong to the abelian ideal of $\mathfrak{gl}(N^2 - 1, \mathbb{R}) \oplus_s \mathbb{R}^{N^2-1}$, are regarded as (infinitesimal) translations acting on $\mathbb{R}^{N^2-1}$. To carry over this picture to $\mathfrak{g}^E$ we define $\tau \in \mathfrak{g}^E$ to be an *(infinitesimal) translation* if

$$\tau\big|_{\mathfrak{her}_0(N)} \equiv 0 \quad \text{and} \quad \tau(\mathbb{1}_N) \in \mathfrak{her}_0(N)\,. \tag{1.48}$$

[1] Note that $\mathfrak{gl}\big(\mathfrak{her}(N)\big)$ is a real vector space, while $\mathfrak{gl}(N^2, \mathbb{C})$ is regarded as complex vector space.

Note, that the second condition is always fulfilled since we assume $\tau \in \mathfrak{g}^E$. Then, for any Hermitian matrix $\rho := \mathbb{1}_N + \rho_0$ with $\rho_0 \in \mathfrak{her}_0(N)$ one has

$$\exp(\tau)(\mathbb{1}_N + \rho_0) = \mathbb{1}_N + \rho_0 + \tau(\mathbb{1}_N)\,, \tag{1.49}$$

where $\exp(\tau) := \sum_{k=0}^{\infty} \frac{\tau^k}{k!}$, i.e., $\exp(\tau)$ acts as a translation on the hyperplane $\mathbb{1}_N + \mathfrak{her}_0(N)$, which does explain the above terminology. Moreover, denote by $\mathfrak{i}^E \subset \mathfrak{g}^E$ the set of all infinitesimal translations. We now provide a few simple results to provide some insight to the characteristics of these operators.

The first result will be expressed here in its general form, but proved later in Section 2.3.2 as part of Theorem 5.

**Lemma 1.3.5.** *The set $\mathfrak{i}^E$ of all infinitesimal translations is an abelian ideal of $\mathfrak{g}^E$, which splits $\mathfrak{g}^E$ into a semi-direct sum $\mathfrak{g}^E = \mathfrak{g}_0^E \oplus_s \mathfrak{i}^E$.*

**Lemma 1.3.6.** *Let $\tau_1$ and $\tau_2$ be infinitesimal translations acting on $\mathfrak{her}(N_1)$ and $\mathfrak{her}(N_2)$, respectively. Then $\tau_1 \otimes \tau_2$ is an infinitesimal translation acting on $\mathfrak{her}(N_1) \otimes \mathfrak{her}(N_2) \stackrel{\text{iso}}{=} \mathfrak{her}(N_1 N_2)$. However, for non-trivial $\tau_1$ and $\tau_2$, the local operators $\mathrm{id}_1 \otimes \tau_2$ and $\tau_1 \otimes \mathrm{id}_2$ do not yield infinitesimal translations.*

*Proof.* We have that $(\tau_1 \otimes \tau_2)(H_1 \otimes H_2) = \tau_1(H_1) \otimes \tau_2(H_2) = 0$ for $H_1 \in \mathfrak{her}(N_1)$ and $H_2 \in \mathfrak{her}_0(N_2)$ or vice versa. Similarly, $(\tau_1 \otimes \tau_2)(\mathrm{id}_1 \otimes \mathrm{id}_2) \in \mathfrak{her}_0(N_1 N_2)$. Thus $\tau_1 \otimes \tau_2$ is an infinitesimal translation acting on $\mathfrak{her}(N_1 N_2)$. Now let $H_2 \in \mathfrak{her}(N_2)$. Then $(\tau_1 \otimes \mathrm{id}_2)(\mathrm{id}_1 \otimes H_2) = \tau_1(\mathrm{id}_1) \otimes H_2 = H_1 \otimes H_2 \neq 0$ for some non-zero $H_1 \in \mathfrak{her}(N_1)$ and therefore $\tau_1 \otimes \mathrm{id}_2$ is not an infinitesimal translation. The same argument shows that neither is $\mathrm{id}_1 \otimes \tau_2$. □

Based on the above conventions, the image of $\mathfrak{i}^E$ under the $\widehat{(\cdot)}$-operation (superoperator representation) is denoted by $\hat{\mathfrak{i}}^E$ and (by abuse of terminology) elements in $\hat{\mathfrak{i}}^E$ are again called infinitesimal translations.

# Chapter 2

# The Lindblad-Kossakowski Lie Algebra

## 2.1 Introduction

As a starting point for any type of mathematical analysis, which in particular involves explicit representations of the objects which we are to consider, it is of fundamental importance to know the structural details of the underlying Lie algebra. In our case, this body of work will focus on various properties of linear operators which act on the convex set of density matrices. As thoroughly discussed in the introduction, here we will be concerned with geometric structures known as Lie wedges which can be *associated* to a closed subsemigroup in (almost) the same way a Lie algebra can be associated to a Lie group. Since we are computing Lie wedges for a given quantum control system $(\Sigma)$ to determine properties of the corresponding subsemigroup of quantum channels it generates, it is imperative that for any kind of explicit computations a parametrisation of the vector space which the wedge is contained in must be chosen.

Recall that the Lie wedges associated to quantum control systems are contained in the so-called Lindblad-Kossakowski Lie algebra (cf. Eqn. (1.15)) given by

$$\mathfrak{g}^{LK} := \langle \mathfrak{w}^{LK} \rangle_{\text{Lie}} \;, \quad \text{and} \quad \mathfrak{g}_0^{LK} := \langle \mathfrak{w}_0^{LK} \rangle_{\text{Lie}} \;, \tag{2.1}$$

where $\mathfrak{w}^{LK}$ and $\mathfrak{w}_0^{LK}$ are the Lindblad-Kossakowski Lie wedges for non-unital and unital systems, respectively. Lemma 1.3.4 then introduced the Lie algebras $\mathfrak{g}^E$ and $\mathfrak{g}_0^E$ and proved that $\mathfrak{g}^{LK} \subseteq \mathfrak{g}^E$ and $\mathfrak{g}_0^{LK} \subseteq \mathfrak{g}_0^E$. The central results of this chapter are Theorems 4 and 5 which will provide the equalities

$$\mathfrak{g}^{LK} = \mathfrak{g}^E = \big\{ \Phi \in \mathfrak{gl}\big(\mathfrak{her}(N)\big) \;\big|\; \operatorname{Im}\Phi \subset \mathfrak{her}_0(N) \big\} \tag{2.2}$$

and

$$\mathfrak{g}_0^{LK} = \mathfrak{g}_0^E = \big\{ \Phi \in \mathfrak{gl}\big(\mathfrak{her}(N)\big) \;\big|\; \operatorname{Im}\Phi \subset \mathfrak{her}_0(N),\ \mathbb{1}_N \in \ker\Phi \big\} \,. \tag{2.3}$$

In fact, we prove the above equalities in a different, more approachable representation to the interested physicist - one which immediately follows from the types of operators physically allowed to make up Lindblad generators $\mathcal{L}$ (cf. Eqn. (1.17)). Due to these equalities, the extended version of Lemma 1.3.4 in Appendix A then provides the isomorphisms

$$\mathfrak{g}^{LK} \overset{\text{iso}}{=} \mathfrak{gl}(N^2-1,\mathbb{R}) \oplus_s \mathbb{R}^{N^2-1} \;, \quad \text{and} \quad \mathfrak{g}_0^{LK} \overset{\text{iso}}{=} \mathfrak{gl}(N^2-1,\mathbb{R}) \;, \tag{2.4}$$

which demonstrate that the LK-Lie algebra and its unital subalgebra are isomorphic to the Lie algebras which contain Lindblad-Kossakowski generators in coherence vector representation (also known as Bloch sphere representation). For describing single qubit open quantum systems, the coherence vector representation is highly intuitive since the system dynamics can be envisaged as compressions, rotations and translations of a three-dimensional sphere. The classic work of Altafini [2] focused on this representation and provided analysis of notions of controllability, accessibility and reachability of quantum systems subject to additional dissipative noise induced by the environment. There he also provided an explicit basis of $\mathfrak{gl}(3, \mathbb{R}^3) \oplus_s \mathbb{R}^3$ which the Lindblad-Kossakowski generators could be decomposed into for a *single* qubit system.

Here we take a different approach by focusing on an alternative representation of $\mathfrak{g}^{LK}$ and $\mathfrak{g}_0^{LK}$. As noted earlier by Eqn. (1.43), we can represent an abstract operator as a matrix via the isomorphism

$$\widehat{(\cdot)} : \mathfrak{gl}\big(\mathbb{C}^{N\times N}\big) \to \mathfrak{gl}(N^2, \mathbb{C}) \,, \tag{2.5}$$

which yields the superoperator representation of a linear operator via the "vec"-operator. The focus of this Section will be to provide a complete parametrisation of the $n$-qubit Lindblad-Kossakowski Lie algebra in this representation. As we will show, we use a parametrisation based upon the Pauli matrices since many standard examples of quantum operations use them in both open and closed quantum systems, and their commutation properties allow for several nice algebraic results to emerge.

First, in Section 2.2 we provide the parametrisations of $\mathfrak{g}_0^{LK}$ and $\mathfrak{g}^{LK}$ for single qubit unital and non-unital systems (cf. Propositions 2.2.1 and 2.2.2, respectively). We also make an important connection regarding the structure of the non-unital part of the dissipative dynamics - that which is responsible for the affine shift of the identity (density) matrix. We provide a matrix representation of these operators which is immediately visible from the decomposition of non-unital Lindblad-Kossakowski generators in operator form. This matrix representation can be envisaged as the matrix operator version of the "usual" vector in $\mathbb{R}^3$ from the coherence vector picture that describes the identity shift for non-unital systems. The operators can be explicitly obtained by the new decomposition of the Lindblad generator given in Lemma 1.3.3.

Section 2.3 then extends this representation to $n$-qubit systems. The structure of the unital Lindblad-Kossakowski Lie algebra given by Theorem 4 remains relatively simple, however, the operators which are the direct generalisations of those which make up the non-unital part of the single qubit LK-Lie algebra require special treatment for multi-qubit systems. Namely, these so called *quasi-translation* operators arise in even simple non-unital Lindblad-Kossakowski generators and they induce both unital *and* non-unital dynamics jointly. This is in striking contrast to their behaviour in single qubit systems. After investigation these new types of operators, Theorem 5 in Section 2.3.2 provides a complete representation of the $n$-qubit non-unital Lindblad-Kossakowski Lie algebra. Knowing the explicit structure of this Lie algebra allows one to see, for example, the interplay between symmetric dissipative components and how they can interact via commutation relations to give new skew-symmetric operators - either a coherent-type Hamiltonian generator, or more exotic skew-symmetric operators (for multi-qubit systems).

Finally, Section 2.4 then considers some of the deeper structural aspects related to the purely non-unital part of this Lie algebra and is a fundamental part of this thesis. We provide a novel parametrisation of its basis elements, which we refer to as translation operators. Explicitly, we parametrise the different possible translation operators by an associated *direction*. This parametrisation is extremely useful throughout the thesis since it provides a geometric picture of which "direction" the identity is shifted along by the non-unital noise. Using this concept, we therefore obtain a new intuition of non-unital dynamics which we use in the following chapter to design purely dissipative non-unital noise to drive a system to a target fixed point.

## 2.2 Single-Qubit Systems

Define the sets $I := \{1, x, y, z\}$ and fix the following ordering $1 < x < y < z$. Then the Pauli basis for $\mathfrak{her}(2)$, the set of all Hermitian $2\times 2$-matrices, is given by $\mathcal{B} := \{\sigma_p \mid p \in I\}$ where $\sigma_1 := \mathbb{1}_2$ and

$$\sigma_x := \begin{bmatrix} 0 & 1 \\ 1 & 0 \end{bmatrix}, \quad \sigma_y := \begin{bmatrix} 0 & -\mathrm{i} \\ \mathrm{i} & 0 \end{bmatrix}, \quad \sigma_z := \begin{bmatrix} 1 & 0 \\ 0 & -1 \end{bmatrix}. \tag{2.6}$$

Likewise, for $I_0 := \{x, y, z\}$ we obtain $\mathcal{B}_0 := \{\sigma_p \mid p \in I_0\}$ as a basis for the traceless Hermitian $2 \times 2$-matrices denoted by $\mathfrak{her}_0(2)$. Moreover, we introduce the shortcuts

$$\hat{\sigma}_\nu := \tfrac{1}{2}(\mathbb{1}_2 \otimes \sigma_\nu - \sigma_\nu^\top \otimes \mathbb{1}_2) \tag{2.7}$$

$$\hat{\sigma}_\nu^+ := \tfrac{1}{2}(\mathbb{1}_2 \otimes \sigma_\nu + \sigma_\nu^\top \otimes \mathbb{1}_2) \tag{2.8}$$

for $\widehat{\mathrm{ad}}_{\frac{\sigma_\nu}{2}}$ and $\widehat{\mathrm{ad}}^+_{\frac{\sigma_\nu}{2}}$, respectively. In the remaining of this work we will omit the "hat" when dealing with such matrix representations of operators other than those in Eqns. (2.7) and (2.8). Only unless it is necessary will we explicitly use it. Due to the prefactor $\frac{1}{2}$ in Eqn. (2.7) one easily recovers the $\mathfrak{su}(2)$ commutation relations

$$[\mathrm{i}\,\hat{\sigma}_p, \mathrm{i}\,\hat{\sigma}_q] = -\varepsilon_{pqr}\,\mathrm{i}\,\hat{\sigma}_r \tag{2.9}$$

where $(p, q, r)$ is any permutation of $(x, y, z)$ and $\varepsilon_{pqr}$ is the Levi-Civita symbol, i.e. $\varepsilon_{pqr} = +1$ if $(p, q, r)$ is an even permutation, $\varepsilon_{pqr} = -1$ if $(p, q, r)$ is an odd permutation, and $\varepsilon_{pqr} = 0$ in any other case. Thus,

$$\mathrm{ad}_{\mathfrak{su}(2)} := \langle \mathrm{i}\hat{\sigma}_x, \mathrm{i}\hat{\sigma}_y, \mathrm{i}\hat{\sigma}_z \rangle \overset{\text{iso}}{=} \mathfrak{su}(2)\,. \tag{2.10}$$

For a single open qubit system in the above super-operator representation, the group lift of the controlled master equation (cf. Eqn. 1.7) takes the form

$$\dot{X}(t) = -\Big(\mathrm{i}\big(\,\mathrm{ad}_{H_0} + \sum_j u_j(t)\,\mathrm{ad}_{H_j}\,\big) + \Gamma\Big)\,X(t)\,. \tag{2.11}$$

Here, $X(t)$ may be regarded as a qubit quantum channel represented in $GL(4, \mathbb{C}\,)$.

Subsequently, we refer to the well-known generators of unital and non-unital single qubit quantum channels as *standard single-qubit* generators. For these systems, the Hamiltonians $H_j$ and the components $C_k$, $D_k$ of the Lindblad terms $V_k$ (cf. Eqn. (1.9)) take a particular simple form in the sense that $H_j, C_k, D_k$ are given as scalar multiples of the Pauli matrices $\sigma_x$, $\sigma_y$ and $\sigma_z$. This implies $\{C_k, D_k\}_+ = 0$ for $C_k \neq D_k$ and therefore by Lemma 1.3.3, we obtain the form

$$\dot{X}(t) = -\Big(\mathrm{i}\big(\hat{\sigma}_d + \sum_{j\in I_0} u_j(t)\hat{\sigma}_j\big) + \Gamma\Big)\,X(t)\,. \tag{2.12}$$

with

$$\Gamma = 2 \sum_{\substack{p,q\in I_0 \\ p\neq q}} \gamma_{p,q}\big(\hat{\sigma}_p^2 + \hat{\sigma}_q^2 + 2\mathrm{i}\hat{\sigma}_p\hat{\sigma}_q^+\big) \in \mathfrak{gl}(4, \mathbb{C}\,)\,. \tag{2.13}$$

and if each Lindblad operator $V_k$ is Hermitian, and hence $D_k = 0$, then Eqn. (2.13) reduces to the unital dissipation term

$$\Gamma = 2 \sum_{p \in I_0} \gamma_p \hat{\sigma}_p^2 \in \mathfrak{gl}(4, \mathbb{C}\,)\,, \tag{2.14}$$

which is well known and sometimes called "of double commutator form" in related literature.

Afer these preliminary considerations, we start to characterise the Lie algebraic structure of Lindblad-Kossakowski algebra for single qubit systems in detail. We distinguish to cases: unital and the non-unital systems. The ideas presented in the following will serve us a guideline for the $n$-qubit case.

### 2.2.1 Unital Single-Qubit Systems

For unital single qubit systems, the Lindblad-Kossakowski algebra $\hat{\mathfrak{g}}_0^{LK}$ clearly contains the standard generators $\mathrm{i}\hat{\sigma}_\nu$ and $\hat{\sigma}_\nu^2$ due to Eqns. (2.12) and (2.14). Moreover, it has to embrace all commutators of the form $[\mathrm{i}\hat{\sigma}_\nu, \hat{\sigma}_\mu^2]$. Hence the relation

$$[\mathrm{i}\hat{\sigma}_\nu, \hat{\sigma}_\mu^2] = -\varepsilon_{\nu\mu\lambda}\{\hat{\sigma}_\mu, \hat{\sigma}_\lambda\}_+\,, \tag{2.15}$$

where $(\nu, \mu, \lambda)$ is any permutation of $(x, y, z)$, implies that actually the span

$$\big\langle \mathrm{i}\hat{\sigma}_\nu, \hat{\sigma}_\nu^2, \{\hat{\sigma}_\nu, \hat{\sigma}_\mu\}_+ \,|\, \nu, \mu \in I_0 \big\rangle\,, \tag{2.16}$$

belongs to $\hat{\mathfrak{g}}_0^{LK}$. Then the commutation relations in Appendix E, and the fact that $\hat{\sigma}_\nu^2$ and $\hat{\sigma}_\mu^2$ commute suggest the following result.

**Proposition 2.2.1.** *The unital single qubit Lindblad-Kossakowski algebra* $\hat{\mathfrak{g}}_0^{LK}$ *is a* 9-*dimesional* real *Lie subalgebra of* $\mathfrak{gl}(4, \mathbb{C}\,)$ *given by*

$$\hat{\mathfrak{g}}_0^{\mathrm{LK}} = \big\langle \mathrm{i}\hat{\sigma}_\nu, \hat{\sigma}_\nu^2, \{\hat{\sigma}_\nu, \hat{\sigma}_\mu\}_+ \,|\, \nu, \mu \in I_0, \nu < \mu \big\rangle\,, \tag{2.17}$$

*where the nine elements listed in Eqn.* (2.17) *form a basis of* $\hat{\mathfrak{g}}_0^{LK}$. *Moreover,* $\hat{\mathfrak{g}}_0^{\mathrm{LK}}$ *is isomorphic to* $\mathfrak{gl}(3, \mathbb{R}\,)$ *and admits a Cartan decomposition into skew-Hermitian and Hermitian components, i.e.* $\hat{\mathfrak{g}}_0^{LK} := \hat{\mathfrak{k}}_0 \oplus \hat{\mathfrak{p}}_0$, *where*

$$\hat{\mathfrak{k}}_0 := \langle \mathrm{i}\hat{\sigma}_x, \mathrm{i}\hat{\sigma}_y, \mathrm{i}\hat{\sigma}_y \rangle \overset{\mathrm{iso}}{=} \mathfrak{so}(3) \tag{2.18}$$

$$\hat{\mathfrak{p}}_0 := \langle \hat{\sigma}_\nu^2, \{\hat{\sigma}_\nu, \hat{\sigma}_\mu\}_+ \,|\, \nu, \mu \in I_0, \nu < \mu \rangle \tag{2.19}$$

*with maximal abelian subalgebra*

$$\hat{\mathfrak{a}}_0 := \langle \hat{\sigma}_\nu^2 \,|\, \nu \in I_0 \rangle \subset \mathfrak{p}\,. \tag{2.20}$$

*Furthermore,* $\hat{\mathfrak{a}}_0$ *contains the one-dimensonal center* $\hat{\mathfrak{z}}_o = \langle C_0 \rangle$ *of* $\hat{\mathfrak{g}}_0^{LK}$ *and splits into an orthogonal sum*

$$\hat{\mathfrak{a}}_0 := \langle \hat{\sigma}_\nu^2 - \hat{\sigma}_\mu^2 \,|\, \nu \in I_0, \nu < \mu \rangle \oplus \langle C_0 \rangle\,, \tag{2.21}$$

*with* $C_0 := \hat{\sigma}_x^2 + \hat{\sigma}_y^2 + \hat{\sigma}_z^2$.

*Proof.* The above considerations show that the span in Eqn. (2.16) has to be contained in $\hat{\mathfrak{g}}_0^{LK}$. Moreover, a straightforward computation shows that the set $\{\mathrm{i}\hat{\sigma}_\nu, \hat{\sigma}_\nu^2, \{\hat{\sigma}_\nu, \hat{\sigma}_\mu\}_+ \,|\, \nu, \mu \in I_0, \nu < \mu\}$ is an basis of Eqn. (2.16) . Therefore, Lemma 1.3.4 already implies Eqn. (2.17). And in fact, the commutation relations of Appendix E, Tabs. E.1-E.4 demonstrate that the span in Eqn. (2.16) coincides with its Lie closure and admits the specified Cartan decomposition and isomorphy. For Eqn. (2.18), we simply refer to Eqn. (2.10). □

**Remark 4.** *Note that the basis*

$$\{\mathrm{i}\hat{\sigma}_\nu, \hat{\sigma}_\nu^2, \{\hat{\sigma}_\nu, \hat{\sigma}_\mu\}_+ \,|\, \nu, \mu \in I_0, \nu < \mu\} \tag{2.22}$$

*given in Proposition 2.2.1 is "almost" orthogonal in the sense that all elements are mutually orthogonal except $\hat{\sigma}_x^2$, $\hat{\sigma}_y^2$, and $\hat{\sigma}_z^2$ among each other. Clearly, one could orthogonalize $\hat{\sigma}_x^2$, $\hat{\sigma}_y^2$, and $\hat{\sigma}_z^2$ – but this is in geneal not useful. It is better to pass from $\hat{\sigma}_x^2$, $\hat{\sigma}_y^2$, $\hat{\sigma}_z^2$ to $\hat{\sigma}_x^2 - \hat{\sigma}_y^2$, $\hat{\sigma}_y^2 - \hat{\sigma}_z^2$, $C_0 := \hat{\sigma}_x^2 + \hat{\sigma}_y^2 + \hat{\sigma}_z^2$ as this basis provides further insight into the Lie algebraic structure of $\hat{\mathfrak{g}}_0^{LK}$, cf. Eqn. (2.21).*

### 2.2.2 Non-Unital Single-Qubit Systems

In contrast, for non-unital single-qubit systems Eqn. (2.13) shows that additional terms of the type $\mathrm{i}\hat{\sigma}_\nu\hat{\sigma}_\mu^+$ with $\mu \neq \nu$ must be taken into account. Thus for non-unital systems we consider the linear span

$$\left\langle \mathrm{i}\hat{\sigma}_\nu, \hat{\sigma}_\nu^2, \{\hat{\sigma}_\nu, \hat{\sigma}_\mu\}_+, \mathrm{i}\hat{\sigma}_\nu\hat{\sigma}_\mu^+, \,|\, \nu, \mu \in I_0 \right\rangle .$$

By the commutation relations in Appendix E, Tabs. E.5 and E.6, and noting that $\mathrm{i}\hat{\sigma}_\nu\hat{\sigma}_\mu^+ = -\mathrm{i}\hat{\sigma}_\mu\hat{\sigma}_\nu^+$, we obtain a 12-dimensional Lie-subalgebra of $\mathfrak{gl}(4,\mathbb{C})$ which contains all possible dimensions a single qubit Lie wedge may explore. More precisely, one finds the following:

**Proposition 2.2.2.** *The non-unital single qubit Lindblad-Kossakowski algebra $\hat{\mathfrak{g}}^{LK}$ is a 12-dimesional* real *Lie subalgebra of $\mathfrak{gl}(4,\mathbb{C})$ given by*

$$\hat{\mathfrak{g}}^{LK} = \left\langle \mathrm{i}\hat{\sigma}_\nu, \hat{\sigma}_\nu^2, \{\hat{\sigma}_\nu, \hat{\sigma}_\mu\}_+, \mathrm{i}\hat{\sigma}_\nu\hat{\sigma}_\mu^+ \,|\, \nu, \mu \in I_0, \nu < \mu \right\rangle, \tag{2.23}$$

*where the twelve elements listed in Eqn. (2.23) form an orthogonal basis of $\hat{\mathfrak{g}}^{LK}$. Moreover, $\hat{\mathfrak{g}}^{\mathrm{LK}}$ takes the form of a semidirect sum $\hat{\mathfrak{g}}^{LK} = \hat{\mathfrak{g}}_0^{LK} \oplus_s \hat{\mathfrak{i}}$, where*

$$\hat{\mathfrak{i}} := \langle \mathrm{i}\hat{\sigma}_\nu\hat{\sigma}_\mu^+ \,|\, \nu, \mu \in I_0, \nu < \mu \rangle \tag{2.24}$$

*is an abelian ideal of $\hat{\mathfrak{g}}^{LK}$.*

*Proof.* As mentioned before, the Lie subalgebra structure of the right-hand side of Eqn. (2.23) can be read off the commutation tables in Appendix E. Again, Lemma 1.3.4 and the orthogonality of the listed elements show that one has equality in Eqn. (2.23). Moreover, $\hat{\mathfrak{g}}^{LK}$ is obviously the direct sum of the Lie subalgebra $\hat{\mathfrak{g}}_0^{LK}$ and vector space $\hat{\mathfrak{i}}$. Hence, for establishing the semidirect sum structure of $\hat{\mathfrak{g}}^{LK}$, we still have to show that $\hat{\mathfrak{i}}$ is an abelian ideal of $\hat{\mathfrak{g}}^{LK}$. Yet, this follows again from the commutation relations given in Appendix E and the fact that $[\mathrm{i}\,\hat{\sigma}_\nu\hat{\sigma}_\mu^+, \mathrm{i}\,\hat{\sigma}_\mu\hat{\sigma}_\lambda^+] = 0$ for all $\nu, \mu, \lambda \in I_0$. □

**Example 1.** *Consider the dynamics of a system governed by a Lindblad-Kossakowski operator of the form* $\mathcal{L} = \mathrm{i}\,\mathrm{ad}_{H_d} + \Gamma$*, where* $H_d = \sigma_z$ *and the only Lindblad term of* $\Gamma$ *is given by*

$$V = \begin{bmatrix} 1 & 1 \\ 0 & -1 \end{bmatrix} . \tag{2.25}$$

*Then writing* $V = C + \mathrm{i}D$*, where* $C = \frac{1}{2}\sigma_x + \sigma_z$ *and* $D = \frac{1}{2}\sigma_y$ *and changing to the "vec"-representation of* $\mathcal{L}$*, we obtain*

$$\begin{aligned} \mathcal{L} &= \mathrm{i}\hat{\sigma}_z + \widehat{\mathrm{ad}}_C^2 + \widehat{\mathrm{ad}}_D^2 + \mathrm{i}(\widehat{\mathrm{ad}}_C \widehat{\mathrm{ad}}_D^+ - \widehat{\mathrm{ad}}_D \widehat{\mathrm{ad}}_C^+) \\ &= \mathrm{i}\hat{\sigma}_z + 2\left(\tfrac{1}{4}\hat{\sigma}_x^2 + \hat{\sigma}_z^2 + \tfrac{1}{2}\{\hat{\sigma}_x, \hat{\sigma}_z\}_+ + \tfrac{1}{4}\hat{\sigma}_y^2 + \mathrm{i}\big(\tfrac{1}{2}\hat{\sigma}_x\hat{\sigma}_y^+ + \hat{\sigma}_z\hat{\sigma}_y^+\big)\right) . \end{aligned}$$

*Since any single qubit* $V$ *can be expressed as a sum of a Hermitian matrix and skew-Hermitian matrix, the above type of splitting into basis elements consisting of Pauli elements is always possible. In this example, it's immediately obvious how the operators which make up* $\mathcal{L}$ *relate to the basis elements of* $\hat{\mathfrak{g}}^{LK}$ *as expressed in Proposition 2.2.2.*

The above results allow us directly see the relationship to the coherence vector representation. Recall that the coherence vector/Bloch sphere representation of single qubit dynamics consists of a unital part (which leaves unity invariant) and a non-unital part (which is described by a vector in $\mathbb{R}^3$) such that the overall generator generates a Markovian semigroup of quantum maps. The system $(\Sigma)$ given by Eqn. (1.16) is then lifted to a bilinear control system on $GL(3,\mathbb{R}) \rtimes \mathbb{R}^3$, whose Lie algebra elements are of the general form

$$\left[\begin{array}{c|c} A & a \\ \hline 0 & 0 \end{array}\right] \in \mathfrak{gl}(3,\mathbb{R}) \oplus_s \mathbb{R}^3 , \quad \text{such that} \quad A \in \mathfrak{gl}(3,\mathbb{R}) \quad \text{and} \quad a \in \mathbb{R}^3 , \tag{2.26}$$

where we say (in shorthand notation) that for elements $(A,a),(B,b) \in \mathfrak{gl}(3,\mathbb{R}) \oplus_s \mathbb{R}^3$, the Lie bracket is given by $[(A,a),(B,b)] = ([A,B], Ab - Ba)$. Thus, in order to generate a Markovian semigroup of quantum maps, elements $(A,a) \in \mathfrak{gl}(3,\mathbb{R}) \oplus_s \mathbb{R}^3$ obviously must have a special form [2]. As an alternative point of view to this intuitive single qubit picture, we now know that

$$\hat{\mathfrak{g}}^{LK} = \hat{\mathfrak{g}}_0^{LK} \oplus_s \hat{\mathfrak{i}} \overset{\text{iso}}{=} \mathfrak{gl}(3,\mathbb{R}) \oplus_s \mathbb{R}^3 , \tag{2.27}$$

thus making elements of the ideal $\hat{\mathfrak{i}}$ isomorphic to elements of the form $(0,a) \in \mathfrak{gl}(3,\mathbb{R}) \oplus_s \mathbb{R}^3$ with $a \in \mathbb{R}^3$.

Therefore, the unique Lie subgroup $\hat{\mathbf{G}}^{\mathrm{LK}} \subset GL(4,\mathbb{C})$ with Lie subalgebra $\hat{\mathfrak{g}}^{\mathrm{LK}}$ can be thought of as a semidirect product on the group level

$$\hat{\mathbf{G}}^{\mathrm{LK}} = \hat{\mathbf{G}}_0^{\mathrm{LK}} \otimes_s \hat{\mathbf{T}}^{\mathrm{LK}} \overset{\text{iso}}{=} GL(3,\mathbb{R}) \rtimes \mathbb{R}^3 , \tag{2.28}$$

where $\hat{\mathbf{G}}_0^{\mathrm{LK}}$ and $\hat{\mathbf{T}}^{\mathrm{LK}}$ denote the unique Lie subgroups which correspond to $\hat{\mathfrak{g}}_0^{\mathrm{LK}}$ and $\hat{\mathfrak{i}}$, respectively.

In the next section, we focus on expanding these notions to $n$-qubit systems. We will see that the direct generalised versions of the basis elements of the single qubit ideal $\hat{\mathfrak{i}}$ are *not* elements of the $n$-qubit ideal $\hat{\mathfrak{i}}$. Instead, the generalisation of operators of the form $\mathrm{i}\sigma_p\hat{\sigma}_q^+$ with $p,q \in \{x,y,z\}$ such that $p \neq q$ not only contain a non-unital part, but *also a unital part.* This extremely interesting property then forces us to use a projection operation onto these generalised operations in order to determine a basis for the $n$-qubit ideal $\hat{\mathfrak{i}}$.

## 2.3 $n$-Qubit Systems

### 2.3.1 Unital $n$-Qubit Systems

Let $I^n := \{1, x, y, z\}^n$ and $I_0^n := I^n \setminus \{(1, 1, \ldots, 1)\}$. Moreover, we extend the ordering $1 < x < y < z$ of $I$, which proved very useful in the single qubit case, lexicographically to $I^n$. For compact notation, we make use of the multi-index $\mathbf{p} := (p_1, p_2, ..., p_n) \in I^n$ to define

$$\sigma_{\mathbf{p}} := \sigma_{p_1} \otimes \sigma_{p_2} \otimes \cdots \otimes \sigma_{p_{n-1}} \otimes \sigma_{p_n} \, , \tag{2.29}$$

so that

$$\mathcal{B}^n := \{\sigma_{\mathbf{p}} \mid \mathbf{p} \in I^n\} \, , \text{ and } \mathcal{B}_0^n := \{\sigma_{\mathbf{p}} \mid \mathbf{p} \in I_0^n\} \tag{2.30}$$

are basis for $\mathfrak{her}(2^n)$ and $\mathfrak{her}_0(2^n)$, respectively. Moreover, the natural extensions of Eqn. (2.7) are given by

$$\hat{\sigma}_{\mathbf{p}} := \tfrac{1}{2}(\mathbb{1}_{2^n} \otimes \sigma_{\mathbf{p}} - \sigma_{\mathbf{p}}^\top \otimes \mathbb{1}_{2^n}) \tag{2.31}$$

$$\hat{\sigma}_{\mathbf{p}}^+ := \tfrac{1}{2}(\mathbb{1}_{2^n} \otimes \sigma_{\mathbf{p}} + \sigma_{\mathbf{p}}^\top \otimes \mathbb{1}_{2^n}) \, . \tag{2.32}$$

In the following three Lemmas, we collect a few straightforward results which will be quite helpful in the subsequent proofs.

**Lemma 2.3.1.** *Let $\boldsymbol{p} := (p_1, p_2, ..., p_n)$, $\boldsymbol{q} := (q_1, q_2, ..., q_n)$ and $\sigma_{\boldsymbol{p}}$, $\sigma_{\boldsymbol{q}}$ be defined as in Eqn.* (2.29)*. Then*

$$\sigma_{\boldsymbol{p}}^2 = \mathbb{1}_{2^n} \quad \textit{and} \quad \sigma_{\boldsymbol{p}}\sigma_{\boldsymbol{q}} = (-1)^\varepsilon \sigma_{\boldsymbol{q}}\sigma_{\boldsymbol{p}} \, , \tag{2.33}$$

*where $\varepsilon$ is the number of indices $k$ in $p_k, q_k \in I_0$ with $p_k \neq q_k$ and $p_k \neq 1, q_k \neq 1$. (NB: $\sigma_{\boldsymbol{p}}^2 = \mathbb{1}_{2^n}$, but in the ad-representation $\hat{\sigma}_{\boldsymbol{p}}^2 \neq \mathbb{1}_{4^n}$.)*

*Proof.* Both statements are immediate consequences of well-known properties of the Pauli matrices $\sigma_x$, $\sigma_y$, and $\sigma_z$. □

**Lemma 2.3.2.** *Let $\sigma_{\boldsymbol{p}}, \sigma_{\boldsymbol{q}} \in \mathcal{B}^n$. Then*

1. *$[\sigma_{\boldsymbol{p}}, \sigma_{\boldsymbol{q}}] = 0$ if and only if $\{\sigma_{\boldsymbol{p}}, \sigma_{\boldsymbol{q}}\}_+ \neq 0$ ,*

2. *For a fixed $\sigma_{\boldsymbol{p}} \neq \mathbb{1}$, $\frac{|\mathcal{B}^n|}{2} = \frac{4^n}{2}$ elements of $\mathcal{B}^n$ commute with $\sigma_{\boldsymbol{p}}$, and the remaining half of the elements anti-commute with $\sigma_{\boldsymbol{p}}$.*

*Proof.* 1) By Lemma 2.3.1, its clear that if $\varepsilon$ is even then $[\sigma_{\mathbf{p}}, \sigma_{\mathbf{q}}] = 0$ and $\{\sigma_{\mathbf{p}}, \sigma_{\mathbf{q}}\}_+ = 2\sigma_{\mathbf{p}}\sigma_{\mathbf{q}}$ whereas if $\varepsilon$ is odd then $[\sigma_{\mathbf{p}}, \sigma_{\mathbf{q}}] = 2\sigma_{\mathbf{p}}\sigma_{\mathbf{q}}$ and $\{\sigma_{\mathbf{p}}, \sigma_{\mathbf{q}}\}_+ = 0$. Hence $[\sigma_{\mathbf{p}}, \sigma_{\mathbf{q}}] = 0$ if and only if $\{\sigma_{\mathbf{p}}, \sigma_{\mathbf{q}}\}_+ \neq 0$.
2) We prove by induction. Clearly, the statement holds for $n = 1$. Now let $\sigma_{\mathbf{p}}, \sigma_{\mathbf{q}} \in \mathcal{B}^n$ such that $\sigma_{\mathbf{p}} \neq \mathbb{1}$. Without loss of generality, choose the element $\sigma_{\mathbf{p}} \otimes \mathbb{1} \in \mathcal{B}^{n+1}$. Then for any $\sigma_{\mathbf{q}} \otimes \sigma_\alpha$, with $\alpha \in \{\mathbb{1}, x, y, z\}$ we have that $[\sigma_{\mathbf{p}} \otimes \mathbb{1}, \sigma_{\mathbf{q}} \otimes \sigma_\alpha] = [\sigma_{\mathbf{p}}, \sigma_{\mathbf{q}}] \otimes \sigma_\alpha$. By the induction hypothesis, there are $\frac{|\mathcal{B}^n|}{2} = \frac{4^n}{2}$ elements $\sigma_{\mathbf{q}} \in \mathcal{B}^n$ such that $[\sigma_{\mathbf{p}}, \sigma_{\mathbf{q}}] = 0$ and since $\alpha \in \{\mathbb{1}, x, y, z\}$ we get that there are $\frac{4^n}{2} \cdot 4 = \frac{4^{n+1}}{2}$ elements in $\mathcal{B}^{n+1}$ which commute with the fixed $\sigma_{\mathbf{p}} \otimes \mathbb{1} \in \mathcal{B}^{n+1}$. Since $[\sigma_{\mathbf{p}}, \sigma_{\mathbf{q}}] = 0$ if and only if $\{\sigma_{\mathbf{p}}, \sigma_{\mathbf{q}}\}_+ \neq 0$, the remaining half of the basis elements of $\mathcal{B}^{n+1}$ must anti-commute with the fixed element. □

**Lemma 2.3.3.** *Let $\hat\sigma_{\boldsymbol\mu}$, $\hat\sigma_{\boldsymbol\nu}$, $\hat\sigma_{\boldsymbol p}$, $\hat\sigma_{\boldsymbol q}$ be defined as in Eqn.* (2.31) *and let* $\langle A, B\rangle_{\mathrm{tr}} := \operatorname{tr} A^\dagger B$ *denote the Hilbert-Schmidt inner product between matrices A and B. Then*

*(a) The set $\{\hat\sigma_{\boldsymbol p}^2 \mid \boldsymbol p \in I_0^n\}$ is linearly independent.*

*(b) The matrices $\{\hat\sigma_{\boldsymbol\mu}, \hat\sigma_{\boldsymbol\nu}\}_+$ and $\{\hat\sigma_{\boldsymbol p}, \hat\sigma_{\boldsymbol q}\}_+$ are orthogonal, i.e.*

$$\big\langle \{\hat\sigma_{\boldsymbol\mu}, \hat\sigma_{\boldsymbol\nu}\}_+, \{\hat\sigma_{\boldsymbol p}, \hat\sigma_{\boldsymbol q}\}_+ \big\rangle_{\mathrm{tr}} = 0 \tag{2.34}$$

*if and only if one of the following conditions is met*

*(i)* $\boldsymbol\mu = \boldsymbol\nu$, $\boldsymbol p \neq \boldsymbol q$,

*(ii)* $\boldsymbol\mu \neq \boldsymbol\nu$, $\boldsymbol p = \boldsymbol q$

*(iii)* $\boldsymbol\mu \neq \boldsymbol\nu$, $\boldsymbol p \neq \boldsymbol q$, $(\boldsymbol\mu, \boldsymbol\nu) \neq (\boldsymbol p, \boldsymbol q)$, $(\boldsymbol\mu, \boldsymbol\nu) \neq (\boldsymbol q, \boldsymbol p)$.

*Proof.* (a) From the equality

$$\hat\sigma_{\mathbf p}^2 = \tfrac{1}{2}(\mathbb{1}_{4^n} - \sigma_{\mathbf p}^T \otimes \sigma_{\mathbf p}) \tag{2.35}$$

one easily sees that $\hat\sigma_{\mathbf p}^2$ and $\sigma_{\mathbf q}^T \otimes \sigma_{\mathbf q}$ are orthogonal for all $\mathbf p, \mathbf q \in I_0^n$ with $\mathbf p \neq \mathbf q$. This clearly implies that the set $\{\hat\sigma_{\mathbf p}^2 \mid \mathbf p \in I_0^n\}$ is linearly independent.

(b) "$\Longleftarrow$": Here, we exemplify only the case $\boldsymbol\mu = \boldsymbol\nu$, $\mathbf p \neq \mathbf q$, because the same arguments can be applied in all other cases. Note that

$$\{\hat\sigma_{\mathbf p}, \hat\sigma_{\mathbf q}\}_+ = \tfrac{1}{4}\big(\mathbb{1}_{2^n} \otimes \{\sigma_{\mathbf p}, \sigma_{\mathbf q}\}_+ + \{\sigma_{\mathbf p}^\top, \sigma_{\mathbf q}^\top\}_+ \otimes \mathbb{1}_{2^n}\big) - 2\big(\sigma_{\mathbf p}^\top \otimes \sigma_{\mathbf q} + \sigma_{\mathbf q}^\top \otimes \sigma_{\mathbf p}\big)\ . \tag{2.36}$$

Then by Eqn. (2.35) and the trace identity $\operatorname{tr}(A \otimes B) = \operatorname{tr} A \cdot \operatorname{tr} B$ for square matrices $A$, $B$, one has

$$\begin{aligned}\langle \hat\sigma_{\boldsymbol\mu}^2, \{\hat\sigma_{\mathbf p}, \hat\sigma_{\mathbf q}\}_+\rangle_{\mathrm{tr}} &= \tfrac{1}{8}\operatorname{tr}\big(\mathbb{1}_{2^n} \otimes \{\sigma_{\mathbf p}, \sigma_{\mathbf q}\}_+\big) - \tfrac{1}{8}\operatorname{tr}\big(\sigma_{\boldsymbol\mu}^\top \otimes \sigma_{\boldsymbol\mu}\{\sigma_{\mathbf p}, \sigma_{\mathbf q}\}_+\big) + \text{“six more terms”}\\ &= \tfrac{1}{8}\operatorname{tr}(\mathbb{1}_{2^n})\operatorname{tr}(\{\sigma_{\mathbf p}^\top, \sigma_{\mathbf q}^\top\}_+) - \tfrac{1}{8}\operatorname{tr}(\sigma_{\boldsymbol\mu}^\top)\operatorname{tr}(\sigma_{\boldsymbol\mu}\{\sigma_{\mathbf p}, \sigma_{\mathbf q}\}_+) + \text{“six more terms”}\ .\end{aligned}$$

Since the Pauli matrices are traceless, it easily follows from Lemma 2.3.1 that each of the above terms vanishes and thus $\langle \hat\sigma_{\boldsymbol\mu}^2, \{\hat\sigma_{\mathbf p}, \hat\sigma_{\mathbf q}\}_+\rangle_{\mathrm{tr}} = 0$.

"$\Longrightarrow$": If none of the above conditions (i)–(iii) is met then one has $\boldsymbol\mu = \boldsymbol\nu$, $\mathbf p = \mathbf q$ and therefore orthogonality fails by Eqn. (2.35). □

With these technical Lemmas out of the way, we are now ready to prove our first main result. Recall that Lemma 1.3.4 introduced the Lie algebra $\mathfrak{g}_0^E$ which was given by

$$\mathfrak{g}_0^E = \big\{\Phi \in \mathfrak{gl}\big(\mathfrak{her}(N)\big) \bigm| \operatorname{Im}\Phi \subset \mathfrak{her}_0(N),\, \mathbb{1}_N \in \ker\Phi\big\}\,, \tag{2.37}$$

and proved that $\mathfrak{g}_0^{LK} \subseteq \mathfrak{g}_0^E$. Then, Proposition 2.2.1 proved that for a single qubit we have $\mathfrak{g}_0^{LK} = \mathfrak{g}_0^E$. The following result then proves this equality for general $n$-qubit systems.

**Theorem 4.** *The unital n-qubit Lindblad-Kossakowski Lie algebra $\hat{\mathfrak{g}}_0^{LK} \subset \mathfrak{gl}(4^n, \mathbb{C})$ is a $(4^n - 1)^2$-dimensional* real *Lie algebra given by*

$$\hat{\mathfrak{g}}_0^{LK} = \hat{\mathfrak{g}}_0^E = \langle \mathrm{i}\hat\sigma_{\boldsymbol p}, \hat\sigma_{\boldsymbol p}^2, \{\hat\sigma_{\boldsymbol\alpha}, \hat\sigma_{\boldsymbol\mu}\}_+ \,|\, \boldsymbol p, \boldsymbol\alpha < \boldsymbol\mu \in I_0^n\rangle_{\mathsf{Lie}}\ . \tag{2.38}$$

*Moreover, it admits a Cartan decomposition into Hermitian and skew-Hermitian elements $\hat{\mathfrak{g}}_0^{LK} = \hat{\mathfrak{k}}_0 \oplus \hat{\mathfrak{p}}_0$ where*

$$\hat{\mathfrak{k}}_0 \ := \ \langle [\{\hat\sigma_{\boldsymbol p}, \hat\sigma_{\boldsymbol q}\}_+, \{\hat\sigma_{\boldsymbol\alpha}, \hat\sigma_{\boldsymbol\mu}\}_+] \,|\, \boldsymbol p \le \boldsymbol q, \boldsymbol\alpha \le \boldsymbol\mu \in I_0^n\rangle \tag{2.39}$$

$$\hat{\mathfrak{p}}_0 \quad := \quad \langle \{\hat{\sigma}_{\boldsymbol{\alpha}}, \hat{\sigma}_{\boldsymbol{\mu}}\}_+ \,|\, \boldsymbol{\alpha} \leq \boldsymbol{\mu} \in I_0^n \rangle \;, \tag{2.40}$$

*with maximally abelian subalgebra* $\hat{\mathfrak{a}}_0 \subset \hat{\mathfrak{p}}_0$ *given by*

$$\hat{\mathfrak{a}}_0 := \langle \hat{\sigma}_{\mathbf{p}}^2 \,|\, \mathbf{p} \in I_0^n \rangle \;. \tag{2.41}$$

*Proof.* It is clear that $\hat{\mathfrak{g}}_0^{LK}$ must contain operators $\mathrm{i}\hat{\sigma}_{\mathbf{p}}$ and $\mathrm{i}\hat{\sigma}_{\mathbf{p}}^2$ due to the generalizations of Eqns. (2.12) and (2.14). Moreover, the extension of Eqn. (2.15) to multi qubit systems is given by

$$[\mathrm{i}\hat{\sigma}_{\mathbf{p}}, \hat{\sigma}_{\boldsymbol{\mu}}^2] = \{[\mathrm{i}\hat{\sigma}_{\mathbf{p}}, \hat{\sigma}_{\boldsymbol{\mu}}], \hat{\sigma}_{\boldsymbol{\mu}}\}_+ \;, \tag{2.42}$$

which follows from the identity $[A, B^2] = \{[A, B], B\}_+$ for arbitrary square matrices $A$, $B$. Therefore, we have the inclusion $\tilde{\mathfrak{g}}_0 \subseteq \hat{\mathfrak{g}}_0^{LK} \subseteq \hat{\mathfrak{g}}_0^{E}$, where

$$\tilde{\mathfrak{g}}_0 := \langle \mathrm{i}\hat{\sigma}_{\mathbf{p}}, \hat{\sigma}_{\mathbf{p}}^2, \{\hat{\sigma}_{\boldsymbol{\alpha}}, \hat{\sigma}_{\boldsymbol{\mu}}\}_+ \,|\, \mathbf{p}, \boldsymbol{\alpha} < \boldsymbol{\mu} \in I_0^n \rangle_{\mathsf{Lie}} \;. \tag{2.43}$$

Next, define $\tilde{\mathfrak{p}}_0 := \tilde{\mathfrak{g}}_0 \cap \mathfrak{her}(4^n)$ and recall $\hat{\mathfrak{p}}_0^E := \hat{\mathfrak{g}}_0^E \cap \mathfrak{her}(4^n)$, cf. Eqn. (A.16). Then one has obviously the inclusion $\tilde{\mathfrak{p}}_0 \subset \hat{\mathfrak{p}}_0^E$. Moreover, a straightforward computation shows that all elements of $\hat{\mathfrak{p}}_0$, cf. Eqn. (2.40), are Hermitian and therefore $\hat{\mathfrak{p}}_0 \subset \tilde{\mathfrak{p}}_0$. Finally, Lemma 2.3.3 and Corollary A.0.1 yield the following estimates

$$\tfrac{(4^n-1)(4^n)}{2} \leq \dim_{\mathbb{R}}\, \hat{\mathfrak{p}}_0 \leq \dim_{\mathbb{R}}\, \tilde{\mathfrak{p}}_0 \leq \dim_{\mathbb{R}}\, \hat{\mathfrak{p}}_0^E = \tfrac{(4^n-1)(4^n)}{2} \;,$$

which implies $\hat{\mathfrak{p}}_0 = \tilde{\mathfrak{p}}_0 = \hat{\mathfrak{p}}_0^E$ and thus $\hat{\mathfrak{p}}_0^E \subset \tilde{\mathfrak{g}}_0$.
Now, according to Lemma 1.3.4 the Lie algebra $\hat{\mathfrak{g}}_0^E$ is isomorphic to $\mathfrak{gl}(4^n-1, \mathbb{R})$ and therefore one has $\hat{\mathfrak{k}}_0^E = [\hat{\mathfrak{p}}_0^E, \hat{\mathfrak{p}}_0^E]$ or, equivalenty, $\hat{\mathfrak{g}}_0^E = [\hat{\mathfrak{p}}_0^E, \hat{\mathfrak{p}}_0^E] \oplus \hat{\mathfrak{p}}_0^E$. Hence, by the inclusion $\hat{\mathfrak{p}}_0^E \subset \tilde{\mathfrak{g}}_0$ we conclude $\hat{\mathfrak{k}}_0^E \subset \tilde{\mathfrak{g}}_0$ and thus $\tilde{\mathfrak{g}}_0 = \hat{\mathfrak{g}}_0^{LK} = \hat{\mathfrak{g}}_0^E$.
So finally, we have to show that

$$\hat{\mathfrak{a}}_0 := \langle \hat{\sigma}_{\mathbf{p}}^2 \,|\, \mathbf{p} \in I_0^n \rangle \;, \tag{2.44}$$

is a maximal abelian subalgebra of $\hat{\mathfrak{p}}_0$. The fact that $\hat{\mathfrak{a}}_0$ is abelian easily follows from Eqn. (2.35) and Lemma 1.3.4. Furthermore, Lemma 2.3.3 implies that $\hat{\mathfrak{a}}_0$ is $4^n-1$-dimensional and therefore maximal due to the isomorphy of $\hat{\mathfrak{g}}_0^{LK}$ to $\mathfrak{gl}(4^n-1, \mathbb{R})$. □

**Corollary 2.3.1.** *The Lie algebra* $\hat{\mathfrak{k}}_0 = [\hat{\mathfrak{p}}_0, \hat{\mathfrak{p}}_0]$ *given by Eqn.* (2.39)*, can be alternatively generated as* $\hat{\mathfrak{k}}_0 = [\hat{\mathfrak{a}}_0, \hat{\mathfrak{p}}_0]$ *and therefore,*

$$\hat{\mathfrak{k}}_0 = \langle [\hat{\sigma}_{\boldsymbol{p}}^2, \{\hat{\sigma}_{\boldsymbol{\alpha}}, \hat{\sigma}_{\boldsymbol{\mu}}\}_+] \,|\, \boldsymbol{p}, \boldsymbol{\alpha} < \boldsymbol{\mu} \in I_0^n \rangle \;. \tag{2.45}$$

*Proof.* Since $\hat{\mathfrak{g}}_0^{LK}$ is isomorphic to $\mathfrak{gl}(4^n-1, \mathbb{R})$ by Theorem 4 we know that $\hat{\mathfrak{k}}_0$ and $\hat{\mathfrak{p}}_0$ are isomorphic to $\mathfrak{so}(4^n-1)$ and $\mathfrak{sym}(4^n-1)$, respectively. Here, $\mathfrak{sym}(4^n-1)$ denotes the set of all symmetric matrices in $\mathfrak{gl}(4^n-1, \mathbb{R})$. Now, for $\mathfrak{gl}(4^n-1, \mathbb{R})$ the relation $\mathfrak{so}(4^n-1) = [\mathfrak{a}, \mathfrak{sym}(4^n-1)]$, where $\mathfrak{a}$ is a maximal abelian subalgebra of $\mathfrak{sym}(4^n-1)$, is well-known. Therefore, by the above isomorphy, we conclude $\hat{\mathfrak{k}}_0 = [\hat{\mathfrak{a}}_0, \hat{\mathfrak{p}}_0]$. □

By the following Lemma, we are able to provide a necessary and sufficient condition on how to generate elements in $\mathrm{ad}_{\mathfrak{su}(2^n)} \subset \hat{\mathfrak{k}}_0$ from basis elements of $\hat{\mathfrak{p}}_0$.

**Lemma 2.3.4.** *Let* $\mathbf{p}, \boldsymbol{\alpha}, \boldsymbol{\mu}, \in I_0^n$ *such that* $\boldsymbol{\alpha} \neq \boldsymbol{\mu}$. *Then* $[\hat{\sigma}_{\boldsymbol{p}}^2, \{\hat{\sigma}_{\boldsymbol{\alpha}}, \hat{\sigma}_{\boldsymbol{\mu}}\}_+] \in \mathrm{ad}_{\mathfrak{su}(2^n)}$ *if and only if either* $\boldsymbol{p} = \boldsymbol{\alpha}$ *or* $\boldsymbol{p} = \boldsymbol{\mu}$.

*Proof.* First, we remark that $A := [\hat{\sigma}_{\mathbf{p}}^2, \{\hat{\sigma}_{\boldsymbol{\alpha}}, \hat{\sigma}_{\boldsymbol{\mu}}\}_+]$ can be expressed as $A = B - B^\dagger - C$ with

$$\begin{aligned} B &:= \tfrac{1}{4}\big((\sigma_{\boldsymbol{\alpha}}\sigma_{\mathbf{p}})^\top \otimes \sigma_{\mathbf{p}}\sigma_{\boldsymbol{\mu}} - (\sigma_{\mathbf{p}}\sigma_{\boldsymbol{\mu}})^\top \otimes \sigma_{\boldsymbol{\alpha}}\sigma_{\mathbf{p}}\big)\ , \quad \text{and} \\ C &:= \tfrac{1}{4}\big(\sigma_{\mathbf{p}}^\top \otimes [\sigma_{\mathbf{p}}, \{\sigma_{\boldsymbol{\alpha}}, \sigma_{\boldsymbol{\mu}}\}_+] - [\sigma_{\mathbf{p}}, \{\sigma_{\boldsymbol{\alpha}}, \sigma_{\boldsymbol{\mu}}\}_+]^\top \otimes \sigma_{\mathbf{p}}\big)\ . \end{aligned} \tag{2.46}$$

Note that $B$ is either Hermitian or skew-Hermitian and therefore $B - B^\dagger$ is either zero or $2B$. With these definitions at hand we are prepared to proof our statement.

The "$\Longleftarrow$" direction is straightforward. If $\mathbf{p} = \boldsymbol{\alpha}$ (or $\mathbf{p} = \boldsymbol{\mu}$) then $C = 0$ in Eqn. (2.46) and $A = (B - B^\dagger) \in \mathrm{ad}_{\mathfrak{su}(2^n)}$, since $\sigma_{\mathbf{p}}\sigma_{\boldsymbol{\alpha}} = \mathbb{1}_{2^n}$ (or $\sigma_{\mathbf{p}}\sigma_{\boldsymbol{\mu}} = \mathbb{1}_{2^n}$).

The "$\Longrightarrow$" direction is deduced from the decomposition $A = B - B^\dagger - C$ as follows. As in the proof of Lemma 2.3.3 (b), and the fact that the Pauli matrices are traceless, one has $\langle B, C\rangle_{\mathrm{tr}} = \langle B^\dagger, C\rangle_{\mathrm{tr}} = \langle \mathrm{i}\hat{\sigma}_{\mathbf{p}}, C\rangle_{\mathrm{tr}} = 0$ for all $\mathbf{p} \in I_0^n$. Therefore, $C = 0$ is necessary for $A \in \mathrm{ad}_{\mathfrak{su}(2^n)}$. Using the same techniques, we see that $B - B^\dagger$ is orthogonal to $\mathrm{ad}_{\mathfrak{su}(2^n)}$ whenever $\mathbf{p} \neq \boldsymbol{\alpha}$ and $\mathbf{p} \neq \boldsymbol{\mu}$. Therefore, we have shown that $A \in \mathrm{ad}_{\mathfrak{su}(2^n)}$ implies $\mathbf{p} = \boldsymbol{\alpha}$ or $\mathbf{p} = \boldsymbol{\mu}$. □

**Corollary 2.3.2.**

$$\mathrm{ad}_{\mathfrak{su}(2^n)} = \langle [\hat{\sigma}_{\boldsymbol{p}}^2, \{\hat{\sigma}_{\boldsymbol{p}}, \hat{\sigma}_{\boldsymbol{\mu}}\}_+] \,|\, \boldsymbol{p} \neq \boldsymbol{\mu} \in I_0^n \rangle\ . \tag{2.47}$$

*Proof.* The inclusion "$\supset$" follows from Lemma 2.3.4. On the other hand, for $\mathbf{p} = \boldsymbol{\alpha}$ one has $\sigma_{\mathbf{p}}\sigma_{\boldsymbol{\alpha}} = \mathbb{1}_{2^n}$ and $C = 0$ and therefore it is easy to see that by an suitable choice of $\mathbf{p}$ and $\boldsymbol{\mu}$ the matrix $B - B^\dagger$ can represent (up to a sign factor) any $\mathrm{i}\hat{\sigma}_{\mathbf{q}}$ where $\mathbf{q} \in I_0^n$. Hence, we obtain the desired equality. □

### 2.3.2 Non-Unital $n$-Qubit Systems

In Section 2.2 we considered the Lindblad generators for "standard" single qubit unital and non-unital noise. Recall that the corresponding Lindblad terms were of the form $V_k = C_k + \mathrm{i}D_k$, where $C_k$ and $D_k$ are scalar multiples of Pauli matrices. We showed that by considering these especially simple, yet common, Lindblad terms we could provide an elegant representation of the Lindblad-Kossakowski Lie algebra $\hat{\mathfrak{g}}^{LK}$. In that situation, it turned out that the non-unital part of the dissipative dynamics could easily be interpreted as elements of an (abelian) ideal $\hat{\mathfrak{i}}$ of $\hat{\mathfrak{g}}^{LK}$ given by the (real) span

$$\hat{\mathfrak{i}} = \langle \mathrm{i}\hat{\sigma}_y\hat{\sigma}_z^+, \mathrm{i}\hat{\sigma}_z\hat{\sigma}_x^+, \mathrm{i}\hat{\sigma}_x\hat{\sigma}_y^+ \rangle \tag{2.48}$$

Furthermore, these elements which make up the ideal are immediately visible from the decomposition given by Lemma 1.3.3 applied to a Lindblad generator which describes standard single qubit noise i.e.

$$\Gamma = 2 \sum_{\substack{p,q \in I_0 \\ p \neq q}} \gamma_{p,q}\big(\hat{\sigma}_p^2 + \hat{\sigma}_q^2 + 2\mathrm{i}\hat{\sigma}_p\hat{\sigma}_q^+\big) \in \mathfrak{gl}(4, \mathbb{C})\,, \tag{2.49}$$

where we recall that $I_0 = \{x, y, z\}$.

To address the situation of non-unital $n$-qubit systems, we can generalize Eqn. (2.49) and we will show that there again exists an abelian ideal $\hat{\mathfrak{i}}$ such that $\hat{\mathfrak{g}}^{LK}$ admits a semidirect sum decomposition $\hat{\mathfrak{g}}^{LK} = \hat{\mathfrak{g}}_0^{LK} \oplus_s \hat{\mathfrak{i}}$, where $\hat{\mathfrak{g}}_0^{LK}$ denotes the unital n-qubit Lindblad-Kossakowski Lie algebra cf. Eqn. (2.38).

A standard multi-qubit dissipative process can be separated into a *unital* component and an additional *mixed* unital and non-unital component in the following sense. For Lindblad terms of the form $V_k = \sqrt{\gamma_k}(\sigma_{\mathbf{p}_k} + \mathrm{i}\sigma_{\mathbf{q}_k})$, such that $\mathbf{p}_k \neq \mathbf{q}_k$ and $\gamma_k \in \mathbb{R}^+$ for all $k$, we can use the decomposition of $\Gamma$ given by Lemma 1.3.3 to provide the splitting $\Gamma = \Gamma_0 + \Gamma_m$ where

$$\Gamma_0 := 2\sum_k \gamma_k \left(\hat{\sigma}^2_{\mathbf{p}_k} + \hat{\sigma}^2_{\mathbf{q}_k}\right) \;, \quad \text{and} \quad \Gamma_m := 2\mathrm{i}\sum_k \gamma_k \left(\hat{\sigma}_{\mathbf{p}_k}\hat{\sigma}^+_{\mathbf{q}_k} - \hat{\sigma}_{\mathbf{q}_k}\hat{\sigma}^+_{\mathbf{p}_k}\right) \;. \tag{2.50}$$

Again by Lemma 1.3.3, if $\{\sigma_{\mathbf{p}_k}, \sigma_{\mathbf{q}_k}\}_+ = 0$ for all $k$ we obtain

$$\Gamma_m = 4\mathrm{i}\sum_k \gamma_k \hat{\sigma}_{\mathbf{p}_k}\hat{\sigma}^+_{\mathbf{q}_k} \;. \tag{2.51}$$

Since we are focused on the operators which make up Eqn. (2.51), unless explicitly stated, we will omit the factor of four for simplicity. Furthermore, denoting the *real* linear span of operators $\Gamma_m$ given in Eqn. (2.50) as $\hat{\mathfrak{m}} \subset \mathfrak{gl}(4^n, \mathbb{C})$, i.e.

$$\hat{\mathfrak{m}} := \langle \mathrm{i}(\hat{\sigma}_{\mathbf{p}}\hat{\sigma}^+_{\mathbf{q}} - \hat{\sigma}_{\mathbf{q}}\hat{\sigma}^+_{\mathbf{p}}) \mid \mathbf{p}, \mathbf{q} \in I_0^n \rangle \;,$$

we arrive at the following decomposition result.

**Proposition 2.3.1.** *The subspace $\hat{\mathfrak{m}} \subset \mathfrak{gl}(4^n, \mathbb{C})$ can be decomposed as $\hat{\mathfrak{m}} = \hat{\mathfrak{m}}_{\mathrm{qt}} \oplus \hat{\mathfrak{m}}_{\mathrm{s}}$, where*

$$\hat{\mathfrak{m}}_{\mathrm{qt}} := \langle \mathrm{i}\hat{\sigma}_{\boldsymbol{p}}\hat{\sigma}^+_{\boldsymbol{q}} \mid [\sigma_{\boldsymbol{p}}, \sigma_{\boldsymbol{q}}] \neq 0 \rangle \;, \quad \text{and} \quad \hat{\mathfrak{m}}_{\mathrm{s}} = \langle \mathrm{i}(\hat{\sigma}_{\boldsymbol{p}}\hat{\sigma}^+_{\boldsymbol{q}} - \hat{\sigma}_{\boldsymbol{q}}\hat{\sigma}^+_{\boldsymbol{p}}) \mid [\sigma_{\boldsymbol{p}}, \sigma_{\boldsymbol{q}}] = 0 \rangle \;,$$

*and in particular, $\hat{\mathfrak{m}}_{\mathrm{s}} \subseteq \hat{\mathfrak{k}}_0 \subseteq \hat{\mathfrak{g}}_0^{LK}$, where $\hat{\mathfrak{k}}_0$ is the skew-Hermitian part of the Cartan decomposition of $\hat{\mathfrak{g}}_0^{LK} = \hat{\mathfrak{k}}_0 \oplus \hat{\mathfrak{p}}_0$.*

*Proof.* The set $\hat{\mathfrak{m}}_{\mathrm{qt}}$ follows by Lemma 1.3.3 and clearly by Lemma 2.3.2, $\hat{\mathfrak{m}}_{\mathrm{qt}} \cap \hat{\mathfrak{m}}_{\mathrm{s}} = \{0\}$, thus the direct sum follows. Furthermore, note that when $k = 1$ in Eqn. (2.50) we obtain $\Gamma_m = 2\mathrm{i}\gamma(\hat{\sigma}_{\mathbf{p}}\hat{\sigma}^+_{\mathbf{q}} - \hat{\sigma}_{\mathbf{q}}\hat{\sigma}^+_{\mathbf{p}})$ and hence

$$\Gamma_m \;=\; \mathrm{i}\gamma\left(\sigma_{\mathbf{q}}^\top \otimes \sigma_{\mathbf{p}} - \sigma_{\mathbf{p}}^\top \otimes \sigma_{\mathbf{q}}\right) + \tfrac{\mathrm{i}\gamma}{2}\widehat{\mathrm{ad}}^+_{[\sigma_{\mathbf{p}}, \sigma_{\mathbf{q}}]} \;, \tag{2.52}$$

where as usual $\gamma \in \mathbb{R}^+$. Thus, if $[\sigma_{\mathbf{p}}, \sigma_{\mathbf{q}}] = 0$, then $\Gamma_m \in \hat{\mathfrak{k}}_0 \subset \hat{\mathfrak{g}}_0^{LK}$. □

**Remark 5.** *If $\sigma_{\boldsymbol{q}} = \mathbb{1}_{2^n}$, then the skew-Hermitian element $\Gamma_m \in \hat{\mathfrak{m}}_{\mathrm{s}}$ is contained in $\mathrm{ad}_{\mathfrak{su}(2^n)}$, whereas in a generic case, $\Gamma_{\mathrm{m}}$ is a more exotic type of skew-symmetric operation. This skew-symmetry motivates the definition of the set $\hat{\mathfrak{m}}_{\mathrm{s}}$.*

For reasons which will become clear later, elements in $\hat{\mathfrak{m}}_{\mathrm{qt}}$ will be called *quasi translations.* The following Lemma provides a useful fact which will be relevant in the remaining of the thesis.

**Lemma 2.3.5.** $(\mathrm{i}\hat{\sigma}_{\boldsymbol{p}}\hat{\sigma}^+_{\boldsymbol{q}})^2 = 0$ *for all elements* $\mathrm{i}\hat{\sigma}_{\boldsymbol{p}}\hat{\sigma}^+_{\boldsymbol{q}} \in \hat{\mathfrak{m}}_{\mathrm{qt}}$.

*Proof.* By direct computation we obtain $\hat{\sigma}_{\mathbf{p}}\hat{\sigma}_{\mathbf{q}}^{+} = \frac{1}{4}(\mathbb{1} \otimes \sigma_{\mathbf{p}}\sigma_{\mathbf{q}} + \sigma_{\mathbf{q}}^{\top} \otimes \sigma_{\mathbf{p}} - \sigma_{\mathbf{p}}^{\top} \otimes \sigma_{\mathbf{q}} - (\sigma_{\mathbf{q}}\sigma_{\mathbf{p}})^{\top} \otimes \mathbb{1})$ and therefore

$$\hat{\sigma}_{\mathbf{p}}\hat{\sigma}_{\mathbf{q}}^{+}\sigma_{\mathbf{p}} = \tfrac{1}{4}(\mathbb{1} \otimes \sigma_{\mathbf{p}}\sigma_{\mathbf{q}}\sigma_{\mathbf{p}} - \sigma_{\mathbf{p}}^{\top} \otimes \sigma_{\mathbf{p}}\sigma_{\mathbf{q}} + \sigma_{\mathbf{q}}^{\top} \otimes \sigma_{\mathbf{p}}^{2} - (\sigma_{\mathbf{p}}\sigma_{\mathbf{q}})^{\top} \otimes \sigma_{\mathbf{p}} - \sigma_{\mathbf{p}}^{\top} \otimes \sigma_{\mathbf{q}}\sigma_{\mathbf{p}} + (\sigma_{\mathbf{p}}^{\top})^{2} \otimes \sigma_{\mathbf{q}} - (\sigma_{\mathbf{q}}\sigma_{\mathbf{p}})^{\top} \otimes \sigma_{\mathbf{p}} + (\sigma_{\mathbf{q}}\sigma_{\mathbf{p}})^{\top} \otimes \mathbb{1}) \,.$$

Clearly $\sigma_{\mathbf{p}}^2 = \sigma_{\mathbf{q}}^2 = \mathbb{1}$ and by 1) of Lemma 2.3.2 (and its proof), we know that $\sigma_{\mathbf{p}}\sigma_{\mathbf{q}} = -\sigma_{\mathbf{q}}\sigma_{\mathbf{p}}$ since $[\sigma_{\mathbf{p}}, \sigma_{\mathbf{q}}] \neq 0$. Thus we obtain $\hat{\sigma}_{\mathbf{p}}\hat{\sigma}_{\mathbf{q}}^{+}\sigma_{\mathbf{p}} = 0$ and hence $(\mathrm{i}\hat{\sigma}_{\mathbf{p}}\hat{\sigma}_{\mathbf{q}}^{+})^2 = 0$. □

The above Lemma makes intuitive sense when we consider the single qubit case. That is, in Proposition 2.2.2 we proved that operators of the form $\mathrm{i}\hat{\sigma}_p\hat{\sigma}_q^{+}$ with $p, q \in I_0 = \{x, y, z\}$ such that $p \neq q$ (and noting that $\mathrm{i}\hat{\sigma}_p\hat{\sigma}_q^{+} = -\mathrm{i}\hat{\sigma}_q\hat{\sigma}_p^{+}$) are a basis for the abelian ideal $\hat{\mathfrak{i}}$. That is, elements $\Gamma_m$ can *only* be elements of $\hat{\mathfrak{m}}_{\mathrm{qt}}$ in this single qubit case, and by counting degrees of freedom we see that in fact $\hat{\mathfrak{m}} = \hat{\mathfrak{m}}_{\mathrm{qt}} = \hat{\mathfrak{i}}$. Furthermore, recall that these ideal elements can be represented in the coherence vector representation as elements of the form $(0, a) \in \mathfrak{gl}(3, \mathbb{R}) \oplus_s \mathbb{R}^3$ with $a \in \mathbb{R}^3$ (cf. Eqn. (2.26)). Thus, it is immediately apparent that these ideal elements are nilpotent matrices since $(0, a) \cdot (0, a) = (0, 0)$ for all $a \in \mathbb{R}^3$. Finally, we note that unlike the single qubit case, for multi-qubit systems, the product of two quasi-translation operators is *not always zero.*

Now we finally arrive at a crucial distinction between operators $\Gamma_m \in \hat{\mathfrak{m}}_{\mathrm{qt}}$ when we are comparing the single qubit scenario to multi-qubit systems. Appendix C provides a detailed description of the kernel and range of such an operator which then leads us to the following result.

**Proposition 2.3.2.** *For multi-qubit systems, the elements $\Gamma_m = \mathrm{i}\hat{\sigma}_{\boldsymbol{p}}\hat{\sigma}_{\boldsymbol{q}}^{+} \in \hat{\mathfrak{m}}_{\mathrm{qt}}$ are not contained in the ideal $\hat{\mathfrak{i}}$, nor are they contained in the unital subalgebra $\hat{\mathfrak{g}}_0^{LK}$. That is,*

$$\mathrm{i}\hat{\sigma}_{\boldsymbol{p}}\hat{\sigma}_{\boldsymbol{q}}^{+} \in \hat{\mathfrak{g}}^{LK} = \hat{\mathfrak{g}}_0^{LK} \oplus_s \hat{\mathfrak{i}} \tag{2.53}$$

*such that $\mathrm{i}\hat{\sigma}_{\boldsymbol{p}}\hat{\sigma}_{\boldsymbol{q}}^{+} \notin \hat{\mathfrak{i}}$ and $\mathrm{i}\hat{\sigma}_{\boldsymbol{p}}\hat{\sigma}_{\boldsymbol{q}}^{+} \notin \hat{\mathfrak{g}}_0^{LK}$.*

*Proof.* First we show that the operators $\mathrm{i}\hat{\sigma}_{\mathbf{p}}\hat{\sigma}_{\mathbf{q}}^{+} \in \hat{\mathfrak{m}}_{\mathrm{qt}}$ are non-unital and therefore and not elements of $\hat{\mathfrak{g}}_0^{LK}$ by proving that $\mathrm{i}\hat{\sigma}_{\mathbf{p}}\hat{\sigma}_{\mathbf{q}}^{+}(\mathbb{1}) \neq 0$. Since $[\sigma_{\mathbf{p}}, \sigma_{\mathbf{q}}] \neq 0$, then by (the proof of) Lemma 2.3.2, we know that $[\sigma_{\mathbf{p}}, \sigma_{\mathbf{q}}] = 2\sigma_{\mathbf{p}}\sigma_{\mathbf{q}}$. Thus, in operator representation we immediately get $\mathrm{i}\,\mathrm{ad}_{\sigma_{\mathbf{p}}} \circ \mathrm{ad}_{\sigma_{\mathbf{q}}}^{+}(\mathbb{1}) = 2\mathrm{i}[\sigma_{\mathbf{p}}, \sigma_{\mathbf{q}}] = 4\mathrm{i}\sigma_{\mathbf{p}}\sigma_{\mathbf{q}}$ and thus the operator is not in $\hat{\mathfrak{g}}_0^{LK}$.

Now we prove that these operators are not contained in the ideal $\hat{\mathfrak{i}} \subseteq \hat{\mathfrak{i}}^E$. Recall that an element $\tau \in \mathfrak{i}^E$ is called an (infinitesimal) translation element and by definition it satisfies

$$\tau\big|_{\mathfrak{her}_0(N)} \equiv 0 \quad \text{and} \quad \tau(\mathbb{1}_N) \in \mathfrak{her}_0(N) \,, \tag{2.54}$$

(for more details see the paragraphs following Lemma 1.3.4 in Section 1.3). Proposition C.0.1 in Appendix C provides the range of such an operator and proves it's always the case that $\mathrm{i}\,\mathrm{ad}_{\mathbf{p}}\,\mathrm{ad}_{\mathbf{q}}^{+}(\sigma_{\mathbf{m}}) = 4\mathrm{i}\sigma_{\mathbf{p}}\sigma_{\mathbf{q}}\sigma_{\mathbf{m}}$ whenever $[\sigma_{\mathbf{p}}, \sigma_{\mathbf{m}}] = [\sigma_{\mathbf{q}}, \sigma_{\mathbf{m}}] = 0$. The operator clearly violates the first (infinitesimal) translation element condition which implies it is not contained in $\hat{\mathfrak{i}}^E$ and therefore not in $\hat{\mathfrak{i}}$ either. □

The above result implies that for multi-qubit systems, operators of the form $\mathrm{i}\hat{\sigma}_{\mathbf{p}}\hat{\sigma}_{\mathbf{q}}^{+} \in \hat{\mathfrak{m}}_{\mathrm{qt}}$ describe a dynamic on the system which has a unital and non-unital component to it.

Furthermore, since they are not contained within the ideal $\hat{\mathfrak{i}}$, they are *not* infinitesimal translations as they are in the single qubit case. We therefore aim to introduce a projection operator to "project out" the unital part of these *quasi-translation* operators thereby leaving only the translational component which is contained in the ideal.

Recall Lemma 1.3.4 where we introduced the Lie algebra $\mathfrak{g}^E$ which contains $\mathfrak{g}^{LK}$. Furthermore, in the introduction to this chapter we noted that we will in fact prove the equality $\mathfrak{g}^{LK} = \mathfrak{g}^E$. This is essential because it proves that $\mathfrak{g}^{LK}$ is isomorphic to the Lie algebra $\mathfrak{gl}(N^2-1,\mathbb{R}) \oplus_s \mathbb{R}^{N^2-1}$ which is the Lie algebra which contains Lindblad-Kossakowski operators $\mathcal{L}$ in the coherence vector representation.

**Proposition 2.3.3.** *Recall that $\hat{\mathfrak{g}}^E$ can be decomposed into the semidirect sum $\hat{\mathfrak{g}}^E = \mathfrak{g}_0^E \oplus_s \hat{\mathfrak{i}}^E$, where $\hat{\mathfrak{i}}^E$ is an abelian ideal of $\hat{\mathfrak{g}}^E$. Moreover, define the operator $\chi : \hat{\mathfrak{g}}^E \longrightarrow \hat{\mathfrak{g}}^E$, as $\chi(A) = [C_0, A]$, where*

$$C_0 := \frac{1}{2^{2n-1}} \sum_{p \in I_0^n} \hat{\sigma}_p^2 \,. \tag{2.55}$$

*Then $\chi : \hat{\mathfrak{g}}^E \longrightarrow \hat{\mathfrak{g}}^E$ is an orthogonal projection such that $\chi(\hat{\mathfrak{g}}^E) \subseteq \hat{\mathfrak{i}}^E$.*

Before we prove Proposition 2.3.3, we establish a Lemma which shows that the projection operator does indeed project out the unital parts of the quasi-translations and leaves only the (infinitesimal) translation part.

**Lemma 2.3.6.** *Let $\chi$ be the operator defined in Proposition 2.3.3. Then*

$$\chi(\hat{\mathfrak{g}}_0^E) = \{0\} \quad \textit{and} \quad \chi(\hat{\mathfrak{m}}_{\mathrm{qt}}) \neq \{0\} \,, \tag{2.56}$$

*and hence $\chi(\hat{\mathfrak{m}}_{\mathrm{qt}}) \subseteq \hat{\mathfrak{i}}^E$*

*Proof.* The result is proved in operator representation in Lemma B.0.1 in Appendix B. □

*Proof of Proposition 2.3.3.* First we state the obvious that since $\hat{\sigma}_{\mathbf{p}}^2 \in \hat{\mathfrak{g}}^E$ for all $\mathbf{p} \in I_0^n$, then $\chi(\hat{\mathfrak{g}}^E) \in \hat{\mathfrak{g}}^E$.

Throughout the remaining of the proof we will exploit the fact that $\hat{\mathfrak{g}}^E \overset{\text{iso}}{=} \mathfrak{gl}(4^n-1,\mathbb{R}) \oplus_s \mathbb{R}^{4^n-1}$ and hence we change to the coherence vector representation. Recalling Eqn. (2.26), a generic element $A \in \hat{\mathfrak{g}}^E$ can be represented as

$$\left[\begin{array}{c|c} A' & a \\ \hline 0 & 0 \end{array}\right] \in \mathfrak{gl}(4^n-1,\mathbb{R}) \oplus_s \mathbb{R}^{4^n-1} \,, \tag{2.57}$$

such that $A' \in \mathfrak{gl}(4^n-1,\mathbb{R})$ and $a \in \mathbb{R}^{4^n-1}$. Thus, the product of two elements $(A',a),(B',b) \in \mathfrak{gl}(4^n-1,\mathbb{R}) \oplus_s \mathbb{R}^{4^n-1}$ is given by $(A',a)\cdot(B',b) = (A'B', A'b)$.

To prove that $\chi$ is an orthogonal projection, we will use the fact that $C_0$ in Eqn. (2.55) is itself an orthogonal projection. This follows immediately by Remark 15 in Appendix B where we see $C_0\big|_{\mathfrak{hers}_0(2^n)} = \mathbb{1}_{2^n}$ and therefore in the coherence vector representation we see that (under a slight abuse of notation) $C_0 = (\mathbb{1}_{4^n-1}, 0)$. Thus, clearly we have that $C_0^2 = C_0^\dagger = C_0$ and hence $\chi^\dagger = \chi$.

Now we show that $\chi^2 = \chi$ and $\chi(\hat{\mathfrak{g}}^E) \subseteq \hat{\mathfrak{i}}^E$. For any $A \in \hat{\mathfrak{g}}_0^E$ and $B \in \hat{\mathfrak{i}}^E$, we have

$$\chi^2(A+B) = [C_0,[C_0,A]] + [C_0,[C_0,B]] = [C_0,[C_0,B]] \,, \tag{2.58}$$

since $\chi(A) = 0$ by Lemma 2.3.6. In the coherence vector representation it is immediately obvious that $BC_0 = 0$ and $C_0 B = B$ and hence $\chi^2(A + B) = [C_0, [C_0, B]] = C_0^2 B = C_0 B = B = \chi(A + B) \in \hat{\mathfrak{i}}^E$. □

Using this projection operator, we can now describe the structure of the abelian ideal $\hat{\mathfrak{i}}$ and the (infinitesimal) translation operators which it's made up of. In doing so, we now have all the tools to prove a central result of this chapter - that we have in fact the equality $\hat{\mathfrak{g}}^{LK} = \hat{\mathfrak{g}}^E$ and hence we have obtained a complete description of the $n$-qubit non-unital Lindblad-Kossakowski Lie algebra $\hat{\mathfrak{g}}^{LK}$.

**Theorem 5.** *The non-unital n-qubit Lindblad-Kossokowski Lie algebra $\hat{\mathfrak{g}}^{LK} \subset \mathfrak{gl}(4^n, \mathbb{C})$ decomposes into a semidirect sum $\hat{\mathfrak{g}}^{LK} = \hat{\mathfrak{g}}_0^{LK} \oplus_s \hat{\mathfrak{i}} = \hat{\mathfrak{g}}^E$, where $\hat{\mathfrak{g}}_0^{LK}$ is given by Eqn. (2.38) and $\hat{\mathfrak{i}} \subset \hat{\mathfrak{g}}^{LK}$ is a $4^n - 1$ dimensional abelian ideal defined by*

$$\hat{\mathfrak{i}} := \chi(\hat{\mathfrak{m}}_{\mathrm{qt}}) = \langle \chi(\mathrm{i}\hat{\sigma}_{\boldsymbol{p}}\hat{\sigma}_{\boldsymbol{q}}^+) \mid [\sigma_{\boldsymbol{p}}, \sigma_{\boldsymbol{q}}] \neq 0 \rangle = \hat{\mathfrak{i}}^E \, . \tag{2.59}$$

*Proof.* We already know the following inclusions/equalities: $\hat{\mathfrak{g}}^{LK} \subseteq \hat{\mathfrak{g}}^E$, $\hat{\mathfrak{g}}_0^{LK} = \hat{\mathfrak{g}}_0^E$, $\hat{\mathfrak{g}}^E = \hat{\mathfrak{g}}_0^E \oplus_s \hat{\mathfrak{i}}^E$ and $\hat{\mathfrak{i}} \subseteq \hat{\mathfrak{i}}^E$, where the last inclusion is given in Proposition 2.3.3 since $\hat{\mathfrak{i}} := \chi(\hat{\mathfrak{m}}_{\mathrm{qt}})$. Hence, we only need to prove $\hat{\mathfrak{i}} = \hat{\mathfrak{i}}^E \subset \hat{\mathfrak{g}}^{LK}$. Let $A \in \hat{\mathfrak{g}}^{LK}$ and $\mathrm{i}\hat{\sigma}_{\mathbf{p}}\hat{\sigma}_{\mathbf{q}}^+ \in \hat{\mathfrak{m}}_{\mathrm{qt}}$. By the Jacobi identity,

$$\begin{aligned} \chi([A, \chi(\mathrm{i}\hat{\sigma}_{\mathbf{p}}\hat{\sigma}_{\mathbf{q}}^+)]) &= [A, \chi(\chi(\mathrm{i}\hat{\sigma}_{\mathbf{p}}\hat{\sigma}_{\mathbf{q}}^+))] - [\chi(\mathrm{i}\hat{\sigma}_{\mathbf{p}}\hat{\sigma}_{\mathbf{q}}^+), \chi(A)] && (2.60)\\ &= [A, \chi^2(\mathrm{i}\hat{\sigma}_{\mathbf{p}}\hat{\sigma}_{\mathbf{q}}^+)] && (2.61)\\ &= [A, \chi(\mathrm{i}\hat{\sigma}_{\mathbf{p}}\hat{\sigma}_{\mathbf{q}}^+)] \in \hat{\mathfrak{i}}^E \cap \hat{\mathfrak{g}}^{LK} \, , && (2.62) \end{aligned}$$

which follows from the fact that $\hat{\mathfrak{i}}^E$ is abelian and so $[\chi(\mathrm{i}\hat{\sigma}_{\mathbf{p}}\hat{\sigma}_{\mathbf{q}}^+), \chi(A)] = 0$ and $\chi^2 = \chi$ by Proposition 2.3.3. Therefore $[\hat{\mathfrak{g}}^{LK}, \hat{\mathfrak{i}}] \subseteq \hat{\mathfrak{i}}^E \cap \hat{\mathfrak{g}}^{LK} = \hat{\mathfrak{i}}^E \cap (\hat{\mathfrak{g}}_0^{LK} \oplus_s \hat{\hat{\mathfrak{i}}}) = \hat{\mathfrak{i}}$ and hence $\hat{\mathfrak{i}} \subseteq \hat{\mathfrak{i}}^E$ is an ideal.

Since $\mathfrak{gl}(4^n - 1, \mathbb{R})$ acts transitively on $\mathbb{R}^{4^n-1}$, and $\mathfrak{gl}(4^n - 1, \mathbb{R}) \oplus_s \mathbb{R}^{4^n-1} \overset{\text{iso}}{=} \hat{\mathfrak{g}}_0^E \oplus_s \hat{\mathfrak{i}}^E$ by Lemma 1.3.4, then $\hat{\mathfrak{i}}^E$ contains no other ideals except for itself and 0. By Proposition 2.3.3, $\hat{\mathfrak{i}} \neq \{0\}$ and thus $\hat{\mathfrak{i}} = \hat{\mathfrak{i}}^E \subseteq \hat{\mathfrak{g}}^{LK}$ which finally shows that $\hat{\mathfrak{g}}^{LK} = \hat{\mathfrak{g}}^E$. □

Here and henceforth, we call $\hat{\mathfrak{i}}$ the Lindblad-Kossakowski (LK) ideal. As we will shall see throughout this thesis, this operator representation of the LK-Lie algebra will provide us with new insights into the interplay between the coherent and incoherent operators/parts of an open systems dynamics. In particular, the next section will further investigate the exact structure of the (infinitesimal) translation operators which make up the LK-ideal $\hat{\mathfrak{i}}$.

## 2.4 Translation Operators and the Lindblad-Kossakowski Ideal

As exemplified in the last section, the quasi-translation elements which make up the "mixed" part $\Gamma_m$ of $\Gamma = \Gamma_0 + \Gamma_m$ induces both unital and non-unital dissipative dynamics for multi-qubit systems (cf. Proposition 2.3.2). We then showed in Theorem 5 that the unital component of these quasi-translation operators needed to be projected out, thereby leaving only the non-unital (infinitesimal) translation component to define a basis of the Lindblad-Kossakowski ideal. As discussed prior to Lemma 1.3.5 in Section 1.3, elements $\mathcal{T} \in \mathfrak{i}$ are called (infinitesimal) translations due to the following reasoning. For any Hermitian matrix $\rho := \mathbb{1}_{2^n} + \rho_0$ with $\rho_0 \in \mathfrak{her}(2^n)$ and $\exp(\mathcal{T}) := \sum_{k=0}^{\infty} \frac{\mathcal{T}^k}{k!}$, we see that $\exp(\mathcal{T})(\mathbb{1}_{2^n} + \rho_0 + \mathcal{T}(\mathbb{1}_{2^n}))$. Now, since $\mathcal{T}(\mathbb{1}_{2^n}) \in \mathfrak{her}(2^n)$ its clear $\exp(\mathcal{T})$ acts as a translation on the hyperplane $\mathbb{1}_{2^n} + \mathfrak{her}_0(2^n)$. The LK-ideal then consists of (infinitesimal) translation elements and the goal of this section is to parametrise its basis. As we will see, this can be done by determining the form of translation operators $\mathcal{T} \in \mathfrak{i}$ such that

$$\mathcal{T}(\mathbb{1}_{2^n}) = \sigma_{\mathbf{m}} \ , \quad \text{for} \quad \sigma_{\mathbf{m}} \in \mathcal{B}_0^n \ . \tag{2.63}$$

First we must introduce some notation that ultimately will simplify the discussion and then be used throughout this thesis.

Define as usual the local Pauli operators by

$$\sigma_{p,k} := \mathbb{1}_2 \otimes \cdots \otimes \mathbb{1}_2 \otimes \sigma_p \otimes \mathbb{1}_2 \otimes \cdots \otimes \mathbb{1}_2 \ , \tag{2.64}$$

where $\sigma_p$ appears at the $k^{th}$ position. Thus, $\sigma_{p,k} = \sigma_{\mathbf{p}}$, where $\mathbf{p} := (1, \ldots, 1, p, 1, \ldots, 1) \in I_0^n$ with $p \in I_0$ at the $k^{th}$ position. Then the corresponding "localized" extensions of Eqns. (2.31) and (2.32) are given by

$$\hat{\sigma}_{p,k} \ := \ \tfrac{1}{2}\big(\mathbb{1}_{2^n} \otimes \sigma_{p,k} - \sigma_{p,k}^T \otimes \mathbb{1}_{2^n}\big) \tag{2.65}$$

$$\hat{\sigma}_{p,k}^{+} \ := \ \tfrac{1}{2}\big(\mathbb{1}_{2^n} \otimes \sigma_{p,k} + \sigma_{p,k}^T \otimes \mathbb{1}_{2^n}\big) \tag{2.66}$$

Define the sets

$$\mathbb{I} := \{\pm x, \pm y, \pm z, \pm 1\} \quad \text{and} \quad \mathbb{I}_0 := \{\pm x, \pm y, \pm z\} \ , \tag{2.67}$$

and note that for some $p \in I$, then $-p \in \mathbb{I}$ and we define $-\sigma_p := \sigma_{-p}$. Moreover, for any $p, q \in \mathbb{I}$ we define a "product" as follows

$$p \star q := \begin{cases} 1 & \text{if } p = q \ , \\ \pm p & \text{if } q = \pm 1 \ , \\ \pm q & \text{if } p = \pm 1 \ , \\ \pm r & \text{else, where } \pm r \text{ is determined by } \sigma_p \sigma_q = \pm \mathrm{i} \sigma_r \ . \end{cases} \tag{2.68}$$

Furthermore, we can extend the above star-product over $n$-indices for $\mathbf{p}, \mathbf{q} \in \mathbb{I}^n$ as

$$\mathbf{m} := \mathbf{p} \star \mathbf{q} \ , \quad \text{where} \quad m_k = p_k \star q_k \quad \text{for all } k = 1, \ldots, n. \tag{2.69}$$

**Remark 6.** *Note that by Eqn.* (2.68)*, each $m_k \in \{1, \pm p_k, \pm q_k, \pm r_k\}$ and thus we can define a notion of positivity of the product $\boldsymbol{m} = \boldsymbol{p} \star \boldsymbol{q}$. We say that $\boldsymbol{p} \star \boldsymbol{q} > 0$ ( resp. $\boldsymbol{p} \star \boldsymbol{q} < 0$) whenever there are an* even *(resp.* odd*) number of negative indices of $\boldsymbol{m}$. Moreover, we remark that this star-product does* not *induce a group structure on the set $\mathbb{I}$ since it is clearly not an associative operation.*

The motivation for introducing such a product is the following. For any $\sigma_{\mathbf{p}}, \sigma_{\mathbf{q}} \in \mathcal{B}_0^n$ such that $[\sigma_{\mathbf{p}}, \sigma_{\mathbf{q}}] \neq 0$, then by Lemma 2.3.1 we get $[\sigma_{\mathbf{p}}, \sigma_{\mathbf{q}}] = 2\sigma_{\mathbf{p}}\sigma_{\mathbf{q}} = 2\mathrm{i}\sigma_{\mathbf{m}}$, where $\mathbf{m} = \mathbf{p} \star \mathbf{q}$. Therefore, this star-product serves as a shortcut to identify the index of the Pauli matrix resulting from a commutation between two other Pauli matrices and the positivity/negativity of the product is simply the corresponding sign of the commutator. This is extremely useful for our purposes throughout the paper.

Now we are ready to apply this concept to the single qubit ideal elements. For a single qubit we know that for $p, q \in \{x, y, z\}$ such that $p \neq q$ then $\mathrm{i}\hat{\sigma}_q\hat{\sigma}_p^+ \in \hat{\mathfrak{i}}$ since $[\sigma_q, \sigma_p] \neq 0$ and hence in operator representation we see

$$\tfrac{\mathrm{i}}{4}\,\mathrm{ad}_{\sigma_q} \circ \mathrm{ad}^+_{\sigma_p}(\mathbb{1}_2) = \tfrac{\mathrm{i}}{2}[\sigma_q, \sigma_p] = \mathrm{i}\sigma_q\sigma_p = -\mathrm{i}\sigma_p\sigma_q = \sigma_m \ , \tag{2.70}$$

since $\sigma_p\sigma_q = \mathrm{i}\sigma_m$ for $m = p \star q$ as defined above.

For $p, q \in I_0$ such that $[\sigma_q, \sigma_p] \neq 0$ we can therefore define a *translation operator* in the $m^{th}$ direction as

$$\mathcal{T}_m := \mathrm{i}\hat{\sigma}_q\hat{\sigma}_p^+ \ , \quad \text{where} \quad m = p \star q \tag{2.71}$$

and hence via the star-product, $m = \pm r$ where $\mathcal{T}_{-r} := -\mathcal{T}_r$. As we have seen in Proposition 2.2.2 in Section 2.2.2, operators of this type form the basis of the single qubit Lindblad-Kossakowski ideal.

**Remark 7.** *The ordering of the product which the translation is defined as may look like it should be reversed due to the fact that we have defined the direction $m$ to be given by $m = p \star q$. However, due to Eqn.* (2.70) *we see that (in operator representation) $\mathcal{T}_m(\mathbb{1}_2) = \sigma_m$ and hence if for example $p \star q = -r$ we get $\mathcal{T}_{-r}(\mathbb{1}_2) = -\mathcal{T}_r(\mathbb{1}_2) = -\sigma_r$ as expected.*

This brings us to a fundamental concept of this thesis which is highlighted in the following example.

**Example 2.** *Consider a single qubit Lindblad generator whose only dissipative term is given by $V = \frac{1}{2}(\sigma_x + \mathrm{i}\sigma_y)$. It is well known that this noise generator describes amplitude damping noise, see for instance [39, 6]. Following Eqn.* (2.13) *by writing* $\Gamma$ *in super-operator representation we obtain*

$$-\mathcal{L} = -\Gamma = -\tfrac{1}{2}(\hat{\sigma}_x^2 + \hat{\sigma}_y^2) - \mathrm{i}\hat{\sigma}_x\hat{\sigma}_y^+ = -\tfrac{1}{2}(\hat{\sigma}_x^2 + \hat{\sigma}_y^2) + \mathcal{T}_z \ , \tag{2.72}$$

*where $\mathcal{T}_z = -\mathrm{i}\hat{\sigma}_x\hat{\sigma}_y^+ = \mathrm{i}\hat{\sigma}_y\hat{\sigma}_x^+$. This leads to the following intuitive picture. It's also well known that in the Bloch sphere representation, amplitude damping noise contracts the Bloch sphere and then* translates *the states upwards along the* z-axis *towards the north pole (the $|0\rangle$-state) [39]. Usually, this translation is described by a vector in $\mathbb{R}^3$ - however here, we have provided the operator picture which describes this affine shift* and *its corresponding direction.*

*Since the translation direction drives the system towards its unique fixed point, a main result of this thesis will be to use this intuition to our advantage. Namely, in Chapter 3 we will use generalisations of these translation operators to engineer purely dissipative noise that drives $n$-qubit systems to desired unique fixed points.*

Thus, we now aim to generalise the single qubit translation operators to $n$-qubit systems. The immediate extension is then to introduce a "local quasi-translation" in the $m^{th}$ direction on the $k^{th}$ qubit as

$$\mathcal{T}_{m,k} := \mathrm{i}\hat{\sigma}_{q,k}\hat{\sigma}_{p,k}^+ \ , \quad \text{for} \quad [\sigma_q, \sigma_p] \neq 0 \ , \quad \text{and} \quad m = p \star q \tag{2.73}$$

where we again identify $\mathcal{T}_{-r,k} := -\mathcal{T}_{r,k}$. We stress that by Proposition 2.3.2 these are indeed quasi-translations and are not translation elements (i.e. not contained in $\hat{\mathfrak{i}}$) although they carry the tau-notation. The notation has a benefit which we will see in the following.

**Proposition 2.4.1.** *Denote the span of all local quasi-translations by* $\hat{\mathfrak{m}}_{\mathrm{loc}} \subseteq \hat{\mathfrak{m}}_{\mathrm{qt}} \subseteq \hat{\mathfrak{m}}$. *Then* $\hat{\mathfrak{m}}_{\mathrm{loc}}$ *is an abelian Lie subalgebra and furthermore, for a single qubit system we have*

$$\hat{\mathfrak{i}} = \hat{\mathfrak{m}}_{\mathrm{loc}} = \hat{\mathfrak{m}}_{\mathrm{qt}} = \hat{\mathfrak{m}}\ . \tag{2.74}$$

*Proof.* For a single qubit system the equality $\hat{\mathfrak{i}} = \hat{\mathfrak{m}}_{\mathrm{loc}} = \hat{\mathfrak{m}}_{\mathrm{qt}} = \hat{\mathfrak{m}}$ follows immediately from the fact that no Pauli matrices commute and hence there are no quasi-translations.

Now we consider the multi-qubit case. Local quasi-translations on separate qubits commute. We need to only consider the case of two local quasi-translation on the same qubit. Then on a fixed qubit $k$ and for all $r_1, r_2 \in \mathbb{I}_0$, we get that $\mathcal{T}_{r_1,k}\mathcal{T}_{r_2,k} = (\hat{\sigma}_{p_1,k}\hat{\sigma}^+_{q_1,k}\hat{\sigma}_{p_2,k})\hat{\sigma}^+_{q_2,k}$. Now if $p_2 = p_1$ we get that $(\hat{\sigma}_{p_1,k}\hat{\sigma}^+_{q_1,k}\hat{\sigma}_{p_1,k}) = 0$ by the proof of Proposition 2.3.1. If $p_2 = q_1$ then $(\hat{\sigma}_{p_1,k}\hat{\sigma}^+_{q_1,k}\hat{\sigma}_{q_1,k}) = 0$ since by direct computation one sees that $\sigma^+_{q_1,k}\hat{\sigma}_{q_1,k} = 0$. Finally, if $p_2 \neq p_1$ and $p_2 \neq q_1$ then $\sigma_{p_1,k}\sigma_{q_1,k} = \mathrm{i}\varepsilon_{p_1q_1p_2}\sigma_{p_2,k}$ and the direct computation shows that again $(\hat{\sigma}_{p_1,k}\hat{\sigma}^+_{q_1,k}\hat{\sigma}_{p_2,k}) = 0$ and hence $[\mathcal{T}_{r_1,k}, \mathcal{T}_{r_2,k}] = 0$. □

**Remark 8.** *Consider* $n$ *local quasi-translations* $\mathcal{T}_{m_1,1}, \mathcal{T}_{m_2,2}, \ldots, \mathcal{T}_{m_n,n}$ *each acting on different qubits. Then*

$$\prod_{k=1}^{n} \mathcal{T}_{m_k,k} \in \hat{\mathfrak{i}}\ , \tag{2.75}$$

*where* $\hat{\mathfrak{i}}$ *is now the* $n$*-qubit LK-ideal. This is a generalized statement of Lemma 1.3.6 that for* $n$ *translation operators acting on* $n$ *separate single qubit systems, taking the tensor product* $(n-1)$ *times we have that* $\mathcal{T} \otimes \mathcal{T} \otimes \cdots \otimes \mathcal{T}$ *is again an infinitesimal translation but now acting on* $\mathfrak{her}(2^n)$*. See Lemma B.0.2 in Appendix B for additional details.*

With these new parameterisations, we then arrive at two important Corollaries of Theorem 5. The first will be of fundamental importance in this thesis and is a straightforward consequence of the star-product defined above.

**Corollary 2.4.1.** *For any quasi-translation operator* $\mathrm{i}\hat{\sigma}_{\boldsymbol{p}}\hat{\sigma}^+_{\boldsymbol{q}} \in \hat{\mathfrak{m}}_{\mathrm{qt}}$ *define*

$$\mathcal{T}_{\mathbf{m}} := \chi(\mathrm{i}\hat{\sigma}_{\boldsymbol{q}}\hat{\sigma}^+_{\boldsymbol{p}})\ , \quad \textit{where} \quad \boldsymbol{m} = \boldsymbol{p} \star \boldsymbol{q}\ . \tag{2.76}$$

*Then the* $4^n - 1$ *dimensional* $n$*-qubit LK-ideal is given by*

$$\hat{\mathfrak{i}} = \left\langle \mathcal{T}_{\mathbf{m}} \mid \mathbf{m} \in \mathbb{I}_0^n \right\rangle\ . \tag{2.77}$$

Corollary 2.4.1 is powerful since it provides the direct connection between the quasi-translation elements $\mathrm{i}\hat{\sigma}_{\mathbf{q}}\hat{\sigma}_{\mathbf{p}} \in \hat{\mathfrak{m}}_{\mathrm{qt}}$ which are visible in the Lindblad generators of non-unital noise, and the associated translation operator along the "direction" $\mathbf{m}$. In light of Remark 8, we can give an alternative representation of the translation elements $\mathcal{T}_{\mathbf{m}}$ given by Eqn. (2.76) in terms of *local* quasi-translations $\mathcal{T}_{r,k} \in \hat{\mathfrak{m}}_{\mathrm{loc}}$ as defined by Eqn. (2.73). In a sense, this is a finer decomposition since $\hat{\mathfrak{m}}_{\mathrm{loc}} \subseteq \hat{\mathfrak{m}}_{\mathrm{qt}}$.

Recall that $\mathbf{m} \in \mathbb{I}_0^n$ is expressed as $\mathbf{m} = (m_1, m_2, \ldots, m_n)$ where $m_k \in \{x, y, z, 1\}$ such that $\mathbf{m} \neq (1, 1, \ldots, 1)$.

**Corollary 2.4.2.** *For a fixed $\boldsymbol{m} \in \mathbb{I}_0^n$ we have the equality*

$$\tau_{\boldsymbol{m}} = \chi(\prod_k \tau_{m_k,k}) \ , \quad \textit{for} \quad \boldsymbol{m} = (m_1, m_2, \dots, m_n) \ , \tag{2.78}$$

*where $k$ varies over the index numbers of $\boldsymbol{m}$ which have $m_k \neq 1$. Furthermore, for a fixed $\boldsymbol{m} \in \mathbb{I}_0^n$ which has no $k^{th}$ element $m_k$ equal to one, Eqn. (2.78) simplifies to*

$$\tau_{\boldsymbol{m}} = \prod_{k=1}^{n} \tau_{m_k,k} \ , \quad \textit{for} \quad \boldsymbol{m} = (m_1, m_2, \dots, m_n) \ . \tag{2.79}$$

*Proof.* The proof is simpler in operator representation and thus is relegated to Appendix B. □

**Example 3.** *Consider the quasi-translation operator for a three-qubit system given by $\mathrm{i}\hat{\sigma}_{\boldsymbol{q}}\hat{\sigma}_{\boldsymbol{p}}^{+} \in \hat{\mathfrak{m}}_{\mathrm{qt}}$, where $\boldsymbol{p} = (1, x, z)$ and $\boldsymbol{q} = (x, x, y)$. Then $\boldsymbol{p} \star \boldsymbol{q} = \boldsymbol{m} = (x, 1, -x)$ and therefore*

$$-\tau_{x1x} = \chi(\mathrm{i}\hat{\sigma}_{\boldsymbol{q}}\hat{\sigma}_{\boldsymbol{p}}^{+}) = -\chi(\tau_{x,1}\tau_{x,3}) \ , \tag{2.80}$$

*Furthermore, if instead $\boldsymbol{p} = (z, 1, 1)$, and $\boldsymbol{q} = (y, 1, x)$, then we obtain $\boldsymbol{p} \star \boldsymbol{q} = \boldsymbol{m} = (-x, 1, x)$ and hence again $-\tau_{x1x} = \chi(\mathrm{i}\hat{\sigma}_{\boldsymbol{q}}\hat{\sigma}_{\boldsymbol{p}}^{+}) = -\chi(\tau_{x,1}\tau_{x,3})$. Thus, there exist multiple operators in $\hat{\mathfrak{m}}_{\mathrm{qt}}$ which, post-projection, give the same basis element of the LK-ideal $\hat{\mathfrak{i}}$.*

Now we give the number of degenerate choices of the pairs $(\mathbf{p}, \mathbf{q}) \in \mathbb{I}_0^n \times \mathbb{I}_0^n$ which give a single $\mathbf{m} \in \mathbb{I}_0^n$.

**Lemma 2.4.1.** *Define the set of pairs of $\boldsymbol{p}, \boldsymbol{q} \in I_0^n$ which under the star-product give the same $\boldsymbol{m} \in \mathbb{I}_0^n$ element as*

$$S_{\boldsymbol{m}} := \{(\boldsymbol{p}, \boldsymbol{q}) \in \mathbb{I}_0^n \times \mathbb{I}_0^n \mid \boldsymbol{p} \star \boldsymbol{q} = \boldsymbol{m} \neq \boldsymbol{q} \star \boldsymbol{p}\} \ . \tag{2.81}$$

*Then $|S_{\boldsymbol{m}}| = 4^{n-1}$.*

*Proof.* We prove by induction. For $n = 1$, its clear that $|S_{\mathbf{m}}| = 1$ for $\mathbf{m} \in \mathbb{I}_0$ since $\sigma_p \sigma_q = \mathrm{i}\varepsilon_{pqr}\sigma_r$. Now assume that for general $n$, (and hence $\mathbf{m} \in \mathbb{I}_0^n$), $|S_{\mathbf{m}}| = 4^{n-1}$. Therefore for $n' = n + 1$ ( and $\mathbf{m}' \in \mathbb{I}_0^{n+1}$) we want to show that $|S_{\mathbf{m}'}| = 4^n$. Without loss of generality we choose $\mathbf{m}' \in \mathbb{I}_0^{n+1}$ such that $\sigma_{\mathbf{m}'} = \sigma_{\mathbf{m}} \otimes \mathbb{1}$ where $\mathbf{m} \in \mathbb{I}_0^n$. By assumption, there are $4^{n-1}$ pairs $(\mathbf{p}, \mathbf{q}) \in \mathbb{I}_0^n \times \mathbb{I}_0^n$ such that $\sigma_{\mathbf{p}}\sigma_{\mathbf{q}} = \mathrm{i}\sigma_{\mathbf{m}}$ and clearly for any $\sigma_p \in \{\mathbb{1}, \sigma_x, \sigma_y, \sigma_z\}$ we have that $(\sigma_{\mathbf{p}} \otimes \sigma_p)(\sigma_{\mathbf{q}} \otimes \sigma_p) = \mathrm{i}\sigma_{\mathbf{m}} \otimes \mathbb{1}$ and hence $|S_{\mathbf{m}'}| = 4^{n-1} \cdot 4 = 4^n$. □

Section 3.3 in Chapter 3 applies this degeneracy of associated quasi-translation terms (for a fixed translation) to fixed point engineering. Namely, we show that for a fixed translation direction $\tau_{\mathbf{m}}$, the degeneracy in its conversion to the multiple quasi-translation operators $\mathrm{i}\hat{\sigma}_{\mathbf{q}}\hat{\sigma}_{\mathbf{p}}^{+} \in \hat{\mathfrak{m}}_{\mathrm{qt}}$ such that $\mathbf{m} = \mathbf{p} \star \mathbf{q}$ results in the fact that the fixed point sets of different Lindblad generators with $V = \frac{1}{2}(\sigma_{\mathbf{p}} + \mathrm{i}\sigma_{\mathbf{q}})$ *are all equivalent.* Thereby showing that although the unital part of $\Gamma$ differs between different realisations of $V$ in terms of $\sigma_{\mathbf{p}}, \sigma_{\mathbf{q}}$ Pauli matrices, the fact that the associated translation direction is the same implies equality of the fixed point sets. We then expand on this notion by constructing sets of Lindblad terms which overall result in specific translation directions to obtain unique target fixed points.

# Chapter 3

# Purely Dissipative State Engineering

## 3.1 Introduction

This chapter will focus on various aspects relating to fixed points of Markovian semigroups of quantum channels which are generated purely dissipative Lindblad-Kossakowski operators. We prove a variety of results which allow us to provide a new connection which relates the geometric interpretation of a Lindblad-Kossakowski operator and the corresponding fixed points of the Markovian semigroup it generates. Explicitly, this geometric interpretation is a result of the ideal structure introduced in Section 2.4 as we can consider the elements of the abelian ideal of the Lindblad-Kossakowski Lie algebra (cf. Theorem 5) as *translation* directions. These operators describe the overall affine shifts the initial state undergoes as it evolves towards the fixed point state(s) of the systems dynamics. Using this intuition, we are able to provide several new insights and results to this very active current field of research.

In Section 3.2 we introduce the general theory of fixed points of these Markovian semigroups. We first establish important notions and concepts such as invariant subspaces of the underlying Hilbert space, special forms of a generator which is restricted to such subspaces and necessary and sufficient conditions for when these subspaces support fixed points (cf. Proposition 3.2.1, Theorem 6 and Corollary 3.2.2). The core of these results can be found both implicitly and explicitly in the existing literature [49, 32, 10, 8, 9, 50, 45, 51, 48, 52]. With these insights, we provide a complete classification of subspaces which support pure state fixed points - so-called "generalised " dark state spaces - and are those which are a direct generalisation of the dark state spaces defined in [32]. We then provide a unique decomposition of the support space of the fixed point set which shows that fixed points belong to two distinct sets whose supports are orthogonal to one another - the vector space composed of generalised dark state spaces and that which supports "intrinsic" higher rank fixed points (cf. Theorem 7). The latter are a class we are seemingly the first to consider in this manner and are those which cannot be decomposed into pure states which themselves are fixed points. The Hilbert space decomposition implied by the fixed point decomposition is of a similar form to that considered in [9], but here the splitting we use has an immediate consequence in simplifying the problem of engineering arbitrary pure or mixed state fixed points. That is, by showing that there are two distinct classes of fixed points - generalised dark state fixed points and intrinsic higher rank fixed points - each of which have orthogonal sup-port to one another, we can provide several useful results showing when either of these classes do or do not exist, and most importantly, when the fixed point is unique (cf. Corollary 3.2.3, Propositions 3.2.2, 3.2.3, 3.2.4, and 3.2.5).

In Section 3.3 we apply the previous results to purely dissipative systems which undergo a general class of noise processes - those which have Lindblad terms $\{V_k\}$ which are of *Canonical Form* (cf. Eqn. (3.36)). These are matrices which are generalisations of the usual atomic raising and lowering operators $\sigma^+$ and $\sigma^-$. We prove that the fixed point sets of noise processes which are described by Lindblad terms of this form are simple to determine and simplify many of the general results of Section 3.2. In particular, Theorem 8 provides a necessary and sufficient condition for when any given pure state fixed point is the systems unique fixed point. We finish this section by providing a general construction scheme which relates the centraliser of the target pure state fixed point to a Lindblad-Kossakowski operator which generates a Markovian semigroup of quantum channels with the target state as its unique fixed point (cf. Algorithm 1).

In Section 3.4 we then apply all the previous results by providing illustrative examples of how to obtain several states which are useful in quantum information processing. Namely, we make use of Algorithm 1 and show how one can obtain a plethora of purely dissipative noise processes which drive the system uniquely to: the $n$-qubit GHZ state, W State, Stabiliser/Graph States and the Toric Code Subspace. As an outlook, we show how to obtain the 4-qubit Symmetric Dicke State and outline how to generalise the procedure to $n$-qubits. We compare and contrast our findings to those existing in the current literature and show that many of the existing techniques are but special cases of this general construction.

Finally, in Section 3.5 we show that using a slight generalisation of the Lindblad terms in canonical form presented in Section 3.3 and used in Section 3.4, we are able to engineer dissipative processes which have a unique *mixed* state fixed point. These mixed state fixed points are exactly the "intrinsic" higher rank fixed points, the second and last class of fixed points categorised in Section 3.2. We then provide several sufficient conditions for obtaining any diagonal mixed state (with its non-zero eigenvalues non-degenerate) fixed point as the unique fixed point of the system (see for instance Theorems 9 and 10). By providing a scheme to obtain entire classes of Lindblad-Kossakowski operators which have these diagonal states as the systems unique fixed point, we provide a *constructive* method to obtain any target state (which has its non-zero eigenvalues non-degenerate) as a systems *unique* fixed point (cf. Theorem 11). We conclude the section with several non-trivial examples and an algorithm which provides the Lindblad term construction to obtain these diagonal mixed state fixed points as the unique fixed points.

## 3.2 Fixed Point Sets and Invariant Subspaces

In Section 1.3 we rigorously discussed the basic concepts and notions of a Lindblad-Kossakowski operator (henceforth referred to as a LK-operator)

$$\mathcal{L} := \mathrm{i}\,\mathrm{ad}_{\mathrm{H}} + \Gamma \qquad \text{with} \qquad \Gamma(\rho) = \tfrac{1}{2}\sum_{k=1}^{m}\left(V_k^\dagger V_k\rho + \rho V_k^\dagger V_k - 2V_k\rho V_k^\dagger\right), \tag{3.1}$$

which is the infinitesimal generator of a completely positive semigroup of linear operators $t \mapsto \Phi(t)$, $t \geq 0$. We called a LK-operator "purely dissipative " whenever it induced *no* Hamiltonian dynamics on the system which geometrically can be described as

$$\langle \mathcal{L}, \mathrm{ad}_H \rangle = 0\ , \qquad \text{for all} \qquad H \in \mathfrak{su}(2^n)\ . \tag{3.2}$$

Furthermore Lemma 1.3.1 provided the sufficient condition that $\Gamma$ given in Eqn. (3.1) is itself purely dissipative if each of the Lindblad terms are traceless i.e. $V_1, V_2, \dots, V_m \in \mathfrak{sl}(N, \mathbb{C})$. Proposition 1.3.1 then strengthened this condition by proving that $\Gamma$ is purely dissipative *if and only if* the vector $\alpha := (\operatorname{tr} V_1, \operatorname{tr} V_2, \dots, \operatorname{tr} V_m)$ is contained in the kernel of the operator $\kappa$, where $\kappa(\alpha) := \frac{\mathrm{i}}{2} \sum_{k=1}^{m} \left( \alpha_k V_k^\dagger - \overline{\alpha}_k V_k \right)$. By Theorem 3, we can decompose the LK-operator uniquely into a purely dissipative Lindblad term $\Gamma$ and "intrinsic " Hamiltonian term via

$$\mathcal{L} = \mathrm{i}\ \mathrm{ad}_{\mathrm{H}_0} + \Gamma \ , \qquad \text{where} \qquad H_0 \in \mathfrak{her}_0(N). \tag{3.3}$$

With these preliminaries out of the way we can discuss concepts relating to the fixed point set of a Markovian semigroup of quantum channels generated by a purely dissipative LK-operator. Since a purely dissipative LK-operator is equivalent to a purely dissipative Lindblad generator i.e. $\mathcal{L} = \Gamma$, we will often say that $\Gamma$ is a purely dissipative LK-operator.

We express a Lindblad generator $\Gamma$ which is associated to multiple Lindblad terms $\{V_k\}$ as

$$\Gamma = \sum_k \Gamma_{V_k} \ , \tag{3.4}$$

where $\Gamma_{V_k}$ is the Lindblad generator corresponding to a single Lindblad term $V_k$. Let $\Phi_\Gamma(t) = e^{t\Gamma}$ be the Markovian semigroup of quantum channels generated by $\Gamma$ and define its fixed point set to be

$$\mathcal{F}(\Phi_\Gamma) \ := \ \{\rho \in \mathfrak{pos}_1(N) \mid \Phi_\Gamma(t)\rho = \rho \text{ for all } t \geq 0\} = \ker(\Gamma) \cap \mathfrak{pos}_1(N) \ . \tag{3.5}$$

In a slight abuse of language, we will often say that $\rho$ is a fixed point of $\Gamma$.

An important concept we will make use of is the support of a density matrix. Let $\rho \in \mathfrak{her}(\mathcal{H})$ be any Hermitian operator acting on $\mathcal{H}$. Its range is called support and denoted by $\operatorname{supp} \rho$. Since $\rho = \sum_k \lambda_k P_k$ can be uniquely written as a linear combination of orthogonal projections $P_k$ having mutually orthogonal ranges $\operatorname{Im} P_k$, one has the equality $\operatorname{supp} \rho = \bigoplus_k \operatorname{Im} P_k$. Furthermore, let $\mathcal{S} \subset \mathcal{H}$ be any subspace of $\mathcal{H}$ and let $\rho$ be a density operator with support in $\mathcal{S}$. Since $\rho$ can be uniquely identified with an element in $\mathfrak{her}(\mathcal{S})$, in the following we do not distinguish between $\mathfrak{her}(\mathcal{S})$ and the set of all Hermitian operators acting on $\mathcal{H}$ but being supported in $\mathcal{S}$. Certainly, the same applies to $\mathfrak{pos}_1(\mathcal{S})$ and the set of all density operators acting on $\mathcal{H}$ but being supported in $\mathcal{S}$.

We also define a reduced Lindblad generator as $\Gamma\big|_{\mathcal{S}} : \mathfrak{her}(\mathcal{S}) \to \mathfrak{her}(\mathcal{S})$ which is given by

$$\Gamma\big|_{\mathcal{S}}(\rho) := P_{\mathcal{S}} \circ \Gamma(\rho) \circ P_{\mathcal{S}} \ , \tag{3.6}$$

where $P_{\mathcal{S}}$ denotes the orthogonal projection onto $\mathcal{S}$. In particular, if $\mathcal{S}$ is equal to the span of the first $r$ standard basis vectors of $\mathbb{C}^n$ then there exists basis in which $\rho$ and $\{V_k\}$ have block decompositions (by choosing an appropriate unitary transformation) of the form

$$V_k := \begin{bmatrix} A_k & B_k \\ C_k & D_k \end{bmatrix} \quad \text{and} \quad \rho := \begin{bmatrix} \rho_A & 0 \\ 0 & 0 \end{bmatrix} \ , \tag{3.7}$$

where $\rho_A$ is an $r \times r$ full rank matrix and the reduced $\Gamma\big|_{\mathcal{S}}$ is then given by

$$\Gamma\big|_{\mathcal{S}}(\rho) \ = \ \sum_k \Gamma_{A_k}(\rho_A) - \tfrac{1}{2}(C_k^\dagger C_k \rho_A + \rho_A C_k^\dagger C_k) \ , \tag{3.8}$$

where $\Gamma_{A_k}$ is the Lindblad generator with the single Lindblad term $A_k$.

With these introductory concepts established, we first consider the relationship between invariant subspaces of the Lindblad generator, the corresponding Markovian semigroup of quantum channels and the Lindblad terms $\{V_k\}$ which are associated to each generator. Let $\mathcal{S} \subseteq \mathcal{H} = \mathbb{C}^{2^n}$ be a subspace of the underlying Hilbert space. We say that $\mathcal{S}$ is an invariant subspace of a $2^n \times 2^n$ matrix $A$ if $Av \in \mathcal{S}$ for all $v \in \mathcal{S}$ (and hence $A\mathcal{S} \subseteq \mathcal{S}$). Clearly then, the span of an eigenvector of the matrix $A$ is a one dimensional invariant subspace and the range of the matrix $A$ is another example of an invariant subspace of $A$. We then have the following Proposition.

**Proposition 3.2.1.** *Let $\Gamma$ be a (purely dissipative) Lindblad-Kossakowski operator with Lindblad terms $\{V_k\}$ and let $\mathcal{S}$ be a subspace of $\mathcal{H}$. Then the following statements are equivalent:*

1. *One has the inclusions*

$$V_k\mathcal{S} \subseteq \mathcal{S} \quad \text{for all } k \text{ and } \quad \sum_k V_k^\dagger V_k \mathcal{S} \subseteq \mathcal{S}\,. \tag{3.9}$$

2. *$\mathfrak{her}(\mathcal{S})$ is an invariant subspace of $\Gamma$.*

3. *$\mathfrak{pos}_1(\mathcal{S})$ are invariant under $\Phi_\Gamma(t)$, $t \geq 0$.*

*Proof.* (1) $\Longrightarrow$ (2): For any $M \in \mathfrak{her}(\mathcal{S})$ we have that range$(V_kM) \subseteq \mathcal{S}$, range$(MV_k^\dagger) \subseteq \mathcal{S}$, range$(\sum_k V_k^\dagger V_k M) \subseteq \mathcal{S}$, and range$(M\sum_k V_k^\dagger V_k) \subseteq \mathcal{S}$ for all $k$ so clearly $\mathfrak{her}(\mathcal{S})$ is an invariant subspace of $\Gamma$.

(2) $\Longrightarrow$ (1): If $\mathcal{S} = \mathcal{H}$ the implication is trivial so therefore we assume that $\dim\mathcal{S} < \dim\mathcal{H}$. For any matrix $M \in \mathfrak{her}(\mathcal{S})$, there exists a block decomposition as in Eqn. (3.7) of each $V_k$ and $M$ such that $M = \mathrm{diag}(M_A, 0)$, where $M_A$ has support in $\mathcal{S}$ and is full rank. Now if supp$(\Gamma(M)) \subseteq \mathcal{S}$, then this imposes the off-diagonal block and lower diagonal block conditions 1) $\sum_k M_A(A_k^\dagger B_k + C_k^\dagger D_k) - 2(A_k M_A C_k^\dagger) = 0$ and 2) $\sum_k C_k M_A C_k^\dagger = 0$. Since $M_A$ is a rank $r$, $r\times r$ matrix then $\sum_k C_k M_A C_k^\dagger = 0$ if and only if $C_k = 0$ for all $k$ which implies $V_k\mathcal{S} \subseteq \mathcal{S}$ for all $k$. Substituting $C_k = 0$ for all $k$ into condition 1), we obtain the condition $M_A(\sum_k A_k^\dagger B_k) = 0$ but since $M_A$ has full block rank then $\sum_k A_k^\dagger B_k = 0$. Using the relation

$$\sum_k V_k^\dagger V_k = \begin{bmatrix} \sum_k A_k^\dagger A_k & \sum_k A_k^\dagger B_k \\ \sum_k B_k^\dagger A_k & \sum_k B_k^\dagger B_k + D_k^\dagger D_k \end{bmatrix}\,,$$

we see that $\sum_k A_k^\dagger B_k = 0$ if and only if $\sum_k V_k^\dagger V_k \mathcal{S} \subseteq \mathcal{S}$.

(2) $\Longrightarrow$ (3): Note that the power series of the exponential function is given by $e^\Gamma = \sum_k^\infty \frac{1}{k!}\Gamma^k$, for $k \in \mathbb{N}$ and since $\mathfrak{her}(\mathcal{S})$ is an invariant subspace of $\Gamma$, and hence that $\mathfrak{her}(\mathcal{S})$ is invariant under $\Phi_\Gamma(t)$ for $t \geq 0$. Moreover, since $\mathfrak{pos}_1(\mathcal{H})$ is trivially invariant under $\Phi_\Gamma(t)$ for $t \geq 0$ then we have that the intersection $\mathfrak{pos}_1(\mathcal{S}) = \mathfrak{her}(\mathcal{S}) \cap \mathfrak{pos}_1(\mathcal{H})$ is as well.

(3) $\Longrightarrow$ (2): Since $\mathfrak{pos}_1(\mathcal{S})$ is invariant under $\Phi_\Gamma(t)$, $t \geq 0$, then so is $\mathfrak{her}(\mathcal{S}) = \mathfrak{pos}_1(\mathcal{S}) - \mathfrak{pos}_1(\mathcal{S})$ and hence we have that $e^{t\Gamma}(M) \in \mathfrak{her}(\mathcal{S})$ for all $M \in \mathfrak{her}(\mathcal{S})$. The derivative of the flow at time $t = 0$ must also be invariant and hence $\frac{d}{dt}(e^{t\Gamma})\big|_{t=0}(M) = \Gamma(M) \in \mathfrak{her}(\mathcal{S})$.

□

Due to Proposition 3.2.1, a subspace $\mathcal{S} \subseteq \mathcal{H}$ is called an invariant subspace of $\Gamma$ if it satisfies one and therefore all of the above conditions.

**Remark 9.** *In general, the invariance of $\mathcal{S}$ does not imply that $S^{\perp}$ is an invariant subspace of $\Gamma$. Consider $k = 1$, $V_1 := \sigma_x + \mathrm{i}\sigma_y$ and $\mathcal{S} := \mathrm{span}_{\mathbb{C}}\{e_1\}$.*

It is a well known fact that as a standard application of Brouwer's fixed point theorem (see for example Chap. V, Prop. 22.13 in [4]) we obtain the following.

**Corollary 3.2.1.** *If $\mathcal{S} \subset \mathcal{H}$ is an invariant subspace of $\Gamma$ then $\Gamma$ (resp. $\Phi_\Gamma$) has at least one fixed point in $\mathfrak{pos}_1(\mathcal{S})$.*

Using the invariant subspaces considerations presented above, a more detailed analysis shows that one can obtain necessary and sufficient conditions for the existence of a rank $r$ fixed point. This can essentially be found implicitly in terms of attractivity of invariant subspaces in [49] but here we present it in the (new) manner which explicitly makes use of the invariant subspace inclusions given by Eqn. (3.9) in Proposition 3.2.1.

**Theorem 6.** *Let $\Gamma$ be a (purely dissipative) Lindblad-Kossakowski operator. A density operator $\rho \in \mathfrak{pos}_1(\mathcal{H})$ is a fixed point of $\Gamma$ if and only if the following conditions are satisfied:*

1. *$\mathcal{S} := \mathrm{supp}\,\rho$ is an invariant subspace of $\Gamma$.*
2. *$\rho|_{\mathcal{S}}$ is a fixed point of $\Gamma\big|_{\mathcal{S}}$.*

*Proof.* Again for full rank fixed points the result is obvious. We prove it for existence of a fixed point $\rho \in \mathfrak{pos}_1(\mathcal{H})$ with rank $r < d$. Block decomposing $\rho$ and each $V_k$ as in Eqn. (3.7), and noting that $\rho$ is a fixed point if and only if $\Gamma(\rho) = 0$, each of the 4-blocks of $\Gamma(\rho) = 0$ must be zero and therefore we obtain the necessary and sufficient conditions that

$$\sum_k \Gamma_{A_k}(\rho_A) - \tfrac{1}{2}(C_k^\dagger C_k \rho_A + \rho_A C_k^\dagger C_k) = 0$$
$$\sum_k \rho_A(A_k^\dagger B_k + C_k^\dagger D_k) - 2(A_k \rho_A C_k^\dagger) = 0$$
$$\sum_k C_k \rho_A C_k^\dagger = 0\ .$$

The second and third equations are true if and only if $V_k\mathcal{S} \subseteq \mathcal{S}$ for all $k$ and $\sum_k V_k^\dagger V_k \mathcal{S} \subseteq \mathcal{S}$ as shown in the proof of Proposition 3.2.1 which is true if and only if condition 1. holds. Moreover, the first equation is precisely the reduced Lindblad generator given in Eqn. (3.8).

□

When $\dim(\mathcal{S}) = 1$, this result gives the purely dissipative analogue of Theorem 1 of [32] as an immediate Corollary.

**Corollary 3.2.2.** *For a (purely dissipative) Lindblad-Kossakowski operator $\Gamma$, a pure state $\rho = |\psi\rangle\langle\psi|$ is a fixed point if and only if the following conditions are satisfied:*

*1. $|\psi\rangle$ is a simultaneous right eigenvector of all $V_k$, i.e.*

$$V_k|\psi\rangle = \lambda_k|\psi\rangle \quad \text{for all } k \text{ and some } \lambda_k \in \mathbb{C}. \tag{3.10}$$

*2. $|\psi\rangle$ is a left eigenvector of $V_\sum := \sum_k \overline{\lambda}_k V_k$ to the eigenvalue $\lambda_\sum := \sum_k |\lambda_k|^2$, i.e.*

$$\langle\psi|V_\sum = \langle\psi|\lambda_\sum\,. \tag{3.11}$$

**Remark 10.** *The above conditions 1. and 2. are obviously equivalent to 1. and 2'. which is given by*

*2'. $|\psi\rangle$ is a right eigenvector of $\sum_k V_k^\dagger V_k$ to the eigenvalue $\lambda_\sum := \sum_k |\lambda_k|^2$, i.e.*

$$\Big(\sum_k V_k^\dagger V_k\Big)|\psi\rangle = \lambda_\sum|\psi\rangle\,. \tag{3.12}$$

*Proof.* Theorem 6 gives necessary and sufficient conditions for existence of a fixed point of rank $r$ so here we consider the case where $r = 1$ and the subspace $\mathcal{S} = \text{span}_\mathbb{C}\{|\psi\rangle\}$. Condition (1) in Theorem 6 implies that the invariance conditions given by Eqn. (3.9) take the form of $V_k|\psi\rangle = \lambda_k|\psi\rangle$ for all $k$ and some $\lambda_k \in \mathbb{C}$ and $\sum_k V_k^\dagger V_k|\psi\rangle = \sum_k \lambda_k V_k^\dagger|\psi\rangle = \tilde{\lambda}|\psi\rangle$ and $\tilde{\lambda} \in \mathbb{C}$. Right multiplying this equation by $\langle\psi|$ we get that $\sum_k |\lambda_k|^2 = \tilde{\lambda}$. Note that the second condition in Theorem 6 becomes trivial since for a rank one fixed point, the same block decomposition of $\rho$ used in the proof shows that $\rho_A = 1$ and thus the reduced Lindblad term is trivial. □

Let $\Gamma$ be a (purely dissipative) Lindblad-Kossakowski operator. For $\Lambda := (\lambda_1, \dots, \lambda_m) \in \mathbb{C}^m$, we define the set $\mathcal{D}_\Lambda$ as follows

$$\mathcal{D}_\Lambda := \Big\{|\psi\rangle \in \mathcal{H} \;\Big|\; V_k|\psi\rangle = \lambda_k|\psi\rangle \text{ for all } k = 1, \dots, m \text{ and } \langle\psi|V_\sum = \langle\psi|\lambda_\sum\Big\}\,, \tag{3.13}$$

where $V_\sum$ and $\lambda_\sum$ are defined as in Corollary 3.2.2. Moreover, we define the *complete dark space* as

$$\mathcal{D} := \bigoplus_{\Lambda\in\mathbb{C}^m} \mathcal{D}_\Lambda\,. \tag{3.14}$$

Note that for almost all $\Lambda$ one has $\mathcal{D}_\Lambda = \{0\}$. But, whenever $\mathcal{D}_\Lambda \neq \{0\}$ it is called a *generalized dark space* of $\Gamma$. Therefore, every pure state fixed point gives rise to a generalized dark space. Conversely, if $\mathcal{D}_\Lambda$ is a generalized dark space, then any density operator in $\mathfrak{pos}_1(\mathcal{D}_\Lambda)$ is a fixed point of $\Lambda$.

The following result provides a special decomposition of any fixed point into two distinct types of fixed points. Those whose supports are contained within the generalised darks state spaces - and therefore can be decomposed into a convex combination of pure state fixed points - and, "intrinsic" higher rank fixed points which are those whose supports are orthogonal to the generalised dark state space. This splitting provides an orthogonal decomposition of the total Hilbert space subspace which supports all the systems fixed points. The details regarding these types of subspace splittings were thoroughly worked out in [8, 9] and discussed relative to fixed point engineering in [45]. In Theorem 3 of [9] the authors prove that the Hilbert space subspace which consists of the supports of all the systems fixed points can be represented as an orthogonal decomposition where each subspace supports one and only one fixed point. Their Theorem 7 then provides a further explicit decomposition of this support space and they prove that any arbitrary fixed point of the system can be expressed as a special linear combination of fixed points whose individual supports are related to the special subspace decomposition.

Here we present an alternate formulation of these two results by using the notions of generalised dark state spaces and intrinsic higher rank fixed points - thus providing a different overall decomposition of the total fixed point support space which is more suited for engineering arbitrary pure and mixed state fixed points.

**Theorem 7.** *Let $\Gamma$ be a (purely dissipative) Lindblad-Kossakowski operator and let $\rho \in \mathfrak{pos}_1(\mathcal{H})$ be any fixed point of $\Gamma$. Moreover, let $\mathcal{D}_{\Lambda_1}, \ldots, \mathcal{D}_{\Lambda_r}$ be the generalized dark spaces of $\Gamma$. Then the following assertions are fulfilled:*

1. *$\mathcal{D}_{\Lambda_1}, \ldots, \mathcal{D}_{\Lambda_r}$ are mutually orthogonal.*
2. *$\rho$ can be decomposed as a convex combination of the form*

$$\rho = \alpha_0 \rho_0 + \sum_{j=1}^{r} \alpha_j \sigma_j, \tag{3.15}$$

*where $\rho_0, \sigma_1, \ldots, \sigma_r \in \mathfrak{pos}_1(\mathcal{H})$ satify the conditions:*

*(a) For $j = 1, \ldots, r$, the inclusion $\operatorname{supp} \sigma_j \subseteq \mathcal{D}_{\Lambda_j}$ holds and hence each $\sigma_j$ is a convex combination of rank-1 fixed points and thus itself a fixed point of $\Gamma$.*

*(b) The operator $\rho_0$ is an intrinsic higher rank fixed point of $\Gamma$ with $\operatorname{supp} \rho_0 \perp \mathcal{D}$, i.e. $\rho_0$ cannot be decomposed as a convex combination of a rank-1 fixed point and another fixed point $\rho_0'$.*

The above decomposition then immediately yields the generalisation of Theorem 2 in [32].

**Corollary 3.2.3.** *Let $\mathcal{D}_\Lambda$ be a generalized dark space of $\Gamma$. If there exists no subspace $\mathcal{S} \subseteq \mathcal{H}$ such that $\mathcal{S} \perp \mathcal{D}_\Lambda$ and $V_k \mathcal{S} \subseteq \mathcal{S}$ for all $k$, then the only fixed points of $\Gamma$ are those with support in $\mathcal{D}_\Lambda$.*

*Proof.* This is a direct consequence of Eqn. (3.15) and Theorem 6. □

For the proof of Theorem 7 we need two technical Lemmas.

**Lemma 3.2.1.** *Let $\rho$ be a fixed point and let $\mathcal{D}_\Lambda$ be an arbitrary generalized dark space. Moreover, assume $\dim\big(\operatorname{supp}(\rho) \cap \mathcal{D}_\Lambda\big) = r \geq 1$. Then there exists fixed points $\sigma$ and $\rho'$ such that $\operatorname{supp}(\sigma) \subset \mathcal{D}_\Lambda$ and $\operatorname{supp}(\rho') \perp \mathcal{D}_\Lambda$ (and hence $\operatorname{supp}(\rho') \perp \operatorname{supp}(\sigma)$) such that $\rho = \mu\sigma + (1-\mu)\rho'$ with $0 \leq \mu \leq 1$.*

*Proof.* Assume w.l.o.g. that $\rho$ has rank $r'$ and that $\operatorname{supp}(\rho) \cap \mathcal{D}_\Lambda = \operatorname{span}_{\mathbb{C}} \{e_1, \ldots, e_r\}$ such that $\sigma_k := e_k e_k^\dagger$ are generalised dark state fixed points. By a change of basis, we obtain the block decompositions in Eqn. (3.7) of each $V_k$ and $\rho$, where $C_k = 0$ for all $k$. By assumption, since $e_k \in \operatorname{supp}(\rho)$ and $\sigma_k = e_k e_k^\dagger$ are pure state fixed points such that $V_k e_k = \lambda_k e_k$ for all $k$, we can further decompose each $A_k$ block and $\rho_A$ as

$$A_k := \begin{bmatrix} \lambda_k I_r & B_k' \\ 0 & D_k' \end{bmatrix} \quad \text{and} \quad \rho_A := \begin{bmatrix} \rho_{11} & \rho_{12} \\ \rho_{12}^\dagger & \rho_{22} \end{bmatrix}, \tag{3.16}$$

and thus we have to show that $\rho_{12} = 0$. By Theorem 6, the reduced $\Gamma$ has $\rho_A$ as a full rank fixed point and hence $\sum_k \Gamma_{A_k}(\rho_A) = 0$ which implies each of the 4 blocks of $\sum_k \Gamma_{A_k}(\rho_A) = 0$ must be zero. For the upper-left block, the condition is then given by

$$\sum_k -\overline{\lambda}_k B'_k \rho_{12}^\dagger - \lambda_k \rho_{12} B_k^{'\dagger} - 2B'_k \rho_{22} B_k^{'\dagger} = 0 \ . \tag{3.17}$$

Since $|\psi\rangle$ is a pure state fixed point, the conditions in Corollary 3.2.2 imply that $\langle\psi| \sum_k \overline{\lambda}_k A_k = \langle\psi| \sum_k |\lambda_k|^2 = (\sum_k |\lambda_k|^2, 0)$. Also note that $\langle\psi| \sum_k \overline{\lambda}_k A_k = (\sum_k |\lambda_k|^2, \sum_k \overline{\lambda}_k B'_k)$ for $|\psi\rangle = e_i$ for $i = 1, \ldots, r$, which implies that $\sum_k \overline{\lambda}_k B'_k = 0$. Therefore Eqn. (3.17) reduces to

$$\sum_k B'_k \rho_{22} B_k^{'\dagger} = 0 \ , \tag{3.18}$$

and since $\rho_A$ is full rank then so is $\rho_{22}$ which implies that $B'_k = 0$ for all $k$. Using this fact then the upper right block of $\sum_k \Gamma_{A_k}(\rho_A) = 0$ yields

$$0 = \sum_k 2\lambda_k \rho_{12} D_k'^\dagger - |\lambda_k|^2 \rho_{12} - \rho_{12} D_k'^\dagger D'_k \ , \tag{3.19}$$

right multiplying by $\rho_{12}^\dagger$, and taking the trace gives

$$0 = \sum_k 2\,\mathrm{Re}\,\mathrm{tr}\left(\lambda_k \rho_{12} D_k'^\dagger \rho_{12}^\dagger\right) - |\lambda_k|^2 \|\rho_{12}^\dagger\|^2 - \|D'_k \rho_{12}^\dagger\|^2 = -\sum_k \|\lambda_k \rho_{12}^\dagger - D'_k \rho_{12}^\dagger\|^2 \tag{3.20}$$

Thus one has $D'_k \rho_{12}^\dagger = \lambda_k \rho_{12}^\dagger$ for all $k$. Now we see that each $j^{th}$ row of the (rectangular) matrix $\rho_{12}$ is a generalised dark state of the same $\Lambda$-type as $|\psi\rangle$ and therefore each row must be in $\mathrm{span}_{\mathbb{C}}\{e_1, \ldots, e_r\}$. However, this is only possible when each row consists of zeros and hence the entire block matrix $\rho_{12}$ is zero. This finally implies that the matrix $\sigma := \mathrm{diag}(\rho_{11}, 0)$ is a fixed point with support in $\mathcal{D}_\Lambda$. Furthermore, defining $\rho' = \mathrm{diag}(0, \rho_{22})$ we obtain $\Gamma(\rho) = \mu\Gamma(\sigma) + (1-\mu)\Gamma(\rho') = \Gamma(\rho') = 0$ for $0 \le \mu \le 1$ and hence $\rho'$ is again a fixed point such that $\mathrm{supp}(\rho') \perp \mathcal{D}_\Lambda$. □

This result provides an immediate Corollary which is interesting in its own right.

**Corollary 3.2.4.** *Let $\sigma = |\psi\rangle\langle\psi|$ be a pure state fixed point such that $|\psi\rangle \in D_\Lambda$. Moreover, let $\rho$ be any other fixed point of $\Gamma$ such that $\mathrm{supp}(\rho) \cap \mathcal{D}_\Lambda$ is spanned by $|\psi\rangle$. Then $|\psi\rangle$ is an eigenvector of $\rho$.*

*Proof.* Without loss of generality, assume that the pure state fixed point is given by $\sigma = e_1 e_1^\dagger$ and that $\rho$ is rank $r$ with $1 \le r \le 2^n$. Then we have that $\mathrm{supp}(\rho) \cap \mathcal{D}_\Lambda = \mathrm{span}_{\mathbb{C}}\{e_1\}$ and hence by Lemma 3.2.1, $\rho = e_1 e_1^\dagger + \rho'$ (w.l.o.g neglecting the scalar coefficients) such that $\mathrm{supp}(\rho') \perp \mathrm{span}_{\mathbb{C}}\{e_1\}$. Therefore $\rho e_1 = \lambda e_1$ for $\lambda \in \mathbb{R}$ since $e_1^\dagger e_1 = 1$ and $\rho' e_1 = 0$. □

**Lemma 3.2.2.** *Let $\sigma = |\psi\rangle \in D_\Lambda$ correspond to a pure state fixed point and let $\rho$ be any other fixed point of $\Gamma$. Moreover, let $P_{\mathcal{S}}$ and $P_{\mathcal{S}^\perp}$ be the orthogonal projectors onto $\mathcal{S} := \mathrm{supp}(\rho)$ and $\mathcal{S}^\perp$. Then each of the vectors*

$$|\psi_1\rangle := P_{\mathcal{S}}|\psi\rangle \quad \textit{and} \quad |\psi_2\rangle := P_{\mathcal{S}^\perp}|\psi\rangle \tag{3.21}$$

*is either zero or gives rise to a generalised dark state in $D_\Lambda$.*

*Proof.* As usual, for $\rho$ being rank $r$ with $1 \le r \le 2^n$, we can block decompose the Lindblad terms and $\rho$ as in Eqn. (3.7) and we can decompose $|\psi\rangle = |\psi_1\rangle + |\psi_2\rangle$, where $|\psi_1\rangle := P_{\mathcal{S}}|\psi\rangle = (|\psi_1'\rangle, 0)^\top$ and $|\psi_2\rangle := P_{\mathcal{S}^\perp}|\psi\rangle = (0, |\psi_2'\rangle)^\top$. Furthermore, by Theorem 6, we know that $\sum_k V_k^\dagger V_k \mathcal{S} \subseteq \mathcal{S}$ for $\mathcal{S} = \operatorname{supp}(\rho)$ and hence $\sum_k A_k^\dagger B_k = 0$. Corollary 3.2.2 implies that $V_k|\psi\rangle = \lambda_k|\psi\rangle$ for all $k$ and $\sum_k V_k^\dagger V_k|\psi\rangle = \lambda_\sum|\psi\rangle$, where $\lambda_\sum = \sum_k |\lambda_k|^2$ which gives the conditions

$$V_k|\psi\rangle = \begin{bmatrix} A_k|\psi_1'\rangle + B_k|\psi_2'\rangle \\ D_k|\psi_2'\rangle \end{bmatrix} = \lambda_k \begin{bmatrix} |\psi_1'\rangle \\ |\psi_2'\rangle \end{bmatrix} \ , \quad \text{for all } \ k \ , \tag{3.22}$$

and

$$\sum_k V_k^\dagger V_k|\psi\rangle = \begin{bmatrix} \sum_k A_k^\dagger A_k|\psi_1'\rangle \\ \sum_k (B_k^\dagger B_k + D_k^\dagger D_k)|\psi_2'\rangle \end{bmatrix} = \lambda_\sum \begin{bmatrix} |\psi_1'\rangle \\ |\psi_2'\rangle \end{bmatrix} \ . \tag{3.23}$$

Eqn. (3.22) implies that $D_k|\psi_2'\rangle = \lambda_k|\psi_2'\rangle$ for all $k$ and hence Eqn. (3.23) gives the condition $\sum_k (B_k^\dagger B_k + \lambda_k D_k^\dagger)|\psi_2'\rangle = \lambda_\sum|\psi_2'\rangle$. Multiplying from the left by $\langle\psi_2'|$ gives $\sum_k \|B_k|\psi_2'\rangle\|^2 + \lambda_\sum\langle\psi_2'|\psi_2'\rangle = \lambda_\sum\langle\psi_2'|\psi_2'\rangle$ and hence $\sum_k \|B_k|\psi_2'\rangle\|^2 = 0$ which implies $B_k|\psi_2'\rangle = 0$ for all $k$ and therefore $|\psi_1\rangle$ and $|\psi_2\rangle$ also satisfy the conditions of Corollary 3.2.2. □

*Proof of Theorem. 7.* 1) Without loss of generality, assume that there are only two generalised dark state spaces, $\mathcal{D}_\Lambda$ and $\mathcal{D}_{\Lambda'}$ where $\Lambda \ne \Lambda'$ and consider the fixed points $\sigma = |\psi\rangle\langle\psi|$ and $\rho = |\psi'\rangle\langle\psi'|$ where $|\psi\rangle \in \mathcal{D}_\Lambda$ and $|\psi'\rangle \in \mathcal{D}_{\Lambda'}$. Note that clearly $\operatorname{supp}(\sigma) \subseteq \mathcal{D}_\Lambda$ and $S := \operatorname{supp}(\rho) \subseteq \mathcal{D}_{\Lambda'}$. By Lemma 3.2.2, we know that $|\psi_1\rangle := \mathcal{P}_{\mathcal{S}}|\psi\rangle$ is either zero or again corresponds to a pure state fixed point of the same $\Lambda$-type. Since $|\psi_1\rangle \in D_\Lambda$, then if $|\psi_1\rangle \ne 0$ this would imply that $\operatorname{supp}(\rho) \nsubseteq \mathcal{D}_{\Lambda'}$ which is a contradiction and therefore $|\psi_1\rangle = 0$. Then we have shown that $|\psi'\rangle \perp |\psi\rangle$ and so $\mathcal{D}_{\Lambda'} \perp \mathcal{D}_\Lambda$.

2a) Holds by construction of the generalised dark state spaces.

2b) If $\mathcal{D}_\Lambda = \{0\}$ for all $\Lambda \in \mathbb{C}^m$ i.e. if there are no generalised dark spaces of $\Gamma$ then clearly every fixed point $\rho$ is an intrinsic higher rank fixed point. Therefore, we can assume w.l.o.g. that there exists some $\Lambda_1 \in \mathbb{C}^m$ such that $\mathcal{D}_{\Lambda_1} \ne \{0\}$. Now, consider any fixed point $\rho$ of $\Gamma$ whose support is not contained in $\mathcal{D}_{\Lambda_1}$. We will first show that $\rho$ can be decomposed into the two fixed points

$$\rho = \sigma_1 + \rho' \tag{3.24}$$

with $\operatorname{supp}(\sigma_1) \in \mathcal{D}_{\Lambda_1}$ and $\operatorname{supp}(\sigma_1) \perp \operatorname{supp}\rho'$. Let $\mathcal{S} := \operatorname{supp}(\rho)$ and define the subspace $\tilde{\mathcal{D}}_{\Lambda_1} := P_{\mathcal{S}}(\mathcal{D}_{\Lambda_1})$, where $P_{\mathcal{S}}$ is the orthogonal projector onto the support of $\rho$. Assume that $\tilde{\mathcal{D}}_{\Lambda_1} \ne \{0\}$. By Lemma 3.2.2, every non-zero element in $\tilde{\mathcal{D}}_{\Lambda_1}$ corresponds to a pure state fixed point of the same $\Lambda_1$-type and moreover, by Lemma 3.2.1, we have the splitting $\rho = \sigma_1 + \rho'$. Since $\rho'$ is fixed, we can again ask if now $\operatorname{supp}(\rho') \subseteq \mathcal{D}_{\Lambda_2}$, for some $\Lambda_2 \ne \Lambda_1$. If it is contained then we know that $\rho$ is decomposed into pure state fixed points and we are done. Assume that it is not contained. Then using the same argument as before it follows that $\rho = \sigma_1 + \sigma_2 + \rho''$, with $\operatorname{supp}(\rho'') \subseteq \mathcal{D}_{\Lambda_2}$. Iterating this procedure, we eventually exhaust all fixed points belonging to the generalised dark state spaces and are left with an intrinsic higher rank fixed point $\rho_0$ such that $\operatorname{supp}(\rho_0) \perp \mathcal{D}$. □

As an application of the ideas presented above, the following result concerns pure state fixed points of a purely dissipative Lindblad generator which consists of a single Lindblad term.

**Proposition 3.2.2.** *Let $\Gamma$ be a (purely dissipative) Lindblad-Kossakowski operator with a single $V$-term. Then the following results on pure state fixed points hold.*

1. *$\sigma := |\psi\rangle\langle\psi|$ is the only pure state fixed point (possibly among other mixed state fixed points) if and only if one of the following conditions is satisfied:*
   - *(a) $\ker V \subseteq \operatorname{span}_{\mathbb{C}}\{|\psi\rangle\}$ and $\operatorname{span}_{\mathbb{C}}\{|\psi\rangle\}$ is the only simultaneous eigenspace of $V$ and $V^\dagger$.*
   - *(b) $\ker V = \operatorname{span}_{\mathbb{C}}\{|\psi\rangle\}$ and there are no simultaneous eigenvectors of $V$ and $V^\dagger$.*

*Moreover, the "only-if" direction in 1. can be strengthened as follows:*

2) *If a) is satisfied then a second (mixed state) fixed point $\rho$ with $\operatorname{supp}\rho = \{|\psi\rangle\}^\perp$ is given by*

$$\rho := \frac{(V_\perp^\dagger V_\perp)^+}{\operatorname{tr}\{(V_\perp^\dagger V_\perp)^+\}} \tag{3.25}$$

*with $V_\perp := P_\perp V P_\perp^\dagger$ and $P_\perp := \mathbb{1} - |\psi\rangle\langle\psi|$. Here, $(\,\cdot\,)^+$ denotes the Moore-Penrose inverse.*

3) *$\sigma$ is the unique fixed point of $\Gamma$ if and only if case b) is satisfied and $V$ has no eigenvector $x \in \operatorname{span}_{\mathbb{C}}\{|\psi\rangle\}^\perp$.*

*Proof.* 1) Since there is only a single Lindblad term $V$, the necessary and sufficient conditions on the existence of a pure state fixed point given by Eqns. (3.10) and (3.11) in Corollary 3.2.2, simplify to $V|\psi\rangle = \lambda|\psi\rangle$ for $\lambda \in \mathbb{C}$ and $\langle\psi|\overline{\lambda}V = \langle\psi||\lambda|^2$, respectively. Clearly then the only time these equalities can be satisfied are with either $\lambda = 0$ or $|\psi\rangle$ is a simultaneous left and right eigenvector of $V$.

2) Without loss of generality assume that $|\psi\rangle = e_1$ and hence $V_\perp = \operatorname{diag}(0, D)$ where $D$ is a full rank $(d-1)\times(d-1)$ matrix. Then $\sigma := (\operatorname{tr}\{(V_\perp^\dagger V_\perp)^+\})^{-1}\cdot(V_\perp^\dagger V_\perp)^+$ is given by the state $\sigma := (\operatorname{tr}(\tilde{D}))^{-1}\cdot\operatorname{diag}(0,\tilde{D})$ where $\tilde{D} := (D^\dagger D)^{-1}$ and its clear that $\Gamma(\sigma) = 0$.

3) If case 1b) is satisfied and there is no eigenvector $x \in \operatorname{span}_{\mathbb{C}}\{|\psi\rangle\}^\perp$ then there cannot exist an invariant subspace $\mathcal{S} \subseteq \mathcal{H}$ such that $\mathcal{S} \perp \operatorname{span}_{\mathbb{C}}\{|\psi\rangle\}$ such that $V\mathcal{S} \subseteq \mathcal{S}$ and hence by Corollary 3.2.3 it is unique. If the fixed point is unique then clearly case 1b) must be satisfied since if case 1a) was satisfied then the fixed point is not unique by 2). Furthermore, since the fixed point is unique then by Corollary 3.2.3 there cannot exist such an orthogonal subspace and therefore neither an eigenvector contained in it. □

In light of the decomposition given in Theorem 7, we can now make the crucial class distinction between the fixed points whose supports are contained in generalised dark state spaces and those which are intrinsic higher rank fixed points. The set of fixed points associated to the generalised dark state spaces is given by

$$\mathcal{F}_{\mathcal{D}} := \{\rho \in \mathfrak{pos}_1(N) \mid \Gamma(\rho) = 0,\ \text{such that } \operatorname{supp}(\rho) \subseteq \mathcal{D}\}\ , \tag{3.26}$$

where $\mathcal{D}$ is defined by Eqn. (3.14) and hence this fixed point set consists of all density matrices that are either pure states or mixed states which are a convex combination of pure states which are themselves fixed points. Furthermore, we can define the set of intrinsic higher rank fixed points as

$$\mathcal{F}_{\mathcal{D}^\perp} := \{\rho \in \mathfrak{pos}_1(N) \mid \Gamma(\rho) = 0, \text{ such that } \operatorname{supp}(\rho) \subseteq \mathcal{D}^\perp\} \;, \tag{3.27}$$

which consists of density matrices that are mixed states that cannot be decomposed into a convex combination which contains any pure state which is also a fixed point. Clearly then have the set inclusions $\mathcal{F}_{\mathcal{D}} \subseteq \mathcal{F}(\Phi_\Gamma)$ and $\mathcal{F}_{\mathcal{D}^\perp} \subseteq \mathcal{F}(\Phi_\Gamma)$ and hence the entire fixed point set is given by

$$\mathcal{F}(\Phi_\Gamma) = \operatorname{conv}\{\mathcal{F}_{\mathcal{D}} \cup \mathcal{F}_{\mathcal{D}^\perp}\} \;. \tag{3.28}$$

Therefore, the goal of constructing arbitrary unique pure state fixed points boils down to finding a Lindblad generator with appropriate Lindblad terms $\{V_k\}_{k=1}^r \in \mathfrak{sl}(N, \mathbb{C})$ such that

$$1) \quad \mathcal{F}_{\mathcal{D}^\perp} = \varnothing \quad \text{and} \quad 2) \quad \mathcal{F}_{\mathcal{D}} = \{|\psi\rangle\langle\psi|\} \;, \tag{3.29}$$

and hence

$$\mathcal{F}(\Phi_\Gamma) = \mathcal{F}_{\mathcal{D}} = \{|\psi\rangle\langle\psi|\} \;. \tag{3.30}$$

In general, it is useful to have conditions which guarantee the necessary condition 1). Once condition 1) is satisfied, adding additional Lindblad terms (while maintaining condition 1) ) to reduce the fixed point set $\mathcal{F}(\Phi_\Gamma) = \mathcal{F}_{\mathcal{D}}$ to a single unique element can be used to engineer desired fixed point states. Thus, we have the following immediate but useful result which is a consequence of Theorem 7.

**Proposition 3.2.3.** *$\mathcal{F}_{\mathcal{D}^\perp} = \varnothing$ and hence $\mathcal{F}(\Phi_\Gamma) = \mathcal{F}_{\mathcal{D}}$ if there exists no invariant subspace $\mathcal{S}$ of $\Gamma$ such that $S \subseteq \mathcal{D}^\perp$.*

Note that the existence of an additional invariant subspace $S \subseteq \mathcal{D}^\perp$ of $\Gamma$ requires that $V_k\mathcal{S} \subseteq \mathcal{S}$ for all $k$ and $\sum_k V_k^\dagger V_k \mathcal{S} \subseteq \mathcal{S}$ by Proposition 3.2.1. The problem of determining whether a set of matrices $A_1, \ldots, A_m$ have a common invariant subspace of dimension $r < N$ is a non-trivial problem. Motivated by the fact that a quantum channel is irreducible if and only if its Kraus operators do not have a non-trivial common invariant subspace, Jamiolkowski and Pastuszak have made very recent progress on this problem [27]. Namely, they provided a necessary and sufficient condition to check whether a set of matrices have a common eigenvector (a one dimensional invariant subspace) by providing a subspace which will always be a common invariant subspace (which may be trivial) for each matrix $A_k$. They also provide a computable criteria for the *existence* of a simultaneous invariant subspace of dimension larger than one in the case the matrices $A_1, \ldots A_m$ each have pairwise different eigenvalues. For an excellent summary of this subject we refer the reader to [27] and the references therein. Nonetheless, the general problem of determining whether a common $1 \leq r \leq N$ dimensional invariant subspace *exists* for a set of matrices was solved entirely by Arapura and Peterson using techniques from algebraic geometry and Gröbner base theory [5]. However, since by Proposition 3.2.3 we are looking for the non-existence of a subspace which is *orthogonal* to an already existing invariant subspace $\mathcal{D}$, the application of these results to our scenario remains an open problem.

To conclude this section we now present two useful results which provide partial solutions to the invariant subspace problem outlined above. They provide sufficient conditions guaranteeing no such invariant subspace $\mathcal{S} \perp \mathcal{D}$ exists and hence serve as key tools allowing us to make use of Proposition 3.2.3 to engineer unique pure state fixed points.

**Proposition 3.2.4.** *Let $\mathcal{L}$ be a Lindblad-Kossakowki generator with associated (square matrix) Lindblad terms $\{V_k\}_{k=1}^r$. The fixed point set is given by*

$$\mathcal{F}(\Phi_\Gamma) = \mathcal{F}_\mathcal{D}$$

*if there exists no eigenvector $|\psi\rangle \in \mathcal{D}^\perp$ of the matrix $\sum_k^r V_k$.*

*Proof.* It is a standard linear algebra result that for a finite-dimensional matrix $A$, if there exists a subspace $\mathcal{S} \subseteq \mathcal{H}$ such that $A\mathcal{S} \subseteq \mathcal{S}$ then $\mathcal{S}$ must contain an eigenvector of A. Thus, if there is no eigenvector $|\psi\rangle \in \mathcal{D}^\perp$ of $\sum_k^r V_k$, then there exists no subspace $\mathcal{S} \subseteq \mathcal{D}^\perp$ such that $\sum_k^r V_k\mathcal{S} \subseteq \mathcal{S}$ and therefore there exists no subspace $\mathcal{S} \subseteq \mathcal{D}^\perp$ such that $V_k\mathcal{S} \subseteq \mathcal{S}$ for all $k$. By Theorem 6, $V_k\mathcal{S} \subseteq \mathcal{S}$ for all $k$ is a necessary condition for $\mathfrak{pos}_1(\mathcal{S})$ to support a fixed point and therefore by Proposition 3.2.3, $\mathcal{F}(\Phi_\Gamma) = \mathcal{F}_\mathcal{D}$. □

A main application of these results in this paper will focus on Lindblad-Kossakowski operators whose Lindblad terms are nilpotent. Since the eigenvalues of each Lindblad term are zero, Corollary 3.2.2 implies that the pure state fixed points are composed from vectors which are simultaneous nullvectors of each Lindblad term. Therefore, the generalised dark state spaces $\mathcal{D}_\Lambda$ as defined by Eqn. (3.13) reduce to that of the "usual" dark state space [32] given by

$$\mathcal{D}_0 := \Big\{ |\psi\rangle \in \mathcal{H} \;\Big|\; V_k|\psi\rangle = 0 \text{ for all } k = 1, \dots, m \Big\}, \tag{3.31}$$

and hence since $\Lambda = (\lambda_1, \dots, \lambda_m) = (0, \dots, 0) \in \mathbb{C}^m$ we get that the complete dark state space generically defined by Eqn. (3.14) simplifies to

$$\mathcal{D} = \mathcal{D}_0 = \bigcap_k \ker(V_k)\ . \tag{3.32}$$

**Proposition 3.2.5.** *Let $\Gamma$ be a (purely dissipative) Lindblad-Kossakowski operator with nilpotent Lindblad terms $\{V_k\}_{k=1}^r$. Then if $r = 1$ or if for $r \geq 2$, $[V_i, V_j] = 0$ for all $1 \leq i, j \leq r$ then the total fixed point set is given by*

$$\mathcal{F}(\Phi_\Gamma) = \mathcal{F}_{\mathcal{D}_0}\ , \tag{3.33}$$

*and hence the only fixed points are those which are convex combinations of pure dark state fixed points.*

*Proof.* For the $r = 1$ case, all the eigenvectors of a single Lindblad term are null vectors and hence the result follows by Proposition 3.2.4. For the $r \geq 2$ case, recall that Proposition 3.2.3 states that $\mathcal{F}(\Phi_\Gamma) = \mathcal{F}_{\mathcal{D}_0}$ if there exists no subspace $\mathcal{S} \subseteq \mathcal{D}_0^\perp$ which satisfies the conditions of Theorem 6. Assume by contradiction there does exist such a subspace and hence $V_k\mathcal{S} \subseteq \mathcal{S}$ for all $k$. Note that since $\mathcal{D}_0 = \cap_k \ker(V_k)$, then if for some $x \in \mathcal{D}_0^\perp$, $V_k x = 0$ for all $k$ this would imply $x \in \mathcal{D}_0$ which would be a contradiction. We will show this will always eventually occur.

Clearly, there must exist a vector $x \in \mathcal{D}_0^\perp$ such that $V_k x \in \mathcal{S}$ and $V_k x \notin \mathcal{D}_0$ for all $k$. If at this stage, $V_k x = 0$ for all $k$ then we obtain our contradiction. Assume there are non-zero elements of this form and define the subspace $\mathcal{S}_1 \subseteq \mathcal{S}$ as

$$\mathcal{S}_1 := \langle x, V_k x \mid \forall\, V_k \text{ s.t } V_k x \neq 0 \rangle\ . \tag{3.34}$$

By assumption $V_k\mathcal{S}_1 \subseteq \mathcal{S}$ for all $k$ and hence iterating this process by either obtaining a contradiction and if not, adding the non-zero elements into the successive subsets of $\mathcal{S}$ we eventually obtain the element $s_0 \in \mathcal{S}$ given by

$$s_0 := V_1^{n_1-1} V_2^{n_2-1} \dots V_r^{n_r-1} x \neq 0\ , \tag{3.35}$$

where $\{n_k\}_{k=1}^{r}$ are the degrees of nilpotency of each Lindblad term $V_k$. Then $V_k s_0 = 0$ for all $k$ which gives the final possible contradiction. □

## 3.3 Engineering Pure State Fixed Points

### 3.3.1 Motivation and Lindblad Terms of Canonical Form

This subsection will present a collection of results on the structure of fixed point sets which arise from choosing the Lindblad terms of a Lindblad generator to be of a special form. As we will see, they greatly simplify the discussion of the previous subsections on the theory of engineering fixed points. Specifically, we will use these special forms of Lindblad terms to show one can completely characterise the fixed point set. Later, in Section 3.4 we will use these results by providing explicit examples showing how to obtain several well known pure states as unique fixed points.

Let $\mathbf{p}, \mathbf{q} \in I^n$ be three $n$-tuples of indices which define the Kronecker product of Pauli matrices $\sigma_{\mathbf{p}}, \sigma_{\mathbf{q}}, \sigma_{\mathbf{m}} \in \mathcal{B}_0^n$ as defined in Eqns. (2.29) and (2.30). Then define a Lindblad term to be of *Non-Unital Canonical Form* when it is given by

$$V = \tfrac{\sqrt{\gamma}}{2}(\sigma_{\mathbf{p}} + \mathrm{i}\sigma_{\mathbf{q}}) \;, \quad \text{such that} \quad [\sigma_{\mathbf{p}}, \sigma_{\mathbf{q}}] \neq 0 \;, \tag{3.36}$$

for $\gamma \in \mathbb{R}^+$. We can now immediately provide a useful Lemma which shows Lindblad terms of this form admit a special kind of decomposition which we will later generalise for mixed state fixed point engineering.

**Lemma 3.3.1.** *Let $V$ be in canonical form. Then it can be decomposed as*

$$V = \sqrt{\gamma}\sigma_{\boldsymbol{p}} P \;, \quad \textit{where} \quad P = \tfrac{1}{2}(\mathbb{1} - \sigma_{\boldsymbol{m}}) \;, \tag{3.37}$$

*is an orthogonal projection and $\boldsymbol{m} = \boldsymbol{p} \star \boldsymbol{q}$ for the star-product defined by Eqns.* (2.68) *and* (2.69).

*Proof.* Note first that $V = \frac{\sqrt{\gamma}}{2}(\sigma_{\mathbf{p}} + \mathrm{i}\sigma_{\mathbf{q}}) = \frac{\sqrt{\gamma}}{2}\sigma_{\mathbf{p}}(\mathbb{1} + \mathrm{i}\sigma_{\mathbf{p}}\sigma_{\mathbf{q}}) = \frac{\sqrt{\gamma}}{2}\sigma_{\mathbf{p}}(\mathbb{1} - \sigma_{\mathbf{m}})$ where $\mathbf{m} = \mathbf{p} \star \mathbf{q}$ since $\sigma_{\mathbf{p}}\sigma_{\mathbf{q}} = \mathrm{i}\sigma_{\mathbf{m}}$. Clearly $P$ is Hermitian, and furthermore, since $\sigma_{\mathbf{m}}^2 = \mathbb{1}$ we see that $P^2 = \frac{1}{4}(2\mathbb{1} - 2\sigma_{\mathbf{m}}) = P$. □

These types of matrices are the natural generalisation of atomic raising and lowering operators given by $\sigma^{\pm} := \frac{1}{2}(\sigma_x \pm \mathrm{i}\sigma_y)$ and are known to describe a non-unital dissipative process. For the remaining of the thesis we will suppress the terminology “non-unital” and simply call them *canonical Lindblad terms.* Moreover, by Lemma 2.3.2, $[\sigma_{\mathbf{p}}, \sigma_{\mathbf{q}}] \neq 0$ if and only if $\{\sigma_{\mathbf{p}}, \sigma_{\mathbf{q}}\}_+ = 0$ and thus by Lemma 1.3.3 the corresponding Lindblad generator is given by

$$\Gamma_V = \tfrac{\gamma}{2}\big(\hat{\sigma}_{\mathbf{p}}^2 + \hat{\sigma}_{\mathbf{q}}^2\big) + \mathrm{i}\gamma\hat{\sigma}_{\mathbf{p}}\hat{\sigma}_{\mathbf{q}}^{+} \in \mathfrak{gl}(4^n, \mathbb{C}\,) \;, \tag{3.38}$$

whose coefficient follows from the fact that $\Gamma_{\sigma_{\mathbf{p}}} = 2\hat{\sigma}_{\mathbf{p}}^2$ due to the scaling of $\hat{\sigma}_{\mathbf{p}}$ (see Eqn. (2.31) and the subsequent plus commutator). Thus, Eqn. (3.38) is the generalisation of the single qubit Lindblad generator given by Eqn (2.13) whose Lindblad terms were said to have been in *single qubit standard form* - which is non-other than the canonical form restricted to a single qubit. This decomposition of the Lindblad generator allows one to directly see the connection to the representation theory of the Lindblad-Kossakowski Lie algebra provided in Chapter 2 - and in particular the quasi-translation operators $\mathrm{i}\hat{\sigma}_{\mathbf{p}}\hat{\sigma}_{\mathbf{q}}^{+} \in \hat{\mathfrak{m}}_{\mathrm{qt}}$ (cf. Proposition 2.3.1). We now present a Lemma which will be of utmost importance in simplifying pure state fixed point engineering. Recall that a square matrix $V$ which is nilpotent is one which satisfies $V^a = 0$ for some minimal positive integer $a$. We call this smallest number $a$ the degree of nilpotency of $V$.

**Lemma 3.3.2.** *Let $V$ be in canonical form. Then $V$ is nilpotent of degree 2.*

*Proof.* First note that $\sigma_{\mathbf{p}}^2 = \sigma_{\mathbf{q}}^2 = \mathbb{1}$ and by the definition of a Lindblad term of canonical form we know $\{\sigma_{\mathbf{p}}, \sigma_{\mathbf{q}}\}_+ = 0$ since $[\sigma_{\mathbf{p}}, \sigma_{\mathbf{q}}] \neq 0$ by Lemma 2.3.2. Then $V^2 = (\frac{\sqrt{\gamma}}{2}(\sigma_{\mathbf{p}} + \mathrm{i}\sigma_{\mathbf{q}}))^2 = \frac{\gamma}{4}(\sigma_{\mathbf{p}}^2 - \sigma_{\mathbf{q}}^2 + 2\mathrm{i}\{\sigma_{\mathbf{p}}, \sigma_{\mathbf{q}}\}_+) = 0$. □

Since every Lindblad term which is in canonical form is nilpotent (and therefore every eigenvalue is zero), Corollary 3.2.2 implies that a pure state $\rho = |\psi\rangle\langle\psi|$ is a fixed point of $\Gamma = \sum_k \Gamma_{V_k}$ if and only if $V_k|\psi\rangle = 0$ for all $k$. Therefore, the fixed point set of the associated Markovian semigroup is given by

$$\mathcal{F}(\Phi_\Gamma) = \mathrm{conv}\,\{\mathcal{F}_{\mathcal{D}_0} \cup \mathcal{F}_{\mathcal{D}_0^\perp}\}\ , \tag{3.39}$$

where by Theorem 7, $\mathcal{F}_{\mathcal{D}_0}$ is the fixed point set containing the pure state fixed points and their convex combinations (cf. Eqn. (3.26)) whereas $\mathcal{F}_{\mathcal{D}_0^\perp}$ is the set of mixed state fixed points which are not associated to the eigenvalue zero and hence are intrinsic higher rank fixed points (cf. Eqn. (3.27)).

Clearly, by analysing the fixed point set of a Lindblad generator which has a *single* Lindblad term (in canonical form), later, we can consider the more general problem of when a Lindblad generator has multiple Lindblad terms (in canonical form). Essentially, this "piecewise" understanding of a Lindblad generator allows us to selectively add additional Lindblad terms to a pre existing set that ultimately narrow down the fixed point set until it yields the correct target fixed point state. We will now show that we can obtain a complete classification of the fixed point set $\mathcal{F}(\Phi_\Gamma)$ when the Lindblad generator has a single Lindblad term and is of canonical form.

Recall the star-product $\mathbf{p}\star\mathbf{q} = \mathbf{m}$ as defined by Eqns. (2.68) and (2.69). Furthermore, Remark 6 provided a notion of positivity/negativity of the product $\mathbf{p}\star\mathbf{q}$ since for example we can have cases where $\mathbf{m} = (-x, 1, x) \in \mathbb{I}_0^3$. That is, $\mathbf{p}\star\mathbf{q} > 0$ whenever there are an even number of negative index terms of $\mathbf{m}$, and $\mathbf{p}\star\mathbf{q} < 0$ whenever there is an odd number of negative index terms of $\mathbf{m}$. In this example, $\sigma_{\mathbf{m}}$ is *not* a Pauli matrix since not every index element is positive and thus we cannot discuss its matrix properties. To get around this small technicality, consider the following.

For $r \in \{x, y, z\}$ we defined $\sigma_{-r} := -\sigma_r$ and thus for a generic $\mathbf{m} \in \mathbb{I}_0^n$ we can define a "factored" version of $\mathbf{m}$ in the sense that $\sigma_{\mathbf{m}} = -\sigma_{\mathbf{m}'}$ where $\mathbf{m}'$ now has all the same index letters (i.e. $x, y, z, 1$) as $\mathbf{m}$ except now they all come as positive entries, i.e. $\mathbf{m}' \in I_0^n$. That is,

$$\sigma_{\mathbf{m}} = \begin{cases} \sigma_{\mathbf{m}'} & \text{if } \mathbf{p}\star\mathbf{q} > 0 \\ -\sigma_{\mathbf{m}'} & \text{if } \mathbf{p}\star\mathbf{q} < 0\ . \end{cases} \tag{3.40}$$

Since $\mathbf{m}' \in I_0^n$ then $\sigma_{\mathbf{m}'} \in \mathcal{B}_0^n$ is indeed a valid Pauli matrix. With this concept in mind, we now have the following result.

**Lemma 3.3.3.** *For $V$ in canonical form we have*

$$\ker(V) = \begin{cases} E_1(\sigma_{\mathbf{m}'}) & \text{if} \quad \mathbf{p}\star\mathbf{q} > 0 \\ E_{-1}(\sigma_{\mathbf{m}'}) & \text{if} \quad \mathbf{p}\star\mathbf{q} < 0\ , \end{cases} \tag{3.41}$$

*where $E_1(\sigma_{\mathbf{m}})$ and $E_{-1}(\sigma_{\mathbf{m}})$ are the $+1$ and $-1$ eigenspaces of $\sigma_{\mathbf{m}'}$, respectively.*

*Proof.* By Lemma 3.3.1 we know that a Lindblad term in canonical form (and recall the additional condition that $[\sigma_{\mathbf{p}}, \sigma_{\mathbf{q}}] \neq 0$) can be decomposed as

$$V = \begin{cases} \frac{\sqrt{\gamma}}{2}\sigma_{\mathbf{p}}(\mathbb{1} - \sigma_{\mathbf{m'}}) & \text{if} \quad [\sigma_{\mathbf{p}}, \sigma_{\mathbf{q}}] = 2\mathrm{i}\sigma_{\mathbf{m'}} \\ \frac{\sqrt{\gamma}}{2}\sigma_{\mathbf{p}}(\mathbb{1} + \sigma_{\mathbf{m'}}) & \text{if} \quad [\sigma_{\mathbf{p}}, \sigma_{\mathbf{q}}] = -2\mathrm{i}\sigma_{\mathbf{m'}} \ , \end{cases} \tag{3.42}$$

and therefore $\ker(V) = \ker(\mathbb{1} \pm \sigma_{\mathbf{m'}})$ and the result follows. □

Recall from the introduction of Section 2.2 that we defined the index set $I_0 = \{x, y, z\}$. For any $m \in I_0$, let $|0\rangle_m$ and $|1\rangle_m$ be the $+1$-eigenvector and $-1$-eigenvector, respectively, of $\sigma_m$. Moreover, we denote $|0\rangle := (1,0)^\top = |0\rangle_z$, $|1\rangle := (0,1)^\top = |1\rangle_z$ and set $|0\rangle_1 := |0\rangle$ and $|1\rangle_1 := |1\rangle$. Now, for a fixed $\mathbf{m'} \in I_0^n$, define a generic vector of the form

$$|\phi_{\mathbf{m'}}\rangle := |\phi(1)\rangle_{m_1} \otimes |\phi(2)\rangle_{m_2} \otimes \cdots \otimes |\phi(n)\rangle_{m_n} \in (\mathbb{C}^2)^{\otimes n} \ , \tag{3.43}$$

where $\phi(k) \in \{0, 1\}$ for all $k$. Since every $\sigma_{\mathbf{m'}}$ has $2^{n-1}$ plus one and minus one eigenvalues, a simple counting argument shows that the eigenspaces of $\sigma_{\mathbf{m'}}$ are given by

$$E_1(\sigma_{\mathbf{m'}}) \quad = \quad \mathrm{span}_{\mathbb{C}}\,\{\, |\phi_{\mathbf{m'}}\rangle \mid \sum_k^n \phi(k) \text{ is even }\} \ \text{ and} \tag{3.44}$$

$$E_{-1}(\sigma_{\mathbf{m'}}) \quad = \quad \mathrm{span}_{\mathbb{C}}\,\{\, |\phi_{\mathbf{m'}}\rangle \mid \sum_k^n \phi(k) \text{ is odd }\} \ , \tag{3.45}$$

where $\sum_k^n \phi(k) = 0$ is considered even. This observation along with Lemma 3.3.3 gives a complete account of the set of fixed points of the Markovian semigroup generated by the associated Lindblad generator. The discussion can be summarised by the following Proposition.

**Proposition 3.3.1.** *Let $\Gamma$ be a (purely dissipative) Lindblad Kossakowski operator with a single Lindblad term $V$ which is of canonical form. Then the fixed point set is given by*

$$\mathcal{F}(\Phi_\Gamma) \quad = \quad \{\rho \in \mathfrak{pos}_1(N) \mid \mathrm{supp}(\rho) \subseteq \ker(V)\} \ , \quad \textit{where} \tag{3.46}$$

$$\ker(V) \quad = \quad \begin{cases} \mathrm{span}_{\mathbb{C}}\,\{\, |\phi_{\boldsymbol{m'}}\rangle \mid \sum_k^n \phi(k) \textit{ is even}\,\} & \textit{if } \boldsymbol{p} \star \boldsymbol{q} > 0 \\ \mathrm{span}_{\mathbb{C}}\,\{\, |\phi_{\boldsymbol{m'}}\rangle \mid \sum_k^n \phi(k) \textit{ is odd}\,\} & \textit{if } \boldsymbol{p} \star \boldsymbol{q} < 0 \ , \end{cases} \tag{3.47}$$

*and where $\sum_k^n \phi(k) = 0$ is considered even.*

*Proof.* As noted in the preceding paragraphs, since $V$ is nilpotent then $\mathcal{F}(\Phi_\Gamma) = \mathrm{conv}\,\{\mathcal{F}_{\mathcal{D}_0} \cup \mathcal{F}_{\mathcal{D}_0^\perp}\}$. Moreoever, by Proposition 3.2.5, we in fact have $\mathcal{F}(\Phi_\Gamma) = \mathcal{F}_{\mathcal{D}_0} = \{\rho \in \mathfrak{pos}_1(N) \mid \mathrm{supp}(\rho) \subseteq \ker(V) = \mathcal{D}_0\}$. Lemma 3.3.3 combined with the observation in Eqn. (3.44) provides the kernel of the Lindblad term $V$. □

**Remark 11.** *It's important to note that $\dim(\ker(V)) = 2^{n-1}$ and therefore the fixed point set $\mathcal{F}(\Phi_\Gamma)$ above is not a unique fixed point.*

For a complete analysis of how these fixed points relate to the decomposition of the Lindblad generator into its unital and non-unital components we refer the reader to Appendix C. We also show there that specific (possibly mixed) fixed points related to a Lindblad generators associated *translational* component are always contained in the fixed point set (See Proposition C.0.2).

Recall that Lemma 2.4.1 introduced the set of pairs of element $\mathbf{p}, \mathbf{q} \in I_0^n$ which under the star-product give the same (fixed) $\mathbf{m} \in \mathbb{I}_0^n$ element as

$$S_{\mathbf{m}} := \{(\mathbf{p}, \mathbf{q}) \in \mathbb{I}_0^n \times \mathbb{I}_0^n \mid \mathbf{p} \star \mathbf{q} = \mathbf{m} \neq \mathbf{q} \star \mathbf{p}\} . \tag{3.48}$$

and stated that there are $4^{n-1}$ such pairs. Note that the inequality on the right hand side implies that $[\sigma_{\mathbf{p}}, \sigma_{\mathbf{q}}] \neq 0$. Since the fixed point set of a Markovian semigroup of quantum channels generated by a Lindblad generator of canonical form depends solely on $\mathbf{m}$ by Proposition 3.3.1, we immediately obtain the following result.

**Proposition 3.3.2.** *Let $V_1, V_2, \ldots, V_{4^{n-1}}$ be Lindblad terms of canonical form such that $\boldsymbol{p}_k \star \boldsymbol{p}_k = \boldsymbol{m}$ for a fixed $\boldsymbol{m} \in \mathbb{I}_0^n$ for all $k$. Then the fixed point sets of the Markovian semigroup of quantum channels generated by each $\Gamma_{V_k}$ are equal, i.e.*

$$\mathcal{F}(\Phi_{\Gamma_{V_1}}) = \mathcal{F}(\Phi_{\Gamma_{V_2}}) = \cdots = \mathcal{F}(\Phi_{\Gamma_{V_{4^{n-1}}}}) . \tag{3.49}$$

Thus, we have shown the relationship between the kernel of a Lindblad term in canonical form and the fixed point set depends only on the associated $\mathbf{m}$ index. One can also interpret this as a dependence on the associated *translation* direction, i.e. the associated ideal element $\tau_{\mathbf{m}} \in \hat{\mathfrak{i}}$, in the following way.

A Lindblad generator whose (single) Lindblad term is of canonical form is given by Eqn. (3.38). Now, recall that in Section 2.3.2, we proved that each quasi-translation operator $\mathrm{i}\hat{\sigma}_{\mathbf{p}}\hat{\sigma}_{\mathbf{q}}^+ \in \hat{\mathfrak{m}}_{\mathrm{qt}}$ (cf. Proposition 2.3.1) is associated to an ideal element in the sense that $\chi(\mathrm{i}\hat{\sigma}_{\mathbf{p}}\hat{\sigma}_{\mathbf{q}}^+) = \chi(\tau_{\mathbf{m}}) \in \hat{\mathfrak{i}}$, where $\chi$ is a projection operator (cf. Proposition 2.3.3 and Corollary 2.4.1). This observation, combined with Proposition 3.3.2, provide the following conclusion.

**Proposition 3.3.3.** *For a fixed translation element $\tau_{\boldsymbol{m}} \in \hat{\mathfrak{i}}$, there are $4^{n-1}$ different Lindblad generators $\Gamma_{V_k} \in \mathfrak{w}^{LK}$, each of which have the same fixed point set, and are associated to $\tau_{\boldsymbol{m}} \in \hat{\mathfrak{i}}$.*

As a final result, we consider the case when we have *multiple* Lindblad terms, each of which are in canonical form. Since the fixed point set is dependent on the existence of simultaneous invariant subspaces of each Lindblad term i.e. $V_k\mathcal{S} \subseteq \mathcal{S}$ for all $k$ and $\mathcal{S} \subseteq \mathcal{H}$ (cf. Proposition 3.2.1 ), and not just the kernel of the single Lindblad term anymore, in general we have that

$$\mathcal{F}(\Phi_\Gamma) = \mathrm{conv}\,\{\mathcal{F}_{\mathcal{D}_0} \cup \mathcal{F}_{\mathcal{D}_0^\perp}\} \ , \text{ where } \ \mathcal{D}_0 = \cap_k \ker(V_k) \ . \tag{3.50}$$

However, we do know the exact structure of the kernels of each individual Lindblad term $V_k$ and therefore we obtain the following.

**Theorem 8.** *Let $\Gamma$ be a (purely dissipative) Lindblad-Kossakowski operator with Lindblad terms $\{V_k\}_{k=1}^r$ in canonical form for some $r \in \mathbb{N}$. Then $\rho = |\psi\rangle\langle\psi|$ is the unique fixed point if and only if $|\psi\rangle \in \mathcal{H}$ is the only vector which satisfies*

$$\sigma_{\boldsymbol{m}'_k}|\psi\rangle = \pm|\psi\rangle \ , \quad \forall\, \sigma_{\boldsymbol{m}'_k} \ , \tag{3.51}$$

*where $\boldsymbol{m}'_k$ has only positive index elements of $\boldsymbol{m}_k = \boldsymbol{p}_k \star \boldsymbol{q}_k$ and there exists no subspace $\mathcal{S} \perp \mathrm{span}_{\mathbb{C}}\{|\psi\rangle\}$ with $\dim(\mathcal{S}) \geq 2$ such that $V_k\mathcal{S} \subseteq \mathcal{S}$ for all $k$.*

*Proof.* The “ if ” direction. If $\rho$ is the unique fixed point then $\mathrm{span}_{\mathbb{C}}\{|\psi\rangle\} =: \mathcal{S}$ is the only (non-trivial) invariant subspace of $\Gamma$ and hence $\mathcal{D}_0 = \mathcal{S}$. Trivially, this also implies that there exists no other invariant subspace $\mathcal{S}' \perp \mathbb{C}|\psi\rangle$ which has dimension greater than or equal to two since $\mathcal{D}_0^\perp = (\mathrm{span}_{\mathbb{C}}\{|\psi\rangle\})^\perp$ supports intrinsic higher rank fixed points by Theorem 7. Now, since $\mathcal{D}_0 = \mathrm{span}_{\mathbb{C}}\{|\psi\rangle\}$, then by Proposition 3.3.1 this implies that $V_k|\psi\rangle = 0$ and therefore $\sigma_{\mathbf{m}'_k}|\psi\rangle = \pm|\psi\rangle$ for all $k$ by Proposition 3.3.1.

The “ only if ” direction. Since $\mathcal{D}_0 = \cap_k \ker(V_k)$ and $|\psi\rangle \in \mathcal{H}$ is the only simultaneous eigenvector of each $\sigma_{\mathbf{m}'_k}$, then by the structure of the kernel of each Lindblad term given in Proposition 3.3.1 this implies that in fact $\mathcal{D}_0 = \cap_k \ker(V_k) = \mathrm{span}_{\mathbb{C}}\{|\psi\rangle\}$ and therefore $\rho = |\psi\rangle\langle\psi|$ is the only pure state fixed point. By Theorem 7, every other fixed point must have support $\mathcal{S} \subseteq \mathcal{D}_0^\perp = \mathrm{span}_{\mathbb{C}}\{|\psi\rangle\}$. If there exists no subspace $\mathcal{S} \subseteq \mathcal{D}_0^\perp$ such that $V_k\mathcal{S} \subseteq \mathcal{S}$ for all $k$, then by Proposition 3.2.1, $\mathfrak{her}(\mathcal{S})$ is not an invariant subspace of $\Gamma$ and therefore all the fixed points have support in $\mathcal{D}_0$ and therefore the fixed point is unique. □

### 3.3.2 Summary and Pure State Symmetry Considerations

We remark that Theorem 8 implies that we have to prove the non-existence of an invariant subspace contained in $\mathcal{D}_0^\perp = (\mathrm{span}_{\mathbb{C}}\{|\psi\rangle\})^\perp$ in order for $\rho = |\psi\rangle\langle\psi|$ to be the unique fixed point. Propositions 3.2.4 and 3.2.5 in Section 3.2, provide two useful conditions which guarantee this holds. We use these three key results in Section 3.4 where we provide a plethora of useful examples in which we engineer many well-known quantum states useful in quantum information processing as unique fixed points. The following illustrative example demonstrates the relationship between invariant subspaces, choices of canonical Lindblad terms, and the uniqueness of a pure state fixed point. It should be considered as a toy-model since as we will later show in Section 3.4 there is a much easier and intuitive solution.

**Example 4.** *Let $\Gamma := \Gamma_{V_1} + \Gamma_{V_2}$ be the Lindblad-Kossakowski operator with Lindblad terms $V_1$ and $V_2$ such that the corresponding translation directions (upon projecting the unital component of the dynamics out cf. Proposition 2.3.3) are given by $\chi(\Gamma_{V_1}) = \chi(\tau_{z\mathbb{1}})$ and $\chi(\Gamma_{V_2}) = \chi(\tau_{zz})$, respectively. By Proposition 3.3.3 we can realise the operators $V_1$ and $V_2$ each in four different ways, and thus for example $V_1 := \sigma_{x1} + \mathrm{i}\sigma_{y1}$ and $V_2 := \sigma_{x1} + \mathrm{i}\sigma_{yz}$, without loss of generality by setting the normalisation factors to one. Note that $[V_1, V_2] \neq 0$ and hence we cannot apply Proposition 3.2.5 to show that $\mathcal{F}_{\mathcal{D}_0^\perp} = \varnothing$ (and hence to show there are no intrinsic higher rank fixed points).*

*Since $\sigma_{\mathbf{m}_1} = \sigma_{z1}$ and $\sigma_{\mathbf{m}_2} = \sigma_{zz}$, have $|00\rangle$ as their only simultaneous eigenvector (even to the same eigenvalue of one), then $\mathcal{D}_0 = \cap\ker(V_k) = \mathrm{span}_{\mathbb{C}}\{|00\rangle\}$ and hence the first necessary condition of Theorem 8 is satisfied. Furthermore it can be verified that $V_1 + V_2$ has two eigenvectors perpendicular to $|00\rangle$ given by $v_1 := |01\rangle + |11\rangle$ and $v_2 := -|01\rangle + |11\rangle$, thus we cannot use Proposition 3.2.4 to show that $\mathcal{F}_{\mathcal{D}_0^\perp} = \varnothing$. In fact, $\mathcal{S} := \mathrm{span}_{\mathbb{C}}\{v_1, v_2\} = \mathrm{span}_{\mathbb{C}}\{|01\rangle, |11\rangle\}$ is an invariant subspace in $\mathcal{D}_0^\perp$ and therefore by Theorem 8 the target fixed point is non-unique since there exists an intrinsic higher rank fixed point (given by $\rho = \frac{1}{2}(|01\rangle\langle 01| + |11\rangle\langle 11|)$).*

*In order to make the pure state the unique fixed point, we can use an alternative realisation of the Lindblad terms. That is, keeping $V_1$ the same but instead choosing $V_2' := \sigma_{1x} + \mathrm{i}\sigma_{zy}$, clearly the first necessary condition of Theorem 8 is satisfied and hence $\mathcal{D}_0 = \cap\ker(V_k) = \mathrm{span}_{\mathbb{C}}\{|00\rangle\}$. Furthermore, the only non-trivial eigenvector of $V_1 + V_2'$ is $|00\rangle$ and therefore by Proposition 3.2.4 there exists no invariant subspace $\mathcal{S} \subseteq \mathcal{D}_0^\perp$ and*

*hence by Theorem 8, the fixed point is unique i.e.*

$$\mathcal{F}(\Phi_\Gamma) = \mathcal{F}_{\mathcal{D}_0} = \{|00\rangle\langle 00|\} \ .$$

As a final culminating result, we now outline a systematic method in which one can engineer the purely dissipative dynamics to drive the system to the target pure state fixed point. In many scenarios, this general method produces elegant solutions which reproduce and generalise the current state-of-the-art known solutions. Moreover, it connects the entirety of Chapter 2 and Sections 3.2-3.3 by considering the symmetries of the target pure state fixed point and their relations to translation directions, invariant subspaces and in many cases, Lindblad terms of canonical form.

Let $\rho = |\psi\rangle\langle\psi|$ be a target pure state fixed point. Moreover, the centraliser of $\rho$ (*with respect to* the $\mathfrak{su}(2^n)$ Lie algebra) is given by

$$\mathfrak{s}_\rho := \{s \in \mathfrak{su}(2^n) \,|[s, \rho] = 0\} \ , \tag{3.52}$$

and is a Lie subalgebra of $\mathfrak{su}(2^n)$. Note that if $[s, \rho] = 0$ then $|\psi\rangle$ is an eigenvector of $s \in \mathfrak{su}(2^n)$, i.e. $s|\psi\rangle = \lambda|\psi\rangle$ for $\lambda \in \mathbb{C}$. We now want to compute non-normal Lindblad terms from this centraliser and thus obtain a non-unital noise operation.

For $\lambda \neq 0$, define a *shifted* centraliser element $P^S$ as

$$P^S := |\lambda|\mathbb{1} \pm \mathrm{i}s \ , \quad \text{such that} \quad P^S|\psi\rangle = 0 \ , \tag{3.53}$$

which will always exist since the eigenvalue $\lambda$ to $|\psi\rangle$ has the form $\lambda = \mathrm{i}\tilde{\lambda}$ where $\tilde{\lambda} \in \mathbb{R}$. Clearly it's still true that $[P^S, \rho] = 0$, except now $P^S \notin \mathfrak{su}(2^n)$. By including the extra complex factor within $P^S$ this implies that in fact $P^S$ is Hermitian. We can now define a Lindblad term which is *centraliser generated* as one which has strength coefficient $\gamma \in \mathbb{R}^+$ and is of the form

$$V := \sqrt{\gamma}\sigma_{\mathbf{p}}P^S = \sqrt{\gamma}\sigma_{\mathbf{p}}(|\lambda|\mathbb{1} \pm \mathrm{i}s) \ , \text{ where } \ s \in \mathfrak{s}_\rho \quad \text{such that} \quad [\sigma_{\mathbf{p}}, s] \neq 0 \ , \quad \text{and} \quad \mathrm{tr}(\sigma_{\mathbf{p}}s) = 0 \tag{3.54}$$

A few remarks are in order. First note that the condition $[\sigma_{\mathbf{p}}, s] \neq 0$ guarantees that $V \neq V^\dagger$ since $V^\dagger = \sqrt{\gamma}(|\lambda|\sigma_{\mathbf{p}} \pm \mathrm{i}s\sigma_{\mathbf{p}})$ which follows from the fact that $s \in \mathfrak{s}_\rho$ is skew-Hermitian. If $V = V^\dagger$ then the Lindblad term would describe a unital noise process - which we want to avoid. Furthermore, the last condition that $\mathrm{tr}(\sigma_{\mathbf{p}}s) = 0$ imposes that $V$ is traceless and hence describes a purely dissipative process. Finally, its clear that Lindblad terms constructed in this fashion satisfy $V|\psi\rangle = 0$. Therefore, constructing a set of Lindblad terms this way guarantees that $|\psi\rangle \in \mathcal{D}_0$, the dark space of $\Gamma$, and thus $\rho = |\psi\rangle\langle\psi|$ is indeed a fixed point.

As we will see soon, Lindblad terms of canonical form (and hence are nilpotent) are indeed a special type of these centraliser generated Lindblad terms. However, in general, it is easy to provide examples where general Lindblad terms constructed in this fashion are *not* nilpotent, let alone nilpotent of degree two.

**Remark 12.** *A general $s \in \mathfrak{s}_\rho$ is given by $s := \mathrm{i}\sum_i r_i\sigma_{\boldsymbol{m}_i}$ with $r_i \in \mathbb{R}$. The index $\boldsymbol{m}_i$ of each element in the sum* does not *always equal every induced translation direction i.e. if $s = \mathrm{i}(\sigma_{z1} + \sigma_{1z})$ then one cannot immediately deduce that the projection of the total Lindblad generator $\Gamma$ onto the ideal of translation directions yields only the translations $\tau_{z1}$ and $\tau_{1z}$. That is, there may be induced translation directions $\tau_{\boldsymbol{m}}$ whose $\boldsymbol{m}$ index does not equal one of those of the elements of the sum $s = \mathrm{i}\sum_i r_i\sigma_{\boldsymbol{m}_i}$.*

*For example, let $\rho = |\psi\rangle\langle\psi|$ be a target pure state fixed point and suppose there is a centraliser element given by $s = \mathrm{i}(\sigma_{z1} + \sigma_{1z})$ such that $s|\psi\rangle = -\mathrm{i}|\psi\rangle$ and hence $\lambda = -\mathrm{i}$.*

*Then $|\lambda| = 1$, and therefore $P^S|\psi\rangle = 0$ where $P^S = \mathbb{1} - \mathrm{i}s$ is the shifted centraliser element. By choosing $\sigma_{\boldsymbol{p}} = \sigma_{x1}$ and without loss of generality set the overall strength coefficient of the noise $\gamma = 1$, we get a Lindblad term of the form*

$$V = \sigma_{\boldsymbol{p}} P^S = \sigma_{\boldsymbol{p}}(\mathbb{1} - \mathrm{i}s) = \sigma_{x1}(\mathbb{1} - \mathrm{i}(\mathrm{i}\sigma_{z1} + \mathrm{i}\sigma_{1z})) = \sigma_{x1} - \mathrm{i}\sigma_{y1} + \sigma_{xz} = (\sigma_{x1} + \sigma_{xz}) - \mathrm{i}\sigma_{y1} \quad (3.55)$$

*Decomposing as $V = C + \mathrm{i}D$ with $C := \sigma_{x1} + \sigma_{xz}$ and $D := -\sigma_{y1}$, the associated quasi-translation operator $\mathrm{ad}_C \circ \mathrm{ad}_D^+ - \mathrm{ad}_D \circ \mathrm{ad}_C^+$ (as expressed in the decomposition of $\Gamma$ in Eqn. (1.38)) can be constructed, from which, one can identify the translation elements associated to each quasi-translation operator within the summation. That is, the first term gives*

$$\mathrm{ad}_C \circ \mathrm{ad}_D^+ = \mathrm{ad}_{\sigma_{x1}+\sigma_{xz}} \circ \mathrm{ad}_{-\sigma_{y1}}^+ = -\,\mathrm{ad}_{\sigma_{x1}} \circ \mathrm{ad}_{\sigma_{y1}}^+ - \mathrm{ad}_{\sigma_{xz}} \circ \mathrm{ad}_{\sigma_{y1}}^+ \;, \quad (3.56)$$

*whereas the second term gives $\mathrm{ad}_D \circ \mathrm{ad}_C^+ = -\,\mathrm{ad}_{\sigma_{y1}} \circ \mathrm{ad}_{\sigma_{x1}}^+ - \mathrm{ad}_{\sigma_{y1}} \circ \mathrm{ad}_{\sigma_{xz}}^+$. On each piecewise element of the total quasi-translation term we can perform a star-product calculation to determine the induced translation direction. For example, setting $\boldsymbol{p}_1 := (x,1)$ and $\boldsymbol{q}_1 := (y,1)$ to correspond to the indices from the first term we get that $\boldsymbol{p}_1 \star \boldsymbol{q}_1 = (x,1)\star(y,1) = (z,1) =: \boldsymbol{m}_1$. Continuing in this fashion, we see that the translation directions induced by the total quasi-translation term are $\mathcal{T}_{z1}$ and $\mathcal{T}_{zz}$ (each with a pre-factor of $-2$).*

However, this non one-to-one correspondence between centraliser elements and translation directions is *not* always the case. It turns out for several useful states in quantum information processing we are able to simplify the discussion greatly and are able to immediately identify the associated translation directions. Consider the following.

Suppose there are "bare" generators of the algebra $\mathfrak{su}(2^n)$ contained in the centraliser of a target pure state fixed point in the sense that $\mathrm{i}\sigma_{\mathbf{m}} \in \mathfrak{s}_\rho$. Since this implies $\mathrm{i}\sigma_{\mathbf{m}}|\psi\rangle = \pm\mathrm{i}|\psi\rangle$, then $\lambda = \pm\mathrm{i}$ and we obtain $P^S = (\mathbb{1} \pm \sigma_{\mathbf{m}})$ such that $P^S|\psi\rangle = 0$. Assume that for example $P^S = (\mathbb{1} + \sigma_{\mathbf{m}})$, then the Lindblad term is given by

$$V = \sqrt{\gamma}\sigma_{\mathbf{p}} P^S = \sqrt{\gamma}\sigma_{\mathbf{p}}(\mathbb{1} + \sigma_{\mathbf{m}}) = \begin{cases} \sqrt{\gamma}(\sigma_{\mathbf{p}} + \mathrm{i}\sigma_{\mathbf{q}'}) & \text{if} \quad \mathbf{p} \star \mathbf{m} > 0 \\ \sqrt{\gamma}(\sigma_{\mathbf{p}} - \mathrm{i}\sigma_{\mathbf{q}'}) & \text{if} \quad \mathbf{p} \star \mathbf{m} < 0 \;, \end{cases} \quad (3.57)$$

where $\mathbf{q}' \in I_0^n$ is the positive version of $\mathbf{q} = \mathbf{p} \star \mathbf{m}$ and hence $V$ is a Lindblad term of *canonical form* (mod the normalising coefficient) as introduced in Section 3.3. As we will see in the next section, in many cases one can select such a "bare" element of the centraliser which is often even an element of a maximally abelian subalgebra which we denote $\mathfrak{a}_\rho \subseteq \mathfrak{s}_\rho$.

In conclusion, one can distill an algorithm from the above considerations which provides sets of Lindblad terms that drive the system to a unique target pure state fixed point. The procedure is given by Algorithm 1.

**Algorithm 1:** Unique Target Pure State Fixed Point
Via State Symmetries

*Input:* Target state $\rho = |\psi\rangle\langle\psi|$
*Output:* Set(s) of Lindblad terms
1. Calculate the centraliser $\mathfrak{s}_\rho$
2. Determine maximally abelian subalgebra $\mathfrak{a}_\rho \subseteq \mathfrak{s}_\rho$
If possible,
  2a. Identify translation directions $\tau_{\mathbf{m}_k}$ from $\mathrm{i}\sigma_{\mathbf{m}_k} \in \mathfrak{a}_\rho$
  2b. Construct $V_k := \sigma_{\mathbf{p}_k}(\mathbb{1} \pm \sigma_{\mathbf{m}_k})$ such that $V_k|\psi\rangle = 0$
  2c. Ensure $\rho$ is unique fixed point via Theorem 8
  2d. If $\rho$ not unique, chose another of the $4^{n-1}$ choices of $\sigma_{\mathbf{p}_k}$ such that $[\sigma_{\mathbf{p}_k}, \sigma_{\mathbf{m}_k}] \neq 0$
  2e. Repeat 2b-2d
Else,
3. Determine eigenvalue $\lambda_k \neq 0$ via $s_k|\psi\rangle = \lambda_k|\psi\rangle$ for $s_k \in \mathfrak{s}_\rho$
4. Construct $V_k := \sigma_{\mathbf{p}_k}(|\lambda_k|\mathbb{1} \pm \mathrm{i}s_k)$ such that $V_k|\psi\rangle = 0$
  4a. Ensure $\mathcal{D}_0 = \cap_k \ker(V_k) = \mathrm{span}_{\mathbb{C}}\{|\psi\rangle\}$
  4b. Ensure $\rho$ is unique fixed point via Proposition 3.2.3
  4c. If $\rho$ not unique, go to 3. and choose new $s_k \in \mathfrak{s}_\rho$
5. Return Lindblad term solution set(s) $\{V_k\}$

## 3.4 Applications

This section will focus on applying the general theory of pure state fixed point existence and uniqueness discussed in Section 3.2, and in particular, make use of of the results in Section 3.3 where we presented the the problem in detail. The goal here is to provide the reader with a summary of the previous results which culminate in a simple overarching procedure to determine sets of Lindblad terms $\{V_k\}$ which drive a purely dissipative system into a given desired target fixed-point state $\rho = \rho_\infty$. The full algorithm is given by Algorithm 1 in Section 3.3.2, however here we present it again in its most basic form. We recall that it is just based on the symmetries of $\rho$ (i.e. its centraliser) and on exploiting the structure of canonical Lindblad terms ensuring uniqueness of the fixed point. In the canonical case, its basic steps are the following:

(0) fix target state $\rho$

(1) characterize $\rho$ by its *symmetries* given by the centraliser of $\rho$ in $\mathfrak{su}(N)$,

$$\mathfrak{s}_\rho := \{ s \in \mathfrak{su}(N) \,|\, [s, \rho] = 0 \}$$

(2) choose a convenient *maximally abelian subalgebra* $\mathfrak{a}_\rho \subset \mathfrak{s}_\rho$

(3) pick an appropriate set of translations $\{\tau_{\mathbf{m}}^{(k)}\}$ according to the $\{a_{\mathbf{m}}\} \subseteq \mathfrak{a}_\rho$

(4) convert selected translations $\{\tau_{\mathbf{m}}^{(k)}\}$ into a set of canonical Lindblad terms

$$\{V_k := \sigma_{\mathbf{p}}^{(k)}(\mathbb{1} \pm \sigma_{\mathbf{m}}^{(k)}) = (\sigma_{\mathbf{p}}^{(k)} + \mathrm{i}\,\sigma_{\mathbf{q}}^{(k)})\} \;,$$

(5) ensure uniqueness of fixed point $\rho_\infty$ by satisfying Theorem 8 or Proposition 3.2.3: Typically $n$ nilpotent terms $V_k$ with $[V_k, V_{k'}] = 0$ are needed (see Propositions 3.2.4 and 3.2.5).

In fact, Section 3.5 will later provide the details outlining how this algorithm can be generalised to also encompass the mixed state scenario. This shows that, most remarkably, this scheme provides a unified frame for both pure-state and mixed-state fixed-point engineering at the same time encompassing the stabilizer formalism for graph states and topological states. Clearly, the centraliser $\mathfrak{s}_\rho$ of the target state generates the corresponding *stabilizer group* $\mathbf{S}_\rho := \exp \mathfrak{s}_\rho$. Therefore, to every set of Lindblad terms $\mathcal{V} = \{V_k\}$ driving a purely dissipative system into the (unique) target state $\rho$, one immediately finds (infinitely many) equivalent sets of Lindblad terms (driving the system also into $\rho$) by coordinate transformation under the stabilizer group in the sense $\mathcal{V}' = S\mathcal{V}S^\dagger$ with $S \in \mathbf{S}_\rho$. This paves the way to a systematic way of simplifying Lindblad terms.

We start the examples with a method to engineer dissipation which drives the system to the ground state of an $n$-qubit system uniquely. Though well known, the connection to translation directions will give a new intuition for later examples.

### 3.4.1 $n$-Qubit Ground State

Trivially, a purely dissipative $n$-qubit system with $n$ local amplitude damping Lindblad terms $\{V_k\}_{k=1}^n = \{\sigma_k^+\}$ drives into the ground state $\rho_0 := \operatorname{diag}(1, 0, \ldots, 0)$. Here we use the shorthand notions $\sigma_k^+ := \frac{1}{2}\,(\sigma_x^{(k)} + \mathrm{i}\sigma_y^{(k)})$ with $\sigma_x^{(k)} := \mathbb{1}_2^{\otimes(k-1)} \otimes \sigma_x \otimes \mathbb{1}_2^{\otimes(n-k)}$ so that $\sigma_x$ appears at the $k^{\text{th}}$ place. — A geometric way to see this will be useful in the sequel.

The centralizer to the ground state $\rho_0$ in $\mathfrak{su}(N)$ can be viewed as an embedding of $\mathfrak{u}(N-1)$ in $\mathfrak{su}(N)$ of the form

$$\mathfrak{s}_{\rho_0} = \{(a \oplus M) \in \mathfrak{su}(N) \,|\, M \in \mathfrak{u}(N-1) \text{ and } a = -\operatorname{tr} M\}\,, \tag{3.58}$$

consisting of $(2^n - 1)^2$ basis elements and where a convenient choice for a basis to the maximal abelian subalgebra (torus) is given by all the $n$-fold products of $\mathbb{1}_2$ and $\sigma_z$ except $\mathbb{1}_2^{\otimes n}$

$$\mathfrak{a}_{\rho_0} = \operatorname{span}_{\mathbb{R}} \{\mathrm{i}\sigma_{\mathbf{m}} \in \mathfrak{su}(N) \,|\, m_k \in \{z, 1\}\}\,, \tag{3.59}$$

which is generated from $2^n - 1$ basis elements in total.

We note that the state vector $|\psi\rangle = |00\ldots0\rangle$ is contained in each of the $+1$ eigenspaces of each basis element of the maximally abelian subalgebra $\mathfrak{a}_{\rho_0}$ in Eqn. (3.59). By the procedure outlined following Remark 12 in Section 3.3.2, steps (3) and (4) of the above algorithm yield the known $n$ Lindblad terms of the form $V_k := \frac{1}{2}(\sigma_x^{(k)}(\mathbb{1} - \sigma_z^{(k)})) = \sigma_k^+$ such that $V_k|\psi\rangle = 0$ for all $k$. The geometric interpretation is now obvious. Each Lindblad term induces a local $z$-translation towards the north pole of that states Bloch sphere - the $|0\rangle$ state. Remarkably, each of these translation directions are one-to-one with the parametrisation of the abelian subalgebra basis $\mathfrak{a}_{\rho_0}$. The connection between the individual translation directions, the abelian subalgebra element and each amplitude damping term is

$$V_k := \tfrac{1}{2}(\sigma_x^{(k)}(\mathbb{1} - \sigma_z^{(k)})) = \sigma_k^+ \quad\longrightarrow\quad \mathcal{T}_z^{(k)}\,. \tag{3.60}$$

Recall that in Section 3.3 we introduced the concept of Lindblad terms in *canonical form*, which were those of the form $V_k = \frac{\sqrt{\gamma}}{2}(\sigma_{\mathbf{p}} + \mathrm{i}\sigma_{\mathbf{q}})$, where $[\sigma_{\mathbf{p}}, \sigma_{\mathbf{q}}] \neq 0$ and $\gamma \in \mathbb{R}^+$ is the damping coefficient. Since the Lindblad terms obtained here are precisely of this form, we can readily apply the fixed point theory provided in the same section to ensure the fixed point is unique and hence satisfy step (5) of the algorithm above. Precisely, since every Lindblad term is nilpotent and commutes with every other, then by Proposition 3.2.5 we see that $\rho_0$ is the only pure state fixed point. Moreover, it is unique (i.e. there are no mixed state fixed points) by Theorem 8.

**Algorithm 2:** Determining the Centraliser of a Target Fixed Point

*Input:* Target state $\rho = |\psi\rangle\langle\psi|$ and $\{\sigma_{\mathbf{m}}\} \in \mathcal{B}_0^n$ basis of $\mathfrak{su}(2^n)$
*Output:* Centraliser $\mathfrak{s}_\rho$ of $\rho$

1. Compute $\widehat{\mathrm{ad}}_\rho$ relative to $\mathfrak{su}(2^n)$ basis by
   1a. Computing matrix element $M_{ij} := \mathrm{tr}([\rho, \sigma_{\mathbf{m}_i}], \sigma_{\mathbf{m}_j})$, for $\sigma_{\mathbf{m}_i}, \sigma_{\mathbf{m}_j} \in \mathcal{B}_0^n$
   1b. Constructing the matrix $M \equiv \widehat{\mathrm{ad}}_\rho$
2. Determine $\ker(M)$ with $(2^n-1)^2$ basis elements $\{n_k\}$
3. Determine basis elements $s_k := \sum_{j=1}^{2^n-1} n_k[j] \cdot \sigma_{\mathbf{m}_j}$
4. Return centraliser basis elements $\{s_k \mid k = 1, \ldots, (n^2-1)^2\}$

### 3.4.2 GHZ States

For a purely dissipative $n$-qubit system, we wish to determine sets of Lindblad terms driving any initial state uniquely to the $n$-qubit GHZ state $\rho := |\psi\rangle\langle\psi|$ with $|\psi\rangle := \frac{1}{\sqrt{2}}(|00\ldots0\rangle + |11\ldots1\rangle)$. Following the algorithm presented above, step (1) and (2) requires we determine the centraliser and its maximally abelian sub algebra denoted by $\mathfrak{s}_{\mathrm{GHZ}}$ and $\mathfrak{a}_{\mathrm{GHZ}} \subset \mathfrak{s}_{\mathrm{GHZ}}$, respectively. By unitary similarity to the ground state centraliser, clearly the two structures will again have dimensions $(2^n-1)^2$ and $(2^n-1)$, respectively. Furthermore, although an analytic representation of the $n$-qubit centraliser is difficult to obtain, one can easily compute its basis by Algorithm 2.

We can however provide a closed form expression for the associated maximally abelian subalgebra as

$$\mathfrak{a}_{\mathrm{GHZ}} = \mathrm{span}_{\mathbb{R}} \{ \mathrm{i}\sigma_{xx\ldots x}, \mathrm{i}\sigma_{\mathbf{m}}, \mathrm{i}\sigma_{\mathbf{n}} \in \mathfrak{su}(N) \mid m_k \in \{z, 1\}\ ,\ n_k \in \{x, y\}\ , \\ \text{with even } \#\text{ of both } m_k = z \text{ and } n_k = y \text{ terms} \}\ , \tag{3.61}$$

where we note that there are $n$ choose $k$ for all even $k \leq n$ terms of the form $\sigma_{\mathbf{m}}$ as well as $\sigma_{\mathbf{n}}$ and thus including the $\sigma_{xx\ldots x}$ term there are a total of $(2^n-1)$ basis elements as expected.

By step (3) of the algorithm, we now want to identify basis elements of the abelian subalgebra which act locally on a joint system of qubits. However, by choosing all the elements which depend on $z$ and unities one would see that the fixed point fails to be unique by Theorem 8 as there would exist more than one, one dimensional invariant subspace of the dynamics. Therefore, here we choose $n$ abelian subalgebra elements such that by step (4) we obtain canonical Lindblad terms given by

$$\tau_{x\ldots xxxx} \quad \longrightarrow \quad V_1 = \tfrac{1}{2}(\sigma_{\mathbf{p}_1}(\mathbb{1} - \sigma_{xx\ldots x}))$$

$$\begin{aligned}
\tau_{1\ldots 11zz} &\longrightarrow V_2 = \tfrac{1}{2}(\sigma_{\mathbf{p}_2}(\mathbb{1} - \sigma_{1\ldots 11zz})) \\
\tau_{1\ldots 1zz1} &\longrightarrow V_3 = \tfrac{1}{2}(\sigma_{\mathbf{p}_3}(\mathbb{1} - \sigma_{1\ldots 1zz1})) \\
\vdots &\longrightarrow \vdots \\
\tau_{zz1\ldots 11} &\longrightarrow V_n = \tfrac{1}{2}(\sigma_{\mathbf{p}_n}(\mathbb{1} - \sigma_{zz1\ldots 11}))\ ,
\end{aligned}$$

where $\sigma_{\mathbf{p}_k}$ terms can each be chosen in $4^n - 1$ ways such that $[\sigma_{\mathbf{p}_k}, \sigma_{\mathbf{m}_k}] \neq 0$ (and thus making each $V_k$ a canonical Lindblad term). The goal is to now choose the $\sigma_{\mathbf{p}_k}$ terms appropriately to ensure uniqueness of the target state fixed point by establishing the conditions of Theorem 8 are satisfied. Since the target state is the only pure state simultaneously in every eigenspace of each $\sigma_{\mathbf{m}_k}$ term, we see that it is the only pure state fixed point. Now we have to show there are no "intrinsic" higher rank fixed points by showing there exists exists no invariant subspace $\mathcal{S} \subseteq \mathcal{H}$ such that $V_k\mathcal{S} \subseteq \mathcal{S}$ for all $k$. It can easily verified that even in the 3-qubit case, there does not exist a set of $\{\sigma_{\mathbf{p}_k}\}$ terms which make $\{V_k\}$ commute with one another and thus we cannot use Proposition 3.2.5 as in the ground state scenario. Since there are $n$ Lindblad terms and $4^{n-1}$ choices of realising each Lindblad term (via the choice of $\sigma_{\mathbf{p}_k}$ such that $[\sigma_{\mathbf{p}_k}, \sigma_{\mathbf{m}_k}] \neq 0$), there are $4^{n(n-1)}$ possible sets of $\{\sigma_{\mathbf{p}_k}\}$ terms to choose from in order to make the fixed point unique. Below we will show some concrete examples proving it is very easy to obtain the fixed point uniquely by invoking Proposition 3.2.4.

By Eqn. (3.61), the maximally abelian subalgebra for the 2-qubit system is given by

$$\mathfrak{a}_{\mathrm{GHZ}} = \mathrm{span}_{\mathbb{R}}\,\{\mathrm{i}\sigma_{xx}, \mathrm{i}\sigma_{yy}, \mathrm{i}\sigma_{zz}\}\ . \tag{3.62}$$

A few possible sets of Lindblad terms which drive the system to the target unique fixed point are given in Table 3.1.

**Table 3.1:** Three Solution Sets of Lindblad Terms for Unique 2-Qubit Bell State Fixed Point

| Translation Directions | Lindblad Terms | Factored Lindblad Terms |
|---|---|---|
| $\tau_{xx}$ | $V_1 = \frac{1}{2}(\sigma_{y1} + \mathrm{i}\sigma_{zx})$ | $V_1 = \frac{1}{2}\sigma_{y1}(\mathbb{1}_4 - \sigma_{xx})$ |
| $\tau_{zz}$ | $V_2 = \frac{1}{2}(\sigma_{1x} + \mathrm{i}\sigma_{zy})$ | $V_2 = \frac{1}{2}\sigma_{1x}(\mathbb{1}_4 - \sigma_{zz})$ |
| $\tau_{xx}$ | $V_1 = \frac{1}{2}(\sigma_{y1} + \mathrm{i}\sigma_{zx})$ | $V_1 = \frac{1}{2}\sigma_{y1}(\mathbb{1}_4 - \sigma_{xx})$ |
| $-\tau_{yy}$ | $V_2 = \frac{1}{2}(\sigma_{1x} + \mathrm{i}\sigma_{yz})$ | $V_2 = \frac{1}{2}\sigma_{1x}(\mathbb{1}_4 + \sigma_{yy})$ |
| $\tau_{zz}$ | $V_1 = \frac{1}{2}(\sigma_{x1} + \mathrm{i}\sigma_{yz})$ | $V_1 = \frac{1}{2}\sigma_{x1}(\mathbb{1}_4 - \sigma_{zz})$ |
| $-\tau_{yy}$ | $V_2 = \frac{1}{2}(\sigma_{xy} + \mathrm{i}\sigma_{z1})$ | $V_2 = \frac{1}{2}\sigma_{xy}(\mathbb{1}_4 + \sigma_{yy})$ |

Some remarks are in order to relate this method to known solutions which exist in the literature. Another method of obtaining a unique pure state fixed points is via unitary conjugation of the ground state to the target state [32, 51]. Since there exists a unitary conjugation which rotates the ground state into the target state, we can apply such a conjugation to each Lindblad term that drives the system to the ground state such that the new system will have the target state as its unique fixed point. Here we will show that for this particular example, the change of basis approach results in the same translation directions than if we were to engineer the Lindblad terms directly from the centraliser method.

From the unique ground state fixed point example, we know that choosing the Lindblad terms $V_k = \sigma_k^+$ for $k = 1, 2$ results in the ground state as the unique fixed point. Defining the unitary matrix

$$U := \exp(-\mathrm{i}\tfrac{\pi}{8}(\sigma_{xy} + \sigma_{yx}))\ , \tag{3.63}$$

we see that $U|00\rangle = |\psi\rangle$ and hence the Lindblad generator with Lindblad terms $V_1' = U\sigma_1^+ U^\dagger$ and $V_2' = U\sigma_2^+ U^\dagger$ will drive the system to the 2-qubit GHZ state uniquely. Using Algorithm 3 (given at the end of this Section), the associated translation directions are

$$\Gamma' \longrightarrow \tau_{xx} - \tau_{yy} \tag{3.64}$$

Comparing with Table 3.1, we see that although the Lindblad terms are different from those obtained via unitary conjugation, the overall translation directions are *equivalent* to the second solution set in the table. This equality of the translation directions from these two inequivalent engineering schemes does not hold in general. As we will now show, in the 3-qubit scenario, this same unitary rotation approach does not reproduce the simple solutions obtained by the centraliser scheme. Using Algorithm 3 again, we can calculate the associated translation directions to a solution set of Lindblad terms. Table 3.2 gives a sample of the output sets of translation directions and their corresponding canonical Lindblad terms which give the 3-qubit GHZ-state as the unique fixed point.

**Table 3.2:** Lindblad Terms for Unique 3-Qubit GHZ Fixed Point

| Translation Directions | Lindblad Terms | Factored Lindblad Terms |
|---|---|---|
| $-\tau_{yyx}$ | $V_1 = \frac{1}{2}(\sigma_{x11} + \mathrm{i}\sigma_{zyx})$ | $V_1 = \frac{1}{2}\sigma_{x11}(\mathbb{1}_8 + \sigma_{yyx})$ |
| $-\tau_{yxy}$ | $V_2 = \frac{1}{2}(\sigma_{yzy} + \mathrm{i}\sigma_{1y1})$ | $V_2 = \frac{1}{2}\sigma_{yzy}(\mathbb{1}_8 + \sigma_{yxy})$ |
| $-\tau_{xyy}$ | $V_3 = \frac{1}{2}(\sigma_{xyx} + \mathrm{i}\sigma_{11z})$ | $V_3 = \frac{1}{2}\sigma_{xyx}(\mathbb{1}_8 + \sigma_{xyy})$ |
| $\tau_{1zz}$ | $V_1 = \frac{1}{2}(\sigma_{11x} + \mathrm{i}\sigma_{1zy})$ | $V_1 = \frac{1}{2}\sigma_{11x}(\mathbb{1}_8 - \sigma_{1zz})$ |
| $-\tau_{yyx}$ | $V_2 = \frac{1}{2}(\sigma_{xyx} + \mathrm{i}\sigma_{z11})$ | $V_2 = \frac{1}{2}\sigma_{xyx}(\mathbb{1}_8 + \sigma_{yyx})$ |
| $\tau_{xxx}$ | $V_3 = \frac{1}{2}(\sigma_{1y1} + \mathrm{i}\sigma_{xzx})$ | $V_3 = \frac{1}{2}\sigma_{1y1}(\mathbb{1}_8 - \sigma_{xxx})$ |
| $\tau_{xxx}$ | $V_1 = \frac{1}{2}(\sigma_{11y} + \mathrm{i}\sigma_{xxz})$ | $V_1 = \frac{1}{2}\sigma_{11y}(\mathbb{1}_8 - \sigma_{xxx})$ |
| $\tau_{zz1}$ | $V_2 = \frac{1}{2}(\sigma_{x1\alpha} + \mathrm{i}\sigma_{yz\alpha})$ | $V_2 = \frac{1}{2}\sigma_{x1\alpha}(\mathbb{1}_8 - \sigma_{zz1})$ |
| $\tau_{1zz}$ | $V_3 = \frac{1}{2}(\sigma_{\alpha x1} + \mathrm{i}\sigma_{\alpha yz})$ | $V_3 = \frac{1}{2}\sigma_{\alpha x1}(\mathbb{1}_8 - \sigma_{1zz})$ |

where $\alpha \in \{1, x, z\}$.

Using the same change of basis approach as in the 2-qubit case, we obtain the trans-

lation directions

$$\Gamma' \longrightarrow \tfrac{3}{4}(\tau_{xxx} - \tau_{zzz} - \tau_{xyy} - \tau_{yxy} - \tau_{yyx}) + \tfrac{1}{4}(\tau_{11z} + \tau_{1z1} + \tau_{z11}) \ , \tag{3.65}$$

which clearly is a much more complicated set of translation elements as compared to the solution set number three in the table. Thus, using a simple change of basis approach may, in a sense, complicate the dynamics which the system undergoes as it is driven towards the unique fixed point. Although the form of a Lindblad term in canonical form looks somewhat non-standard, as we will now show, they can be factored so they can resemble typical noise.

First we need to discuss the notion of "quasi-local" Lindblad terms. Quasi-local Lindblad terms are those which act as an identity on at least one of the individual qubit subsystems [32]. Notably, in [51] the authors proved that GHZ states and $W$-states cannot be obtained by only using such quasi-local Lindblad terms. Then in their latest work, [52], they provided an engineering scheme where they showed that by choosing Lindblad terms in quasi-local form *and* restricting the initial Hilbert space to a specific subspace - they could drive every system with support in the subspace to the target GHZ state. It turns out that the specific quasi-local Lindblad terms they present are in fact in *canonical* Lindblad form. The difference in our method, is that we allow for *global* dissipative processes i.e. they do not allow for canonical Lindblad terms which induce translation directions of the form $\tau_{\mathbf{m}}$ with $m_i \neq \mathbb{1}$ for each $m_i \in \mathbf{m}$. Moreover, the subspace in which every density matrix can be driven to the target state exactly corresponds to the $+1$ eigenspace of the "global" element $\sigma_{xx\ldots x}$ in the abelian subalgebra used to construct the non-local dissipative component. That being said, it seems that our approach can provide information on which subspaces we can restrict to in using their subspace-restricted quasi-local scheme.

Specifically, the (quasi-local) Lindblad terms that were presented were $V_1 = D \otimes \mathbb{1}$ and $V_2 = \mathbb{1} \otimes D$, where $D := \frac{1}{2}(\sigma^+ \otimes (\sigma_z + \mathbb{1}) - \sigma^- \otimes (\sigma_z - \mathbb{1}))$. In fact, by expanding we see that $D = \frac{1}{2}(\sigma_{x\mathbb{1}} + \mathrm{i}\sigma_{yz})$, which is of canonical Lindblad form and thus so are $V_1$ and $V_2$. This solution is given by the third solution set using the centraliser method in Table 3.2 setting $\alpha = \mathbb{1}$.

### 3.4.3 Stabiliser States

The following basic remarks on stabiliser states which can be found in [19]. Define the $n$-qubit Pauli-group $\mathcal{G}_{\mathcal{B}^n}$ as the group generated by the basis $\mathcal{B}^n$ given by Eqn. (2.30) of tensor products of Pauli matrices and identities, along with multiplicative factors $\pm 1$, $\pm\mathrm{i}$. A quantum state $|\psi\rangle$ is called a stabiliser state if it is the only simultaneous state vector to the eigenvalue $+1$ of a special subgroup of $\mathcal{G}_{\mathcal{B}^n}$ - a so-called *stabiliser* group. Given a stabiliser state $|\psi\rangle$ and its corresponding stabiliser group

$$S(|\psi\rangle) = \{S_k \in \mathcal{G}_{\mathcal{B}^n} \mid S_k|\psi\rangle = |\psi\rangle\} \ , \tag{3.66}$$

Therefore, for constructing a set of Lindblad terms by the algorithm presented at the start of this section, steps (1)-(3) are essentially completed. For any stabiliser state, taking the generators of its stabiliser group and multiplying by a complex factor of i results in a basis set for an abelian subalgebra of the centraliser of $\rho = |\psi\rangle\langle\psi|$. Then, these elements can then be identified with the corresponding necessary translation directions. The only step remaining is to construct the canonical Lindblad terms and ensure uniqueness of the fixed point by steps (4) and (5). To provide concrete examples, we consider a special subclass of stabiliser states well studied in the literature known as Graph States [21, 20].

Consider an undirected simple finite graph $G = (V, E)$ which is associated to a set of vertices $V := \{1, 2, \ldots, n\}$ and edges $E \subset V \times V$. It is common in quantum information processing to only consider simple graphs which are those that do not have edges which connect a node to itself. Given some vertex $k \in V$, the neighbourhood of $a$ is the set of vertices $b \in V$ such that an edge element $(k, b) \in E$ connects the two. We denote such a neighbourhood as $N_k$.

We can associate a graph to a quantum state as follows. To each graph $G = (V, E)$, each node represents a qubit ($|V| = n$) and hence the total Hilbert space represented by the entire graph is $\mathcal{H}^n = (\mathbb{C}^2)^{\otimes n}$. Next, we define the so-called stabilizer operators which are associated to each vertex of the graph as

$$S_k := \sigma_{x,k} \prod_{b \in N_k} \sigma_{z,b} \ , \tag{3.67}$$

for all $k \in V$ where we recall that $\sigma_{p,k}$ for $p \in \{x, y, z\}$ is an operator acting locally on the $k^{th}$ qubit (cf. Eqn. (2.64)). The stabiliser group $S(|\psi\rangle)$ is therefore the group generated by these $n$ stabiliser operators and the *graph state* $|G\rangle \in \mathcal{H}^n$ is the unique common eigenvector to the eigenvalue $+1$ of each of the independent stabiliser operators in the sense that

$$S_k|G\rangle = |G\rangle \ , \quad \text{for all } k \in V \ . \tag{3.68}$$

It is known that certain types of stabiliser states - notably many graph states - can be obtained as unique fixed points by using purely dissipative means [53, 32]. Specifically, the authors of [32] showed how to construct quasi-local Lindblad terms to obtain certain types of graph states as unique fixed points. The operators they constructed to accomplish this task were actually in *canonical form* and as we will see now, fit into our centraliser generated construction method.

Step (4) of the algorithm then dictates we construct canonical Lindblad terms by choosing elements of a maximally abelian subalgebra of the centraliser. In this special scenario, there is no need to compute a maximally abelian subalgebra as we have already obtained an $n$-dimensional subalgebra of the centraliser. We then construct the Lindblad terms as

$$V_k = \tfrac{1}{2}\sigma_{\mathbf{p}}(\mathbb{1} - S_k) \equiv \tfrac{1}{2}\sigma_{\mathbf{p}}(\mathbb{1} - \sigma_{\mathbf{m}}) = \tfrac{1}{2}(\sigma_{\mathbf{p}} \pm \mathrm{i}\sigma_{\mathbf{q}}) \ . \tag{3.69}$$

We now have a direct geometric interpretation of the stabiliser group elements in terms of the systems translation directions induced on each qubits Bloch sphere in the sense that

$$S_k \equiv \sigma_{\mathbf{m}_k} \longrightarrow \tau_{\mathbf{m}_k} \longrightarrow V_k \ , \quad \text{for all } k \in V \ . \tag{3.70}$$

Therefore, we once again have shown that the current state-of-the-art known solutions for a large class of paradigmatic quantum states used in quantum information processing are in fact a special case of the general theory presented here using Lindblad terms of canonical form. Specific examples are provided for $2, 3$ and $4$ qubit systems in Table 4.3.

**Table 3.3:** Example Graph States

| Graph | Abelian Subalgebra $\mathfrak{a}$ | Translation Operators | Lindblad Terms |
|---|---|---|---|
|  | $\langle \mathrm{i}\sigma_{xz}, \mathrm{i}\sigma_{zx} \rangle$ | $\tau_{xz}$ | $V_1 = \sigma_{y1} + \mathrm{i}\,\sigma_{zz}$ |
| | | $\tau_{zx}$ | $V_2 = \sigma_{1y} + \mathrm{i}\,\sigma_{zz}$ |
|  | $\langle \mathrm{i}\sigma_{xz1}, \mathrm{i}\sigma_{zxz}, \mathrm{i}\sigma_{1zx} \rangle$ | $\tau_{xz1}$ | $V_1 = \sigma_{y11} + \mathrm{i}\,\sigma_{zz1}$ |
| | | $\tau_{zxz}$ | $V_2 = \sigma_{1y1} + \mathrm{i}\,\sigma_{zzz}$ |
| | | $\tau_{1zx}$ | $V_3 = \sigma_{11y} + \mathrm{i}\,\sigma_{1zz}$ |
| 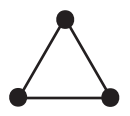 | $\langle \mathrm{i}\sigma_{xzz}, \mathrm{i}\sigma_{zxz}, \mathrm{i}\sigma_{zzx} \rangle$ | $\tau_{xzz}$ | $V_1 = \sigma_{y11} + \mathrm{i}\,\sigma_{zzz}$ |
| | | $\tau_{zxz}$ | $V_2 = \sigma_{1y1} + \mathrm{i}\,\sigma_{zzz}$ |
| | | $\tau_{zzx}$ | $V_3 = \sigma_{11y} + \mathrm{i}\,\sigma_{zzz}$ |
| 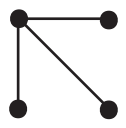 | $\langle \mathrm{i}\sigma_{xzzz}, \mathrm{i}\sigma_{zx11}, \mathrm{i}\sigma_{z1x1}, \mathrm{i}\sigma_{z11x} \rangle$ | $\tau_{xzzz}$ | $V_1 = \sigma_{y11z} + \mathrm{i}\,\sigma_{zzz1}$ |
| | | $\tau_{zx11}$ | $V_2 = \sigma_{1y11} + \mathrm{i}\,\sigma_{zz11}$ |
| | | $\tau_{z1x1}$ | $V_3 = \sigma_{11y1} + \mathrm{i}\,\sigma_{z1z1}$ |
| | | $\tau_{z11x}$ | $V_4 = \sigma_{111y} + \mathrm{i}\,\sigma_{z11z}$ |
| 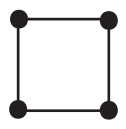 | $\langle \mathrm{i}\sigma_{xz1z}, \mathrm{i}\sigma_{zxz1}, \mathrm{i}\sigma_{1zxz}, \mathrm{i}\sigma_{z1zx} \rangle$ | $\tau_{xz1z}$ | $V_1 = \sigma_{y111} + \mathrm{i}\,\sigma_{zz1z}$ |
| | | $\tau_{zxz1}$ | $V_2 = \sigma_{1y11} + \mathrm{i}\,\sigma_{zzz1}$ |
| | | $\tau_{1zxz}$ | $V_3 = \sigma_{11y1} + \mathrm{i}\,\sigma_{1zzz}$ |
| | | $\tau_{z1zx}$ | $V_4 = \sigma_{111y} + \mathrm{i}\,\sigma_{z1zz}$ |

### 3.4.4 Topological-State Subspaces: Toric Code

In this section we will focus on special kinds of stabiliser states - ones which form an invariant subspace which is usually called a *toric code* space. Thus instead of wanting to engineer Lindblad terms which drive the system to a unique pure state fixed point, we now want to drive the system to a unique invariant *subspace.* As a brief introduction, we provide a basic summary of key points in [39, 1, 26] and [42].

Consider a two dimensional graph which forms an $L \times L$ square lattice with periodic boundary conditions and hence forms a torus. Here, we place a qubit on each edge such that the system comprises of $n = 2L^2$ qubits. Now define the star and plaquette stabiliser operators

$$A_s := \bigotimes_{i \in \mathrm{star}(s)} \sigma_{x,i}\ , \quad \text{and} \quad B_p := \bigotimes_{i \in \mathrm{boundary}(p)} \sigma_{z,i}\ , \tag{3.71}$$

where star($s$) is the set of (the 4) neighbouring qubits connected by one edge at each vertex $s$ and boundary($p$) is the set of (the 4) qubits on each edge of a plaquette face $p$.

Define the graph code space $\mathcal{C}_n \subset \mathcal{H}^n$ as

$$\mathcal{C}_n := \mathrm{span}_{\mathbb{C}}\left\{|\psi\rangle \mid A_s|\psi\rangle = |\psi\rangle \quad \text{and} \quad B_p|\psi\rangle = |\psi\rangle \ , \ \text{for all } s,p\right\} \ , \tag{3.72}$$

and furthermore, noting that $\prod_s A_s = \mathbb{1}^{\otimes n}$ and $\prod_p B_p = \mathbb{1}^{\otimes n}$, it common knowledge that the dimension of the code space $\mathcal{C}_n$ is four. To determine the basis states of this subspace it is useful to define the *logical* operators which act on the subspace $\mathcal{C}_n$ as

$$\overline{X} := \prod_{k\in c} \sigma_{x,k} \quad \text{and} \quad \overline{Z} := \prod_{k\in c'} \sigma_{z,k} \ , \tag{3.73}$$

where $c$ is a loop on the lattice and $c'$ is a cut on the dual lattice. Since the toric code is a 4-dimensional subspace, we see that there are 4 logical operators $\overline{X}_1, \overline{X}_2, \overline{Z}_1, \overline{Z}_2$ which act on two two qubit encoded states in $\mathcal{C}_n$. Furthermore, it can be shown that a basis for $\mathcal{C}_n$ is given by

$$|\psi_1\rangle := \prod_s \tfrac{1}{\sqrt{2}}(\mathbb{1} + A_s)|00\ldots 0\rangle \ , \tag{3.74}$$

and $|\psi_2\rangle := \overline{X}_1|\psi_1\rangle$, $|\psi_3\rangle := \overline{X}_2|\psi_1\rangle$, and $|\psi_4\rangle := \overline{X}_1\overline{X}_2|\psi_1\rangle$.

Another way to describe the code space is by defining a system Hamiltonian as

$$H = \tfrac{1}{2}\sum_s^{n/2}(\mathbb{1} - A_s) + \sum_p^{n/2}(\mathbb{1} - B_p) = \tfrac{1}{2}\sum_k^{n}(\mathbb{1} - S_k) \tag{3.75}$$

where $S_k$ is any stabiliser operator and therefore the code space is then given by the kernel of $H$ i.e. $\ker(H) = \mathcal{C}_n$. As in the graph state case, we can again easily obtain a purely dissipative way to now drive any initial state to the protected code space $\mathcal{C}_n \subset \mathcal{H}^n$ by associating the stabiliser group of the subspace to an abelian subalgebra of the now, *joint* centraliser of the code subspace. The relationship can be summarised as

$$S_k \equiv \sigma_{\mathbf{m}_k} \longrightarrow \tau_{\mathbf{m}_k} \longrightarrow V_k \ , \quad \text{for all } S_k \ , \tag{3.76}$$

and thus we obtain steps (1)-(4) of the algorithm presented at the beginning of this section and all that remains is to prove there are no other fixed points (and hence no other subspace is invariant). Since the corresponding fixed point space associated to pure state fixed points is given by $\mathcal{D} = \cap_k \ker(V_k) = \ker(\sum_k V_k^\dagger V_k) = \ker(\sum_k(\mathbb{1} - S_j)^2) = \ker(\sum_k 2(\mathbb{1} - S_j)) = \ker(H) = \mathcal{C}_n$, step (5) then requires we would have to show that there is no other invariant subspace contained in $\mathcal{D}^\perp$ via Theorem 7 and Corollary 3.2.3. Depending on the specific example and size of the lattice, this can be proven by invoking Propositions 3.2.4 or 3.2.5 using the degeneracy of the canonical Lindblad terms cf. Propositions 3.3.2 and 3.3.3.

### 3.4.5 W States

Suppose we want to obtain the $n$-qubit W-state $\rho = |\psi\rangle\langle\psi|$, where

$$|\psi\rangle = \tfrac{1}{\sqrt{n}}(|100\ldots 00\rangle + |010\ldots 00\rangle + \cdots + |000\ldots 01\rangle) \tag{3.77}$$

as the unique fixed point. Furthermore, it is a classic result that the $W$-state is *not* a graph/stabiliser state (see for example [20]) and thus the considerations of the previous examples cannot be applied here. Nonetheless, the algorithm presented in the introduction to this section allows us to obtain many solution sets of Lindblad terms which allows us to obtain the target state as the unique fixed point.

First, we note that Proposition 3.3.1 shows that the only Lindblad generator with a single Lindblad term of canonical form which contains the W-state in its fixed point set is one which is associated to the centraliser element $\sigma_{zz\ldots z}$ (with the target state being contained in the $-1$ eigenspace of $\sigma_{zz\ldots z}$). Since there does not exist a further $n-1$ Lindblad terms of canonical form its clear that we cannot obtain it as the unique fixed point by only Lindblad terms of canonical form. Here we present one particularly simple method of obtaining the fixed point uniquely based on choosing special operators contained in the W-states centraliser (which are not contained in the maximally abelian subalgebra). Thus this example (and the following symmetric Dicke state example) will serve to illustrate that by knowing the centraliser, one can use a wide variety of possible solutions including those of the flavour presented here.

Either by calculating the centraliser directly by Algorithm 2, or by noticing the fact that

$$\begin{aligned} (\mathbb{1}-\sigma_z)\otimes\sigma_x^{\otimes n-1}|\psi\rangle &= |11\ldots1\rangle\ , \\ \sigma_x\otimes(\mathbb{1}-\sigma_z)\otimes\mathbb{1}^{\otimes n-2}|\psi\rangle &= |11\ldots1\rangle\ , \\ &\vdots \\ \sigma_x^{\otimes n-1}\otimes(\mathbb{1}-\sigma_z)|\psi\rangle &= |11\ldots1\rangle\ , \end{aligned}$$

and hence operators made by differences of the form

$$s_{(1,2)} := (\mathbb{1}-\sigma_z)\otimes\sigma_x^{\otimes n-1} - \sigma_x\otimes(\mathbb{1}-\sigma_z)\otimes\mathbb{1}^{\otimes n-2}\ , \tag{3.78}$$

are contained in the centraliser (by including the complex factor i) $\mathfrak{s}_{\mathrm{W}}$ of the target state such that $s_{(1,2)}|\psi\rangle = 0$. Henceforth we let $s_{(i,j)} \in \mathfrak{s}_{\mathrm{W}}$ denote the difference operator of the above form where $\frac{1}{2}(\mathbb{1}-\sigma_z)$ acts on the $i^{th}$ and $j^{th}$ qubit. By the procedure outlined prior to Remark 12 in Section 3.3.2, we can associate these elements to Lindblad terms. That is, in this case the "shifted " centraliser terms are actually trivial in the sense that

$$P_{s_{(i,j)}} := |\lambda|\mathbb{1} \pm s_{(i,j)} = s_{(i,j)}\ , \quad \text{since} \quad s_{(i,j)}|\psi\rangle = 0\ , \quad \text{and hence } \lambda = 0\ . \tag{3.79}$$

Therefore, the Lindblad terms are given by $V_k := \sigma_{\mathbf{p}_k} P_{s_{(i,j)}} = \sigma_{\mathbf{p}}(s_{(i,j)})$ such that $[\sigma_{\mathbf{p}}, s_{(i,j)}] \neq 0$ and $\mathrm{tr}(\sigma_{\mathbf{p}} s_{(i,j)}) = 0$. The easiest choice of $\sigma_{\mathbf{p}_k}$ terms is to set $\sigma_{\mathbf{p}_k} = \sigma_{xx,\ldots,x}$ for all $k$ and hence each Lindblad term is of the form (in a slight abuse of notation)

$$V_k := \sigma_{xx,\ldots,x}(s_{(i,j)}) = \sigma_i^+ - \sigma_j^+\ , \tag{3.80}$$

where we recall that $\sigma_i^+$ and $\sigma_j^+$ are the amplitude damping terms acting locally on qubits $i$ and $j$, respectively, such that $i \neq j$. Choosing $V_1$ to be of canonical Lindblad form such that the associated global translation direction is given by $-\tau_{zz,\ldots,z}$ and the remaining $n-1$ Lindblad terms of the local form of Eqn. (3.80) acting on separate qubit subsystems one then has completed steps (1)-(4) of the algorithm (without requiring a maximally abelian subalgebra) and therefore all we need to show is that the fixed point is unique (via step (5)).

Choosing the $n-1$ Lindblad terms $\{V_k\}_{k=2}^n$ which are of the form of Eqn. (3.80) such that $i = 1, \ldots, n-1$ and $j = i+1$ and therefore $V_2 = \sigma_1^+ - \sigma_2^+$, $V_2 = \sigma_2^+ - \sigma_3^+$,

etc. we have that

$$\bigcap_{k=2}^{n} \ker(V_k) = \mathrm{span}_{\mathbb{C}} \{|00\dots 0\rangle, |\psi\rangle\} , \tag{3.81}$$

and since $|00\dots 0\rangle$ is not contained in the nullspace of the canonical Lindblad term $V_1 = \frac{1}{2}\sigma_{\mathbf{p}_1}(\mathbb{1} + \sigma_{zz\dots z})$, the target state is the *unique pure state* fixed point since

$$\bigcap_{k=1}^{n} \ker(V_k) = \mathrm{span}_{\mathbb{C}} \{|\psi\rangle\} . \tag{3.82}$$

By choosing an appropriate $\sigma_{\mathbf{p}_1}$ term, Proposition 3.2.4 can be used to show that there exists *no other* invariant subspace perpendicular to $|\psi\rangle$ and hence by Proposition 3.2.3 the fixed point would be unique.

Alternatively, we provide one more example of a non-trivial solution for a 3-qubit system. First note that there exists a (non-trivial) three dimensional abelian subalgebra of the centraliser spanned by (here ignoring their complex coefficients of i)

$$\begin{aligned} a_1 &:= \sigma_{zzz} , \\ a_2 &:= \sigma_{xx1} + \sigma_{x1x} + \sigma_{1xx} + \sigma_{yy1} + \sigma_{y1y} + \sigma_{1yy} , \quad \text{and} \\ a_3 &:= \sigma_{xxz} + \sigma_{xzx} + \sigma_{zxx} + \sigma_{yyz} + \sigma_{yzy} + \sigma_{zyy} . \end{aligned}$$

Using the same construction as before by shifting the centraliser elements (cf. Eqns. (3.79) and (3.80)) we see that step (4) of the algorithm gives

$$V_1 := \tfrac{1}{2}(\sigma_{x11}(\mathbb{1} + a_1)) = \tfrac{1}{2}(\sigma_{x11} - \mathrm{i}\sigma_{yzz}) , \quad V_2 := \sigma_{1x1}(4\mathbb{1} - a_2) , \quad \text{and} \quad V_3 := \sigma_{11x}(4\mathbb{1} - a_3) \tag{3.83}$$

which suffice to drive the system to the unique target state.

To conclude, some remarks on the relation of these methods to known solutions are in order. As the final details of this solution method were being completed, the authors of [52] provided two (quasi-local) engineering schemes to obtain the W-state as the unique fixed point (1) by purely dissipative means while restricting the initial states to a particular subspace and (2) by using an additional quasi-local Hamiltonian term and no longer restricting the initial support space. Here our method differs in the sense that we allow for environment dynamics to act globally on the system and hence do not restrict any of our Lindblad terms to act as an identity on any number of individual qubits (which would make them quasi-local). We briefly note that scheme (2) involved quasi-local Lindblad terms which were precisely of the form used in our first method here which were the difference of neighbouring qubit atomic raising operators. By including the global Lindblad term $V_1$ we were able to obtain the target fixed point purely dissipatively without resorting to adding additional Hamiltonian drift dynamics. Finally, the short proof of principle 3-qubit case example solution given by Eqn. (3.83) is one example of a new solution this centraliser method yields.

### 3.4.6 Outlook: Symmetric Dicke State

Suppose we would like to obtain the target state $\rho = |\psi\rangle\langle\psi|$, where

$$|\psi\rangle = \binom{n}{n/2}^{-\frac{1}{2}} (|11..10\dots 0\rangle + \text{all permutations}) ,$$

**Algorithm 3:** Determining the Associated Translation Directions

*Input:* Lindblad terms $\{V_k\}_{k=1}^{r}$ and $\{\sigma_{\mathbf{m}}\} \in \mathcal{B}_0^n$ basis
*Output:* Translation direction and scaling factor

1. Calculate $\widehat{\Gamma}$ in vec-representation
2. Change the basis to coherence vector representation by calculating $\Gamma' := U\widehat{\Gamma}U^\dagger$, with $U := (\mathrm{vec}\,(\sigma_{\mathbf{m}_1}), \ldots, \mathrm{vec}\,(\sigma_{\mathbf{m}_{n-1}}), \mathrm{vec}\,\mathbb{1}_n)$
3. For $1 \leq k < n$, if the matrix elements $[\Gamma']_{k,n} \neq 0$, then return "Scaling factor of $[\Gamma']_{k,n}$ in the $\mathbf{m}_k$ direction" where there are an even number of both 1 and 0 states. Exactly as in the W-state case, it can be shown that the only Pauli basis matrices which have $|\psi\rangle$ in their eigenspaces are $\sigma_{zz\ldots z}$, $\sigma_{xx\ldots x}$, and $\sigma_{yy\ldots y}$ thus determining the maximally abelian subalgebra of the centraliser in order to use its "bare" Pauli matrix generators will not suffice for this class of states either.

For one possible solution set we identify the first Lindblad term to be in canonical form and be associated to the centraliser element $\sigma_{xx\ldots x}$ and hence translation direction $\mathcal{T}_{xx\ldots x}$. For the remaining Lindblad terms we use those which are precisely the ones which are generalisations of those chosen to obtain the W-state uniquely. That is, we select centraliser elements which are differences of operators which contain $\frac{n}{2}$ terms of the form $(\mathbb{1} - \sigma_z)$ and the remaining $\frac{n}{2}$ terms in the tensor product being equal to $\sigma_x$. Since $|\psi\rangle$ is then a nullvector of each of these centraliser elements, we can construct the Lindblad terms in step (4) of the algorithm again by simply multiplying each of them by a global $\sigma_{xx\ldots x}$ term to obtain terms of the form (for example)

$$V_k = \sigma^+ \otimes \sigma^+ \otimes \cdots \otimes \sigma^+ \otimes \mathbb{1}^{\otimes(\frac{n}{2})} - \sigma^+ \otimes \sigma^+ \otimes \cdots \otimes \mathbb{1} \otimes \sigma^+ \otimes \mathbb{1}^{\otimes(\frac{n}{2}-1)} \ , \tag{3.84}$$

where in general the Lindblad terms will be composed of the difference of two operators each of which will have $\frac{n}{2}$ amplitude damping terms. To complete the algorithm and satisfy step (5) we must have that $|\psi\rangle$ as the unique element in the joint kernel of each Lindblad term and that there exists no invariant subspace perpendicular to $|\psi\rangle$ which by by Proposition 3.2.3 would imply the fixed point would be unique.

For an explicit example, one possible solution set of Lindblad operators to obtain the 4-qubit symmetric Dicke state is given by (using the previously introduced shorthand notation)

$$V_1 = \tfrac{1}{2}(\sigma_{y111} + \mathrm{i}\sigma_{zxxx}) \ , \quad V_2 = \sigma_1^+\sigma_2^+ - \sigma_1^+\sigma_3^+ \ , \quad V_3 = \sigma_2^+\sigma_3^+ - \sigma_2^+\sigma_4^+ \ , \quad V_4 = \sigma_3^+\sigma_4^+ - \sigma_1^+\sigma_3 \ .$$

It is very simple to arrive at other possible solutions for the four qubit case. However, at the time of this thesis submission, we have yet to determine a simple solution set for the six qubit scenario. The simplest generalisations of the four qubit solution set result in $\dim(\cap_k \ker(V_k)) \geq 8$. Thus it is currently an open problem to determine possible sets of five Lindblad terms which consist of atomic raising operator differences - a simple calculation reveals there are 190 possible Lindblad terms of the desired form.

## 3.5 Engineering Mixed State Fixed Points

### 3.5.1 Motivation and Lindblad Terms of Generalised Canonical Form

This section will focus on techniques to engineer mixed state fixed points using a generalised method of the one used in Section 3.3 to obtain pure state fixed points. We recall that the fixed point set of a Markovian semigroup of quantum channels admits the splitting $\mathcal{F}(\Phi_\Gamma) = \mathrm{conv}\,\{\mathcal{F}_\mathcal{D} \cup \mathcal{F}_{\mathcal{D}^\perp}\}$ where

$$\mathcal{F}_\mathcal{D} := \{\rho \in \mathfrak{pos}_1(N) \mid \Gamma(\rho) = 0,\ \text{such that}\ \mathrm{supp}(\rho) \subseteq \mathcal{D}\}\ ,$$

and

$$\mathcal{F}_{\mathcal{D}^\perp} := \{\rho \in \mathfrak{pos}_1(N) \mid \Gamma(\rho) = 0,\ \text{such that}\ \mathrm{supp}(\rho) \subseteq \mathcal{D}^\perp\}\ , \tag{3.85}$$

where $\mathcal{D}$ is given by Eqn. (3.14) and called the complete dark state space. The fixed point set $\mathcal{F}_\mathcal{D}$ consists of all density matrices that are either pure states or mixed states which are a convex combination of pure states that are themselves fixed points whereas $\mathcal{F}_{\mathcal{D}^\perp}$ consists of so called *intrinsic* higher rank mixed state fixed points - precisely those which can not be decomposed into pure state fixed points. This splitting was of central importance to engineer pure state fixed points (e.g. see Proposition 3.2.3) since the trick was to show that there were no intrinsic higher rank fixed points (and hence $\mathcal{F}_{\mathcal{D}^\perp} = \varnothing$) which implied $\mathcal{F}(\Phi_\Gamma) = \mathcal{F}_\mathcal{D}$. Some preliminary results showing sufficient conditions for when $\mathcal{F}_{\mathcal{D}^\perp} = \varnothing$ were given by Propositions 3.2.5 and 3.2.4. Once these could be established, Corollary 3.2.2 gave a necessary and sufficient condition for pure states to be fixed points and therefore one could engineer precise Lindblad terms which would drive the system to a target pure state fixed point i.e. ensuring that $\mathcal{F}(\Phi_\Gamma) = \mathcal{F}_\mathcal{D} = \{|\psi\rangle\langle\psi|\}$.

In Section 3.3 we were able to show that using Lindblad terms of *canonical* form (cf. Eqn. (3.36)), one could reduce the complexity of the task greatly since each Lindblad term of this form was nilpotent of degree two (cf. Lemma 3.3.2). Recall that a Lindblad term in canonical form was given by

$$V = \tfrac{1}{2}(\sigma_\mathbf{p} + \mathrm{i}\sigma_\mathbf{q}) \quad \text{such that} \quad [\sigma_\mathbf{p}, \sigma_\mathbf{q}] \neq 0\ . \tag{3.86}$$

Furthermore, by recalling the star-product $\mathbf{p}\star\mathbf{q} = \mathbf{m}$ defined by Eqns. (2.68) and (2.69) one arrives at the equivalent factorized form

$$V = \sigma_\mathbf{p} P\ , \quad \text{where} \quad P = \tfrac{1}{2}(\mathbb{1} - \sigma_\mathbf{m}) \tag{3.87}$$

is an orthogonal projection. Now, given a set of these nilpotent Lindblad terms, the generalised dark state spaces $\mathcal{D}_\Lambda$ (cf. Eqn. (3.13)) and subsequently the complete dark state space $\mathcal{D}$ reduced to

$$\mathcal{D} = \mathcal{D}_0 = \bigcap_k \ker(V_k)\,. \tag{3.88}$$

Therefore, the fixed point set is given by $\mathcal{F}(\Phi_\Gamma) = \mathcal{F}_{\mathcal{D}_0}$ whenever the set of intrinsic higher rank fixed points $\mathcal{F}_{\mathcal{D}_0^\perp} = \varnothing$. Now we will show how to engineer Lindblad terms which instead will be used to obtain unique *intrinsic* higher rank fixed points. That is, the fixed point set will be given by

$$\mathcal{F}(\Phi_\Gamma) = \mathcal{F}_{\mathcal{D}_0^\perp} = \{\rho\}\ , \tag{3.89}$$

and we will see that this category of fixed points can be obtained by using a slight generalisation of the canonical Lindblad terms used to obtain unique pure state fixed points.

First we make an important observation. Whenever $\mathbf{m}$ in Eqn. (3.86) is given by $\mathbf{m} = (m_1, m_2, \ldots, m_n)$ such that $m_j \in \{1, z\}$ for $1 \leq j \leq n$, we see that there exists a choice of the pairs $(\mathbf{p}, \mathbf{q})$ such that the $\mathbf{p}$ $n$-tuple is fixed as $\mathbf{p} = (p_1, p_2, \ldots, p_n)$ with $p_j \in \{1, x\}$ for $1 \leq j \leq n$. The Lindblad term in canonical form is then given by

$$V = \tfrac{1}{2}(\sigma_{\mathbf{p}} + \mathrm{i}\sigma_{\mathbf{q}}) = \tfrac{1}{2}(\sigma_{\mathbf{p}}(\mathbb{1} - \sigma_{\mathbf{m}})) = \sigma_{\mathbf{p}} P \ , \tag{3.90}$$

thereby making $P$ an orthogonal projection into the orthogonal complement of the target state. Note that $\mathrm{rank}(P) \geq 1$ with equality if and only if we are considering a single qubit system i.e. $n = 1$ and hence $V = |0\rangle\langle 0|$ or $V = |1\rangle\langle 1|$. We want to generalise this construction of Lindblad term such that the orthogonal projection $P$ can be rank one for more than a single qubit system - this entails that we no longer are restricted to having a single $\sigma_{\mathbf{m}}$ in Eqn. (3.90), but now a summation of such diagonal Pauli matrices.

An important first step that will be useful later is to notice that Pauli matrices of the form $\sigma_{\mathbf{p}} = (p_1, p_2, \ldots, p_n)$ with $p_k \in \{1, x\}$ for $1 \leq j \leq n$ are elements of the group defined by

$$\mathcal{G}_n := \{g_1 \otimes g_2 \otimes \cdots \otimes g_n \mid g_j \in \{\mathbb{1}, \sigma_x\} \text{ for all } 1 \leq j \leq n\} \ . \tag{3.91}$$

Now let $P$ be any diagonal orthogonal projection which has rank $\mu$. Notice that any such projection can be decomposed into a linear combination of the identity and Pauli matrices which has the form

$$P = \frac{1}{2^n}(\mu\mathbb{1} + \sum_i c_i \sigma_{\mathbf{m}_i}) \ , \tag{3.92}$$

where the coefficients $c_i \in \mathbb{Z}$ and each $n$-tuple $\mathbf{m}_i = (m_1, m_2, \ldots, m_n)_i$ only has indices which consist of unities and z's. Finally, identifying an element of the group $G \in \mathcal{G}_n$ as $G \equiv \sigma_{\mathbf{p}}$, we define Lindblad terms of *generalised* canonical form with strength coefficient $\sqrt{\gamma} \in \mathbb{R}^+$ as

$$V := \sqrt{\gamma} G P \tag{3.93}$$

Note that now $\mathrm{rank}(P) \geq 1$ for *any* $n$ as compared to the Lindblad terms in canonical form where even though they admitted a similar factorized form, the projection $P$ was only rank one in the single qubit case.

**Example 5.** *Let* $P = diag(1,1,1,0) = \frac{1}{4}(3\mathbb{1} + \sigma_{z1} + \sigma_{1z} - \sigma_{zz})$ *be a rank three orthogonal projection and let* $G \equiv \sigma_{x1} \in \mathcal{G}_2$ *be a group element. Then we can construct a Lindblad term of generalised canonical form as* $V = \sqrt{\gamma}\sigma_{x1} P = \frac{\sqrt{\gamma}}{4}\sigma_{x1}(3\mathbb{1} + \sigma_{z1} + \sigma_{1z} - \sigma_{zz}) = \frac{\sqrt{\gamma}}{4}(3\sigma_{x1} + \sigma_{xz} + \mathrm{i}(\sigma_{yz} - \sigma_{y1}))$ *which in explicit matrix representation gives*

$$V = \sqrt{\gamma}\begin{bmatrix} 0 & 0 & 1 & 0 \\ 0 & 0 & 0 & 0 \\ 1 & 0 & 0 & 0 \\ 0 & 1 & 0 & 0 \end{bmatrix} \ . \tag{3.94}$$

*We lastly note that* $V = \frac{\sqrt{\gamma}}{4}(3\sigma_{x1} + \sigma_{xz} + \mathrm{i}(\sigma_{yz} - \sigma_{y1}))$ *is truly a type of generalised version of a Lindblad term of canonical form (which was defined as* $V = \frac{1}{2}(\sqrt{\gamma}(\sigma_{\mathbf{p}} + \mathrm{i}\sigma_{\mathbf{q}}))$ *where* $[\sigma_{\mathbf{p}}, \sigma_{\mathbf{q}}] \neq 0$*) since there are now multiple* $\sigma_{\mathbf{p}}$ *and/or* $\sigma_{\mathbf{q}}$ *terms which compose the Lindblad term.*

With this notion in hand we can now make the great connection between Lindblad terms in canonical form, generalised canonical form and those which are centraliser generated. That is, recall that in Section 3.3.2 we provided a systematic way of constructing Lindblad terms from the centraliser (with respect to $\mathfrak{su}(2^n)$) of a target pure state $\rho = |\psi\rangle\langle\psi|$ given by

$$\mathfrak{s}_\rho := \{s \in \mathfrak{su}(2^n) \,|[s,\rho] = 0\} \ , \tag{3.95}$$

which allowed us to give an algorithm to determine sets of Lindblad terms that drove the system to the unique target fixed point. This was accomplished by first choosing a centraliser element $s \in \mathfrak{s}_\rho$ and noticing that since $[s,\rho] = 0$ then $s|\psi\rangle = \lambda|\psi\rangle$ for $\lambda \in \mathbb{C}$. Selecting an element $s \in \mathfrak{s}_\rho$ such that $\lambda \neq 0$ we then defined its *shifted* form as

$$P^S := |\lambda|\mathbb{1} \pm \mathrm{i}s \ , \quad \text{such that} \quad P^S|\psi\rangle = 0 \ . \tag{3.96}$$

A Lindblad term which was said to be *centraliser generated* was then defined as

$$V := \sqrt{\gamma}\sigma_{\mathbf{p}}P^S = \sqrt{\gamma}\sigma_{\mathbf{p}}(|\lambda|\mathbb{1} \pm \mathrm{i}s) \ , \quad \text{such that} \quad [\sigma_{\mathbf{p}}, s] \neq 0 \ , \quad \text{and} \quad \mathrm{tr}(\sigma_{\mathbf{p}}s) = 0 \ , \tag{3.97}$$

and we showed that canonical Lindblad terms were a special case of this form (see for example the discussion following Remark 12).

We now turn our attention to a target state $\rho$ which is a *diagonal mixed state* fixed point. A maximally abelian subalgebra of the centraliser of such a diagonal state specialises to

$$\mathfrak{a}_\rho = \mathrm{span}_{\mathbb{R}} \{\mathrm{i}\sigma_{\mathbf{m}} \mid m_k \in \{1, z\} \text{ for } k = 1, \ldots, n\} \ . \tag{3.98}$$

A Lindblad term of *generalised* canonical form was constructed by first decomposing a diagonal orthogonal projection into its linear combination of the identity matrix and Pauli matrices which are tensor products (for multi-qubit systems) of the identity and $\sigma_z$ terms i.e. $P = \frac{1}{2^n}(\mu\mathbb{1} + \sum_i c_i\sigma_{\mathbf{m}_i})$, where $\mu = \mathrm{Rank}(P)$, the $c_i$ elements are real coefficients and the $\mathbf{m}_i$ $n$-tuples contain only 1's and z's. For any strength coefficient $\gamma \in \mathbb{R}^+$, the Lindblad term was then given by

$$V := \sqrt{\gamma}GP \ , \quad \text{where} \quad P = \frac{1}{2^n}(\mu\mathbb{1} + \sum_i c_i\sigma_{\mathbf{m}_i}) \ , \tag{3.99}$$

where $G \equiv \sigma_{\mathbf{p}}$ was a special Pauli matrix whose elements of the $\mathbf{p}$ $n$-tuple consisted of only 1's and x's. Taking the Pauli matrix decomposition component of the diagonal orthogonal projection $P$ and multiplying by a factor of i we get the new element $s := \mathrm{i}\sum_i c_i\sigma_{\mathbf{m}_i}$. Now it's indeed clear that this element "$s$" *is contained in the maximally abelian subalgebra* (given by Eqn. (3.98)) of the state centraliser of a diagonal target $\rho$! Therefore, in a sense we can identify the projection $P$ used to construct a Lindblad term of *generalised canonical* form (used for diagonal mixed states) as a type of "shifted" centraliser element $P^S$ similar to those used to construct Lindblad terms which were *centraliser generated* (used for pure states).

Just as Lindblad terms which are centraliser generated, not every Lindblad term of generalised canonical form is nilpotent and the induced translation directions are not simply given by the index of the abelian subalgebra elements (See Remark 12) which make up the projection $P$ in $V = \sqrt{\gamma}GP$. The following Lemma establishes the conditions in which the Lindblad term is nilpotent of degree two, just like Lindblad terms of canonical form.

**Lemma 3.5.1.** *Let $V = \sqrt{\gamma}GP$ be a Lindblad term of generalised canonical form where $G \in \mathcal{G}_n$ such that $G \neq \mathbb{1}_n$ and the diagonal orthogonal projection is given by $P = \sum_k P_k$ with $P_k = e_k e_k^\dagger$ such that each $e_k$ is standard basis vector of $\mathbb{R}^{2^n}$. Let $S$ denote the set of basis vectors which compose the rank one projections which make up $P$. If $\langle G(e_i), e_j \rangle = 0$ for all $e_i, e_j \in S$, then $V$ is nilpotent of degree two.*

*Proof.* Clearly $V_k^2 = \gamma(GP)(GP) = \gamma(\sum_i GP_i G)\sum_j P_j = \gamma \sum_i \sum_j G(e_i)(G(e_i))^\dagger e_j e_j^\dagger = \gamma \sum_i \sum_j \langle G(e_i)|e_j\rangle G(e_i)e_j^\dagger = 0$ since $\langle G(e_i), e_j\rangle = 0$ for all $e_i, e_j \in S$. □

As a final remark, we note that in a slight abuse of notation, we will often say that a Lindblad generator is proportional to its associated translation direction(s) and will denote the relationship by $\Gamma_{V_k} \propto \sum_i \mathcal{T}_{\mathbf{m}_i}$. With these considerations, we will show to to construct Lindblad terms in generalized canonical form which allow us to obtain unique mixed state fixed points.

### 3.5.2 Full Rank Fixed Points

As further motivation, we consider the single qubit scenario since it will provide the general structure of the $n$-qubit generalisation.

**Example 6.** *For a purely dissipative system, suppose we want to obtain any unique fixed point of the form $\rho = diag(\lambda_1, \lambda_2)$ where $0 < \lambda_1, \lambda_2 < 1$ such that $\lambda_1 + \lambda_2 = 1$. Defining the two orthogonal projections $P_1 := e_1 e_1^\dagger = \frac{1}{2}(\mathbb{1} + \sigma_z)$ and $P_2 := e_2 e_2^\dagger = \frac{1}{2}(\mathbb{1} - \sigma_z)$, we can construct the Lindblad terms in canonical form*

$$\begin{aligned} V_1 &= \sqrt{\gamma_1}\sigma_x P_1 = \tfrac{\sqrt{\gamma_1}}{2}(\sigma_x(\mathbb{1} + \sigma_z)) = \tfrac{\sqrt{\gamma_1}}{2}(\sigma_x - \mathrm{i}\sigma_y) \quad \text{and} \\ V_2 &= \sqrt{\gamma_2}\sigma_x P_2 = \tfrac{\sqrt{\gamma_2}}{2}(\sigma_x(\mathbb{1} - \sigma_z)) = \tfrac{\sqrt{\gamma_2}}{2}(\sigma_x + \mathrm{i}\sigma_y) \end{aligned}$$

*such that $\Gamma_{V_1}$ and $\Gamma_{V_2}$ each induce an individual translation direction of $-\gamma_1 \mathcal{T}_z$ and $\gamma_2 \mathcal{T}_z$, respectively. Note that for a general diagonal mixed state fixed point of the form $\tilde{\rho} = diag(\tilde{\rho}_{11}, \tilde{\rho}_{22})$ we have that*

$$\begin{aligned} \Gamma(\tilde{\rho}) &= -\frac{1}{2}\sum_k V_k^\dagger V_k \tilde{\rho} + \tilde{\rho} V_k^\dagger V_k - 2V_k \tilde{\rho} V_k^\dagger = \sum_k \gamma_k P_k \tilde{\rho} - V_k \tilde{\rho} V_k^\dagger & (3.100) \\ &= diag(\gamma_1 \tilde{\rho}_{11}, \gamma_2 \tilde{\rho}_{22}) - diag(\gamma_2 \tilde{\rho}_{22}, \gamma_1 \tilde{\rho}_{11}) = 0 & (3.101) \end{aligned}$$

*since $V_k^\dagger V_k = P_k^2 = P_k$ and $[P_k, \tilde{\rho}] = 0$ for $k = 1, 2$. Thus, $\tilde{\rho}_{11} = \frac{\gamma_2}{\gamma_1}\tilde{\rho}_{22}$, which results in the normalised fixed point $\rho' = \frac{1}{\gamma_1 + \gamma_2} diag(\gamma_2, \gamma_1)$ and hence choosing $\gamma_1 = \lambda_1$, $\gamma_2 = \lambda_2$ or $\gamma_1 = \frac{1-\lambda_1}{\lambda_1}$ and $\gamma_2 = 1$ gives the target fixed point $\rho := \rho'$. Since each $V_k$ is nilpotent and $\cap_k \ker(V_k) = \{0\}$ then by Corollary 3.2.2 there are no pure state fixed points and in the proof of the following theorem we will prove that fixed points obtained from this type of construction will always be unique.*

First we make the observation that

$$\begin{aligned} V_1(e_1 + e_2) &= \sqrt{\gamma_1}\sigma_x P_1(e_1 + e_2) = \sqrt{\gamma_1} e_2 \quad \text{and} & (3.102) \\ V_2(e_1 + e_2) &= \sqrt{\gamma_2}\sigma_x P_2(e_1 + e_2) = \sqrt{\gamma_2} e_1 \,, & (3.103) \end{aligned}$$

and hence by considering the group $\mathcal{G}_1 = \{\mathbb{1}, \sigma_x\}$, and defining $G_1 = G_2 = \sigma_x \in \mathcal{G}_1$, it holds that $\{G_k e_k \mid k = 1, 2\} = \{e_k \mid k = 1, 2\}$ and $G_k e_k \neq e_k$ for $k = 1, 2$. These two group conditions were the keys to obtaining the *mixing* of diagonal elements in Eqn. (3.101) which resulted in the desired unique fixed point. We can now use the notion of Lindblad terms in *generalised* canonical form in order to engineer intrinsic higher rank mixed state fixed points for more than one qubit using a straightforward extension of these two group conditions.

Define the index set of all diagonal elements of a (target) $2^n \times 2^n$ matrix as $\mathcal{I}_{\text{tar}} := \{1, \dots, 2^n\}$, and define the set of index subsets $\mathcal{I}_{\mathcal{P}} := \mathcal{P}(\mathcal{I}_{\text{tar}}) \setminus \{\mathcal{I}_{\text{tar}}, \emptyset\}$, where $\mathcal{P}(\mathcal{I}_{\text{tar}})$ is the power set of $\mathcal{I}_{\text{tar}}$. Thus $\mathcal{I}_{\mathcal{P}}$ contains $4^n - 2$ elements (sets). Noting that the group $\mathcal{G}_n$ (cf. Eqn. (3.91)) acts transitively on the set $\{e_k \,|k = 1, \dots, 2^n\}$, we can define two group element conditions which we will show will guarantee uniqueness of the target fixed point.

(1) Existence of Target Subspace:

$$\{G_k e_k \mid k \in \mathcal{I}_{\text{tar}}\} = \{e_k \mid k \in \mathcal{I}_{\text{tar}}\} \tag{3.104}$$

(2) Uniqueness of Target Invariant Subspace:

$$\{G_k e_k \mid k \in S\} \neq \{e_k \mid k \in S\} \text{ for each index set } S \subset \mathcal{I}_{\mathcal{P}} . \tag{3.105}$$

Furthermore, a set of group elements $\{G_k\} \in \mathcal{G}_n$ which as a whole, satisfy Eqns. (3.104) and (3.105) can then define the sets

$$\begin{aligned} Sol(\mathcal{I}_{\text{tar}}) := \{G_1 P_1, \dots, G_{2^n} P_{2^n} \mid \{G_k\}_{k=1}^{2^n} & \text{ satisfies Eqns. (3.104)} \\ & \text{and (3.105) and } P_k := e_k e_k^\dagger \ \forall\, k\} . \end{aligned} \tag{3.106}$$

**Theorem 9.** *Let $\rho = diag(\lambda_1, \dots, \lambda_{2n})$ be any diagonal mixed state of full rank with non-degenerate eigenvalues and let $Sol(\mathcal{I}_{\text{tar}})$ be a solution set of operators as given by Eqn. (3.106) where $\mathcal{I}_{\text{tar}} = \{1, 2, \dots, 2^n\}$. Then $\rho$ is obtained as the unique fixed point of a purely dissipative Lindblad-Kossakowski operator with $2^n$ Lindblad terms given by $V_k := \sqrt{\gamma_k} G_k P_k$, where $G_k P_k \in Sol(\mathcal{I}_{\text{tar}})$ and $\gamma_k = \frac{\lambda_{2^n}}{\lambda_k}$ for all $k$.*

*Proof.* Condition one given by Eqn. (3.104) guarantees that

$$\Gamma(\rho) = \sum_k (\gamma_k P_k(\rho) - \gamma_k G_k P_k(\rho) P_k G_k^\dagger) = \lambda_{2^n} \mathbb{1} - \lambda_{2^n} \sum_k G_k (e_k e_k^\dagger) G_k^\dagger = 0 , \tag{3.107}$$

since $V_k^\dagger V_k = P_k^2 = \gamma_k P_k$ and $[P_k, \tilde{\rho}] = 0$ for all $k$ and therefore $\rho$ is a fixed point. Furthermore, condition two given by Eqn. (3.105) implies that $G_k e_k \perp e_k$ and hence by Lemma 3.5.1 each Lindblad term $V_k$ is nilpotent and thus the dynamics are purely dissipative.

Now we prove uniqueness of the fixed point. First we will show that there cannot exist any rank degenerate fixed point. Suppose there exists a rank degenerate fixed point $\rho'$ and let $\mathcal{S}' := \text{supp}(\rho')$. By Proposition 3.2.1 and Corollary 3.2.1 this would imply $V_k \mathcal{S}' \subseteq \mathcal{S}'$ for all $k$ and $\sum_k V_k^\dagger V_k \mathcal{S}' \subseteq \mathcal{S}'$. Since the second condition simplifies to

$$\sum_k V_k^\dagger V_k = \sum_k \gamma_k P_k = \text{diag}(\gamma_1, \gamma_2, \dots, \gamma_{2^n}) , \tag{3.108}$$

we know that the non-trivial invariant subspaces of $\sum_k V_k^\dagger V_k$ are

$$S_1 := \text{span}_{\mathbb{C}} \{e_1\} , \quad S_2 := \text{span}_{\mathbb{C}} \{e_2\} , \quad S_3 := \text{span}_{\mathbb{C}} \{e_1, e_2\} , \quad \dots \text{etc.} \tag{3.109}$$

and thus are all combinations of spans of standard basis vectors. We will show that condition two given by Eqn. (3.105) guarantees that no such subspace is *simultaneously* invariant for each $V_k$.

Condition two ensures that no group elements which comprise the Lindblad terms satisfy relations such as $\{G_1e_1, G_2e_2\} = \{e_1, e_2\}$ with $G_1e_1 = e_2$ and $G_2e_2 = e_1$. Defining $\mathcal{S} := \mathrm{span}_{\mathbb{R}}\{e_1, e_2\}$, we would then have that $V_1\mathcal{S} \subseteq \mathcal{S}$, $V_2\mathcal{S} \subseteq \mathcal{S}$ and $V_k\mathcal{S} \subseteq \mathcal{S}$ for $k = 3, \ldots, 2^n$ trivially since $\mathcal{S} \subseteq \ker(V_k)$ for $k = 3, \ldots, 2^n$. Therefore, $\mathcal{S}$ would be an invariant subspace which supports a fixed point since it is also a simultaneous invariant subspace of the matrix given by Eqn. (3.108). Condition two then ensures there are no invariant subspaces which are spans of unit vectors. This implies there cannot exist any rank deficient fixed point.

Now suppose that the additional fixed point $\rho'$ is full rank. Taking the affine combination of $\rho$ and $\rho'$ gives $\Gamma(a_1\rho + a_2\rho') = 0$ for all $a_1, a_2 \in \mathbb{R}$ such that $a_1 + a_2 = 1$. Note that there exists some choice of $a_1$ and $a_2$ such that $\sigma := a_1\rho + a_2\rho'$ lies on the boundary of the set of density matrices and hence has rank $r < 2^n$ which implies $\sigma$ would be a rank degenerate fixed point. By the previous discussion, we know this cannot occur and therefore we are done.

□

**Example 7.** *Suppose we want to obtain a unique full rank diagonal (with non-degenerate eigenvalues) 2-qubit mixed state fixed point. Choosing the group elements $G_1, G_2, G_3, G_4 \in \mathcal{G}_2$ as*

$$G_1 = \mathbb{1} \otimes \sigma_x \ , \quad G_2 = \sigma_x \otimes \sigma_x \ , \quad G_3 = \mathbb{1} \otimes \sigma_x \ , \quad G_4 = \sigma_x \otimes \sigma_x \ , \tag{3.110}$$

*since $G_1e_1 = e_2$, $G_2e_2 = e_3$ , $G_3e_3 = e_4$, $G_4e_4 = e_1$ one sees that they satisfy the conditions given by Eqns.* (3.104) *and* (3.105) *therefore we have one particular solution set given by $Sol(\mathcal{I}_{\mathrm{tar}}) = \{G_kP_k \mid P_k = e_ke_k^\dagger \ , k = 1, \ldots, 4\}$ which can then be used to construct the Lindblad terms*

$$V_1 = \sqrt{\tfrac{\lambda_4}{\lambda_1}}(\mathbb{1} \otimes \sigma_x)P_1 = \tfrac{1}{2}\sqrt{\tfrac{\lambda_4}{\lambda_1}}\big((\mathbb{1} + \sigma_z) \otimes \sigma^-\big) \ , \quad V_2 = \sqrt{\tfrac{\lambda_4}{\lambda_2}}(\sigma_x \otimes \sigma_x)P_2 = \sqrt{\tfrac{\lambda_4}{\lambda_2}}\big(\sigma^- \otimes \sigma^+\big) \ , \tag{3.111}$$

$$V_3 = \sqrt{\tfrac{\lambda_4}{\lambda_3}}(\mathbb{1} \otimes \sigma_x)P_3 = \tfrac{1}{2}\sqrt{\tfrac{\lambda_4}{\lambda_3}}\big((\mathbb{1} - \sigma_z) \otimes \sigma^-\big) \ , \quad V_4 = (\sigma_x \otimes \sigma_x)P_4 = \sigma^+ \otimes \sigma^+ \ , \tag{3.112}$$

*where the equalities follows from $P_1 = \frac{1}{4}(\mathbb{1} \otimes \mathbb{1} + \sigma_z \otimes \mathbb{1} + \mathbb{1} \otimes \sigma_z + \sigma_z \otimes \sigma_z)$, $P_2 = \frac{1}{4}(\mathbb{1} \otimes \mathbb{1} + \sigma_z \otimes \mathbb{1} - \mathbb{1} \otimes \sigma_z - \sigma_z \otimes \sigma_z)$, $P_3 = \frac{1}{4}(\mathbb{1} \otimes \mathbb{1} - \sigma_z \otimes \mathbb{1} + \mathbb{1} \otimes \sigma_z - \sigma_z \otimes \sigma_z)$ and $P_4 = \frac{1}{4}(\mathbb{1} \otimes \mathbb{1} - \sigma_z \otimes \mathbb{1} - \mathbb{1} \otimes \sigma_z + \sigma_z \otimes \sigma_z)$.*

It is important to note the connection between the target mixed state fixed point and the associated translation directions which are induced by the individual Lindblad terms. Remark 12 explained how one can obtain the associated translation directions from each Lindblad term. Explicitly, we discussed how one cannot simply deduce the translation directions based upon the index $\mathbf{m}$ of each Pauli matrix which the (in this case) projections $\{P_k\}$ are decomposed into. It's a simple calculation to see that the translation directions associated to each "piecewise" part of the full $\Gamma = \sum_k^4 \Gamma_{V_k}$ are given by of Example 7 are

$$\Gamma_{V_1} \propto \tfrac{1}{2}\tfrac{\lambda_4}{\lambda_1}(-\mathcal{T}_{1z} - \mathcal{T}_{zz}) \ , \quad \Gamma_{V_2} \propto \tfrac{1}{2}\tfrac{\lambda_4}{\lambda_2}(-\mathcal{T}_{z1} + \mathcal{T}_{1z}) \ , \tag{3.113}$$

$$\Gamma_{V_3} \propto \tfrac{1}{2}\tfrac{\lambda_4}{\lambda_3}(-\mathcal{T}_{z1} + \mathcal{T}_{zz}) \ , \quad \Gamma_{V_4} \propto \tfrac{1}{2}(\mathcal{T}_{z1} + \mathcal{T}_{1z}) \ . \tag{3.114}$$

This fixed point engineering method allows us to obtain *multiple* solutions for obtaining the desired mixed state fixed point. For example, to obtain the diagonal mixed state in Example 7 as the unique fixed point we could of alternatively used a *different* solution set, say, $Sol(\mathcal{I}_{\text{tar}}) = \{(\sigma_x \otimes \mathbb{1})P_1, (\mathbb{1} \otimes \sigma_x)P_2, (\mathbb{1} \otimes \sigma_x)P_3, (\sigma_x \otimes \mathbb{1})P_4\}$. This provides an entire new set of Lindblad terms which drive the system to the target unique fixed point. This time however, not only are the individual translations induced by each Lindblad term different than previously, but the overall (i.e. sum) translations for $\Gamma$ are different since

$$\Gamma_{V_1} \propto \tfrac{1}{2}\tfrac{\lambda_4}{\lambda_1}(-\mathcal{T}_{z1} - \mathcal{T}_{zz}) \; , \quad \Gamma_{V_2} \propto \tfrac{1}{2}\tfrac{\lambda_4}{\lambda_2}(\mathcal{T}_{1z} + \mathcal{T}_{zz}) \; , \tag{3.115}$$

$$\Gamma_{V_3} \propto \tfrac{1}{2}\tfrac{\lambda_4}{\lambda_3}(-\mathcal{T}_{1z} + \mathcal{T}_{zz}) \; , \quad \Gamma_{V_4} \propto \tfrac{1}{2}(\mathcal{T}_{z1} - \mathcal{T}_{zz}) \; . \tag{3.116}$$

We now compare this situation with the single qubit case. In the single qubit full rank mixed state scenario in Example 6 there was only one possible solution, $Sol(\mathcal{I}_{\text{tar}}) = \{\sigma_x P_1, \sigma_x P_2\}$ which gave $V_1 := \sqrt{\gamma_1}\sigma^-$ and $V_2 := \sqrt{\gamma_2}\sigma^+$. There we remarked that each separate Lindblad term induced the translation directions $-\gamma_1 \mathcal{T}_z$ and $\gamma_2 \mathcal{T}_z$, respectively, and thus providing an intuitive geometric picture of the translation directions along the $z$-axis of the bloch sphere. It is immediately apparent that the single qubit Bloch sphere interpretation of the translation direction towards the fixed point of the system generalises to multi-qubit systems by allowing for several different direction "paths" one can steer the system through. The multiple paths available for the multi-qubit set of states results in the multiple solution sets of Lindblad terms which describe different trajectory paths towards the fixed point. Thus, this engineering method may prove to be useful for experimental implementation due to the fact that there may be solution sets of Lindblad terms which may easier to realise physically than the others.

### 3.5.3 Rank Deficient Mixed State Fixed Points

We now consider a target diagonal fixed point $\rho$ which has rank $1 < r < 2^n$ which has non-degenerate non-zero eigenvalues. Recall that if the rank of the diagonal target state was one, we would be in the situation described in Section 3.4.1 where we showed how to obtain the ground/excited state as the unique fixed point using Lindblad terms of canonical form. Here we present two results which provide methods of constructing Lindblad generators which drive the system to any diagonal rank deficient (which has its non-zero eigenvalues being non-degenerate) mixed state fixed point. The first is a slight generalisation of Theorem 9 and we show how to obtain the desired fixed point using $2^n$ Lindblad terms of a specific generalised canonical form. The second result then shows that when the rank of the target fixed point is greater than or equal to one half the dimension size (and not equal to $2^n$) we can obtain it uniquely by using $r + 1$ Lindblad terms.

Let $\rho$ be the target diagonal fixed point with rank $1 < r < 2^n$. As in the full rank mixed state case, define the index set of all non-zero diagonal elements as $\mathcal{I}_{\text{tar}} := \{1, 2, \ldots, r\}$ and the set of index subsets $\mathcal{I}_{\mathcal{P}} := \mathcal{P}(\mathcal{I}_{\text{tar}}) \setminus \{\mathcal{I}_{\text{tar}}, \emptyset\}$, where $\mathcal{P}(\mathcal{I}_{\text{tar}})$ is the power set of $\mathcal{I}_{\text{tar}}$. Note that $\mathcal{I}_{\mathcal{P}}$ contains $2^r - 2$ elements (sets). Moreover, define the set of remaining diagonal element indices as $\mathcal{I}_{\text{tar}^\perp} = \{r+1, r+2, \ldots, 2^n\}$. The group element conditions then reduce to

(1) Existence of a Target Invariant Subspace:

$$\{G_k e_k \mid k \in \mathcal{I}_{\text{tar}}\} = \{e_k \mid k \in \mathcal{I}_{\text{tar}}\} \tag{3.117}$$

(2) Non-existence of Pure State Invariant Subspaces:

$$G_k e_k = e_j \quad \text{for each} \quad k \in \mathcal{I}_{tar^\perp} \text{ such that } j \in \mathcal{I}_{tar} \;, \tag{3.118}$$

(3) Uniqueness of Target Invariant Subspace:

$$\{G_k e_k \bigm| k \in S\} \neq \{e_k \bigm| k \in S\} \text{ for every set } S \subset \mathcal{I}_{\mathcal{P}} \;. \tag{3.119}$$

Conditions one and three given by Eqns. (3.117) and (3.119) are equivalent to the full rank mixed state conditions one and two given by Eqns. (3.104) and (3.105) just reduced to a smaller "target " index set $\mathcal{I}_{\rm tar}$. The new condition given by Eqn. (3.118) ensures that for each basis vector $e_k \in \{e_{r+1}, \ldots, e_{2^n}\}$ there is an element of the solution set that shifts $e_k$ into the support of the target rank $r$ fixed point. This eliminates the possibility of pure state fixed points whose support is orthogonal to $\mathcal{S} := \text{supp}(\rho)$ by Corollary 3.2.2. The proof of the following result is along the same lines as that of Theorem 9.

**Theorem 10.** *Let $\rho = diag(\lambda_1, \ldots, \lambda_r, 0, \ldots, 0)$ be any rank $r$ diagonal mixed state with non-degenerate non-zero eigenavalues and let $Sol(\mathcal{I}_{\rm tar})$ be a solution set of operators of the form given by Eqn. (3.106), except now the elements $\{G_k\}_{k=1}^{2^n}$ satisfy Eqns. (3.117)-(3.119) with $\mathcal{I}_{\rm tar} = \{1, 2, \ldots, r\}$. Then $\rho$ is obtained as the unique fixed point of a purely dissipative Lindblad-Kossakowski operator with $2^n$ Lindblad terms given by $V_k := \sqrt{\gamma_k} G_k P_k$, where $G_k P_k \in Sol(\mathcal{I}_{\rm tar})$ such that $\gamma_k = \frac{\lambda_r}{\lambda_k}$ for $1 \leq k \leq r$ and $\gamma_k \in \mathbb{R}^+$ for $r < k \leq 2^n$.*

*Proof.* Condition one given by Eqn. (3.117) guarantees that

$$\Gamma(\rho) = \sum_k^{2^n} (\gamma_k P_k(\rho) - \gamma_k G_k P_k(\rho) P_k G_k^\dagger) = \lambda_r \sum_k^{r} (e_k e_k^\dagger - G_k (e_k e_k^\dagger) G_k^\dagger) = 0 \;, \tag{3.120}$$

and therefore $\rho$ is a fixed point. Furthermore, condition two given by Eqn. (3.118) implies $G_k e_k \perp e_k$ for $r < k \leq 2^n$ and condition three given by Eqn. (3.119) implies that $G_k e_k \perp e_k$ for $1 \leq k \leq r$. By Lemma 3.5.1 this implies that each Lindblad term $V_k$ is nilpotent and thus the dynamics are purely dissipative.

We now prove uniqueness of the target fixed point. For $\mathcal{S} := \text{supp}(\rho) = \text{span}_{\mathbb{C}} \{e_1, e_2, \ldots, e_r\}$, condition two given by Eqn. (3.118) guarantees there are no $e_j \in \mathcal{S}^\perp$ such that $V_k e_j = 0$ for all $k$ and hence there exists no pure state fixed points of the system whose support would be orthogonal to $\mathcal{S}$. Thus we have that $V_k \mathcal{S}^\perp \subseteq \mathcal{S}$ for all $k$ and hence there are no fixed points with support contained in $\mathcal{S}^\perp$.

Suppose there exists another rank degenerate fixed point $\rho'$ and define $\mathcal{S}' := \text{supp}(\rho')$. Then either $\mathcal{S}' \subseteq \mathcal{S}^\perp$ or $\mathcal{S}' \subseteq \mathcal{S}$, or simply $\mathcal{S}' \subseteq \mathcal{H}$ such that it is not completely contained in either $\mathcal{S}$ or $\mathcal{S}'$. From the above discussion we know that the first case cannot occur. Now we consider the second scenario where we assume that $\mathcal{S}' \subseteq \mathcal{S}$. Along the same lines as the proof of Theorem 9, we note that $\sum_k V_k^\dagger V_k$ is diagonal with unique non-zero eigenvalues. This then implies that we know all the invariant subspaces of $\sum_k V_k^\dagger V_k$ thereby showing in the same manner that there cannot exist any invariant subspace $\mathcal{S}' \subseteq \mathcal{S}$ which supports a fixed point.

For the third and final case we now assume instead that the additional rank degenerate fixed point $\rho'$ does not have its support entirely contained in either $\mathcal{S}$ or $\mathcal{S}'$. Then clearly $\mathcal{S}' \cap \mathcal{S} \subseteq \mathcal{S}$ and since $\mathcal{S}$ itself satisfies $V_k \mathcal{S} \subseteq \mathcal{S}$ for all $k$ and $\sum_k V_k^\dagger V_k \mathcal{S} \subseteq \mathcal{S}$ (by Proposition 3.2.1), it holds that

$$V_k \left(\mathcal{S}' \cap \mathcal{S}\right) \subseteq \left(\mathcal{S}' \cap \mathcal{S}\right) \;, \qquad \text{for all } k\text{, and} \quad \sum_k V_k^\dagger V_k \left(\mathcal{S}' \cap \mathcal{S}\right) \subseteq \left(\mathcal{S}' \cap \mathcal{S}\right) \;. \tag{3.121}$$

Therefore $(\mathcal{S}' \cap \mathcal{S}) \subseteq \mathcal{S}$ is another invariant subspace which supports a new fixed point. We know this inclusion cannot occur and hence there exists no other rank degenerate mixed state fixed points.

Finally, if instead we assume there exists a full rank fixed point we can use the exact same argument at the end of Theorem 9 to show that this implies there must exist a rank deficient fixed point - which we know cannot exist. Thus, the target fixed point is unique.

□

We can also obtain a rank $r < 2^n$ fixed point using $r+1$ Lindblad terms if the rank is larger than half the dimension of the total state space.

**Corollary 3.5.1.** *Let $\rho = diag(\lambda_1, \dots, \lambda_r, 0, \dots, 0)$ be any diagonal mixed state of rank $2^{n-1} \leq r < 2^n$ with non-degenerate non-zero eigenvalues. Then $\rho$ is obtained as the unique fixed point of a purely dissipative Lindblad-Kossakowski operator with $r+1$ Lindblad terms given by*

$$V_k = \sqrt{\gamma_k} G_k P_k \quad \text{where } P_k = e_k e_k^\dagger \quad \text{for } k = 1, \dots, r \ , \quad \text{and} \tag{3.122}$$

$$V_{r+1} = \sum_{j=r+1}^{2^n} (G_{r+1}) P_j \quad \text{where } P_j = e_j e_j^\dagger \ , \tag{3.123}$$

*such that Eqns.* (3.117) *and* (3.119) *are satisfied,* $\mathrm{range}(V_{r+1}) \subseteq$ *supp*$(\rho)$ *and* $\gamma_k = \frac{\lambda_r}{\lambda_k}$ *for $k = 1, \dots, r$.*

*Proof.* The structure of $V_{r+1} = \sum_{j=r+1}^{2^n} (G_{r+1}) P_j$ and the condition that $\mathrm{range}(V_{r+1}) \subseteq \mathrm{supp}(\rho)$ just ensures that there are no simultaneous eigenvectors contained in $\mathcal{S}^\perp$ which would be a necessary condition that $\mathcal{S}^\perp$ contains another invariant subspace. The rest of the proof is identical to that of Theorem 10. □

After obtaining a desired diagonal fixed point it's clear that one can apply a unitary rotation to any other density matrix with those desired eigenvalues. Thus the discussion above provides a general result showing how to obtain *any* arbitrary density matrix with non-degenerate non-zero eigenvalues as the unique fixed point. Although Corollary 3.5.1 provides the upper bound of $r+1$ terms required when $r \geq 2^{n-1}$, the following result can be considered as a general "worst case" upper bound on the number of Lindblad terms required.

**Theorem 11.** *Let $\rho$ be any rank $r$ quantum state with non-degenerate non-zero eigenvalues $\{\lambda_k\}_{k=1}^r$. Then there exists multiple sets of $2^n$ Lindblad terms of the form*

$$V_k = \sqrt{\gamma_k} U (G_k(P_k)) U^\dagger \quad \text{for all } k = 1, \dots, 2^n \ , \tag{3.124}$$

*where $\gamma_k = \frac{\lambda_r}{\lambda_k}$ for $1 \leq k \leq r$ and $\gamma_k \in \mathbb{R}^+$ for $r+1 \leq k \leq 2^n$ and $P_k = e_k e_k^\dagger$ for all $k$ such that the associated purely dissipative Lindblad generator has $\rho$ as the unique fixed point state.*

*Proof.* Let $U$ be the unitary which diagonalises $\rho$ as $\rho = \mathrm{diag}(\lambda_1, \lambda_2, \dots, \lambda_r, 0, \dots)$, where clearly if $\rho$ is full rank then there are no zeros on the diagonal. The proof then follows from a basic change of basis and Theorems 9 and 10. □

**Remark 13.** *Theorems 9, 10 and 11 are stated in a manner which assumes the non-zero eigenvalues of the target fixed point state are non-degenerate. The construction method of the Lindblad terms* also *seem to work for degenerate eigenvalues but the final step in this completely general proof required some details to be polished at the time of this thesis submission. The general result will be published in follow up work.*

*Example 8 provides an example showing the same construction of Lindblad terms also works.*

Another remark is in order to connect this result to the current literature. In [48], Ticozzi, Schirmer and Wang also considered the problem of determining a Lindblad-Kossakowski operator which drives the system to a target mixed state fixed point. They showed that it is possible using a single Hamiltonian $H$ and Lindblad term $L$ which were quintdiagonal and tridiagonal, respectively. In [41], Pechen proved a similar result to that of Theorem 11 where he considered a physical implementation of "all to one " controls which are those which can simultaneously transfer every initial state to any desired target state. Using a procedure motivated by the experimental use of incoherent light, it was shown that every general initial state (i.e. those which are rank $r$ and have each matrix element non-zero) can be driven to the target state by using $4^n$ Lindblad terms which are rank one and of the form $L = e_i e_j^\dagger$. Our method does not depend on these types of rank one Lindblad terms which are experimentally realisable via engineered radiation, instead, we have provided a general engineering scheme which extends the pure state construction that used Lindblad terms of canonical form in Sections 3.3 and 3.4 to obtain unique pure state fixed points.

As the following example shows, we expect there is a similar bound to that of Corollary 3.5.1 when the rank of the target mixed state $r$ is less than $2^{n-1}$. Nonetheless, knowing bounds on how many Lindblad terms are required does not necessarily simplify an experimental implementation as they are usually more complicated and perhaps more difficult to implement overall.

**Example 8.** *Consider the 3-qubit target state $\rho_{tar} = \frac{1}{2}diag(1,1,0,0,0,0,0,0)$. Choosing the group elements $G_1, G_2, G_3, G_4 \in \mathcal{G}_3$ as*

$$G_1 = \mathbb{1} \otimes \mathbb{1} \otimes \sigma_x \ , \ G_2 = \mathbb{1} \otimes \mathbb{1} \otimes \sigma_x \ , \ G_3 = \sigma_x \otimes \mathbb{1} \otimes \sigma_x \ , \ G_4 = \sigma_x \otimes \sigma_x \otimes \sigma_x \ , \tag{3.125}$$

*Then for $\mathcal{H} = \mathrm{span}_{\mathbb{R}}\{e_1, \dots, e_8\}$ and a generic element $\tilde{v} := \sum_{k=1}^{8} a_k e_k \in \mathcal{H}$ where each $a_k \in \mathbb{R}$ we have that $G_1 P_1 \tilde{v} = a_1 e_2$, $G_2 P_2 \tilde{v} = a_2 e_1$ and*

$$G_3(\sum_{j=3}^{6} P_j)\tilde{v} = a_6 e_1 + a_5 e_2 + a_4 e_7 + a_3 e_8 \ , \quad G_4(\sum_{j=7}^{8} P_k)\tilde{v} = a_8 e_1 + a_7 e_2 \ , \tag{3.126}$$

*and hence the only invariant subspaces are $\mathcal{S} = \mathrm{span}_{\mathbb{C}}\{e_1, e_2\}$ and $\mathcal{H}$. The Lindblad terms*

$$\begin{aligned} V_1 &= \tfrac{1}{4}\big((\mathbb{1}+\sigma_z) \otimes (\mathbb{1}+\sigma_z) \otimes \sigma^-\big) \ , \\ V_2 &= \tfrac{1}{4}\big((\mathbb{1}+\sigma_z) \otimes (\mathbb{1}+\sigma_z) \otimes \sigma^+\big) \ , \\ V_3 &= \tfrac{1}{2}\Big(\big(\sigma^- \otimes (\mathbb{1}-\sigma_z) \otimes \sigma_x + \sigma^+ \otimes (\mathbb{1}+\sigma_z)\big) \otimes \sigma_x\Big) \ , \quad \textit{and} \\ V_4 &= \sigma^+ \otimes \sigma^+ \otimes \sigma_x \end{aligned}$$

*drive the system to the desired target state fixed point.*

We now provide concrete examples of the implementation of Theorem 10 and Corollary 3.5.1.

**Example 9.** *Suppose we want to obtain the 2-qubit state* $\rho_{tar} = \frac{1}{4}diag(1,3,0,0)$ *as the unique fixed point of a purely dissipative system. Since* $(\mathbb{1}\otimes\sigma_x)e_1 = e_2$, $(\mathbb{1}\otimes\sigma_x)e_2 = e_1$, $(\mathbb{1}\otimes\sigma_x)(e_3) = e_1$ *and* $(\mathbb{1}\otimes\sigma_x)(e_4) = e_2$ *then we can define a solution set given by* $Sol(\mathcal{I}_{\rm tar}) = \{(\mathbb{1}\otimes\sigma_x)P_1, (\mathbb{1}\otimes\sigma_x)P_2, (\sigma_x\otimes\mathbb{1})(P_3+P_4)\}$*. The corresponding Lindblad terms which drive the system to the target fixed point are then given by*

$$\begin{aligned} V_1 &= \sqrt{\gamma_1}(\mathbb{1}\otimes\sigma_x)P_1 = \tfrac{1}{2}\sqrt{\gamma_1}\big((\mathbb{1}+\sigma_z)\otimes\sigma^-\big)\ , \\ V_2 &= (\mathbb{1}\otimes\sigma_x)P_2 = \tfrac{1}{2}\big((\mathbb{1}+\sigma_z)\otimes\sigma^+\big)\ , \quad \textit{and} \\ V_3 &= (\sigma_x\otimes\mathbb{1})(P_3+P_4) = \sigma^+\otimes\mathbb{1}\ , \end{aligned}$$

*where* $\gamma_1 = \frac{\lambda_2}{\lambda_1} = 3$*. By Corollary 3.5.1, the fixed point is unique. Moreover, another possible solution set is given by* $Sol(\mathcal{G}_2) = \{(\mathbb{1}\otimes\sigma_x)P_1, (\mathbb{1}\otimes\sigma_x)P_2, (\sigma_x\otimes\sigma_x)(P_3+P_4)\}$*, where only the third element is different. This then results in the new Lindblad term* $V_3' := (\sigma_x\otimes\sigma_x)(P_3+P_4) = \sigma^+\otimes\sigma_x$ *which can be used as an alternative to* $V_3$ *without sacrificing uniqueness of the target fixed point.*

**Example 10.** *Consider a 3-qubit target fixed point which is rank 5 and is given by* $\rho_{tar} = diag(\lambda_1,\lambda_2,\ldots,\lambda_5,0,\ldots,0)$*. Choosing group elements* $G_k\in\mathcal{G}_3$ *as*

$$\begin{aligned} &G_1 = \sigma_x\otimes\mathbb{1}\otimes\mathbb{1}\ , && G_2 = \mathbb{1}\otimes\mathbb{1}\otimes\sigma_x\ , && G_3 = \mathbb{1}\otimes\sigma_x\otimes\sigma_x\ , && G_4 = \sigma_x\otimes\sigma_x\otimes\sigma_x\ , \\ &G_5 = \sigma_x\otimes\sigma_x\otimes\sigma_x\ , && G_6 = \sigma_x\otimes\sigma_x\otimes\sigma_x\ , && G_7 = \sigma_x\otimes\sigma_x\otimes\sigma_x\ , && G_8 = \sigma_x\otimes\sigma_x\otimes\sigma_x\ , \end{aligned}$$

*since*

$$\begin{aligned} &G_1e_1 = e_5\ , && G_2e_2 = e_1\ , && G_3e_3 = e_2\ , && G_4e_4 = e_3\ , \\ &G_5e_5 = e_4\ , && G_6e_6 = e_3\ , && G_7e_7 = e_2\ , && G_8e_8 = e_1\ , \end{aligned}$$

*and one can immediately see that conditions given by Eqns.* (3.117) *and* (3.118) *are satisfied. Furthermore, one can check that condition given by Eqn.* (3.119) *is also satisfied (e.g. for the index set* $S=\{1,3,7\}$ *we have* $\{G_1e_1, G_3e_3, G_7e_7\} = \{e_5,e_2,e_2\} \neq \{e_1,e_3,e_7\}$*). Therefore* $Sol(\mathcal{I}_{\rm tar}) = \{G_kP_k \mid P_k = e_ke_k^\dagger\ ,\ k=1,\ldots,8\}$ *is a solution set which provides the Lindblad terms*

$$\begin{aligned} V_1 &= \tfrac{1}{4}\sqrt{\gamma_1}\big(\sigma^-\otimes(\mathbb{1}+\sigma_z)\otimes(\mathbb{1}+\sigma_z)\big)\ , \\ V_2 &= \tfrac{1}{4}\sqrt{\gamma_2}\big((\mathbb{1}+\sigma_z)\otimes(\mathbb{1}+\sigma_z)\otimes\sigma^+\big)\ , \\ V_3 &= \tfrac{1}{2}\sqrt{\gamma_3}\big((\mathbb{1}+\sigma_z)\otimes\sigma^+\otimes\sigma^-\big)\ , \\ V_4 &= \sqrt{\gamma_4}(\sigma^-\otimes\sigma^+\otimes\sigma^+)\ , \end{aligned}$$

*and* $V_5 = (\sigma^+\otimes\sigma^-\otimes\sigma^-)$, $V_6 = (\sigma^+\otimes\sigma^-\otimes\sigma^+)$, $V_7 = (\sigma^+\otimes\sigma^+\otimes\sigma^-)$, *and* $V_8 = (\sigma^+\otimes\sigma^+\otimes\sigma^+)$ *with* $\gamma_k = \frac{\lambda_5}{\lambda_k}$ *for* $k=1,\ldots,5$ *and by Theorem 10 will drive the system to the unique target state fixed point. Furthermore, since the rank of the target state is greater than four, we remark that we can also concatenate the Lindblad terms* $V_6, V_7$ *and* $V_8$ *into a single term defined as* $V_6' := (\sigma_x\otimes\sigma_x\otimes\sigma_x)(P_6+P_7+P_8)$*. By Corollary 3.5.1, the set of Lindblad terms* $\{V_1,V_2,V_3,V_4,V_5,V_6'\}$ *will also drive the system to the unique target state fixed point.*

### 3.5.4 Summary and Mixed State Symmetry Considerations

In Section 3.3.2 we provided a systematic procedure of determining sets of Lindblad terms which drive the system to a target *pure state fixed point* $\rho = |\psi\rangle\langle\psi|$ based on the symmetries (centraliser) of $\rho$. Afterwards, in Section 3.4 we showed that it was often the case (for several prototypical classes) that we were able to construct Lindblad terms from the symmetries described by elements of a maximally abelian subalgebra of the centraliser. This led to the motivation in Section 3.5.1 where we defined the notion of Lindblad terms of generalised canonical form for obtaining *diagonal mixed state fixed points* and showed that they too are related to a maximally abelian subalgebra

$$\mathfrak{a}_\rho = \mathrm{span}_{\mathbb{R}} \left\{ \mathrm{i}\sigma_{\mathbf{m}} \in \mathfrak{su}(N) \mid m_k \in \{1, z\} \right\} . \tag{3.127}$$

Therefore, Algorithm 1 in Section 3.3.2 which shows how to obtain sets of Lindblad terms which drive the system to the desired pure state fixed point can be generalised in the following sense. For a diagonal target fixed point of full rank (resp. rank degenerate), Theorem 9 (resp. Theorem 10) provided a means of constructing sets of Lindblad terms which drove the system to the target fixed point uniquely by using sets of specially selected projection operators. By the above discussion, we see that these projection operators are *shifted* (cf. Eqn. (3.96)) elements of the maximally abelian subalgebra given by Eqn. (3.127). The generalisation of Algorithm 1 for diagonal mixed state fixed points is given by Algorithm 3.

---

**Algorithm 3:** Unique Target Diagonal* Mixed State Fixed Point Via State Symmetries

---

*Input:* Diagonal Target state $\rho$ of rank $1 < r \leq 2^n$

*Output:* Set(s) of Lindblad terms

1. Establish $\mathfrak{a}_\rho \subseteq \mathfrak{s}_\rho$ is given by Eqn. (3.127)
2. Construct diagonal orthogonal projections $P_k$ from the max. abelian subalgebra elements
3. For each $P_k$, identify a paired element $G_k := \sigma_{\mathbf{p}_k} \in \mathcal{G}_n$
4. If rank $1 < \rho < 2^n$
   - 4a. Ensure the set of all such pairs satisfy Eqns. (3.117)-(3.119)
   - 4b. If not, return to 3. and select different $G_k$ elements
   - 4c. For all $k$, construct $V_k = \sqrt{\gamma_k} G_k P_k$ from the solution set and $\gamma_k$ according to Theorem 10
   - 4d. Return Lindblad term solution set $\{V_k\}$
5. Else (and the fixed point is full rank)
   - 5a. Ensure the set of all pairs satisfy Eqns. (3.104) and (3.105)
   - 5b. If not, return to 3. and select different $G_k$ elements
   - 5c. For all $k$, construct $V_k = \sqrt{\gamma_k} G_k P_k$ from the solution set and $\gamma_k$ according to Theorem 9
   - 5d. Return Lindblad term solution set $\{V_k\}$

---

* with its non-zero eigenvalues being non-degenerate

# Chapter 4

# Lie Wedges Associated to Open Quantum Systems

## 4.1 Introduction

We first briefly summarise the key concepts introduced in Section 1.2. *Controlled* Markovian quantum dynamics are appropriately addressed as right-invariant *bilinear control systems* of the form [16, 13, 15, 17]

$$\dot{\rho}(t) = -\mathcal{L}_{u(t)}\big(\rho(t)\big) \quad , \quad \rho(0) \in \mathfrak{pos}_1(N) \; , \tag{4.1}$$

where $\mathcal{L}_u$ now depends on some control variable $u \in \mathbb{R}^m$. Here, we focus on *coherently controlled* open systems. This means that $\mathcal{L}_u$ has the following special from

$$\mathcal{L}_u(\rho) = \mathrm{i}\,\mathrm{ad}_{H_u}(\rho) + \Gamma(\rho) \quad \text{with} \quad \mathrm{ad}_{H_u} := \mathrm{ad}_{H_d} + \sum_{j=1}^{m} u_j \,\mathrm{ad}_{H_j} \; . \tag{4.2}$$

Note that the control terms $\mathrm{i}\,\mathrm{ad}_{H_j}$ with *control Hamiltonians* $H_j \in \mathfrak{her}(N)$ are usually switched by piecewise constant *control amplitudes* $u_j(t) \in \mathbb{R}$. The drift term of Eqn. (4.2) is then composed of two parts, (i) the term $\mathrm{i}\,\mathrm{ad}_{H_d}$ (in abuse of language sometimes called 'Hamiltonian' drift) accounting for the coherent time evolution and (ii) a dissipative Lindblad part $\Gamma$. So $\mathcal{L}_u$ denotes the *coherently controlled Lindbladian.* As in the uncontrolled case, the system given by Eqn. (4.1) acts on the vector space of all Hermitian operators leaving the set of all density operators invariant. Furthermore, Eqn. (4.1) allows a group lift to $GL(\mathfrak{her}(N))$ which henceforth is referred to as $(\Sigma)$, i.e.

$$(\Sigma) \qquad \dot{X}(t) = -\mathcal{L}_{u(t)}\,X(t), \qquad X(0) \in GL(\mathfrak{her}(N)) \; , \tag{4.3}$$

where for constant control $u(t) \equiv u$, the solutions of which are of the form $T_u(t) := e^{-t\mathcal{L}_u}$ and therefore are *Markovian quantum maps.* We can then consider the *system semigroup* $\mathbf{P}_\Sigma$ associated to $(\Sigma)$ which is given by

$$\mathbf{P}_\Sigma = \langle T_u(t) = \exp(-t\mathcal{L}_u) \,|\, t \geq 0, u \in \mathbb{R}^m \rangle_S \; . \tag{4.4}$$

Now, recall from Section 1.1 that to any closed subsemigroup $\mathbf{S}$ of a group $\mathbf{G}$ its tangent cone $L(\mathbf{S})$ at the identity $\mathbb{1}$ is given by

$$L(\mathbf{S}) := \{A \in \mathfrak{g} \,|\, \exp(tA) \in \mathbf{S} \;\text{ for all } t \geq 0\} \; . \tag{4.5}$$

By Theorem 2, we obtained the following fundamental result that $\overline{\mathbf{P}}_\Sigma$ is a *Lie* subsemigroup and hence

$$\overline{\mathbf{P}}_\Sigma = \overline{\langle \exp(\mathfrak{w}_\Sigma) \rangle}_S \quad \text{where} \quad \mathfrak{w}_\Sigma := L(\overline{\mathbf{P}}_\Sigma) \; . \tag{4.6}$$

This motivated the terminology that $\mathfrak{w}_\Sigma$ is *the* Lie wedge associated to a control system ($\Sigma$) since it is the smallest (global) Lie wedge which contains all evolution directions of the form $\mathcal{L}_u = \mathrm{i}\,\mathrm{ad}_{H_u} + \Gamma$, $u \in \mathbb{R}^m$.

With these semigroup fundamentals refreshed, we can now diverge into different notions of controllability in open systems. To distinguish between varying degrees of control, we define three Lie algebras: the *control Lie algebra* $\mathfrak{k}_c$, the *extended control Lie algebra* $\mathfrak{k}_d$, and the *system Lie algebra* $\mathfrak{g}_\Sigma$ as follows

$$\begin{aligned} \mathfrak{k}_c &:= \langle \mathrm{i}\,\mathrm{ad}_{H_j} \,|\, j = 1, \dots, m \rangle_{\mathsf{Lie}} \; , \\ \mathfrak{k}_d &:= \langle \mathrm{i}\,\mathrm{ad}_{H_d}, \mathrm{i}\,\mathrm{ad}_{H_j} \,|\, j = 1, \dots, m \rangle_{\mathsf{Lie}} \; , \\ \mathfrak{g}_\Sigma &:= \langle \mathcal{L}_u |\, u_j \in \mathbb{R} \rangle_{\mathsf{Lie}} = \langle \mathrm{i}\,\mathrm{ad}_{H_d} + \Gamma, \mathrm{i}\,\mathrm{ad}_{H_j} \,|\, j = 1, \dots, m \rangle_{\mathsf{Lie}} \; . \end{aligned} \tag{4.7}$$

Note that $\mathfrak{g}_\Sigma$ is different from $\mathfrak{k}_d$, because it contains the total drift term comprised of both the Hamiltonian component are dissipative component ($\mathrm{i}\,\mathrm{ad}_{H_d} + \Gamma$) for the Lie closure, whereas $\mathfrak{k}_d$ only uses the Hamiltonian component $\mathrm{i}\,\mathrm{ad}_{H_d}$. Then ($\Sigma$) is said to fulfill condition (H), (WH), and (A), respectively, if

$$(H) \quad \mathfrak{k}_c = \mathrm{ad}_{\mathfrak{su}(N)} \tag{4.8}$$

$$(WH) \quad \mathfrak{k}_d = \mathrm{ad}_{\mathfrak{su}(N)} \text{ while } \mathfrak{k}_c \neq \mathrm{ad}_{\mathfrak{su}(N)} \tag{4.9}$$

$$(A) \quad \mathfrak{g}_\Sigma = \mathfrak{g}^{LK} \quad (\text{or } \mathfrak{g}_0^{LK} \text{ for unital systems}) \; , \tag{4.10}$$

where $\mathfrak{g}^{LK}$ and $\mathfrak{g}_0^{LK}$ are the Lindblad-Kossakowski Lie algebra and its unital subalgebra which were the focus of Chapter 2.

While condition (A) respects a standard construction of non-linear control theory [28, 31] to express accessibility, conditions (H) and (WH) serve to characterize different types of *controllability* of the Hamiltonian part of ($\Sigma$) in the absence of relaxation: Condition (H) says that the Hamiltonian part is fully controllable even *without* resorting to the drift Hamiltonian, whereas condition (WH) yields full controllability of the Hamiltonian part with the drift Hamiltonian being *necessary.* We refer to the first scenario as *(fully) H-controllable* and to the second as satisfying the (WH)-condition. Generically, open systems ($\Sigma$) given by Eqn. (1.16) meet the accessibility condition (A) [3, 34]. Finally, note that via $\mathrm{e}^{\mathrm{i}\,\mathrm{ad}_H}(\rho) = \mathrm{e}^{\mathrm{i}H} \rho\, \mathrm{e}^{-\mathrm{i}H}$, the Lie algebra $\mathrm{ad}_{\mathfrak{su}(N)}$ generates the Lie group $\mathrm{Ad}_{SU(N)} \overset{\mathrm{iso}}{=} PSU(N)$ here acting on $\mathfrak{her}_0(N)$ by conjugation.

## 4.2 Computing Lie Wedges I: Approximations and Theory

In view of the examples worked out in detail in this section (and in Appendix D) we first sketch how to approximate a Lie wedge of a controlled Markovian system via an *inner approximation* following [36, 16]. It consists of the following steps:

(1) form the smallest closed convex cone $\mathfrak{w}$ containing $\mathrm{i}\,\mathrm{ad}_{H_d} + \Gamma$ i.e. $\mathfrak{w} = \mathbb{R}_0^+(\mathrm{i}\,\mathrm{ad}_{H_d} + \Gamma)$;

(2) compute the edge $E(\mathfrak{w})$ of the wedge and the smallest Lie algebra $\mathfrak{e}$ containing $E(\mathfrak{w})$, i.e. $\mathfrak{e} := \langle E(\mathfrak{w})\rangle_{\text{Lie}}$;

(3) make the wedge invariant under the Ad action of $\mathfrak{e}$ by forming the set $\bigcup_{A\in\mathfrak{e}} \text{Ad}_{\exp A}(\mathfrak{w})$;

(4) update by taking the convex hull conv $\{S\}$ of the set $S$ obtained in step (3) .

The resulting final wedge $\underline{\mathfrak{w}}$ is henceforth referred to as *inner approximation* to the global Lie wedge $\mathfrak{w}_\Sigma$.

Now, the crucial question arises whether the inner approximation $\underline{\mathfrak{w}}$ is global or not. If it is global, Theorem 2 guarantees that $\underline{\mathfrak{w}}$ is equal to $\mathfrak{w}_\Sigma$. Proving that this inner approximation or another type of outer approximation coincide with the associated system Lie wedge is a delicate problem. In general, up until now there has been no such general method to prove equalities. For completeness, we present two previous results which partially solved the problem of determining the associated Lie wedge. The authors of [16] assumed the system was unital and used an outer approximation to show that for unital single qubit systems which satisfy condition $(H)$ (i.e. fully Hamiltonian controllable) the outer approximation was exact.

**Theorem 12** ([16])**.** *Let (Σ) be a unital controlled open system as in Eqn.* (1.16)*. If there exists a* pointed *cone* $\mathfrak{c}_0$ *in the set of all positive semidefinite operators that act on* $\mathfrak{her}_0(N)$ *so that*

*(1)* $\Gamma \in \mathfrak{c}_0$

*(2)* $[\mathfrak{c}_0, \mathfrak{c}_0] \subset \text{ad}_{\mathfrak{su}(N)}$

*(3)* $[\mathfrak{c}_0, \text{ad}_{\mathfrak{su}(N)}] \subset (\mathfrak{c}_0 - \mathfrak{c}_0)$

*(4)* $\text{Ad}_U\, \mathfrak{c}_0\, \text{Ad}_{U^\dagger} \subset \mathfrak{c}_0$ *for all* $U \in SU(N)$*,*

*then the subsemigroup associated to (Σ) follows the inclusion* $\overline{\boldsymbol{P}}_\Sigma \subseteq \text{Ad}_{SU(N)} \cdot \exp(-\mathfrak{c}_0)$ *and hence its Lie wedge obeys the relation* $\mathfrak{w}_\Sigma \subseteq \text{ad}_{\mathfrak{su}(N)} \oplus (-\mathfrak{c}_0)$*, i.e.* $\text{ad}_{\mathfrak{su}(N)} \oplus (-\mathfrak{c}_0)$ *is a global outer approximation to* $\mathfrak{w}_\Sigma$*.*

**Corollary 4.2.1** ([16, 2])**.** *Let (Σ) be a unital single-qubit system satisfying condition (H) with a generic*[1] *Lindblad term Γ. Then the system semigroup is given by* $\overline{\boldsymbol{P}}_\Sigma = \text{Ad}_{SU(2)} \cdot \exp(-\mathfrak{c}_0)$*, where the cone*

$$\mathfrak{c}_0 := \mathbb{R}_0^+ \,\text{conv}\, \{\text{Ad}_U\ \Gamma\ \text{Ad}_{U^\dagger}\ \mid U \in SU(2)\}$$

*is contained in the set of all positive semidefinite elements in* $\mathfrak{gl}(\mathfrak{her}_0)$*. Furthermore* $\mathfrak{w}_\Sigma = \text{ad}_{\mathfrak{su}(2)} \oplus (-\mathfrak{c}_0)$*.*

We now present the final result along these directions. It solves the inner and outer approximation problem by proving the inner approximation is in fact global and is therefore *the* associated Lie wedge to a control system (Σ). A version of the following result was presented with a proof for unital systems in [40]. Here we extend the result by adding a minor change in proof to now additionally accommodate non-unital systems.

[1] In [16] Corollary 4.2.1 is stated under the above genericity assumption; yet one can drop this additional condition.

**Theorem 13** ([40])**.** *Let* $(\Sigma)$ *be a coherently controlled open quantum system and assume that the Lie group* $\boldsymbol{K}$ *which is generated by the control Lie algebra* $\mathfrak{k}_c := \langle \mathrm{i}\,\mathrm{ad}_{H_1}, \mathrm{i}\,\mathrm{ad}_{H_2}, \dots, \mathrm{i}\,\mathrm{ad}_{H_m} \rangle_{\mathsf{Lie}}$ *is closed. Then the Lie wedge associated to the system is given by*

$$\mathfrak{w}_\Sigma = \mathfrak{k}_c \oplus (-\mathfrak{c})\ , \quad where \quad \mathfrak{c} := \mathbb{R}_0^+ \operatorname{conv}\left\{ \mathrm{Ad}_U \left(\mathrm{i}\,\mathrm{ad}_{H_d} + \Gamma\right) \mathrm{Ad}_U^\dagger \ \mid U \in \boldsymbol{K} \right\}\ , \tag{4.11}$$

*and therefore the reachable set of operators is given by*

$$\overline{\boldsymbol{P}}_\Sigma = \overline{\langle \exp(\mathfrak{w}_\Sigma) \rangle}_S\ . \tag{4.12}$$

*Proof.* If the edge of the wedge $E(\mathfrak{w}_\Sigma) = \mathfrak{w}_\Sigma \cap -\mathfrak{w}_\Sigma$ is given by the control algebra then by construction this is a Lie wedge. We prove this equality first. Assume that $A + B \in E(\mathfrak{w}_\Sigma)$ with $A \in \mathfrak{k}_c$ and $B \in \mathfrak{c}$. Hence, there exists some $A' \in \mathfrak{k}_c$ and $B' \in \mathfrak{c}$ such that $-(A+B) = A' + B' \in E(\mathfrak{w}_\Sigma)$ which implies that $A + A' = -B - B' \in E(\mathfrak{w}_\Sigma)$. It can be shown (see for Example [34]) that $\operatorname{tr}(\Gamma) > 0$ and thus $\mathfrak{k}_c \cap \mathfrak{c} = \{0\}$. Hence the equality $A + A' = -B - B' \in E(\mathfrak{w}_\Sigma)$ implies that $A' = -A$ and $B' = -B$. Therefore, it suffices to show that $B = B' = 0$. Well, since $\operatorname{tr}(\Gamma) > 0$ then clearly $\operatorname{tr}(C) > 0$ for all $C \in \mathfrak{c} \setminus \{0\}$ and hence $\operatorname{tr}(B') = -\operatorname{tr}(B)$ which implies $B' = B = 0$ and therefore $E(\mathfrak{w}_\Sigma) = \mathfrak{k}_c$.

The positive trace argument again implies that the cone $\mathfrak{c}$ is pointed and thus in light of Proposition 5.2.1, all we need to show is that $\mathfrak{c}$ is closed. By assumption, $\mathbf{K}$ is closed (and hence is compact) which implies that the set $\operatorname{conv}\{\mathrm{Ad}_U \left(\mathrm{i}\,\mathrm{ad}_{H_d} + \Gamma\right) \mathrm{Ad}_U^\dagger \ \mid U \in \mathbf{K}\}$ is also closed (and compact), which implies that $\mathfrak{c}$ is closed.

We now prove that the Lie wedge is global. First note that Theorem 1 shows that the Lindblad-Kossakowski Lie wedge $\mathfrak{w}^{LK}$ cf. Eqn. (1.14) (which contains every possible individual system $(\Sigma)$ Lie wedge) is *global*. Furthermore, notice that the Lie wedge $\mathfrak{w}_\Sigma$ given by Eqn. (4.11) is contained in $\mathfrak{w}^{LK}$ and its edge is the Lie algebra of the closed subgroup $\mathbf{K}$. Then since its edge satisfies $E(\mathfrak{w}_\Sigma) = E(\mathfrak{w}^{LK}) \cap \mathfrak{w}_\Sigma$, by Corollary 1.1.1 of Chapter 1, $\mathfrak{w}_\Sigma$ is in fact *global*.

Now, if the system group $\mathbf{G}_\Sigma$ is closed, then by Theorem 2 the global Lie wedge $\mathfrak{w}_\Sigma$ (constructed via the inner approximation) is the associated Lie wedge to the system. By Remark 1 of Chapter 1, if $\mathbf{G}_\Sigma$ is not closed, then a simple restatement of Theorem 2 where the closures are now taken with respect to $\mathbf{G}_\Sigma$ shows again that the Lie wedge $\mathfrak{w}_\Sigma$ is the associated Lie wedge to $(\Sigma)$. Finally, Eqn. (4.12) follows immediately by Theorem 2. □

For the remaining of this thesis we will neglect the subscript $\Sigma$ on the associated Lie wedge. Furthermore, if the Lie wedge is associated to a unital system $(\Sigma)$ we will use the subscript "0" as $\mathfrak{w}_0 = E(\mathfrak{w}_0) \oplus \mathfrak{c}_0$, whereas it will be omitted if the system is non-unital.

## 4.3 Unital Single-Qubit Systems

We start out by analysing the structure of the simplest type of Lie wedges - those which are single qubit and unital. Recall that for a single open qubit system, the controlled master equation ( in fact its group lift ) is of the form

$$\dot{X}(t) = -\Big(\mathrm{i}\big(H_d + \sum_j u_j H_j\big) + \Gamma\Big)\, X(t)\,. \tag{4.13}$$

where $X(t)$ may be a considered a density operator (via the so-called vec-representation or a qubit quantum channel represented in $GL(4, \mathbb{C})$. To ensure complete positivity, the relaxation term $\Gamma$ for the standard unital single-qubit systems (with $V_k$ Hermitian) is given by $\Gamma = 2\sum_k \gamma_k\, \hat\sigma_k^2$ to give the nicely structured generator

$$\mathcal{L}_u = \mathrm{i}\big(H_d + \sum_j u_j H_j\big) + \; 2 \sum_{k\in\{x,y,z\}} \gamma_k\, \hat\sigma_k^2\ . \tag{4.14}$$

The generator is of this form because the i$H$ terms are in the $\mathfrak{k}$-part of the Cartan decomposition of $\mathfrak{gl}(4,\mathbb{C})$ into skew-Hermitian ($\mathfrak{k}$) and Hermitian ($\mathfrak{p}$) matrices, whereas the $\hat\sigma_k^2$ terms are in the $\mathfrak{p}$-part.

### 4.3.1 Systems Satisfying Condition (H): Single Lindblad Term

We start out by considering the class of fully Hamiltonian controllable unital single-qubit systems whose dissipation is governed by a *single* Lindblad operator $\hat\sigma_k^2$ for some $k \in \{x,y,z\}$ i.e. two of the three prefactors $\gamma_x, \gamma_y, \gamma_z$ have to vanish. Choosing the controls $\hat\sigma_x$ and $\hat\sigma_y$ such that the system fulfils condition (H) (since $\langle \mathrm{i}\hat\sigma_x, i\hat\sigma_y\rangle_{\mathsf{Lie}} = \mathrm{ad}_{\mathfrak{su}(2)}$), then it is actually immaterial which single Pauli matrix is chosen as the Lindblad operator $\hat\sigma_k^2$ because all of the Pauli matrices are unitarily equivalent. So without loss of generality, one may choose $k = z$, i.e. $\gamma_x = 0$, $\gamma_y = 0$, and $\gamma_z =: \gamma$. Therefore the fully Hamiltonian controllable version of the bit-flip, phase-flip, and bit-phase-flip channels are dynamically equivalent in as much as they have (up to unitary equivalence) a common global Lie wedge

$$\mathfrak{w}_0 := \mathrm{ad}_{\mathfrak{su}(2)} \oplus -\,\mathfrak{c}_0\ , \tag{4.15}$$

with the cone $\mathfrak{c}_0$ being given by

$$\mathfrak{c}_0 := \mathbb{R}_0^+ \mathrm{conv}\big\{\, \mathrm{Ad}_U\ \hat\sigma_z^2\ \mathrm{Ad}_{U^\dagger}\ \big|\, U \in SU(2)\big\} = \mathbb{R}_0^+ \mathrm{conv}\big\{\mathrm{ad}_M^2 \mid M \in \mathcal{O}_{SU(2)}(\sigma_z)\big\} \tag{4.16}$$

where $\mathcal{O}_{SU(2)}(\sigma_z)$ is the SU(2)-unitary orbit of $\sigma_z$.

### 4.3.2 Systems Satisfying Condition (WH): Single Lindblad Term

Now we investigate an important class of standard unital single-qubit systems which are particularly simple in three regards

(i) their dissipative term is governed by a single Lindblad operator, $\Gamma := 2\gamma\hat\sigma_k^2$ for some $k \in \{x,y,z\}$;

(ii) their switchable Hamiltonian control is brought about by a single Hamiltonian $\hat\sigma_c$ for some $c \in \{x,y,z\}$;

(iii) their non-switchable Hamiltonian drift is $\hat\sigma_d$ for some $d \in \{x,y,z\}$.

Applying the algorithm for the inner approximation of the Lie wedge, we get in step (1)

$$\mathfrak{w}_{dk}^c(1) := \mathrm{i}\,\mathbb{R}\hat\sigma_c \oplus -\mathbb{R}_0^+\big(\mathrm{i}\hat\sigma_d + 2\gamma\hat\sigma_k^2\big)\ , \tag{4.17}$$

where again we note the separation by $\mathfrak{k}$-$\mathfrak{p}$ components. In step (2) we identify the span generated by the control $i\hat\sigma_c$ as the edge $E(\mathfrak{w})$ of the wedge. So the conjugation has to be by the control subgroup, i.e. by $e^{-\mathrm{i}2\theta\hat\sigma_c} = e^{+\mathrm{i}\theta\sigma_c^\top} \otimes e^{-\mathrm{i}\theta\sigma_c}$. Thus in step (3) one obtains as $\mathfrak{k}$" component of the conjugated drift

$$K_d^c(\theta) := \quad e^{-\mathrm{i}\theta\hat{\sigma}_c}(\mathrm{i}\hat{\sigma}_d)e^{\mathrm{i}\theta\hat{\sigma}_c} = \begin{cases} \mathrm{i}\,\hat{\sigma}_d & \text{for } c = d \\ \mathrm{i}\,\cos(\theta)\hat{\sigma}_d + \mathrm{i}\,\varepsilon_{cdq}\sin(\theta)\hat{\sigma}_q & \text{else} \end{cases} \tag{4.18}$$

and as $\mathfrak{p}$-component

$$P_k^c(\theta) := \quad e^{-i\theta\hat{\sigma}_c}(2\gamma\,\hat{\sigma}_k^2)e^{i\theta\hat{\sigma}_c} = \begin{cases} 2\gamma\,\hat{\sigma}_k^2 & \text{for } c = k \\ 2\gamma\,\big(\cos(\theta)\hat{\sigma}_k + \,\varepsilon_{ckr}\sin(\theta)\hat{\sigma}_r\big)^2 & \text{else .} \end{cases} \tag{4.19}$$

The last expression (for $c \neq k$) can be further resolved using the anticommutator $\{A, B\}_+ := AB + BA$

$$P_k^c(\theta) = 2\gamma \begin{bmatrix} \cos^2(\theta) \\ \sin^2(\theta) \\ \cos(\theta)\sin(\theta) \end{bmatrix} \cdot \begin{bmatrix} \hat{\sigma}_k^2 \\ \hat{\sigma}_r^2 \\ \varepsilon_{ckr}\{\hat{\sigma}_k,\hat{\sigma}_r\}_+ \end{bmatrix} = \frac{\gamma}{2} \begin{bmatrix} 2 \\ 1+\cos(2\theta) \\ 1-\cos(2\theta) \\ \sin(2\theta) \end{bmatrix} \cdot \begin{bmatrix} \mathbb{1} \\ -(\sigma_k^\top \otimes \sigma_k) \\ -(\sigma_r^\top \otimes \sigma_r) \\ -\varepsilon_{ckr}((\sigma_k^\top \otimes \sigma_r)+(\sigma_r^\top \otimes \sigma_k)) \end{bmatrix}, \tag{4.20}$$

where the latter identity gives a decomposition into mutually orthogonal Pauli-basis elements. To summarize, if the control Hamiltonian neither commutes with the Hamiltonian part nor with the dissipative part of the drift, one obtains in terms of the above $K_d^c(\theta)$ and $P_k^c(\theta)$

$$\mathfrak{c}_{dk}^c := \mathbb{R}_0^+ \operatorname{conv}\{K_d^c(\theta) + P_k^c(\theta) \,|\, \theta \in \mathbb{R}\ \} \tag{4.21}$$

However, if $[\hat{\sigma}_c, \Gamma] = 0$, then the convex cone in Eqn. (4.21) simplifies by $P_k^c(\theta) = \Gamma$ to

$$\mathfrak{c}_{dk}^c = \mathbb{R}_0^+ \operatorname{conv}\left\{ \begin{bmatrix} \cos(\theta) \\ \sin(\theta) \\ 1 \end{bmatrix} \cdot \begin{bmatrix} \mathrm{i}\hat{\sigma}_d \\ \mathrm{i}\varepsilon_{cdq}\hat{\sigma}_q \\ \Gamma \end{bmatrix} \;\middle|\; \theta \in \mathbb{R} \right\} \quad . \tag{4.22}$$

The final Lie wedge admits the orthogonal decomposition

$$\mathfrak{w}_{dk}^c := \mathrm{i}\mathbb{R}\,\hat{\sigma}_c \oplus -\mathfrak{c}_{dk}^c\ . \tag{4.23}$$

**Table 4.1:** Controlled Single-Qubit Channels and Their Lie Wedges

| Channel | Lindblad Terms*) | Kraus Operators | Lie Wedges | |
|---|---|---|---|---|
| | | | WH-Controllable | H-Controllable |
| Bit Flip | $V_1 = \sqrt{a_{11}}\ \sigma_x$ | $E_1 = \sqrt{r_{11}}\ \sigma_x$<br>$E_0 = \sqrt{q_{11}}\ \mathbb{1}$ | $\mathfrak{w}^c_{dx} = \langle \mathrm{i}\hat{\sigma}_c \rangle \oplus -\mathfrak{c}^c_{dx}$<br>[see Eqns. (4.21,4.22)] | $\mathfrak{w}_0 = \mathrm{ad}_{\mathfrak{su}(2)} \oplus -\ \mathfrak{c}_0$<br>[see Eqn. (4.16)] |
| Phase Flip | $V_1 = \sqrt{a_{22}}\ \sigma_z$ | $E_1 = \sqrt{r_{22}}\ \sigma_z$<br>$E_0 = \sqrt{q_{22}}\ \mathbb{1}$ | $\mathfrak{w}^c_{dz} = \langle \mathrm{i}\hat{\sigma}_c \rangle \oplus -\mathfrak{c}^c_{dz}$<br>[see Eqns. (4.21,4.22)] | ——”—— |
| Bit-Phase Flip | $V_1 = \sqrt{a_{33}}\ \sigma_y$ | $E_1 = \sqrt{r_{33}}\ \sigma_y$<br>$E_0 = \sqrt{q_{33}}\ \mathbb{1}$ | $\mathfrak{w}^c_{dy} = \langle \mathrm{i}\hat{\sigma}_c \rangle \oplus -\mathfrak{c}^c_{dy}$<br>[see Eqns. (4.21,4.22)] | ——”—— |
| Depolarizing | $V_1 = \sqrt{a_{11}}\ \sigma_x$<br>$V_2 = \sqrt{a_{22}}\ \sigma_y$<br>$V_3 = \sqrt{a_{33}}\ \sigma_z$ | $E_1 = \sqrt{r_1}\ \sigma_x$<br>$E_2 = \sqrt{r_2}\ \sigma_y$<br>$E_3 = \sqrt{r_3}\ \sigma_z$<br>$E_0 = \sqrt{r_0}\ \mathbb{1}$ | $\mathfrak{w}^c_{d,xyz} = \langle \mathrm{i}\hat{\sigma}_c \rangle \oplus -\mathfrak{c}^c_{d,xyz}$<br>[see Eqns. (4.18,4.27,4.30)] | $\mathfrak{w}_0 = \mathrm{ad}_{\mathfrak{su}(2)} \oplus -\ \mathfrak{c}_{xyz}$<br>[see Eqns. (4.31,4.32)] |
| Amplitude Damping | $V_1 = \sqrt{a_{11}}\ (\sigma_x + \mathrm{i}\sigma_y)$ | $E_1^{**)}$<br>$E_0^{**)}$ | $\mathfrak{w}^c_{d} = \langle \mathrm{i}\hat{\sigma}_c \rangle \oplus -\mathfrak{c}^c_{d}$<br>[see Eqns. (4.40,4.41,4.42)] | $\mathfrak{w}_0 = \mathrm{ad}_{\mathfrak{su}(2)} \oplus -\ \mathfrak{c}_0$<br>[see Eqns. (4.36,4.37)] |

*) Primary operators are for purely dissipative time evolutions (no Hamiltonian drift no control). Then the time dependence of the Kraus operators roots in the GKS matrix $\{a_{ii}\}_{i=1}^{3}$. Define: $\lambda_1 := a_{22} + a_{33}, \lambda_2 := a_{22} - a_{33}$, $\lambda_3 := a_{11} + a_{22}$, and thereby $q_{ii} := \frac{1}{2}(1 + e^{-a_{ii}t})$, $r_{ii} := \frac{1}{2}(1 - e^{-a_{ii}t})$, $r_0 := \frac{1}{4}(1 + e^{-\lambda_1 t} + e^{-\lambda_2 t} + e^{-\lambda_3 t})$, $r_1 := \frac{1}{4}(1 - e^{-\lambda_1 t} + e^{-\lambda_2 t} - e^{-\lambda_3 t})$, $r_2 := \frac{1}{4}(1 + e^{-\lambda_1 t} - e^{-\lambda_2 t} - e^{-\lambda_3 t})$, $r_3 := \frac{1}{4}(1 - e^{-\lambda_1 t} - e^{-\lambda_2 t} + e^{-\lambda_3 t})$.

**) See for instance Section 8.4.1 of [39].

### 4.3.3 Application: Bit-Flip and Phase-Flip Channels

Table 4.1 provides an overview of standard unital noise, their respective Lindblad terms they are described by as well as the corresponding system Lie wedges. In the absence of any *coherent* drift or control brought about by the respective Hamiltonians $H_d = \hat{\sigma}_d$ or $H_j = \hat{\sigma}_j$, the Kraus representations are standard. By allowing for drifts and controls, the Kraus rank $K$ of the channel usually increases to $K{=}4$ with exception of a single $\hat{\sigma}_d$ or $\hat{\sigma}_j$ commuting with the single Lindblad operator $\hat{\sigma}_k^2$ keeping $K{=}2$. Also the time dependences become more involved. Explicit results will be given elsewhere.

Two further remarks are in order. Clearly, when the (H) condition is satisfied, the Lie wedges of all the three channels become equivalent since the Pauli matrices, and thus the corresponding noise generators are unitarily equivalent. Now suppose the (WH)" condition is satisfied for a control system with a Hamiltonian drift term described by $\hat{\sigma}_z$. Upon including relaxation, now there are two different scenarios: if the control Hamiltonian (indexed by $c \in \{x, y, z\}$) commutes with the noise generator (indexed by $k \in \{x, y, z\}$), one finds a situation as in Eqn. (4.22), otherwise the scenario is more general as in Eqn. (4.21).

### 4.3.4 Systems Satisfying Condition (WH): Several Lindblad Terms

Consider a unital qubit system satisfying the (WH)-condition and whose Lindblad generator is associated to $\ell = 2$ or $\ell = 3$ different Lindblad terms $\hat{\sigma}_k^2$. Then one obtains the following generalisations of the symmetric component $P_k^c(\theta) \in \mathfrak{c}_{dk}^c$.

For $\ell = 2$ and $\sigma_c \perp \sigma_k$, $\sigma_c = \sigma_{k'}$

$$P_{kk'}^c(\theta) = 2 \begin{bmatrix} \gamma' \\ \gamma\cos^2(\theta) \\ \gamma\sin^2(\theta) \\ \gamma\cos(\theta)\sin(\theta) \end{bmatrix} \cdot \begin{bmatrix} \hat{\sigma}_{k'}^2 \\ \hat{\sigma}_k^2 \\ \hat{\sigma}_r^2 \\ \varepsilon_{ckr}\{\hat{\sigma}_k,\hat{\sigma}_r\}_+ \end{bmatrix} \tag{4.24}$$

$$= \tfrac{1}{2} \begin{bmatrix} \gamma' \\ 2(\gamma+\gamma') \\ \gamma(1+\cos(2\theta)) \\ \gamma(1-\cos(2\theta)) \\ \gamma\sin(2\theta) \end{bmatrix} \cdot \begin{bmatrix} -(\sigma_{k'}^\top \otimes \sigma_{k'}) \\ \mathbb{1} \\ -(\sigma_k^\top \otimes \sigma_k) \\ -(\sigma_r^\top \otimes \sigma_r) \\ -\varepsilon_{ckr}((\sigma_r^\top \otimes \sigma_k)+(\sigma_k^\top \otimes \sigma_r)) \end{bmatrix}. \tag{4.25}$$

(4.26)

while for $\ell = 3$ and $\sigma_c \perp \sigma_k$, $\sigma_c \perp \sigma_{k'}$, $\sigma_c = \sigma_{k''}$

$$P_{kk'k''}^c(\theta) = 2 \begin{bmatrix} \gamma'' \\ \gamma\cos^2(\theta)+\gamma'\sin^2(\theta) \\ \gamma'\cos^2(\theta)+\gamma\sin^2(\theta) \\ (\gamma-\gamma')\cos(\theta)\sin(\theta) \end{bmatrix} \cdot \begin{bmatrix} \hat{\sigma}_{k''}^2 \\ \hat{\sigma}_k^2 \\ \hat{\sigma}_{k'}^2 \\ \varepsilon_{ckk'}\{\hat{\sigma}_k,\hat{\sigma}_{k'}\}_+ \end{bmatrix} \tag{4.27}$$

$$= \tfrac{1}{2} \begin{bmatrix} \gamma'' \\ 2(\gamma+\gamma'+\gamma'') \\ \gamma+\gamma'+(\gamma-\gamma')\cos(2\theta) \\ \gamma+\gamma'-(\gamma-\gamma')\cos(2\theta) \\ (\gamma-\gamma')\sin(2\theta) \end{bmatrix} \cdot \begin{bmatrix} -(\sigma_{k''}^\top \otimes \sigma_{k''}) \\ \mathbb{1} \\ -(\sigma_k^\top \otimes \sigma_k) \\ -(\sigma_{k'}^\top \otimes \sigma_{k'}) \\ -\varepsilon_{ckk'}((\sigma_{k'}^\top \otimes \sigma_k)+(\sigma_k^\top \otimes \sigma_{k'})) \end{bmatrix}. \tag{4.28}$$

(4.29)

which for $\gamma = \gamma' = \gamma''$ simplifies to

$$P_{kk'k''}^c(\theta) = \Gamma = 2\gamma(\hat{\sigma}_k^2 + \hat{\sigma}_{k'}^2 + \hat{\sigma}_{k''}^2) \; . \tag{4.30}$$

### 4.3.5 Application: Depolarizing Channel

Clearly a system which is fully Hamiltonian controllable and subject to depolarizing noise follows the bit-flip and phase-flip channels in the structure of its associated Lie wedge as

$$\mathfrak{w}_0 = \mathrm{ad}_{\mathfrak{su}(2)} \oplus - \mathfrak{c}_{xyz} \ , \tag{4.31}$$

where the cone $\mathfrak{c}_{xyz}$ is given by

$$\mathfrak{c}_{xyz} := \mathbb{R}_0^+ \mathrm{conv}\big\{ \mathrm{Ad}_U (\gamma_x \hat{\sigma}_x^2 + \gamma_y \hat{\sigma}_y^2 + \gamma_z \hat{\sigma}_z^2)\, \mathrm{Ad}_{U^\dagger} \ \mid U \in SU(2)\big\} \tag{4.32}$$

Again, the edge of the wedge is the entire algebra $E(\mathfrak{w}_0) = \mathrm{ad}_{\mathfrak{su}(2)}$ and note that the Lie wedge in the fully Hamiltonian controllable depolarizing channel with *isotropic* noise takes the structure of a Lie semialgebra as will be discussed in Chapter 5, whereas for *anisotropic* relaxation, however, this feature does not arise.

If only condition (WH) is satisfied, there are two distinctions: if the noise contributions are isotropic (i.e. with equal contribution by all the Paulis through $\gamma_x = \gamma_y = \gamma_z$), one finds a cone expressed by Eqns. (4.18) and (4.30). However, in the generic anisotropic case, the cone can be expressed by Eqns. (4.18) and (4.27), see also Table 4.1.

## 4.4 Non-Unital Single-Qubit Systems

For non-unital systems, the Lie wedge can be constructed using a mild generalization of the concepts introduced in the previous section. Here we outline the basic construction and illustrate some physically relevant examples. First note that by the commutation relations of Table E.1 one obtains

$$[\hat{\sigma}_r, [\hat{\sigma}_r, \mathrm{i}\hat{\sigma}_p \hat{\sigma}_q^+]] = \mathrm{i}\hat{\sigma}_p \hat{\sigma}_q^+ \ , \tag{4.33}$$

which is the precondition for the following useful relation

$$e^{-\mathrm{i}\theta\hat{\sigma}_c} (\mathrm{i}\, \hat{\sigma}_p \hat{\sigma}_q^+) e^{\mathrm{i}\theta\hat{\sigma}_c} = \begin{cases} \mathrm{i}\, \hat{\sigma}_p \hat{\sigma}_q^+ & \text{for } c \neq p \text{ and } c \neq q, \qquad \text{else,} \\ \mathrm{i}\, \cos(\theta) \hat{\sigma}_p \hat{\sigma}_q^+ + \mathrm{i}\, \sin(\theta) \big(\delta_{cq} \hat{\sigma}_{[\mathrm{i}c,p]} \hat{\sigma}_q^+ + \delta_{cp} \hat{\sigma}_p \hat{\sigma}_{[\mathrm{i}c,q]}^+\big) \ , \end{cases} \tag{4.34}$$

where $p, q \in \{x, y, z\}$ and $\delta_{cq}, \delta_{cp}$ are the usual Dirac-Delta functions.

### 4.4.1 Application: Amplitude Damping Channel Satisfying Condition (H)

Suppose the control system $(\Sigma)$ satisfies condition (H) and induces amplitude damping noise i.e. the single Lindblad term being given by $V_1 := \sqrt{\gamma}\,(\sigma_x + \mathrm{i}\sigma_y)$ (neglecting the normalising coefficient for simplicity). By Eqn. (2.13) the corresponding Lindblad generator simplifies to

$$\Gamma = 2\gamma \left(\hat{\sigma}_x^2 + \hat{\sigma}_y^2\right) + 4\gamma \mathrm{i} \left(\hat{\sigma}_x \hat{\sigma}_y^+\right) \ . \tag{4.35}$$

Then the associated Lie wedge is given by

$$\mathfrak{w} = \mathrm{ad}_{\mathfrak{su}(2)} \oplus (-\mathfrak{c}) \ , \tag{4.36}$$

where the cone $\mathfrak{c}$ is given by (by direct calculation)

$$\mathfrak{c} = \mathbb{R}_0^+ \text{conv}\big\{ \text{Ad}_U \; \Gamma_V \; \text{Ad}_{U^\dagger} \;\big|\; U \in SU(2)\big\} = \mathbb{R}_0^+ \text{conv}\big\{\Gamma_{UVU^\dagger} \mid U \in SU(2)\big\} \; , \quad (4.37)$$

and we note that the lack of a "0" index on the wedge $\mathfrak{w}$ and pointed cone $\mathfrak{c}$ is intentionally omitted to emphasise the system is non-unital.

### 4.4.2 Application: Amplitude Damping Channel Satisfying Condition (WH)

As we will now show, the Lie wedge associated to a control system which only has weak H-controllability ( i.e. satisfies condition (WH)) and subject to amplitude damping noise, has a very different structure than the previous scenario. Assume we have only the single $\hat{\sigma}_y$ control Hamiltonian. Then applying the algorithm for the inner approximation of the Lie wedge, step (1) gives

$$\mathfrak{w}_z^y(1) := \text{i}\, \mathbb{R}\hat{\sigma}_y \oplus -\mathbb{R}_0^+ \big(\text{i}\hat{\sigma}_z + 2\gamma(\hat{\sigma}_x^2 + \hat{\sigma}_y^2) + 4\gamma\text{i}(\hat{\sigma}_x\hat{\sigma}_y^+)\big) \; . \quad (4.38)$$

In step (3) one obtains as part of the total $\mathfrak{k}$ component of the wedge the conjugated drift

$$K_z^y(\theta) := e^{-\text{i}\theta\hat{\sigma}_y}(\text{i}\hat{\sigma}_z)e^{\text{i}\theta\hat{\sigma}_y} = \text{i}\,\cos(\theta)\hat{\sigma}_z + \text{i}\,\sin(\theta)\hat{\sigma}_x, \quad (4.39)$$

as in Eqn. (4.18). The conjugation of the dissipative component $\Gamma$ now has two components: a unital contractive part and a non-unital translational part given by the term $\hat{\sigma}_x\hat{\sigma}_y^+$. The details are worked out in Section 2.2.2 where we showed that such a translation operator is isomorphic to a vector in $\mathbb{R}^3$ and thus applying a unitary rotation to the operator is equivalent to rotating a vector in three dimensional space. Computing the effect of the control on the dissipation then yields

$$\begin{aligned} e^{-\text{i}\theta\hat{\sigma}_y}\Gamma e^{\text{i}\theta\hat{\sigma}_y} &= 2\gamma\, e^{-\text{i}\theta\hat{\sigma}_y}(\hat{\sigma}_x^2 + \hat{\sigma}_y^2)e^{\text{i}\theta\hat{\sigma}_y} + 4\gamma\text{i}\, e^{-\text{i}\theta\hat{\sigma}_y}\left(\hat{\sigma}_x\hat{\sigma}_y{}^+\right)e^{\text{i}\theta\hat{\sigma}_y} \\ &=: P(\theta)_x^y + P(\theta)_y^y + N_z^y(\theta) \; , \end{aligned}$$

where $P_k^c(\theta)$ are given by Eqns. (4.19) and (4.20), while $N_z^y$ describes the change in the translation component (originally along $+z$-direction)

$$N_z^y(\theta) = 4\gamma\text{i}e^{-\text{i}\theta\hat{\sigma}_y}(\hat{\sigma}_x\hat{\sigma}_y{}^+)e^{\text{i}\theta\hat{\sigma}_y} = 4\gamma\text{i}\big(\cos(\theta)\hat{\sigma}_x\hat{\sigma}_y^+ + \gamma\sin(\theta)\hat{\sigma}_y\hat{\sigma}_z^+\big) \; .$$

Finally, the conjugation of the unital component follows by Eqn. (4.19) and gives

$$2\gamma e^{-\text{i}\theta\hat{\sigma}_y}(\hat{\sigma}_x^2 + \hat{\sigma}_y^2)e^{\text{i}\theta\hat{\sigma}_y} = P_x^y(\theta) + 2\gamma\hat{\sigma}_y^2 \; ,$$

so that the convex cone $\mathfrak{c}$ is given as

$$\mathfrak{c} := \mathbb{R}_0^+ \;\text{conv}\left\{K_z^y(\theta) + P_x^y(\theta) + 2\gamma\hat{\sigma}_y^2 + N^y(\theta) \,|\, \theta \in \mathbb{R}\;\right\} \quad (4.40)$$

and the convex hull is taken over the rather large expression

$$K_z^y(\theta) + P_x^y(\theta) + 2\gamma\hat{\sigma}_y^2 + N_{xy}^y(\theta) = 2\begin{bmatrix} \frac{1}{2}\cos(\theta) \\ \frac{1}{2}\sin(\theta) \\ 2\gamma\cos(\theta) \\ 2\gamma\sin(\theta) \\ \gamma\cos(\theta)\sin(\theta) \\ \gamma\cos^2(\theta) \\ \gamma\sin^2(\theta) \\ \gamma \end{bmatrix} \cdot \begin{bmatrix} \text{i}\,\hat{\sigma}_z \\ \text{i}\,\hat{\sigma}_x \\ \text{i}\,\hat{\sigma}_x\hat{\sigma}_y^+ \\ \text{i}\,\hat{\sigma}_y\hat{\sigma}_z^+ \\ \{\hat{\sigma}_x,\hat{\sigma}_z\}_+ \\ \hat{\sigma}_x^2 \\ \hat{\sigma}_z^2 \\ \hat{\sigma}_y^2 \end{bmatrix} \; . \quad (4.41)$$

Thus, the associated Lie wedge is given by

$$\mathfrak{w}_z^y := \text{i}\mathbb{R}\,\hat{\sigma}_y \oplus (-\mathfrak{c}) \; . \quad (4.42)$$

## 4.5 Outlook: Lie Wedges Compared to System Algebras

As a final outlook for this Chapter, we discuss an interesting connection between a systems associated Lie wedge and its system Lie algebra. We have pointed out here that the system semigroup of allowed quantum operations which result from the interplay between the coherent controls and the inherent total drift term can be constructed by the system Lie wedge. Furthermore, we have provided a concrete way to determine this algebraic structure from only the provided control set and knowledge of the noise/drift process the quantum system is subject to (cf. Theorem 13). Unlike in closed quantum systems where the system Lie algebra is the main algebraic structure which provides information such as accessibility and controllability of the system, in open quantum systems we can obtain *other* useful information from the *Lie wedge* which cannot be seen from the system Lie algebra.

Consider the following two simple single qubit scenarios. First suppose that the system undergoes isotropic depolarizing noise combined with a non-switchable drift Hamiltonian $\hat{\sigma}_z$. Furthermore, the system is fully Hamiltonian controllable in the sense that there are two controls $\mathrm{i}\hat{\sigma}_x$ and $\mathrm{i}\hat{\sigma}_y$ and thus the control algebra is given by $\mathrm{ad}_{\mathfrak{su}(2)}$. The complete Lindblad generator is then given by

$$\mathcal{L} = \mathrm{i}(\hat{\sigma}_z + u_1\hat{\sigma}_x + u_2\hat{\sigma}_y) + \Gamma \ , \quad \text{where} \quad \Gamma = 2\gamma(\hat{\sigma}_x^2 + \hat{\sigma}_y^2 + \hat{\sigma}_z^2) \ , \tag{4.43}$$

for $u_1, u_2 \in \mathbb{R}$ and $\gamma \in \mathbb{R}^+$. As outlined in Section 4.3.1, the associated system Lie wedge $\mathfrak{w}_\Sigma$ has $\mathrm{ad}_{\mathfrak{su}(2)}$ as its edge and since every element of the edge commutes with $\Gamma$, the wedge is simply given by

$$\mathfrak{w}_\Sigma = \mathrm{ad}_{\mathfrak{su}(2)} \oplus (-\mathbb{R}_0^+\Gamma) \ . \tag{4.44}$$

We also see that the system algebra is 4-dimensional and is given by

$$\mathfrak{g}_\Sigma = \langle \mathrm{i}\hat{\sigma}_z + \Gamma, \mathrm{i}\hat{\sigma}_x, \mathrm{i}\hat{\sigma}_y \rangle_{\mathsf{Lie}} = \mathrm{ad}_{\mathfrak{su}(2)} \oplus \mathbb{R}\Gamma \ . \tag{4.45}$$

Consider a second scenario where the system undergoes the same total drift term and now instead of having *two* control Hamiltonians we restrict to having only a *single* control term $\mathrm{i}\hat{\sigma}_y$. As explained in Section 4.3.5, this would imply condition (WH) is satisfied since $\mathfrak{k}_c = \mathbb{R}(\mathrm{i}\hat{\sigma}_y)$ is the control Lie algebra, whereas $\mathfrak{k}_d = \mathrm{ad}_{\mathfrak{su}(2)}$ is the extended control Lie algebra. Moreover, the associated Lie wedge is given by

$$\mathfrak{w}_\Sigma = \mathrm{i}\,\mathbb{R}\hat{\sigma}_y \oplus -\mathfrak{c} \quad \text{where} \quad \mathfrak{c} = \mathbb{R}_0^+ \,\mathrm{conv}\left\{ \begin{bmatrix} \cos(\theta) \\ \sin(\theta) \\ 1 \end{bmatrix} \cdot \begin{bmatrix} \mathrm{i}\hat{\sigma}_z \\ \mathrm{i}\hat{\sigma}_x \\ \Gamma \end{bmatrix} \;\middle|\; \theta \in \mathbb{R} \right\} \ . \tag{4.46}$$

However, since $[\mathrm{i}\hat{\sigma}_y, \mathrm{i}\hat{\sigma}_z + \Gamma] = [\mathrm{i}\hat{\sigma}_y, \mathrm{i}\hat{\sigma}_z] = -\mathrm{i}\hat{\sigma}_x$, and then $[\mathrm{i}\hat{\sigma}_x, \mathrm{i}\hat{\sigma}_y] = -\mathrm{i}\hat{\sigma}_z$ it's clear that the system algebra is again given by $\mathfrak{g}_\Sigma = \mathrm{ad}_{\mathfrak{su}(2)} \oplus \mathbb{R}\Gamma$. Thus, these two different control systems have the same system Lie algebra but *different* Lie wedges. In Chapter 5 we will see that the Lie wedge from the first example is a very special type of Lie wedge - one which is also a Lie semialgebra.

It is also useful to define a notion of "dimension" of a Lie wedge. In light of the work of Chapter 2 where we provided a simple basis for the entire Lindblad-Kossakowski Lie algebra for both unital and non-unital systems, one may at first glance attempt to use this same basis. However, from the above examples we see that the isotropic depolarizing Lindblad generator $\Gamma = 2\gamma(\hat{\sigma}_x^2 + \hat{\sigma}_y^2 + \hat{\sigma}_z^2)$ was a basis element of the system algebra $\mathfrak{g}_\Sigma$ which itself is composed of three basis elements of $\mathfrak{g}_0^{LK}$. Furthermore, for more complicated scenarios such as those with multiple (even simple) Lindblad terms as discussed in Section 4.3.4 (see for example Eqn. (4.24)) it is slightly more delicate to see how $\Gamma$ itself must count as a "basis" element of the wedge - which it must since we usually take it to be a basis element of the system algebra $\mathfrak{g}_\Sigma$. Therefore, we instead take the approach of determining the dimension of the vector space *spanned* by the Lie wedge itself, i.e. $\dim(\mathfrak{w}_\Sigma - \mathfrak{w}_\Sigma)$, to mean the dimension of the Lie wedge $\mathfrak{w}_\Sigma$.

Table 4.2 provides a comparison of the system Lie algebra and $\dim(\mathfrak{w} - \mathfrak{w})$ (omitting the $\Sigma$ subscript now) for this example as well as several other standard single qubit scenarios. In those special examples we see that $\mathfrak{w} - \mathfrak{w} = \mathfrak{g}_\Sigma$. This is to say that the Lie wedge "explores" every direction of the system algebra it is embedded in. As we will see later on in Appendix D where we extend the wedge construction methods to multi-qubit systems, this is strikingly *not* the case even in simple two-qubit scenarios. This then provides further evidence that even at the two-qubit level, there is a stark difference between an open systems Lie algebra and its associated Lie wedge and that exploring the geometric and algebraic properties of this structure is worth investigating in future work.

**Table 4.2:** Analysis of Single Qubit System Algebras subject to Ising Drift

| Channel | Lindblad Terms | Control | System Algebra | $\dim(\mathfrak{w} - \mathfrak{w})$ |
|---|---|---|---|---|
| Unital Noise | $\sqrt{\gamma_\mu}\sigma_\mu$ | H | $\hat{\mathfrak{g}}_0^{LK}$ | 9 |
| Isotropic Depolarizing | $\sqrt{\gamma}\sigma_x$<br>$\sqrt{\gamma}\sigma_y$<br>$\sqrt{\gamma}\sigma_y$ | H | $\mathrm{ad}_{\mathfrak{su}(2)} \oplus \mathbb{R}\Gamma$ | 4 |
| ————"———— | ——"—— | WH | $\mathrm{ad}_{\mathfrak{su}(2)} \oplus \mathbb{R}\Gamma$ | 4 |
| Amplitude Damping | $\frac{\sqrt{\gamma}}{2}(\sigma_x + i\sigma_y)$ | H | $\hat{\mathfrak{g}}^{LK}$ | 12 |

where $\mu \in \{x, y, z\}$ and $\dim(\hat{\mathfrak{g}}_0^{LK})$=9 and $\dim(\hat{\mathfrak{g}}^{LK})$=12

# Chapter 5

# Lie Semialgebras Associated to Open Quantum Systems

## 5.1 Introduction

Here we focus on answering the problem: *which types of coherently controlled open quantum systems* $(\Sigma)$ *have associated Lie wedges which specialise to Lie semialgebras?* As outlined in Section 1.1, Lie wedges specialising to Lie semialgebras are important to characterise time-independent Markovian quantum channels. In particular, it was proven in [16] that a time-dependent Markovian quantum channel can only be regarded as a time-independent one if the corresponding Lie wedge satisfies the stronger condition that it is also a Lie semialgebra.

That is, a Lie semialgebra allows one to use a *single* time-independent Lindblad generator contained within it - considered as an "effective" Lindblad generator in connection to effective Hamiltonian theory - which generates a one-parameter semigroup that can describe the entire trajectory of the initial state up to some final time $t_f$. Clearly, this is a special scenario since usually the time evolution of an initial state to any final state requires the switching of controls and thus uses a combination of different generators (and products of their respective one-parameter semigroups) to arrive at the desired state at time $t_f$.

This section completely solves this problem by means of Theorem 15. In fact, we prove a stronger result which shows that for a Lie wedge associated to a controlled open quantum system to specialise to a Lie semialgebra it must also specialise to a stronger type of Lie wedge - a so-called relatively invariant wedge. Furthermore, this occurs if and only if every element of the control algebra has no effect on *both* the drift Hamiltonian and the dissipative dynamics induced by the environment i.e.

$$[\mathrm{i}\,\mathrm{ad}_{H_c}, \mathrm{i}\,\mathrm{ad}_{H_d}] = [\mathrm{i}\,\mathrm{ad}_{H_c}, \Gamma] = 0\ , \quad \text{for all} \quad \mathrm{i}\,\mathrm{ad}_{H_c} \in \mathfrak{k}_c := \langle \mathrm{i}\,\mathrm{ad}_{H_j} \mid j = 1, \ldots, m \rangle_{\mathsf{Lie}}\ , \tag{5.1}$$

where $\mathrm{i}\,\mathrm{ad}_{H_j}$ for $j = 1, \ldots, m$ are the control Hamiltonians of the system.

## 5.2 General Theory of Special Forms of Lie Wedges

First we present a collection of results, which in their general form, are either inherently found in [23] or which are applications of their results to scenarios which will be of use in the following section. Concretely, this section will introduce the essential theory and concepts related to the algebraic analysis of Lie wedges which we will later directly apply to solve the problem of determining when a Lie wedge specialises to a Lie semialgebra. Although Lie semialgebras are the main area of interest here, the work of Hilgert, Hofmann, and Lawson [23] shows there is a finer classification of Lie semialgebras, which will be of importance in the course of this section dealing with effectively time independent Markovian quantum channels.

We start with the following series of notions of invariance of wedges $\mathfrak{w}$ that are contained within a finite dimensional Lie algebra $\mathfrak{g}$. Wedges $\mathfrak{w} \subseteq \mathfrak{g}$ (with edge $E(\mathfrak{w}) := \mathfrak{w} \cap -\mathfrak{w}$) which satisfy relations of the form

$$e^{\mathrm{ad}_A}\mathfrak{w} = e^A \mathfrak{w} e^{-A} = \mathfrak{w}\,, \tag{5.2}$$

where $A$ extends over different sets, can be classified into various types of special wedges:

(1) $\mathfrak{w}$ is a *Lie wedge* if $e^A\mathfrak{w}e^{-A} = \mathfrak{w}$, for all $A \in E(\mathfrak{w})$

(2) $\mathfrak{w}$ is a *Lie semialgebra* if (for $\mathfrak{w}$ restricted to a BCH-neighbourhood near the identity), $e^{\mathfrak{w}}e^{-\mathfrak{w}} \subseteq e^{\mathfrak{w}}$

(3) $\mathfrak{w}$ is a *relatively invariant wedge* if $e^A\mathfrak{w}e^{-A} = \mathfrak{w}$, for all $A \in \mathfrak{w}$

(4) $\mathfrak{w}$ is an *invariant wedge* if $e^A\mathfrak{w}e^{-A} = \mathfrak{w}$, for all $A \in \mathfrak{g}$

Since these definitions rely on the convergent power series due to the exponential, it is essential to have an infinitesimal description of these special wedges in terms of the Lie bracket. As a main result in [23] (Scholium II.2.15), the authors prove the useful equivalent characterisations

(1) $\mathfrak{w}$ is a *Lie wedge*, if and only if $[A, E(\mathfrak{w})] \subseteq T_A\mathfrak{w}$ for all $A \in \mathfrak{w}$,

(2) $\mathfrak{w}$ is a *Lie semialgebra*, if and only if $[A, T_A\mathfrak{w}] \subseteq T_A\mathfrak{w}$ or all $A \in \mathfrak{w}$,

(3) $\mathfrak{w}$ is a *relatively invariant wedge*, if and only if $[A, \mathfrak{w}] \subseteq T_A\mathfrak{w}$ or all $A \in \mathfrak{w}$,

(4) $\mathfrak{w}$ is an *invariant wedge*, if and only if $[A, \mathfrak{g}] \subseteq T_A\mathfrak{w}$ or all $A \in \mathfrak{w}$.

By $E(\mathfrak{w}) \subseteq T_A\mathfrak{w} \subseteq (\mathfrak{w} - \mathfrak{w}) \subseteq \mathfrak{g}$, one immediately gets the hierarchy of wedges via the obvious implications:

Lie wedge $\Leftarrow$ Lie semialgebra $\Leftarrow$ relatively invariant wedge $\Leftarrow$ invariant wedge.

With these classes of wedges in hand, we can now focus on various results and aspects of Lie wedges and tangent spaces which will be of use. The first result focuses on a specific type of Lie wedge $\mathfrak{w} \subseteq \mathfrak{g}$, which as later shown in Section 5.3, is precisely the form obtained in the context of quantum control.

**Proposition 5.2.1.** *Let $E \subseteq \mathfrak{g}$ be a Lie subalgebra and let $C_0 \in \mathfrak{g}$ be any element in $\mathfrak{g}$. If the cone given by $\mathfrak{c} := \mathbb{R}_0^+ \operatorname{conv}\{e^x C_0 e^{-x} \mid x \in E\}$ is closed and pointed and $E \cap \mathfrak{c} = \{0\}$ then*

$$\mathfrak{w} = E \oplus (-\mathfrak{c})\,, \tag{5.3}$$

*is a Lie wedge with $E(\mathfrak{w}) = E$.*

*Proof.* By construction, we trivially have that $e^x \mathfrak{w} e^{-x} = \mathfrak{w}$ for all $x \in E$ and thus it is a Lie wedge. Furthermore, since $E$ is a Lie subalgebra $E \subseteq E(\mathfrak{w})$. Assume there exists an elements $A \in E$ and $B \in \mathfrak{c}_0$ such that $A + B \in E(\mathfrak{w})$. Then there are elements $A' \in E$ and $B' \in \mathfrak{c}_0$ such that $-(A+B) = A'+B' \in E(\mathfrak{w})$ and hence $A+A' = -B-B' \in E(\mathfrak{w})$. Since $E \cap \mathfrak{c} = \{0\}$ then this equality implies that $A + A' = 0$ and $B + B' = 0$ and hence $-B = B'$. Since $\mathfrak{c}$ is pointed, $B$ and $-B$ cannot both be in $\mathfrak{c}$ unless they are zero and therefore $E(\mathfrak{w}) = E$. □

For a Lie wedge $\mathfrak{w}$ contained in a Lie algebra $\mathfrak{g}$, another notion which will arise frequently in our discussion is that of a *generating* Lie wedge. We say that $\mathfrak{w}$ is generating whenever $\mathfrak{w} - \mathfrak{w} = \mathfrak{g}$. Since a Lie wedge may be contained within multiple nested Lie algebras, it will be useful later to explicitly fix a notion of a generating Lie wedge with respect to a *fixed* Lie algebra $\mathfrak{g}$. The following result provides an example of a Lie wedge which is generating in the sense that $\mathfrak{w} - \mathfrak{w} = \langle \mathfrak{w} \rangle_{\text{Lie}}$ and therefore setting $\mathfrak{g} := \langle \mathfrak{w} \rangle_{\text{Lie}}$, the Lie wedge satisfies the generating criteria $\mathfrak{w} - \mathfrak{w} = \mathfrak{g}$.

**Proposition 5.2.2.** *Let $E \subseteq \mathfrak{g}$ be a Lie subalgebra and let $C_0 \in \mathfrak{g}$ be any element in $\mathfrak{g}$. If $\mathfrak{w} = E \oplus (-\mathbb{R}_0^+ C_0)$ is a Lie wedge, then $\mathfrak{w} - \mathfrak{w}$ is a Lie algebra, and in fact $\mathfrak{w} - \mathfrak{w} = \langle \mathfrak{w} \rangle_{\text{Lie}}$.*

*Proof.* We have to show that $\mathfrak{w} - \mathfrak{w} = \langle \mathfrak{w} \rangle_{\mathsf{Lie}}$. Note that $\mathfrak{w} - \mathfrak{w} = E \oplus \mathbb{R} C_0$ and since $\frac{d}{dt} e^{\text{ad}_x} C_0 \big|_{t=0} = [x, C_0] \in \mathfrak{w} - \mathfrak{w}$ for all $x \in E$, this implies that $[x, C_0] = A + B$ where $A \in E$ and $B \in \mathbb{R} C_0$ and thus $\mathfrak{w} - \mathfrak{w} = \langle \mathfrak{w} \rangle_{\mathsf{Lie}}$. □

In the following section where we apply these Lie wedge notions to quantum control systems we will *fix* a specific Lie algebra $\mathfrak{g}$ which every Lie wedge is contained in (see the Convention around Eqn. (5.13)). Furthermore, the following proposition is a collection of results which will be of use and are either explicitly stated in [23] or are directly implied by some explanations in and around related results.

**Proposition 5.2.3.** *([23]) Let $\mathfrak{w} \subseteq \mathfrak{g}$ be a Lie wedge. The following collection of results hold*

*(1) If $\mathfrak{w}$ is a Lie semialgebra then $\mathfrak{w} - \mathfrak{w} = \langle \mathfrak{w} \rangle_{\text{Lie}}$*

*(2) If $\mathfrak{w}$ is generating, then $\mathfrak{w}$ is relatively invariant if and only if it is invariant*

*(3) If $\mathfrak{w}$ is an invariant wedge then $E(\mathfrak{w})$ and $\mathfrak{w} - \mathfrak{w}$ are ideals of $\mathfrak{g}$*

*Proof.* (1) The proof is simple but involves properties of the algebraic interior of $\mathfrak{w}$ which we leave out at the moment. See Proposition II.2.13 in [23].
(2) Lemma II.1.4 in the same reference implies that that a wedge $\mathfrak{w} \subseteq \mathfrak{g}$ is relatively invariant if and only if $e^x \mathfrak{w} e^{-x} = \mathfrak{w}$ for all $x \in \mathfrak{w} - \mathfrak{w}$ and therefore if $\mathfrak{w}$ is a wedge such that $\mathfrak{w} - \mathfrak{w} = \mathfrak{g}$ then $\mathfrak{w}$ is relatively invariant if and only if it is invariant. See comments prior to Proposition II.1.10 in [23].
(3) See the first part of Proposition II.1.10 in the same reference. □

To determine whether a Lie wedge further specialises to a Lie semialgebra, relatively invariant wedge, or invariant wedge, in light of the characterisations (1)-(4) prior to Proposition 5.2.1 it will be necessary to know a precise formulation of what we mean by the *tangent space* at a point $A \in \mathfrak{w}$. We have the following (usual) differential result given in [23] which relates elements in the tangent and subtangent space to differentiable curves.

Let V be a vector space. Then a function $\gamma : D \longrightarrow V$ for $D \subseteq \mathbb{R}^+$ with $\gamma(D\cap ]0,\infty[) \subseteq V$ is defined to be *right-differentiable* at 0 if 0 is a cluster point of positive numbers in D and the *right-derivative at 0* given by

$$\dot\gamma_+(0) := \lim_{\substack{t\to 0\\ t\in D}} \tfrac{1}{t}\big(\gamma(t)-\gamma(0)\big)\ , \quad \text{exists}\ . \tag{5.4}$$

With this differential notion in hand, we have to establish some notions on the geometry of wedges in Lie algebras. That is, let $\mathfrak{w}$ be a wedge contained in $\mathfrak{g}$. For any $A \in \mathfrak{g}$, we define the so-called *opposite* wedge of $A$ with respect to $\mathfrak{w}$ as

$$\mathrm{op}(A) := A^\perp \cap \mathfrak{w}^*\ , \tag{5.5}$$

where we recall that the dual wedge $\mathfrak{w}^*$ is given by $\mathfrak{w}^* := \{A \in \mathfrak{g} \,|\, \langle A, B\rangle \geq 0 \text{ for all } B \in \mathfrak{w}\}$. Letting $A \in \mathfrak{w}$, we can define the *subtangent* space of $\mathfrak{w}$ at $A \in \mathfrak{w}$ as

$$L_A\mathfrak{w} := \mathrm{op}(A)^* \tag{5.6}$$

and therefore the *tangent* space of $\mathfrak{w}$ at $A$ is given by

$$T_A\mathfrak{w} = L_A\mathfrak{w} \cap -L_A\mathfrak{w} = (A^\perp \cap \mathfrak{w}^*)^\perp\ . \tag{5.7}$$

We now have the following proposition which relates the tangent and subtangent vectors of a wedge to the differential characterisation of tangent and subtangent vectors.

**Proposition 5.2.4.** *([23], Prop. I.5.3 and Cor. 1.5.4) Let $\mathfrak{w}$ be a wedge in a Lie algebra $\mathfrak{g}$. Then for $A \in \mathfrak{w}$ and $B \in \mathfrak{g}$ we have that*

1. *$B \in L_A\mathfrak{w} = \overline{\mathfrak{w} - \mathbb{R}_0^+ A}$ if and only if there exists a right differentiable function $\gamma$ such that $\gamma(0) = A$ and $B = \dot\gamma_+(0)$*
2. *$B \in T_A\mathfrak{w} = L_A\mathfrak{w} \cap -L_A\mathfrak{w}$ if and only if there exists a right and left differentiable function $\gamma$ such that $\gamma(0) = A$ and $B = \dot\gamma(0)$*

Clearly, it would be useful to know the exact structure of the tangent space $T_A\mathfrak{w} \subseteq \mathfrak{g}$ at every point $A \in \mathfrak{w}$ due to the characterisations of wedges which specialise to stronger Lie-type structures. Unfortunately, a proof of the exact structure of a tangent space $T_A\mathfrak{w} \subseteq \mathfrak{g}$ at any point $A \in \mathfrak{w}$ remains elusive at the time of this thesis submission. However, we do have the following lower bound which proves to be sufficient for our purposes.

**Proposition 5.2.5.** *Let $\mathfrak{w} = E(\mathfrak{w}) \oplus (-\mathfrak{c})$ be a Lie wedge contained in a Lie algebra $\mathfrak{g}$. Then*

$$T_A\mathfrak{w} \supseteq E(\mathfrak{w}) + \mathbb{R}A + [E(\mathfrak{w}), A]\ , \quad \textit{for any } A \in \mathfrak{w}, \textit{ and in particular}\ , \tag{5.8}$$

*$T_A\mathfrak{w} = E(\mathfrak{w})$ if $A \in E(\mathfrak{w})$ and $T_A\mathfrak{w} = \mathfrak{w} - \mathfrak{w}$ if $A \in int_{\mathfrak{w}-\mathfrak{w}}(\mathfrak{w})$, the interior of $\mathfrak{w}$ relative to $\mathfrak{w} - \mathfrak{w}$.*

*Proof.* By Theorem II.1.12 of [23], if $\mathfrak{w}$ is a Lie wedge then $[E(\mathfrak{w}), A] \subseteq T_A\mathfrak{w}$ for all $A \in \mathfrak{w}$. Clearly $\mathfrak{w}^* \supseteq A^\perp \cap \mathfrak{w}^*$ implies $\mathfrak{w}^{*\perp} \subseteq (A^\perp \cap \mathfrak{w}^*)^\perp = T_A\mathfrak{w}$ and using the fact that $\mathfrak{w}^{*\perp} = E(\mathfrak{w})$ (by Proposition I.1.7 in [23]) we have the inclusion $E(\mathfrak{w}) \subseteq T_A\mathfrak{w}$ for all $A \in \mathfrak{w}$. Again since $A^\perp \supseteq A^\perp \cap \mathfrak{w}^*$ then $\mathbb{R}A \subseteq (A^\perp \cap \mathfrak{w}^*)^\perp = T_A\mathfrak{w}$ for all $A \in \mathfrak{w}$. Thus we have proved the lower bound.

Now let $A \in E(\mathfrak{w})$. Well, since $\mathfrak{w}^* \subseteq E(\mathfrak{w})^\perp \subseteq A^\perp$, we have that $x^\perp \cap \mathfrak{w}^* = \mathfrak{w}^*$, and thus $T_A\mathfrak{w} = (x^\perp \cap \mathfrak{w}^*)^\perp = (\mathfrak{w}^*)^\perp$. Again by Proposition $I.1.7$ in [23], $(\mathfrak{w}^*)^\perp = E(\mathfrak{w})^{\perp\perp} = E(\mathfrak{w})$ and therefore $T_A\mathfrak{w} = E(\mathfrak{w})$ for $A \in E(\mathfrak{w})$. The final result that $T_A\mathfrak{w} = \mathfrak{w} - \mathfrak{w}$ is follows trivially from the fact that $A$ is an interior point of $\mathfrak{w}$ (relative to the vector space $\mathfrak{w} - \mathfrak{w}$) and thus its tangent space is everything. □

In Section 5.3, we focus on applying these general tangent space notions to Lie wedges which are associated to controlled quantum systems. We now consider a few more relationships between the tangent space and the subtangent space of a point $A \in \mathfrak{w}$. The following results outline a particularly nice equivalence which, in the end, turns out to not be necessary for our purposes. However, the general theory may be useful for the reader when considering Lie wedges associated to a different scenario then we do in the context of quantum control.

**Lemma 5.2.1.** *Let $A \in \mathfrak{w}$ and let $X \in T_A\mathfrak{w}$ such that $[A, X] \notin T_A\mathfrak{w}$. Then either $[A, X] \notin L_A\mathfrak{w}$ or $-[A, X] \notin L_A\mathfrak{w}$*

*Proof.* Assume $[A, X] \notin T_A\mathfrak{w}_0$ and $[A, X] \in L_A\mathfrak{w}_0$. If $-[A, X]$ were also in $L_A\mathfrak{w}_0$, then $[A, X]$ would actually be in $T_A\mathfrak{w}_0$ contradicting our assumption $[A, X] \notin T_A\mathfrak{w}_0$. Thus we conclude $-[A, X] \notin T_A\mathfrak{w}_0$. □

The above result has the immediate consequence.

**Corollary 5.2.1.** *Let $A \in \mathfrak{w}$. Then one has the equivalence*

$$[A, T_A\mathfrak{w}] \in T_A\mathfrak{w} \quad \Longleftrightarrow \quad [A, T_A\mathfrak{w}] \in L_A\mathfrak{w} \tag{5.9}$$

**Theorem 14.** *Let $\mathfrak{w} \subseteq \mathfrak{g}$ be a Lie wedge contained in a Lie algebra $\mathfrak{g}$. Then the following are equivalent:*

*(1)* $\mathfrak{w}$ *is* not *a Lie semialgebra;*

*(2) There exists $A \in \mathfrak{w}$, $X \in T_A\mathfrak{w}$ and a linear functional $\lambda : \mathfrak{g} \to \mathbb{R}$ such that $\lambda(L_A\mathfrak{w}) \leq 0$ and $\lambda([A, X]) > 0$;*

*(3) There exists $A \in \mathfrak{w}$, $X \in T_A\mathfrak{w}$ and a linear functional $\lambda : \mathfrak{g} \to \mathbb{R}$ such that $\lambda(\mathfrak{w} - \mathbb{R}_0^+ X) \leq 0$ and $\lambda([A, X]) > 0$.*

*Proof.* (2) ⇒ (3) follows from the equality in Eqn. (5.6) which is a consequence of Proposition I.1.9 in [23].
(1) ⇒ (2). If $\mathfrak{w}$ is not a Lie semialgebra then there exists an $A \in \mathfrak{w}$ and $X \in T_A\mathfrak{w}$ such that $[A, X] \notin T_A\mathfrak{w}$ which is equivalent to $[A, X] \notin L_A\mathfrak{w}$ by Corollary 5.2.1. By the Hahn-Banach separation theorem, we know there always exists a linear functional $\lambda : \mathfrak{g} \longrightarrow \mathbb{R}$ such that $\lambda(L_A\mathfrak{w}_0) \leq 0$ and $\lambda(-L_A\mathfrak{w}_0) \leq 0$ which implies $\lambda(T_A\mathfrak{w}_0) = 0$. Therefore $\lambda([A, X]) > 0$ since $[A, X] \notin L_A\mathfrak{w}$.
(3) ⇒ (1). If $\lambda([A, X]) > 0$ then $[A, X] \notin L_A\mathfrak{w}$ and again by Corollary 5.2.1 this implies that $[A, X] \notin T_A\mathfrak{w}$ and hence $\mathfrak{w}$ is not a Lie semialgebra. □

## 5.3 Application to Quantum Control

We first provide a brief overview of Section 4.1 which provided the background details and notation on controlled Markovian quantum dynamics. That is, *controlled* Markovian quantum dynamics are described by right-invariant *bilinear control systems* of the form

$$(\Sigma) \qquad \dot{\rho}(t) = -\mathcal{L}_{u(t)}\big(\rho(t)\big) \quad , \quad \rho(0) \in \mathfrak{pos}_1(N) \; , \tag{5.10}$$

where $\mathcal{L}_u$ depends on a control variable $u \in \mathbb{R}^m$, and we called $\mathcal{L}_u$ the *coherently controlled Lindbladian.* We focus on *coherently controlled* open systems which are defined to be those such that $\mathcal{L}_u$ is given by

$$\mathcal{L}_u(\rho) = \mathrm{i}\,\mathrm{ad}_{H_u}(\rho) + \Gamma(\rho) \quad \text{with} \quad \mathrm{ad}_{H_u} := \mathrm{ad}_{H_d} + \sum_{j=1}^{m} u_j(t)\,\mathrm{ad}_{H_j} \; , \tag{5.11}$$

where the control terms $\mathrm{i}\,\mathrm{ad}_{H_j}$ with *control Hamiltonians* $H_j \in \mathfrak{her}(N)$ are modulated by piecewise constant *control amplitudes* $u_j(t) \in \mathbb{R}$. The drift term of Eqn. (5.11) is then composed of two parts, (i) the 'Hamiltonian' drift term $\mathrm{i}\,\mathrm{ad}_{H_d}$ which describes the coherent time evolution and (ii) a dissipative Lindbladian part $\Gamma$.

Furthermore, to distinguish between varying degrees of control, we defined three Lie algebras: the *control Lie algebra* $\mathfrak{k}_c$, the *extended control Lie algebra* $\mathfrak{k}_d$, and the *system Lie algebra* $\mathfrak{g}_\Sigma$ as

$$\begin{aligned}
\mathfrak{k}_c &:= \langle \mathrm{i}\,\mathrm{ad}_{H_j} \mid j = 1, \ldots, m\rangle_{\mathsf{Lie}} \ , \\
\mathfrak{k}_d &:= \langle \mathrm{i}\,\mathrm{ad}_{H_d}, \mathrm{i}\,\mathrm{ad}_{H_j} \mid j = 1, \ldots, m\rangle_{\mathsf{Lie}} \ , \qquad (5.12)\\
\mathfrak{g}_\Sigma &:= \langle \mathcal{L}_u \mid u_j \in \mathbb{R}\rangle_{\mathsf{Lie}} = \langle \mathrm{i}\,\mathrm{ad}_{H_d} + \Gamma, \mathrm{i}\,\mathrm{ad}_{H_j} \mid j = 1, \ldots, m\rangle_{\mathsf{Lie}} \ ,
\end{aligned}$$

where we specifically make note that $\mathfrak{g}_\Sigma$ is *different* from $\mathfrak{k}_d$, because it contains the total drift term $\mathrm{i}\,\mathrm{ad}_{H_d} + \Gamma$ for the Lie closure, whereas $\mathfrak{k}_d$ is generated by the Hamiltonian drift component $\mathrm{i}\,\mathrm{ad}_{H_d}$. For more details, see Section 4.1.

In Section 2.3, we provided an explicit representation of the Lindblad-Kossakowski Lie algebras for both $n$-qubit unital ($\mathfrak{g}_0^{LK}$) and non-unital ($\mathfrak{g}^{LK}$) systems, cf. Theorems 4 and 5, respectively. These are the largest (physically) possible Lie algebras a Lie wedge associated to a Markovian semigroup of quantum channels may be contained in.

**Convention.** *Let* $(\Sigma)$ *be a coherently controlled open quantum system,* $\mathfrak{g}_\Sigma$ *the corresponding system algebra and* $\mathfrak{w} \subseteq \mathfrak{g}_\Sigma$ *its Lie wedge. Thus, we are working with the overall inclusion*

$$\mathfrak{w} \subseteq (\mathfrak{w} - \mathfrak{w}) \subseteq \mathfrak{g}_\Sigma \subseteq \mathfrak{g}^{LK} \ , \qquad (5.13)$$

*where* $\mathfrak{w} - \mathfrak{w}$ *is the vector space generated by* $\mathfrak{w}$ *and we define the Lie wedge to be generating if* $\mathfrak{w} - \mathfrak{w} = \mathfrak{g}^{LK}$*, (or* $\mathfrak{w} - \mathfrak{w} = \mathfrak{g}_0^{LK}$ *if it is a unital system). Furthermore, as an application of Proposition 5.2.3, if* $\mathfrak{w}$ *is a Lie semialgebra then we have the equality* $\mathfrak{w} - \mathfrak{w} = \mathfrak{g}_\Sigma$.

## 5.3.1 Coherently Controlled Closed Systems

As a motivation for the open system scenario, it is helpful to discuss closed systems first to show that even in this case, it is non-trivial to show which Lie wedges specialise to Lie semialgebras. Suppose that the system $(\Sigma)$ is closed in the sense that the dissipative Lindbladian component $\Gamma \equiv 0$ and hence $\mathcal{L}_u = \mathrm{i}\,\mathrm{ad}_{H_d} + \mathrm{i}\sum_{j=1}^m u_j(t)\,\mathrm{ad}_{H_j}$, for $u_j(t) \in \mathbb{R}$ and note that $\mathfrak{w}_0 \subseteq \mathfrak{g}_\Sigma = \mathfrak{k}_d$. One might ask in what situations does the system Lie wedge satisfy $\mathfrak{w}_0 = \mathfrak{g}_\Sigma = \mathfrak{k}_d$ and thus would specialise to a Lie semialgebra.

First, suppose that there exists an $x \in \mathfrak{k}_c := \langle \mathrm{i}\,\mathrm{ad}_{H_j} \mid j = 1, \ldots, m\rangle_{\mathsf{Lie}}$ such that $e^{\mathrm{ad}_x}(\mathrm{i}\,\mathrm{ad}_{H_d}) = -\lambda \mathrm{i}\,\mathrm{ad}_{H_d}$ for some $\lambda \in \mathbb{R}^+$. Then $\mathrm{i}\,\mathrm{ad}_{H_d}, \mathrm{i}\,\mathrm{ad}_{H_1}, \ldots, \mathrm{i}\,\mathrm{ad}_{H_m} \in E(\mathfrak{w})$ since $E(\mathfrak{w}_0) = \mathfrak{w}_0 \cap -\mathfrak{w}_0$. Moreover, since $E(\mathfrak{w}_0)$ is a Lie algebra we then get that $E(\mathfrak{w}_0) = \mathfrak{w}_0 = \mathfrak{k}_d = \mathfrak{g}_\Sigma$ and hence $[A, T_A\mathfrak{w}_0] \subseteq T_A$ for all $A \in \mathfrak{w}_0$ where $T_A\mathfrak{w}_0 = E(\mathfrak{w}_0) = \mathfrak{g}_\Sigma$ by Proposition 5.2.5 we see that $\mathfrak{w}_0$ is trivially a Lie semialgebra. Furthermore, a classic result (see for example [31] and Prop. V.0.18 in [23]) shows that every compact subsemigroup (with non-empty interior) of $SU(2^n)$ is in fact a subgroup and therefore if the group $\langle \exp \mathfrak{k}_d\rangle_{\mathbf{G}}$ is closed, then $\mathfrak{w}_0$ is a Lie algebra and thus also a Lie semialgebra.

In the context of open quantum systems i.e. when the Lindblad generator $\Gamma$ is non-zero, the situation is much more delicate. For example, the first scenario described above cannot occur since the Lindblad generator which describes the irreversible processes the environment induces on the system *cannot* be inverted. Moreover, the semigroup argument used in the second scenario above also cannot be used. Thus we must use a different tool-set to tackle this problem, namely, the concepts of tangent and subtangent spaces of the systems associated Lie wedge.

### 5.3.2 Coherently Controlled Open Systems

Now we consider the general open system scenario where $(\Sigma)$ is given by Eqn. (5.11) such that the dissipative Lindbladian component $\Gamma \neq 0$ and hence $\mathcal{L}_u = \mathrm{i}(\mathrm{ad}_{H_d} + \sum_{j=1}^{m} u_j(t)\,\mathrm{ad}_{H_j}) + \Gamma$, for $u_j(t) \in \mathbb{R}$ . Theorem 13 of Section 4.1 proved the form of the Lie wedge associated to such a control system. We restate it here for convenience.

**Theorem 13.** *Let $(\Sigma)$ be a coherently controlled open quantum system and assume that the Lie group $\boldsymbol{K}$ which is generated by the control Lie algebra $\mathfrak{k}_c := \langle \mathrm{i}\,\mathrm{ad}_{H_1}, \mathrm{i}\,\mathrm{ad}_{H_2}, \ldots, \mathrm{i}\,\mathrm{ad}_{H_m}\rangle_{\mathsf{Lie}}$ is closed. Then the Lie wedge associated to the system is given by*

$$\mathfrak{w} = \mathfrak{k}_c \oplus (-\mathfrak{c})\ , \quad \textit{where} \quad \mathfrak{c} := \mathbb{R}_0^+ \operatorname{conv}\left\{ \mathrm{Ad}_U\ \left(\mathrm{i}\,\mathrm{ad}_{H_d} + \Gamma\right)\ \mathrm{Ad}_U^\dagger\ \mid U \in \boldsymbol{K} \right\}\ , \tag{5.14}$$

*and therefore the reachable set of operators is given by*

$$\overline{\boldsymbol{P}}_\Sigma = \overline{\langle \exp(\mathfrak{w}_\Sigma)\rangle}_S\ . \tag{5.15}$$

The fact that Lie wedges for coherently controlled open quantum systems take this form where the convex cone $\mathfrak{c}$ is pointed, closed and in a sense generated by the *single* element $\mathrm{i}\,\mathrm{ad}_{H_d} + \Gamma$ allows us to prove several remarkable properties. The first is that we can provide a necessary and sufficient condition for when a Lie wedge is simultaneously a relatively invariant wedge.

**Proposition 5.3.1.** *Let $(\Sigma)$ be a coherently controlled open quantum system with total drift term $\mathcal{D} := \mathrm{i}\,\mathrm{ad}_{H_d} + \Gamma$, where $\Gamma$ can be unital or $non-unital$. Furthermore, let $\mathfrak{g}_\Sigma$ denote the corresponding system algebra, $\mathfrak{w} \subseteq \mathfrak{g}_\Sigma$ its Lie wedge and $\mathfrak{k}_c = E(\mathfrak{w})$ its control Lie algebra. Then $\mathfrak{w}$ is relatively invariant if and only if $\mathfrak{w} = E(\mathfrak{w}) \oplus (-\mathbb{R}_0^+\mathcal{D})$.*

*Proof.* If $\mathfrak{w}$ is a relatively invariant wedge then $[A, \mathfrak{w}] \subseteq T_A\mathfrak{w}$ for all $A \in \mathfrak{w}$ by the condition prior to Proposition 5.2.1. We will show that by the choices of $A \in \mathfrak{w}$ for a Lie wedge $\mathfrak{w}$ of the form given by Eqn. (5.14) will impose that $\mathfrak{w} = E(\mathfrak{w}) \oplus (-\mathbb{R}_0^+\mathcal{D})$.

First let $A \in E(\mathfrak{w}) = \mathfrak{k}_c$. Then $T_A\mathfrak{w} = E(\mathfrak{w})$ by Proposition 5.2.5 and hence the commutation inclusion implies that it must be true that $[E(\mathfrak{w}), \mathfrak{w}] \subseteq E(\mathfrak{w})$. Now since $\mathcal{D} = \mathrm{i}\,\mathrm{ad}_{H_d} + \Gamma \in \mathfrak{w}$ this shows that $[A, \mathcal{D}] \in E(\mathfrak{w})$ for all $A \in E(\mathfrak{w})$ and therefore $\sum_k \frac{1}{k!}\,\mathrm{ad}_A^k(\mathcal{D}) \in E(\mathfrak{w})$ also. By series expansion of the exponential, we then know that $e^{\mathrm{ad}_A}(\mathcal{D}) = \mathcal{D} + \sum_k \frac{1}{k!}\,\mathrm{ad}_A^k(\mathcal{D}) \in E(\mathfrak{w}) \oplus \mathbb{R}\mathcal{D}$ for all $A \in E(\mathfrak{w})$. By Theorem 13 of Section 4.1, Lie wedges for our quantum control scenario are given by $\mathfrak{w} = E(\mathfrak{w}) \oplus (-\mathfrak{c})$ where $\mathfrak{c} := \mathbb{R}_0^+\mathrm{conv}\{e^{\mathrm{ad}_A}(\mathcal{D}) \mid \text{for all } A \in E(\mathfrak{w})\}$. Thus, $\mathfrak{c} \subseteq E(\mathfrak{w}) \oplus \mathbb{R}\mathcal{D}$ which implies that $\mathfrak{w} \subseteq E(\mathfrak{w}) \oplus \mathbb{R}\mathcal{D}$, which finally shows that the Lie wedge must have the form $\mathfrak{w} = E(\mathfrak{w}) \oplus (-\mathbb{R}_0^+\mathcal{D})$.

Alternatively, suppose that $\mathfrak{w} = E(\mathfrak{w}) \oplus (-\mathbb{R}_0^+\mathcal{D})$ is the Lie wedge of the system - we want to prove that the Lie wedge is relatively invariant by showing that $[A, \mathfrak{w}] \subseteq T_A\mathfrak{w}$ for all $A \in \mathfrak{w}$. We now will go through the possible scenarios for the choice of $A \in \mathfrak{w}$.

First let $A \in E(\mathfrak{w})$ and note again that $T_A\mathfrak{w} = E(\mathfrak{w})$ by Proposition 5.2.5. By the structure of $\mathfrak{c}$ its clear that the edge of the wedge leaves the total drift term $\mathcal{D} = \mathrm{i}\,\mathrm{ad}_{H_d} + \Gamma$ invariant and hence $[A, \mathcal{D}] = 0$ for all $A \in E(\mathfrak{w})$. Since $E(\mathfrak{w})$ is a Lie algebra, it's clearly true that $[A, E(\mathfrak{w})] \subseteq E(\mathfrak{w})$ and thus $[A, \mathfrak{w}] \subseteq T_A\mathfrak{w}$ for $A \in E(\mathfrak{w})$. Now for any other $A \in \mathfrak{w}$ such that $A \notin E(\mathfrak{w})$, the subspace inclusion given by Proposition 5.2.5 implies that $T_A\mathfrak{w} = E(\mathfrak{w}) \oplus \mathbb{R}A = \mathfrak{w} - \mathfrak{w}$. Furthermore, by the structure of the Lie wedge, Proposition 5.2.2 then implies that $T_A\mathfrak{w} = \mathfrak{w} - \mathfrak{w} = \langle \mathfrak{w} \rangle_{\text{Lie}}$. It's clearly true that $[A, \mathfrak{w}] \subseteq \langle \mathfrak{w} \rangle_{\text{Lie}}$ for all such $A \in \mathfrak{w}$ and thus the Lie wedge is relatively invariant.

□

In fact, we can prove a something stronger. The following result proves that for coherently controlled (Markovian) quantum systems, a Lie wedge is a Lie semialgebra *if and only if* it is also a relatively invariant wedge - which we completely characterised in the previous result.

**Theorem 15.** *Let* $(\Sigma)$ *be a coherently controlled open quantum system with total drift term* $\mathcal{D} := \mathrm{i}\,\mathrm{ad}_{H_d} + \Gamma$, *where* $\Gamma$ *can be unital or non* $-$ *unital and assume that the group generated by the control Lie algebra* $\mathfrak{k}_c$ *is closed. Furthermore, let* $\mathfrak{g}_\Sigma$ *denote the corresponding system algebra and* $\mathfrak{w} \subseteq \mathfrak{g}_\Sigma$ *its Lie wedge. Then the following are equivalent*

*(1)* $\mathfrak{w}$ *is a Lie semialgebra*

*(2)* $\mathfrak{w}$ *is a relatively invariant wedge*

*(3)* $\mathfrak{w} = \mathfrak{k}_c \oplus (-\mathbb{R}_0^+\mathcal{D})$

*(4)* $[\mathrm{i}\,\mathrm{ad}_{H_c}, \mathcal{D}] = 0$ *for all* $\mathrm{i}\,\mathrm{ad}_{H_c} \in \mathfrak{k}_c$

*Proof.* (1) ⇒ (4). Suppose that $\mathfrak{w}$ is a Lie semialgebra but $[E(\mathfrak{w}), \mathcal{D}] \neq 0$. By assumption we know that $[A, T_A\mathfrak{w}] \subseteq T_A\mathfrak{w}$ for all $A \in \mathfrak{w}$. Thus for $A := \mathcal{D}$ we have that $[A, [E(\mathfrak{w}), A]] \subseteq T_A\mathfrak{w}$ since $[E(\mathfrak{w}), A] \subseteq T_A\mathfrak{w}$ by Proposition 5.2.4. Now let $A' := \mathrm{i}\,\mathrm{ad}_{H_c} + A$ for any $\mathrm{i}\,\mathrm{ad}_{H_c} \in E(\mathfrak{w})$, and note that by including the edge element, the tangent space doesn't change in the sense that $T_{A'}\mathfrak{w} = T_A\mathfrak{w}$ since $L_{A'}\mathfrak{w} = L_A\mathfrak{w} = \overline{\mathfrak{w} - \mathbb{R}_0^+ A}$ by Proposition 5.2.4. Using the tangent space commutation relation again we have that $[A', [E(\mathfrak{w}), A']] = [\mathrm{i}\,\mathrm{ad}_{H_c}, [E(\mathfrak{w}), \mathrm{i}\,\mathrm{ad}_{H_c}]] + [\mathrm{i}\,\mathrm{ad}_{H_c}, [E(\mathfrak{w}), A]] + [A, [E(\mathfrak{w}), \mathrm{i}\,\mathrm{ad}_{H_c}]] + [A, [E(\mathfrak{w}), A]] \subseteq T_A\mathfrak{w}$. Clearly then the first, third and fourth terms are all already contained in $T_A\mathfrak{w}$ (the fourth from the previous discussion) and thus we must also have that the third term $[\mathrm{i}\,\mathrm{ad}_{H_c}, [E(\mathfrak{w}), A]] \subseteq T_A\mathfrak{w}$. We can now check that $[A', [\mathrm{i}\,\mathrm{ad}_{H_c}, [E(\mathfrak{w}), A]]] \subseteq T_A\mathfrak{w}$ which would show that $\mathrm{ad}_x^3(A) \in T_A\mathfrak{w}$ for all $x = \mathrm{i}\,\mathrm{ad}_{H_c} \in E(\mathfrak{w})$. Iterating this procedure, we have that $e^{\mathrm{ad}_x}A = A + \sum_k \frac{1}{k!}\mathrm{ad}_x^k(A) \in T_A\mathfrak{w}$ for all $x \in E(\mathfrak{w})$. This implies that $\mathfrak{w} \subseteq T_A\mathfrak{w}$, and hence $\mathfrak{w} - \mathfrak{w} \subseteq T_A\mathfrak{w}$. Since trivially we know that $T_A\mathfrak{w} \subseteq \mathfrak{w} - \mathfrak{w}$ this implies that $T_A\mathfrak{w} = \mathfrak{w} - \mathfrak{w}$. However, by Proposition 5.2.3, since $\mathfrak{w}$ is a Lie semialgebra, we know that $\mathfrak{w} - \mathfrak{w} = \mathfrak{g}_\Sigma$ and hence $T_A\mathfrak{w} = \mathfrak{g}_\Sigma$. Due to the fact that $A := \mathcal{D}$ is a boundary point of the Lie wedge, this equality of the tangent space and the system algebra cannot occur. Hence, this contradiction implies $[E(\mathfrak{w}), \mathcal{D}] = 0$.

(4) ⇒ (3). If $[E(\mathfrak{w}), \mathcal{D}] = 0$ then the controls have no effect on the total drift term. Then $e^{\mathrm{ad}_x}(\mathcal{D}) = \mathcal{D}$ for all $x \in \mathfrak{k}_c$ and hence by Theorem 13 of Section 4.1 the associated system Lie wedge is given by $\mathfrak{w} = \mathfrak{k}_c \oplus (-\mathbb{R}_0^+\mathcal{D})$.

(3) ⇒ (2). Follows by Proposition 5.3.1.

(2) $\Rightarrow$ (1). This direction follow from the hierarchy of Lie wedges discussed prior to Proposition 5.2.1 in Section 5.2.

□

This result then has some immediate implications.

**Corollary 5.3.1.** *Suppose we have the same preconditions as Theorem 15 and additionally assume that the edge of the wedge is non-zero. Then there exists no Lie wedges which specialise to invariant Lie wedges.*

*Proof.* Recall that an invariant Lie wedge in our context is one which satisfies $[A, \mathfrak{g}^{LK}] \subseteq T_A\mathfrak{w}$ for all $A \in \mathfrak{w}$ for non-unital systems (or $[A, \mathfrak{g}_0^{LK}] \subseteq T_A\mathfrak{w}$ for all $A \in \mathfrak{w}$ for unital systems). By the hierarchy of types of invariance a wedge can take, we know that an invariant wedge must be relatively invariant. Since Theorem 15 shows that all relatively invariant wedges are of the form $\mathfrak{w} = E(\mathfrak{w}) \oplus (-\mathbb{R}_0^+\mathcal{D})$ this implies that invariant wedges are those of the same form which satisfy the tangent space commutation inclusion just described. Letting $A \in E(\mathfrak{w})$ shows that for a Lie wedge to be an invariant Lie wedge it must be true that $[A, \mathfrak{g}^{LK}] \subseteq E(\mathfrak{w})$ (or $[A, \mathfrak{g}_0^{LK}] \subseteq E(\mathfrak{w})$ for unital systems) since $T_A\mathfrak{w} = E(\mathfrak{w})$. Clearly this cannot always occur and thus we are done. □

**Corollary 5.3.2.** *Suppose we have the same preconditions as Theorem 15 and additionally assume that $(\Sigma)$ is accessible (i.e. $\mathfrak{g}_\Sigma = \mathfrak{g}^{LK}$ or $\mathfrak{g}_0^{LK}$ for non-unital or unital $\Gamma$, respectively). Then the corresponding Lie wedge $\mathfrak{w}$ is not a Lie semialgebra, relatively invariant wedge, or invariant wedge.*

*Proof.* If $\mathfrak{g}_\Sigma = \mathfrak{g}^{LK}$ or $\mathfrak{g}_0^{LK}$ then clearly we must have that there exists at least one $\mathrm{i}\,\mathrm{ad}_{H_c} \in E(\mathfrak{w})$ such that $[\mathrm{i}\,\mathrm{ad}_{H_c}, \mathcal{D}] \neq 0$ and thus the result follows from Theorem 15. □

Which then leaves us with a final (now) trivial result.

**Corollary 5.3.3.** *Suppose we have the same preconditions as Theorem 15. If there exists an $\mathrm{i}\,\mathrm{ad}_{H_c} \in \mathfrak{k}_c$ such that $[\mathrm{i}\,\mathrm{ad}_{H_c}, \mathcal{D}] \neq 0$ then $\mathfrak{w}$ is only a Lie wedge in the sense that it does not specialise to a Lie semialgebra, relatively invariant wedge, or invariant wedge.*

## 5.4 Examples

Theorem 15 completely solves the problem of how to determine which Lie wedges specialise to Lie semialgebras. However, it is nonetheless illuminating to consider several illustrative examples which show other means of proving whether a Lie wedge fails to be a Lie semialgebra . Notably, the application of Theorem 14 may be useful in other contexts related to the Lie wedges we considered here.

**Example 11.** *(Two-Qubit Lie Semialgebra)*
*As a simple example of a Lie wedge which specialises to a Lie semialgebra, we consider a two qubit system corresponding to*

$$\dot{\rho} = -\big(\mathrm{i}\hat{\sigma}_d + \sum_j u_j \mathrm{i}\hat{\sigma}_j + \Gamma\big)\rho\ , \ \textit{with}\ \ \Gamma = 2\sum_{\mu,\nu} \hat{\sigma}_{\mu\nu}^2\ , \ \ \textit{and}\ \ u_1 \in \mathbb{R}\ , \tag{5.16}$$

*for all $\mu, \nu \in \{x, z, y\}$ such that the drift Hamiltonian is given by $\mathrm{i}\hat{\sigma}_d := \mathrm{i}(\hat{\sigma}_{z1} + \hat{\sigma}_{1z})$, and the control Hamiltonians are $\hat{\sigma}_j \in \{\hat{\sigma}_{x1}, \hat{\sigma}_{y1}; \hat{\sigma}_{1x}, \hat{\sigma}_{1y}\}$. This system undergoes non-local* *isotropic depolarising and since* $\langle \mathrm{i}\hat{\sigma}_{x1}, \mathrm{i}\hat{\sigma}_{y1}\rangle_{\mathsf{Lie}} = \mathrm{ad}_{\mathfrak{su}_{\mathsf{A}}(2)} \otimes \mathbb{1}_{\mathsf{B}}$*, and* $\langle \mathrm{i}\hat{\sigma}_{1x}, \mathrm{i}\hat{\sigma}_{1y}\rangle_{\mathsf{Lie}} = \mathbb{1}_{\mathsf{A}} \otimes \mathrm{ad}_{\mathfrak{su}_{\mathsf{B}}(2)}$*, there is full H-controllability on qubit A and B individually. The corresponding Lie wedge is given by*

$$\mathfrak{w}_0 = \mathrm{ad}_{\mathfrak{su}_{\mathsf{A}}(2)\widehat{\oplus}\mathfrak{su}_{\mathsf{B}}(2)} \oplus -\mathbb{R}_0^+(\Gamma)\ , \tag{5.17}$$

*since* $\mathrm{i}\hat{\sigma}_d \in E(\mathfrak{w}_0)$ *and* $[E(\mathfrak{w}_0), \Gamma] = 0$*. By Theorem 15,* $\mathfrak{w}_0$ *is a Lie semialgebra (and relatively invariant wedge). The interested reader may see Table D.2 for a comparison between the wedge dimension and system algebra.*

**Example 12.** *(Invariant Dissipative Component & Non-invariant Drift Hamiltonian) Consider the single qubit system described by*

$$\dot{\rho} = -\big(\mathrm{i}\hat{\sigma}_z + u_1\mathrm{i}\hat{\sigma}_y + \Gamma\big)\rho\ ,\ \textit{with}\ \ \Gamma = 2(\hat{\sigma}_x^2 + \hat{\sigma}_y^2 + \hat{\sigma}_z^2)\ ,\quad \textit{and}\ \ u_1 \in \mathbb{R}\ , \tag{5.18}$$

*which is WH-controllable and undergoes isotropic depolarising noise. The system Lie algebra is given by* $\mathfrak{g}_\Sigma = \langle E(\mathfrak{w}_0), \mathrm{i}\hat{\sigma}_z + \Gamma\rangle_{\mathrm{Lie}} = \mathrm{ad}_{\mathfrak{su}(2)} \oplus \mathbb{R}\Gamma$*, and the corresponding Lie wedge is given by* $\mathfrak{w}_0 = \mathbb{R}\mathrm{i}\hat{\sigma}_y \oplus -\mathfrak{c}_0$*, where*

$$\mathfrak{c}_0 = \mathbb{R}_0^+\,\mathrm{conv}\,\{\cos(\theta)\mathrm{i}\hat{\sigma}_z + \sin(\theta)\mathrm{i}\hat{\sigma}_x + \Gamma \mid\ \textit{for all}\ \ \theta \in \mathbb{R}\}\ , \tag{5.19}$$

*since* $[E(\mathfrak{w}_0), \Gamma] = 0$*. Thus by Theorem 15 we* already *know* $\mathfrak{w}_0$ *is no Lie semialgebra, but this particular wedge is* almost *one in the sense that the controls have no effect on the dissipative component* $\Gamma$*. We will show we inevitably obtain a contradiction that* $[A, T_A\mathfrak{w}_0] \subseteq T_A\mathfrak{w}_0$ *for all* $A \in \mathfrak{w}_0$*. Let* $A := \Gamma$*. Since the system algebra has dimension four, the tangent space inclusion given by Proposition 5.2.5 is in fact an equality and hence we obtain*

$$T_A\mathfrak{w}_0 = \mathbb{R}\mathrm{i}\hat{\sigma}_y \oplus \mathbb{R}\Gamma \oplus \mathbb{R}[\mathrm{i}\hat{\sigma}_y, \Gamma]\ , \tag{5.20}$$

*and since* $[\Gamma, [\mathrm{i}\hat{\sigma}_y, \Gamma]] \in \mathbb{R}\mathrm{i}\hat{\sigma}_y$ *(by the Cartan* $\mathfrak{k} - \mathfrak{p}$ *commutation relations) we do have that* $[A, T_A\mathfrak{w}_0] \subseteq T_A\mathfrak{w}_0$*. Now define* $A' = \mathrm{i}\hat{\sigma}_y + \mathrm{i}\hat{\sigma}_z + \Gamma$*. Then* $[E(\mathfrak{w}_0), A'] = -\mathrm{i}\hat{\sigma}_x \in T_A\mathfrak{w}$ *which gives* $[A', [E(\mathfrak{w}_0), A']] = -[\mathrm{i}\hat{\sigma}_y, \mathrm{i}\hat{\sigma}_x] - [\Gamma, \mathrm{i}\hat{\sigma}_x] = -\mathrm{i}\hat{\sigma}_z \notin T_A\mathfrak{w}_0$ *and therefore* $\mathfrak{w}_0$ *is not a Lie semialgebra, relatively invariant wedge, or invariant wedge.*

**Example 13.** *(H-Controllable Standard Single Qubit System)*
*Consider a single qubit controlled open quantum system*

$$\dot{\rho} = (u_1\mathrm{i}\hat{\sigma}_x + u_2\mathrm{i}\hat{\sigma}_y + \mathrm{i}\hat{\sigma}_z)(\rho) + \Gamma(\rho)\ ,\ \textit{with}\ \ \Gamma = 2\hat{\sigma}_y^2\ , \tag{5.21}$$

*where* $u_1, u_2 \in \mathbb{R}$ *and therefore the system is fully H-controllable. In Section 4.3.1, we showed that a system of this type has its corresponding Lie wedge given by* $\mathfrak{w}_0 = \mathrm{ad}_{\mathfrak{su}(2)} \oplus -\ \mathfrak{c}_0$ *where the cone* $\mathfrak{c}_0$ *is defined by*

$$\mathfrak{c}_0 := \mathbb{R}_0^+\mathrm{conv}\big\{\ \mathrm{ad}_M^2\ \mid M \in \mathcal{O}_{SU(2)}(\sigma_z)\big\}\ . \tag{5.22}$$

*Let* $A := \mathrm{i}\hat{\sigma}_z + \Gamma$ *be the total drift term. Then by Proposition 5.2.5, the tangent space at A contains the subspace*

$$T_A\mathfrak{w}_0 \supseteq \mathrm{ad}_{\mathfrak{su}(2)} \oplus \mathbb{R}\Gamma \oplus \mathrm{span}_{\mathbb{R}}\left\{\{\hat{\sigma}_x, \hat{\sigma}_y\}_+, \{\hat{\sigma}_y, \hat{\sigma}_z\}_+\right\}\ ,$$

*where we used the fact that* $[iA, B^2] = \{[iA, B], B\}_+$ *for general square matrices.*

*Defining the linear functional* $\lambda(g) := \langle(2\hat{\sigma}_x^2 - \hat{\sigma}_y^2), g\rangle$ *for* $g \in \hat{\mathfrak{g}}_\Sigma$*, its clear that* $\lambda(T_A\mathfrak{w}_0) = 0$ *since* $2\hat{\sigma}_x^2 - \hat{\sigma}_y^2 \in (T_A\mathfrak{w}_0)^\perp = A^\perp \cap \mathfrak{w}_0^*$*. Checking if* $[A, T_A\mathfrak{w}_0] \subseteq T_A\mathfrak{w}_0$ *we obtain* $[A, -\{\hat{\sigma}_x, \hat{\sigma}_y\}_+] = 2(\hat{\sigma}_y^2 - \hat{\sigma}_x^2)$ *and therefore* $\lambda(2(\hat{\sigma}_y^2 - \hat{\sigma}_x^2)) = -24$ *which by Theorem 14 implies that* $\mathfrak{w}_0$ *is not a Lie semialgebra, relatively invariant wedge or invariant wedge.*

# Appendix A

# The Lindblad-Kossakowski Lie Algebra: Supplementary Proofs

Here we provide some material which had been omitted in Chapter 1 in order to avoid that reader who is familiar with the subject gets bored. Most proofs are straightforward computations combined with some standard arguments from linear algebra. First we restate Lemma 1.3.1.

**Lemma 1.3.1.** *If* $V_1, \dots, V_m \in \mathfrak{sl}(N, \mathbb{C})$, *i.e. if* $V_1, \dots, V_m$ *are traceless, the operator* $\Gamma$ *given by Eqn.* (1.21) *is purely dissipative.*

*Proof.* Operator approach: First, let $\tilde{\mathrm{ad}}_{\mathrm{i}H}$ and $\tilde{\Gamma}$ denote the canonical extensions of $\mathrm{ad}_{\mathrm{i}H}$ and $\Gamma$, respectively, to $\mathfrak{gl}(N, \mathbb{C})$. Then, it is straightforward to show that orthogonality of $\tilde{\mathrm{ad}}_{\mathrm{i}H}$ and $\tilde{\Gamma}$ is equivalent to orthogonality of $\mathrm{ad}_{\mathrm{i}H}$ and $\Gamma$. Therefore, it is sufficient to prove that $\tilde{\Gamma}$ is orthogonal to any $\tilde{\mathrm{ad}}_{\mathrm{i}H}$. Moreover, due to linearity we can assume that $\tilde{\Gamma}$ consists of a single Lindblad term, i.e.

$$\tilde{\Gamma}(X) = 2VXV^\dagger - V^\dagger V X - X V^\dagger V$$

with $\operatorname{tr} V = 0$. With these preliminary consideration, we obtain

$$\begin{aligned}
\langle \tilde{\mathrm{ad}}_{\mathrm{i}H}, \tilde{\Gamma} \rangle &= \sum_{k,l}^{N} \operatorname{tr}\Big( \big(\tilde{\mathrm{ad}}_{\mathrm{i}H}(e_k e_l^\dagger)\big)^\dagger \tilde{\Gamma}(e_k e_l^\dagger) \Big) \\
&= \sum_{k,l}^{N} \operatorname{tr}\Big( \big[\mathrm{i}H, e_k e_l^\dagger\big]^\dagger \tilde{\Gamma}(e_k e_l^\dagger) \Big) = \sum_{k,l}^{N} \operatorname{tr}\Big( \mathrm{i}H \big[e_l e_k^\dagger, \tilde{\Gamma}(e_k e_l^\dagger)\big] \Big) \\
&= \sum_{k,l}^{N} \operatorname{tr}\Big( \mathrm{i}H \big( 2 e_l e_k^\dagger V e_k e_l^\dagger V^\dagger - e_l e_l^\dagger V^\dagger V - e_l e_k^\dagger V^\dagger V e_k e_l^\dagger \big) \Big) \\
&- \sum_{k,l}^{N} \operatorname{tr}\Big( \mathrm{i}H \big( 2 V e_k e_l^\dagger V^\dagger e_l e_k^\dagger - e_k e_l^\dagger V^\dagger V e_l e_k^\dagger - V^\dagger V e_k e_k^\dagger \big) \Big) \\
&= \sum_{k,l}^{N} \Big( 2\mathrm{i} \operatorname{tr}\big( H e_l e_k^\dagger V e_k e_l^\dagger V^\dagger \big) - \mathrm{i} \operatorname{tr}\big( H e_l e_l^\dagger V^\dagger V \big) - \mathrm{i} \operatorname{tr}\big( H e_l e_k^\dagger V^\dagger V e_k e_l^\dagger \big) \Big)
\end{aligned}$$

$$
\begin{aligned}
- & \sum_{k,l}^{N}\Big(2\mathrm{i}\,\mathrm{tr}\left(HVe_ke_l^\dagger V^\dagger e_le_k^\dagger\right) - \mathrm{i}\,\mathrm{tr}\left(He_ke_l^\dagger V^\dagger Ve_le_k^\dagger\right) - \mathrm{i}\,\mathrm{tr}\left(HV^\dagger Ve_ke_k^\dagger\right)\Big)\,, \\
= & \; 2\mathrm{i}\sum_{k,l}^{N} e_k^\dagger Ve_k\cdot e_l^\dagger V^\dagger He_l - \mathrm{i}\sum_{k,l}^{N} e_l^\dagger V^\dagger VHe_l - \mathrm{i}\sum_{k,l}^{N} e_l^\dagger He_l\cdot e_k^\dagger HV^\dagger Ve_k \\
- & \; 2\mathrm{i}\sum_{k,l}^{N} e_k^\dagger HVe_k\cdot e_l^\dagger V^\dagger e_l + \mathrm{i}\sum_{k,l}^{N} e_k^\dagger He_k\cdot e_l^\dagger V^\dagger Ve_l + \mathrm{i}\sum_{k,l}^{N} e_k^\dagger HV^\dagger Ve_k \\
= & \; 2\mathrm{i}\cdot\mathrm{tr}\,V\cdot\mathrm{tr}(V^\dagger H) - \mathrm{i}N\cdot\mathrm{tr}(V^\dagger VH) - \mathrm{i}\cdot\mathrm{tr}\,H\cdot\mathrm{tr}(V^\dagger V) \\
- & \; 2\mathrm{i}\cdot\mathrm{tr}(HV)\cdot\mathrm{tr}(V^\dagger) + \mathrm{i}\cdot\mathrm{tr}\,H\cdot\mathrm{tr}(V^\dagger V) + \mathrm{i}N\,\mathrm{tr}(HV^\dagger V) \\
= & \; 2\mathrm{i}\cdot\mathrm{tr}\,V\big(\mathrm{tr}(V^\dagger H) - \mathrm{tr}(HV)\big) = 0
\end{aligned}
$$

where the last equality follows from the fact that $V$ is assumed be to tracelass. Thus, the proof is complete.

Kronecker approach: Define

$$\widehat{\mathrm{iad}}_H := \mathrm{i}(\mathbb{1}\otimes H - H^\top\otimes\mathbb{1}) \tag{A.1}$$

and

$$\widehat{\Gamma} := 2\overline{V}\otimes V - \mathbb{1}\otimes V^\dagger V - V^\top\overline{V}\otimes\mathbb{1}\,. \tag{A.2}$$

Exploiting the fact that the $\widehat{(\cdot)}$–isomorphism defined by Eqn. (1.43) is actually scalar product preserving with respect to the standard Hilbert-Schmidt scalar product on $\mathfrak{gl}(N^2,\mathbb{C})$ one yields the following computation

$$
\begin{aligned}
\langle\widehat{\mathrm{iad}}_H,\widehat{\Gamma}\rangle = & \;\langle\mathrm{i}(\mathbb{1}\otimes H - H^\top\otimes\mathbb{1}), 2\overline{V}\otimes V - \mathbb{1}\otimes V^\dagger V - V^\top\overline{V}\otimes\mathbb{1}\rangle \\
= & \; -\mathrm{i}\,\mathrm{tr}\Big((\mathbb{1}\otimes H - H^\top\otimes\mathbb{1})^\dagger(2\overline{V}\otimes V - \mathbb{1}\otimes V^\dagger V - V^\top\overline{V}\otimes\mathbb{1})\Big) \\
= & \; -2\mathrm{i}\,\mathrm{tr}(\overline{V}\otimes HV - \overline{HV}\otimes V) + \mathrm{i}\,\mathrm{tr}(\mathbb{1}\otimes HV^\dagger V - H^\top\otimes V^\dagger V) \\
+ & \;\mathrm{i}\,\mathrm{tr}(V^\top\overline{V}\otimes H - H^\top V^\top\overline{V}\otimes\mathbb{1}) \\
= & \; -2\mathrm{i}\,\mathrm{tr}\,\overline{V}\cdot\mathrm{tr}(HV) + 2\mathrm{i}\,\mathrm{tr}(\overline{HV})\cdot\mathrm{tr}\,V + \mathrm{i}N\,\mathrm{tr}(HV^\dagger V) - \mathrm{i}\,\mathrm{tr}\,H^\top\cdot\mathrm{tr}(V^\dagger V) \\
+ & \;\mathrm{i}\,\mathrm{tr}(V^\top\overline{V})\cdot\mathrm{tr}\,H - \mathrm{i}N\,\mathrm{tr}(H^\top V^\top\overline{V}) \\
= & \; -2\mathrm{i}\,\mathrm{tr}\,\overline{V}\cdot\mathrm{tr}(HV) + 2\mathrm{i}\,\mathrm{tr}(\overline{HV})\cdot\mathrm{tr}\,V = 0
\end{aligned}
$$

where the second last equality follows from $\mathrm{tr}(A) = \mathrm{tr}(A^\top)$ and the last one from the fact that $V$ is assumed to be traceless. Thus, again we have $\widehat{\mathrm{iad}}_H \perp \widehat{\Gamma}$ and hence $\mathrm{i}\,\mathrm{ad_H}\perp\Gamma$. □

For the following results which serve as a fundamental background to the representation theory of Lie algebras used in this work, recall that for $n$-qubit quantum systems, the dimension is given as $N = 2^n$.

**Lemma 1.3.4.** *Let $\mathfrak{g}^{LK}$ and $\mathfrak{g}_0^{LK}$ denote the Lindblad-Kossakowski algebra and its unital subalgebra. Moreover, let $\mathfrak{g}^E$ and $\mathfrak{g}_0^E$ denote the following subsets of $\mathfrak{gl}\big(\mathfrak{her}(N)\big)$:*

$$\mathfrak{g}^E := \big\{\Phi\in\mathfrak{gl}\big(\mathfrak{her}(N)\big)\;\big|\;\mathrm{Im}\,\Phi\subset\mathfrak{her}_0(N)\big\} \tag{A.3}$$

*and*

$$\mathfrak{g}_0^E := \big\{\Phi\in\mathfrak{gl}\big(\mathfrak{her}(N)\big)\;\big|\;\mathrm{Im}\,\Phi\subset\mathfrak{her}_0(N),\,\mathbb{1}_N\in\ker\Phi\big\}\,. \tag{A.4}$$

*Then, one has the following results and commutative diagrams:*

*(a)* $\mathfrak{g}^E$ *and* $\mathfrak{g}_0^E$ *are* real *(Lie) subalgebras satisfying the inclusion relations:*

$$\begin{array}{ccc} \mathfrak{g}_0^{LK} & \xrightarrow{\text{inc}} & \mathfrak{g}_0^E \\ \downarrow{\scriptstyle\text{inc}} & & \downarrow{\scriptstyle\text{inc}} \\ \mathfrak{g}^{LK} & \xrightarrow{\text{inc}} & \mathfrak{g}^E \end{array}$$

*(b) Commutative diagrams for* $\mathfrak{g}_0^{LK}$*:*

$$\begin{array}{ccccc} \mathrm{voc}(\mathfrak{g}_0^{LK}) & \xrightarrow{\text{inc}} & \mathfrak{gl}(N^2-1,\mathbb{R}\,) & \xrightarrow{\mathrm{em_b}} & \mathfrak{gl}(N^2,\mathbb{R}\,) \\ \uparrow{\scriptstyle\text{voc}} & & \uparrow{\scriptstyle\text{voc}} & & \uparrow{\scriptstyle\text{voc}} \\ \mathfrak{g}_0^{LK} & \xrightarrow{\text{inc}} & \mathfrak{g}_0^E & \xrightarrow{\text{inc}} & \mathfrak{gl}\big(\mathfrak{her}(N)\big) \end{array}$$

*and*

$$\begin{array}{ccccccc} \mathfrak{g}_0^{LK} & \xrightarrow{\text{inc}} & \mathfrak{g}_0^E & \xrightarrow{\text{inc}} & \mathfrak{gl}\big(\mathfrak{her}(N)\big) & \xrightarrow{\text{em}} & \mathfrak{gl}\big(\mathbb{C}^{\,N\times N}\big) \\ \downarrow{\scriptstyle\widehat{(\cdot)}} & & \downarrow{\scriptstyle\widehat{(\cdot)}} & & \downarrow{\scriptstyle\widehat{(\cdot)}} & & \downarrow{\scriptstyle\widehat{(\cdot)}} \\ \hat{\mathfrak{g}}_0^{LK} & \xrightarrow{\text{inc}} & \hat{\mathfrak{g}}_0^E & \xrightarrow{\text{inc}} & \hat{\mathfrak{gl}}\big(\mathfrak{her}(N)\big) & \xrightarrow{\text{inc}} & \mathfrak{gl}(N^2,\mathbb{C}\,) \end{array}$$

*(c) Commutative diagrams for* $\mathfrak{g}^{LK}$*:*

$$\begin{array}{ccccc} \mathrm{voc}(\mathfrak{g}^{LK}) & \xrightarrow{\text{inc}} & \mathfrak{gl}(N^2-1,\mathbb{R}\,)\oplus_s\mathbb{R}^{\,N^2-1} & \xrightarrow{\mathrm{em_s}} & \mathfrak{gl}(N^2,\mathbb{R}\,) \\ \uparrow{\scriptstyle\text{voc}} & & \uparrow{\scriptstyle\text{voc}} & & \uparrow{\scriptstyle\text{voc}} \\ \mathfrak{g}^{LK} & \xrightarrow{\text{inc}} & \mathfrak{g}^E & \xrightarrow{\text{inc}} & \mathfrak{gl}\big(\mathfrak{her}(N)\big) \end{array}$$

*and*

$$\begin{array}{ccccccc} \mathfrak{g}^{LK} & \xrightarrow{\text{inc}} & \mathfrak{g}^E & \xrightarrow{\text{inc}} & \mathfrak{gl}\big(\mathfrak{her}(N)\big) & \xrightarrow{\text{em}} & \mathfrak{gl}\big(\mathbb{C}^{\,N\times N}\big) \\ \downarrow{\scriptstyle\widehat{(\cdot)}} & & \downarrow{\scriptstyle\widehat{(\cdot)}} & & \downarrow{\scriptstyle\widehat{(\cdot)}} & & \downarrow{\scriptstyle\widehat{(\cdot)}} \\ \hat{\mathfrak{g}}^{LK} & \xrightarrow{\text{inc}} & \hat{\mathfrak{g}}^E & \xrightarrow{\text{inc}} & \hat{\mathfrak{gl}}\big(\mathfrak{her}(N)\big) & \xrightarrow{\text{inc}} & \mathfrak{gl}(N^2,\mathbb{C}\,) \end{array}$$

*(d) In particular, one has the following bounds on the dimensions of* $\mathfrak{g}_0^{LK}$ *and* $\mathfrak{g}^{LK}$*:*

$$\dim_{\mathbb{R}}\mathfrak{g}_0^{LK}\le(N-1)^2\quad\textit{and}\quad\dim_{\mathbb{R}}\mathfrak{g}^{LK}\le(N-1)N\,. \tag{A.5}$$

*Here,* $\mathrm{inc}:*\to*$ *denotes the cannonical inclsion map and*

$$\begin{aligned} \mathrm{em_b}&:\mathfrak{gl}(N^2-1,\mathbb{R}\,)\to\mathfrak{gl}(N^2,\mathbb{R}\,),\\ \mathrm{em_s}&:\mathfrak{gl}(N^2-1,\mathbb{R}\,)\oplus_s\mathbb{R}^{\,N^2-1}\to\mathfrak{gl}(N^2,\mathbb{R}\,)\\ \mathrm{em}&:\mathfrak{gl}\big(\mathfrak{her}(N)\big)\to\mathfrak{gl}\big(\mathbb{C}^{\,N\times N}\big) \end{aligned} \tag{A.6}$$

*are natural embeddings defined by*

$$A\mapsto\mathrm{em_b}(A):=\begin{pmatrix}A&0\\0&0\end{pmatrix} \tag{A.7}$$

$$(A,b)\mapsto\mathrm{em_s}(A,b):=\begin{pmatrix}A&b\\0&0\end{pmatrix} \tag{A.8}$$

*and*

$$\Phi\mapsto\mathrm{em}(\Phi)\,, \tag{A.9}$$

*where* $\mathrm{em}(\Phi)$ *acts on* $V=C+\mathrm{i}C\in\mathbb{C}^{\,N\times N}$ *with* $C,D\in\mathfrak{her}(N)$ *via* $\mathrm{em}(\Phi)(V):=\Phi(C)+\mathrm{i}\Phi(D)$.

**Remark 14.** *By the above commutative diagrams it is obvious that all result on* $\mathfrak{g}^{LK}$ *and* $\mathfrak{g}_0^{LK}$ *can be immediately carried over to* $\hat{\mathfrak{g}}^{LK}$ *and* $\hat{\mathfrak{g}}_0^{LK}$ *and vice versa.*

*Proof.* (Sketch) Due to the fact that the flow of $\Phi$ is trace-preserving one has $\operatorname{Im}\mathcal{L} \subset \mathfrak{her}_0(N)$ for all $\mathcal{L} \in \mathfrak{w}^{\mathrm{LK}}$. This property clearly carries over to all $\mathcal{L} \in \mathfrak{g}^{LK}$ and therefore (by choosing any matrix representation) an easy counting argument shows $\dim \mathfrak{g}^{LK} \leq (N-1)N$.

For all $\mathcal{L} \in \mathfrak{w}_0^{\mathrm{LK}}$ one has the additional property $\mathbb{R}\,\mathbb{1}_N \subset \ker \mathcal{L}$, which again passes to $\mathfrak{g}_0^{LK}$. Consequently, a similar counting argument yields $\dim \mathfrak{g}_0^{LK} \leq (N-1)^2$. □

Consider the standard Cartan-decomposition of

$$\mathfrak{gl}\big(\mathbb{C}^{N\times N}\big) = \mathfrak{gl}_{\text{skew}}\big(\mathbb{C}^{N\times N}\big) \oplus \mathfrak{gl}_{\text{self}}\big(\mathbb{C}^{N\times N}\big)\,, \tag{A.10}$$

into skew- and self-adjont operators (with respect to the Hilbert-Schmidt scalar product) as well as the standard Cartan-decomposition of

$$\mathfrak{gl}\big(N^2, \mathbb{C}\,\big) = \mathfrak{u}(N^2) \oplus \mathfrak{her}(N^2) \tag{A.11}$$

into unitary and Hermitian matrices. Our next Corollary clarifies how these decompositions go along with the $\widehat{(\cdot)}$-operation.

The following straightforward Corollary serves as the basis for making use of Cartan-decompositions throughout this thesis. It provides the dimensions of the $\mathfrak{k}$ and $\mathfrak{p}$ parts of the Lie algebra $\mathfrak{g}^E$ - which we proved was isomorphic to the Lindblad-Kossakowski Lie algebra for open quantum systems.

**Corollary A.0.1.** *Let the notation be as in Lemma 1.3.4 Then the standard Cartan-decompositions of* $\mathfrak{gl}\big(\mathbb{C}^{N\times N}\big)$ *and* $\mathfrak{gl}\big(N^2, \mathbb{C}\,\big)$ *given by Eqns.* (A.10) *and* (A.11)*, respectively, match with the* $\widehat{(\cdot)}$*-operation, i.e.* $\widehat{\mathfrak{gl}}_{\text{skew}}\big(\mathbb{C}^{N\times N}\big) = \mathfrak{u}(N^2)$ *and* $\widehat{\mathfrak{gl}}_{\text{self}}\big(\mathbb{C}^{N\times N}\big) = \mathfrak{her}(N^2)$*. Moreover, they induce Cartan-decompositions of* $\mathfrak{g}_0^E$ *and* $\hat{\mathfrak{g}}_0^E$ *in the following way:*

$$\mathfrak{g}_0^E = \mathfrak{k}_0^E \oplus \mathfrak{p}_0^E \quad \textit{and} \quad \hat{\mathfrak{g}}_0^E = \hat{\mathfrak{k}}_0^E \oplus \hat{\mathfrak{p}}_0^E \tag{A.12}$$

*with*

$$\mathfrak{k}_0^E := \mathfrak{g}_0^E \cap \mathfrak{gl}_{\text{skew}}\big(\mathbb{C}^{N\times N}\big)\,, \tag{A.13}$$

$$\mathfrak{p}_0^E := \mathfrak{g}_0^E \cap \mathfrak{gl}_{\text{self}}\big(\mathbb{C}^{N\times N}\big)\,, \tag{A.14}$$

*and*

$$\hat{\mathfrak{k}}_0^E := \hat{\mathfrak{g}}_0^E \cap \mathfrak{u}(N^2)\,, \tag{A.15}$$

$$\hat{\mathfrak{p}}_0^E := \hat{\mathfrak{g}}_0^E \cap \mathfrak{her}(N^2)\,. \tag{A.16}$$

*The dimensions of the involved subspaces are*

$$\dim_{\mathbb{R}} \mathfrak{k}_0^E = \dim_{\mathbb{R}} \hat{\mathfrak{k}}_0^E = \tfrac{(N^2-1)(N^2-2)}{2} \tag{A.17}$$

*and*

$$\dim_{\mathbb{R}} \mathfrak{p}_0^E = \dim_{\mathbb{R}} \hat{\mathfrak{p}}_0^E = \tfrac{N^2(N^2-1)}{2}\,. \tag{A.18}$$

# Appendix B

# The Lindblad-Kossakowski Ideal: Supplementary Proofs

In this Appendix we prove several results pertaining to Chapter 2 which use the abstract operator representation of elements contained in the Lindblad-Kossakowski Lie algebra $\mathfrak{g}^{LK} = \mathfrak{g}_0^{LK} \oplus_s \mathfrak{i}$.

**Remark 15.** *For any fixed* $\boldsymbol{q} \in I_0^n$*, there exists* $\boldsymbol{p} \in I_0^n$ *such that* $\mathrm{ad}^2_{\sigma_{\boldsymbol{p}}}(\sigma_{\boldsymbol{q}}) = 4\sigma_{\boldsymbol{q}}$*. Moreover, there exists no* $\boldsymbol{r} \in I_0^n$ *such that* $\mathrm{ad}^2_{\sigma_{\boldsymbol{r}}}(\sigma_{\boldsymbol{q}}) = -4\sigma_{\boldsymbol{q}}$ *and hence* $\sum_{\boldsymbol{p}} \mathrm{ad}^2_{\sigma_{\boldsymbol{p}}}\big|_{\mathfrak{her}_0(2^n)} \neq 0$*. More precisely, by Lemma 2.3.1 there are exactly* $2^{2n-1}$ *elements* $\sigma_{\boldsymbol{p}} \in \mathcal{B}_0^n$ *which do not commute with any single fixed* $\sigma_{\boldsymbol{q}} \in \mathfrak{her}_0(2^n)$ *and therefore*

$$C_0\big|_{\mathfrak{her}_0(2^n)} = \mathbb{1}_{2^n} \;, \quad \textit{where} \quad C_0 := \frac{1}{2^{2n-1}} \sum_{\boldsymbol{p}} \mathrm{ad}^2_{\sigma_{\boldsymbol{p}}} \;. \tag{B.1}$$

**Lemma B.0.1.** *For* $C_0$ *defined in Eqn.* (B.1)*, we have that*

$$\mathrm{ad}_{C_0}\big|_{\mathfrak{g}_0^E} = 0 \quad \textit{and} \quad \mathrm{ad}_{C_0}\big|_{\mathfrak{g}^E} \in \mathfrak{i} \;, \tag{B.2}$$

*and in particular,* $\mathrm{ad}_{C_0}\big|_{\mathfrak{m}_{\mathrm{qt}}} \in \mathfrak{i}$.

*Proof.* Note that we already know that $\mathfrak{g}_0^{LK} = \mathfrak{g}_0^E$. Let $A \in \mathfrak{g}_0^{LK}$ and note that $A\big|_{\mathfrak{her}_0(2^n)} \in \mathfrak{gl}\big(\mathfrak{her}_0(2^n)\big)$. Then $[C_0, A] \in \mathfrak{g}_0^{LK}$ and $[C_0, A](\sigma_{\mathbf{p}}) = (C_0A - AC_0)(\sigma_{\mathbf{p}}) = C_0(A(\sigma_{\mathbf{p}})) - A(C_0(\sigma_{\mathbf{p}})) = A(\sigma_{\mathbf{p}}) - A(\sigma_{\mathbf{p}}) = 0$ which follows from Eqn. (B.1) and hence $[C_0, \mathfrak{g}_0^{LK}] = 0$. To prove the second assertion we note that for $B \in \mathfrak{g}^E$ we have that $B\big|_{\mathfrak{her}_0(2^n)} \in \mathfrak{gl}\big(\mathfrak{her}_0(2^n)\big)$ and $B(\mathbb{1}_{2^n}) \in \mathfrak{her}_0(2^n)$. Again using Eqn. (B.1) we see that $[C_0, B](\mathbb{1}_{2^n}) = C_0B(\mathbb{1}_{2^n}) - BC_0(\mathbb{1}_{2^n}) = C_0B(\mathbb{1}_{2^n}) \in \mathfrak{her}_0(2^n)$ since $C_0(\mathbb{1}_{2^n}) = 0$. Thus, $[C_0, B](\sigma_{\mathbf{p}}) = C_0(B(\sigma_{\mathbf{p}})) - B(C_0(\sigma_{\mathbf{p}})) = 0$ and therefore $[C_0, B]$ satisfies the properties of being an infinitesimal translation i.e. $[C_0, \mathfrak{g}^E] \subseteq \mathfrak{i}$. Finally note that the set of quasi-translations $\mathfrak{m}_{\mathrm{qt}}$ are not contained in $\mathfrak{g}_0^{LK}$ by Proposition 2.3.2 and therefore $[C_0, \mathfrak{m}_{\mathrm{qt}}] \subseteq \mathfrak{i}$. □

**Lemma B.0.2.** *For* $\hat{\tau}_{m,k}$ *defined in Eqn.* (2.73) *(which has its "hat" omitted), its corresponding operator representation is given by* $\tau_{m,k} := \frac{\mathrm{i}}{4}\, \mathrm{ad}_{\sigma_{q,k}} \circ \mathrm{ad}^+_{\sigma_{p,k}}$ *with* $m = p \star q$*. Then*

$$\tau_{m,k} = \mathbb{1}_2 \otimes ... \otimes \tau_m \otimes \cdots \otimes \mathbb{1}_2 \;, \tag{B.3}$$

*where $\mathcal{T}_m$ is at the $k^{th}$ position in the tensor products and hence*

$$\prod_{k=1}^{n} \mathcal{T}_{m_k,k} = \mathcal{T}_{m_1} \otimes \mathcal{T}_{m_2} \otimes \cdots \otimes \mathcal{T}_{m_n} \ . \tag{B.4}$$

*Proof.* For notational convenience assume that the translation acts on the first qubit. For any $\sigma_{\mathbf{r}} \in \mathcal{B}^n$ we have that

$$\begin{aligned} \mathcal{T}_{m,1}(\sigma_{\mathbf{r}}) &= \tfrac{\mathrm{i}}{4}\,\mathrm{ad}_{\sigma_q \otimes \mathbb{1}_2 \cdots \otimes \mathbb{1}_2} \circ \mathrm{ad}^{+}_{\sigma_p \otimes \mathbb{1}_2 \cdots \otimes \mathbb{1}_2}(\sigma_{\mathbf{r}}) \\ &= \tfrac{\mathrm{i}}{4}(\mathrm{ad}_{\sigma_q} \circ \mathrm{ad}^{+}_{\sigma_p} \otimes \mathbb{1}_{2^{n-1}})(\sigma_{\mathbf{r}}) \\ &= (\mathcal{T}_m \otimes \mathbb{1}_{2^{n-1}})(\sigma_{\mathbf{r}}) \ . \end{aligned}$$

Taking $n$ products of local operators each on different qubits then yields the equality in Eqn. (B.4). □

Now that the locality of the quasi-local translation operators is explicit, Corollary 2.4.2 from Section 2.4 follows immediately.

**Corollary 2.4.2.** *For a fixed $\boldsymbol{m} \in \mathbb{I}_0^n$ we have the equality*

$$\mathcal{T}_{\boldsymbol{m}} = \chi\Big(\prod_{k} \mathcal{T}_{m_k,k}\Big) \ , \quad \textit{for} \quad \boldsymbol{m} = (m_1, m_2, \ldots, m_n) \ , \tag{B.5}$$

*where $k$ varies over the index numbers of $\boldsymbol{m}$ which have $m_k \neq 1$. Furthermore, for a fixed $\boldsymbol{m} \in \mathbb{I}_0^n$ which has no $k^{th}$ element $m_k$ equal to one, Eqn.* (B.5) *simplifies to*

$$\mathcal{T}_{\boldsymbol{m}} = \prod_{k=1}^{n} \mathcal{T}_{m_k,k} \ , \quad \textit{for} \quad \boldsymbol{m} = (m_1, m_2, \ldots, m_n) \ . \tag{B.6}$$

*Proof.* We prove Eqn. (B.6) in which case Eqn. (B.5) then follows from the same principles. Let $\mathbf{p}, \mathbf{q} \in I_0^n$ such that $\mathbf{m} = \mathbf{p} \star \mathbf{q}$ where $m_k \neq 1$ for all $k \leq n$. Then since $\mathcal{T}_{\mathbf{m}}(\mathbb{1}_2) = \frac{\mathrm{i}}{4}\,\mathrm{ad}_{\sigma_{\mathbf{q}}}\,\mathrm{ad}^{+}_{\sigma_{\mathbf{p}}}(\mathbb{1}_{2^n}) = \sigma_{\mathbf{m}}$, we have that

$$\begin{aligned} \tfrac{\mathrm{i}}{4}\,\mathrm{ad}_{\sigma_{\mathbf{q}}}\,\mathrm{ad}^{+}_{\sigma_{\mathbf{p}}}(\mathbb{1}_{2^n}) &= \sigma_{m_1} \otimes \sigma_{m_2} \otimes \cdots \otimes \sigma_{m_n} \\ &= \mathcal{T}_{m_1}(\mathbb{1}_2) \otimes \mathcal{T}_{m_2}(\mathbb{1}_2) \otimes \cdots \otimes \mathcal{T}_{m_n}(\mathbb{1}_2) \\ &= \big(\mathcal{T}_{m_1} \otimes \mathcal{T}_{m_2} \otimes \cdots \otimes \mathcal{T}_{m_n}\big)(\mathbb{1}_{2^n}) \\ &= \mathcal{T}_{m_1,1}\mathcal{T}_{m_2,2} \ldots \mathcal{T}_{m_n,n}(\mathbb{1}_{2^n}) \\ &= \prod_{k=1}^{n} \mathcal{T}_{m_k,k}(\mathbb{1}_{2^n}) \end{aligned}$$

where the fourth equality follows by Lemma B.0.2. Hence

$$\Big(\tfrac{\mathrm{i}}{4}\,\mathrm{ad}_{\sigma_{\mathbf{q}}}\,\mathrm{ad}^{+}_{\sigma_{\mathbf{p}}} - \prod_{k=1}^{n} \mathcal{T}_{m_k,k}\Big)(\mathbb{1}_{2^n}) = 0 \tag{B.7}$$

and therefore the difference of these two operators is unital. By Lemma B.0.1, $\chi(A) = 0$ for all $A \in \mathfrak{g}_0^{LK}$ (i.e unital A) and hence $\chi(\frac{\mathrm{i}}{4}\,\mathrm{ad}_{\sigma_{\mathbf{q}}}\,\mathrm{ad}^{+}_{\sigma_{\mathbf{p}}} - \prod_{k=1}^{n} \mathcal{T}_{m_k,k}) = 0$. Now since $\mathcal{T}_{\mathbf{m}} := \chi(\frac{\mathrm{i}}{4}\,\mathrm{ad}_{\sigma_{\mathbf{q}}}\,\mathrm{ad}^{+}_{\sigma_{\mathbf{p}}})$ and by Remark 8 in Section 2.4, $\prod_{k=1}^{n} \mathcal{T}_{m_k,k}$ is already contained in the ideal we get that $\mathcal{T}_{\mathbf{m}} = \prod_{k=1}^{n} \mathcal{T}_{m_k,k}$.

If on the other hand, for a fixed $\mathbf{m} \in \mathbb{I}_0^n$ there is some $m_k = 1$ then we can do the same procedure as above except now, by Lemma 1.3.6 we must include the projection operator in the expression $\chi(\prod_k \mathcal{T}_{m_k,k})$ since $\prod_k \mathcal{T}_{m_k,k}$ is not contained in the ideal by itself. □

# Appendix C

# Eigenspaces of Lindblad Generators

We would like to relate the kernel of a Lindblad generator which has a single Lindblad term in canonical form to the kernel of the Lindblad term.

**Lemma C.0.3.** *For $\sigma_{\boldsymbol{p}}, \sigma_{\boldsymbol{q}} \in \mathcal{B}_0^n$, we have that*

$$\dim\ \ker(\sigma_{\boldsymbol{p}} + \mathrm{i}\sigma_{\boldsymbol{q}}) = \begin{cases} 0 & \textit{if} \quad [\sigma_{\boldsymbol{p}}, \sigma_{\boldsymbol{q}}] = 0, \\ 2^{n-1} & \textit{if} \quad \{\sigma_{\boldsymbol{p}}, \sigma_{\boldsymbol{q}}\} = 0. \end{cases} \tag{C.1}$$

*Proof.* By Lemma 2.3.2, these are the only two dimensional possibilities which can occur and they are mutually exclusive. Moreover, note that $\mathrm{rank}(\sigma_{\mathbf{p}} + \mathrm{i}\sigma_{\mathbf{q}}) = \mathrm{rank}((\sigma_{\mathbf{p}} + \mathrm{i}\sigma_{\mathbf{q}})(\sigma_{\mathbf{p}} - \mathrm{i}\sigma_{\mathbf{q}})) = \mathrm{rank}(2\mathbb{1} + \mathrm{i}[\sigma_{\mathbf{p}}, \sigma_{\mathbf{q}}])$. Now if $[\sigma_{\mathbf{p}}, \sigma_{\mathbf{q}}] = 0$, then $\sigma_{\mathbf{p}} + \mathrm{i}\sigma_{\mathbf{q}}$ is full rank, otherwise, $\mathrm{i}[\sigma_{\mathbf{p}}, \sigma_{\mathbf{q}}]$ has $2^{n-1}$ negative two eigenvalues and hence $\mathrm{rank}(2\mathbb{1} + \mathrm{i}[\sigma_{\mathbf{p}}, \sigma_{\mathbf{q}}]) = 2^{n-1}$. □

It will be useful to consider the both the kernel and eigenspaces of both the unital and mixed components of the Lindblad generator as expressed by Eqn. (2.50). The following Proposition gives a complete description of such vector spaces and will allow us to relate the kernel of a Lindblad term in canonical form to the kernel of the entire Lindblad generator.

**Proposition C.0.1.** *Let $\Gamma$ be a Lindblad generator which has a single Lindblad term of canonical form, i.e. $V = \frac{1}{2}(\sigma_{\boldsymbol{p}} + \mathrm{i}\sigma_{\boldsymbol{q}})$ such that $[\sigma_{\boldsymbol{p}}, \sigma_{\boldsymbol{q}}] \neq 0$. Decomposing $\Gamma = \Gamma_u + \Gamma_m$ into unital and mixed components via Eqn. (2.50), the unital component $\Gamma_u$ satisfies*

$$\begin{aligned} \ker(\Gamma_u) &= \langle \sigma_{\boldsymbol{m}} \mid [\sigma_{\boldsymbol{p}}, \sigma_{\boldsymbol{m}}] = [\sigma_{\boldsymbol{q}}, \sigma_{\boldsymbol{m}}] = 0\ , \sigma_{\boldsymbol{m}} \in \mathcal{B}^n \rangle\ , \\ \mathrm{range}(\Gamma_u) &= \langle \sigma_{\boldsymbol{m}} \mid \{\sigma_{\boldsymbol{p}}, \sigma_{\boldsymbol{m}}\}_+ = 0 \textit{ or } \{\sigma_{\boldsymbol{q}}, \sigma_{\boldsymbol{m}}\}_+ = 0\ , \sigma_{\boldsymbol{m}} \in \mathcal{B}^n \rangle \end{aligned}$$

*such that* $\mathrm{range}(\Gamma_u) = E_{-2}(\Gamma_u) \oplus E_{-4}(\Gamma_u)$, *where*

$$\begin{aligned} E_{-2}(\Gamma_u) &:= \langle \sigma_{\boldsymbol{m}} \mid \{\sigma_{\boldsymbol{p}}, \sigma_{\boldsymbol{m}}\}_+ \neq 0 \textit{ and } \{\sigma_{\boldsymbol{q}}, \sigma_{\boldsymbol{m}}\}_+ = 0\ , \\ &\qquad \textit{or}\ , \{\sigma_{\boldsymbol{p}}, \sigma_{\boldsymbol{m}}\}_+ = 0 \textit{ and } \{\sigma_{\boldsymbol{q}}, \sigma_{\boldsymbol{m}}\}_+ \neq 0, \sigma_{\boldsymbol{m}} \in \mathcal{B}^n \rangle\ , \\ E_{-4}(\Gamma_u) &:= \langle \sigma_{\boldsymbol{m}} \mid \{\sigma_{\boldsymbol{p}}, \sigma_{\boldsymbol{m}}\}_+ = \{\sigma_{\boldsymbol{q}}, \sigma_{\boldsymbol{m}}\}_+ = 0\ , \sigma_{\boldsymbol{m}} \in \mathcal{B}^n \rangle\ , \end{aligned}$$

*are the eigenspaces of $\Gamma_u$ corresponding to the eigenvalues $\lambda = -2$ and $\lambda = -4$, respectively. For the mixed component $\Gamma_m$, the following also hold*

$$\ker(\Gamma_m) = \langle \sigma_{\boldsymbol{m}} \mid \{\sigma_{\boldsymbol{p}}, \sigma_{\boldsymbol{m}}\}_+ = 0\ , \textit{or } \{\sigma_{\boldsymbol{q}}, \sigma_{\boldsymbol{m}}\}_+ = 0\ , \sigma_{\boldsymbol{m}} \in \mathcal{B}^n \rangle\ ,$$

$$\mathrm{range}(\Gamma_m) \quad = \quad \langle \sigma_{\boldsymbol{m}} \mid \{\sigma_{\boldsymbol{p}}, \sigma_{\boldsymbol{m}}\}_+ = \{\sigma_{\boldsymbol{q}}, \sigma_{\boldsymbol{m}}\}_+ = 0 \;, \sigma_{\boldsymbol{m}} \in \mathcal{B}^n \rangle$$

*and hence* $\mathrm{range}(\Gamma_m) = E_{-4}(\Gamma_u)$.

*Proof.* It will be helpful to determine the eigenspaces of the operator $\mathrm{ad}^2_{\sigma_{\mathbf{p}}}$ for the proof. By Lemma 2.3.2, $\mathrm{ad}_{\sigma_{\mathbf{p}}}(\sigma_{\mathbf{m}}) = 0$ or $\mathrm{ad}_{\sigma_{\mathbf{p}}}(\sigma_{\mathbf{m}}) = 2\sigma_{\mathbf{p}}\sigma_{\mathbf{m}}$ for $\sigma_{\mathbf{m}} \in \mathcal{B}^n$. Assuming that $\mathrm{ad}_{\sigma_{\mathbf{p}}}(\sigma_{\mathbf{m}}) \neq 0$, we get that $\mathrm{ad}_{\sigma_{\mathbf{p}}}(2\sigma_{\mathbf{p}}\sigma_{\mathbf{m}}) = 2[\sigma_{\mathbf{p}}, \sigma_{\mathbf{p}}\sigma_{\mathbf{m}}] = 4\sigma_{\mathbf{m}}$, and hence $\sigma_{\mathbf{m}}$ is an eigenvector of $\mathrm{ad}^2_{\sigma_{\mathbf{p}}}$ to the eigenvalue zero whenever $\{\sigma_{\mathbf{p}}, \sigma_{\mathbf{q}}\}_+ = 0$, and an eigenvector to the eigenvalue four whenever $[\sigma_{\mathbf{p}}, \sigma_{\mathbf{m}}] \neq 0$. By dimension counting using Lemma 2.3.2, we see that

$$\ker(\tfrac{1}{2}\,\mathrm{ad}^2_{\sigma_{\mathbf{p}}}) \quad = \quad \langle \sigma_{\mathbf{m}} \mid [\sigma_{\mathbf{p}}, \sigma_{\mathbf{m}}] = 0 \rangle \;, \tag{C.2}$$

$$\mathrm{range}(\tfrac{1}{2}\,\mathrm{ad}^2_{\sigma_{\mathbf{p}}}) \quad = \quad \langle \sigma_{\mathbf{m}} \mid \{\sigma_{\mathbf{p}}, \sigma_{\mathbf{m}}\}_+ = 0 \rangle \;. \tag{C.3}$$

Since the $\mathrm{range}(\frac{1}{2}\,\mathrm{ad}^2_{\sigma_{\mathbf{p}}}) = E_2(\frac{1}{2}\,\mathrm{ad}^2_{\sigma_{\mathbf{p}}})$, then its clear that under summation, $\ker(\frac{1}{2}(\mathrm{ad}^2_{\sigma_{\mathbf{p}}} + \mathrm{ad}^2_{\sigma_{\mathbf{q}}})) = \ker(\frac{1}{2}\,\mathrm{ad}^2_{\sigma_{\mathbf{p}}}) \cap \ker(\frac{1}{2}\,\mathrm{ad}^2_{\sigma_{\mathbf{q}}})$ and hence the first claim is proved.

Now $(\ker(\Gamma_u))^\perp = (-\ker(-\frac{1}{2}\,\mathrm{ad}^2_{\sigma_{\mathbf{p}}}) \cap \ker(-\frac{1}{2}\,\mathrm{ad}^2_{\sigma_{\mathbf{q}}}))^\perp$ gives $\mathrm{range}(\Gamma_u) = \mathrm{range}(\frac{1}{2}\,\mathrm{ad}^2_{\sigma_{\mathbf{p}}}) + \mathrm{range}(\frac{1}{2}\,\mathrm{ad}^2_{\sigma_{\mathbf{q}}})$. The eigenspace decomposition of the range follows immediately from the fact that $\frac{1}{2}\,\mathrm{ad}^2_{\sigma_{\mathbf{p}}}(\sigma_{\mathbf{m}}) = 2\sigma_{\mathbf{m}}$ and $\frac{1}{2}\,\mathrm{ad}^2_{\sigma_{\mathbf{q}}}(\sigma_{\mathbf{m}}) = 2\sigma_{\mathbf{m}}$ if either of them are non-zero.

Now we prove statements concerning the kernel and range of $\Gamma_m$. Clearly $\mathrm{ad}_{\sigma_{\mathbf{p}}} \circ \mathrm{ad}^+_{\sigma_{\mathbf{q}}}(\sigma_{\mathbf{m}}) = 0$ when $\{\sigma_{\mathbf{q}}, \sigma_{\mathbf{m}}\}_+ = 0$. Assume that $\mathrm{ad}_{\sigma_{\mathbf{p}}} \circ \mathrm{ad}^+_{\sigma_{\mathbf{q}}}(\sigma_{\mathbf{m}}) = 0$ but $\{\sigma_{\mathbf{q}}, \sigma_{\mathbf{m}}\}_+ \neq 0$. This implies $\mathrm{ad}_{\sigma_{\mathbf{p}}}(2\sigma_{\mathbf{q}}\sigma_{\mathbf{m}}) = 2[\sigma_{\mathbf{p}}, \sigma_{\mathbf{q}}\sigma_{\mathbf{m}}] = 2\{\sigma_{\mathbf{p}}, \sigma_{\mathbf{m}}\}_+\sigma_{\mathbf{q}} = 0$, using the fact that $\{\sigma_{\mathbf{p}}, \sigma_{\mathbf{q}}\}_+ = 0$ and $[\sigma_{\mathbf{q}}, \sigma_{\mathbf{m}}] = 0$ in the second to last equality and hence $\{\sigma_{\mathbf{p}}, \sigma_{\mathbf{m}}\}_+ = 0$.

Defining the subspace $\mathrm{K}(\Gamma_m) \subseteq \ker(\Gamma_m)$ as $\mathrm{K}(\Gamma_m) := \langle \sigma_{\mathbf{m}} \mid \{\sigma_{\mathbf{p}}, \sigma_{\mathbf{m}}\}_+ = 0 \;, \text{or } \{\sigma_{\mathbf{q}}, \sigma_{\mathbf{m}}\}_+ = 0 \;, \sigma_{\mathbf{m}} \in \mathcal{B}^n \rangle$, and taking the orthogonal complement gives $\mathrm{R}(\Gamma_m) := (\mathrm{K}(\Gamma_m))^\perp = \langle \mathrm{i}\sigma_{\mathbf{p}}\sigma_{\mathbf{q}}\sigma_{\mathbf{m}} \mid [\sigma_{\mathbf{p}}, \sigma_{\mathbf{m}}] = [\sigma_{\mathbf{q}}, \sigma_{\mathbf{m}}] = 0 \;, \sigma_{\mathbf{m}} \in \mathcal{B}^n \rangle$ which is a subspace of $\mathrm{range}(\Gamma_m)$. Since $\mathrm{K}(\Gamma_m)$ and $\mathrm{R}(\Gamma_m)$ are use the same preconditions on $\sigma_{\mathbf{m}} \in \mathcal{B}^n$ as the range and kernel of $\Gamma_u$, respectively, and no $\sigma_{\mathbf{m}}$ gives a zero element in any of these four vector spaces then $\mathrm{K}(\Gamma_m) = \ker(\Gamma_m)$ and $\mathrm{R}(\Gamma_m) = \mathrm{range}(\Gamma_m)$. Moreover, since $\Gamma_m \circ \Gamma_m = 0$ and hence is a nilpotent operator, it must be true that $\mathrm{range}(\Gamma_m) \subset \ker(\Gamma_m)$. Applying the constraint for some $\sigma_{\mathbf{m}} \in \ker(\Gamma_m)$ to $\mathrm{i}\sigma_{\mathbf{p}}\sigma_{\mathbf{q}}\sigma_{\mathbf{m}} \in \mathrm{range}(\Gamma_m)$ we see that $\{\sigma_{\mathbf{p}}, \mathrm{i}\sigma_{\mathbf{p}}\sigma_{\mathbf{q}}\sigma_{\mathbf{m}}\}_+ = \{\sigma_{\mathbf{q}}, \mathrm{i}\sigma_{\mathbf{p}}\sigma_{\mathbf{q}}\sigma_{\mathbf{m}}\}_+ = 0$ and hence

$$\mathrm{range}(\Gamma_m) = \langle \sigma_{\mathbf{m}} \mid \{\sigma_{\mathbf{p}}, \sigma_{\mathbf{m}}\}_+ = \{\sigma_{\mathbf{q}}, \sigma_{\mathbf{m}}\}_+ = 0 \;, \sigma_{\mathbf{m}} \in \mathcal{B}^n \rangle \;.$$

□

**Remark 16.** *(Dimensions of Kernel and Ranges)*
*From the above calculation, one can deduce that the following hold*

1. $\dim\ \ker(\Gamma_u) = \dim\ E_{-4}(\Gamma_u) = \dim\ \mathrm{range}(\Gamma_m)$ ,
2. $\dim\ E_{-2}(\Gamma_u) = 2 \cdot \dim\ \ker(\Gamma_u)$ ,
3. $\dim\ \ker(\Gamma_m) = 3 \cdot \dim\ \mathrm{range}(\Gamma_m)$ .

**Example 14.** *For a single qubit, let* $\boldsymbol{p} = (x)$, $\boldsymbol{q} = (y)$ *and hence* $V := \sigma^+ = \frac{1}{2}(\sigma_x + \mathrm{i}\sigma_y)$ *be the atomic raising operator which corresponds to the translation direction* $\tau_z$ *since* $\boldsymbol{p} \star \boldsymbol{q} = \boldsymbol{m} = (z)$. *By Proposition 2.2.2,* $\Gamma_m$ *is an infinitesimal translation and hence by Eqn.* (1.48), $\ker(\Gamma_m) = \langle \sigma_x, \sigma_y, \sigma_z \rangle$ *and* $\mathrm{range}(\Gamma_m) = \langle \sigma_z \rangle$. *Moreover,* $\ker(\Gamma_u) = \langle \mathbb{1} \rangle$ *and* $\mathrm{range}(\Gamma_u) = E_{-2}(\Gamma_u) \oplus E_{-4}(\Gamma_u) = \langle \sigma_x, \sigma_y, \sigma_z \rangle$, *where* $E_{-2}(\Gamma_u) = \langle \sigma_x, \sigma_y \rangle$ *and* $E_{-4}(\Gamma_u) = \langle \sigma_z \rangle$.

*Noting that* $\ker(V) = \langle |e_1\rangle\rangle$, *taking the inverse vectorization of this null vector then gives* $\rho = \frac{1}{2}(\mathbb{1} + \sigma_z) = \frac{1}{2}(\mathbb{1} + \sigma_{\boldsymbol{m}})$. *Under the action of* $\Gamma$, *we obtain*

$$\begin{aligned}\Gamma(\rho) &= \Gamma_u(\rho) + \Gamma_m(\rho) && \text{(C.4)}\\ &= \tfrac{1}{2}(\Gamma_u(\mathbb{1}) + \Gamma_u(\sigma_z)) + \tfrac{1}{2}(\Gamma_m(\mathbb{1}) + \Gamma_m(\sigma_z)) && \text{(C.5)}\\ &= \tfrac{1}{2}(\Gamma_u(\sigma_z)) + \tfrac{1}{2}(\Gamma_m(\mathbb{1})) = -2\sigma_z + 2\sigma_z = 0\ . && \text{(C.6)}\end{aligned}$$

*It is well known that for this single qubit system* $\rho$ *is the unique fixed point and we have made the connection that in fact,*

$$\ker(\Gamma) = \langle \mathrm{vec}^{-1}(v)\ |v \in \ker(V)\rangle\ , \tag{C.7}$$

*where* $\mathrm{vec}^{-1}$ *is the inverse vectorization operator which maps a vector into its matrix representation (see for example [25, 22]).*

One might wonder if this relationship between the kernel of $\Gamma$ and the kernel of the single Lindblad term holds in general for $n$-qubit systems.

**Example 15.** *For a two qubit system, let let* $\boldsymbol{p} = (x, 1), \boldsymbol{q} = (y, z)$ *and hence* $V := \frac{1}{2}(\sigma_x \otimes \mathbb{1} + \mathrm{i}\sigma_y \otimes \sigma_z)$ *which corresponds to the translation direction* $\tau_{zz} = \tau_{\boldsymbol{m}}$ *since* $\boldsymbol{p} \star \boldsymbol{q} = \boldsymbol{m} = (z, z)$. *By Proposition C.0.1 we obtain*

$$\begin{aligned}\ker(\Gamma_u) &= \langle \mathbb{1}\otimes\mathbb{1}, \mathbb{1}\otimes\sigma_z, \sigma_x\otimes\sigma_y, \sigma_x\otimes\sigma_x\rangle && \text{(C.8)}\\ E_{-4}(\Gamma_u) &= \langle \sigma_y\otimes\sigma_y, \sigma_z\otimes\mathbb{1}, \sigma_y\otimes\sigma_x, \sigma_z\otimes\sigma_z\rangle\ , && \text{(C.9)}\end{aligned}$$

*and in general,* $E_{-4}(\Gamma_u) = \mathrm{range}(\Gamma_m) \subset \ker(\Gamma_m)$. *Indeed, again* $\mathrm{vec}^{-1}(\frac{1}{2}(\mathbb{1}\otimes\mathbb{1} + \sigma_z \otimes \sigma_z)) \in \ker(V)$ *and* $\Gamma(\frac{1}{2}(\mathbb{1}\otimes\mathbb{1} + \sigma_z\otimes\sigma_z)) = 0$ *however, by Lemma C.0.3,* $\dim\ker(V) = 2$. *We can obtain the remaining basis elements of* $\ker(\Gamma)$ *by the following. Choosing a fixed element* $k \in \ker(\Gamma_u)$, *then there exists an* $r \in E_{-4}(\Gamma_u) = \mathrm{range}(\Gamma_m)$ *such that* $\Gamma(k \pm r) = \Gamma_u(k \pm r) + \Gamma_m(k \pm r) = \Gamma_u(\pm r) + \Gamma_m(k) = 0$ *since* $r \in \ker(\Gamma_m)$, *and* $\Gamma_m(k) \neq 0$ *and* $\Gamma_u(\pm r) \neq 0$.

From this analysis, we easily obtain the following Proposition which goes without proof.

**Proposition C.0.2.** *Let* $\Gamma$ *be a Lindblad generator which has a single canonical Lindblad term* $V := \frac{\sqrt{\gamma}}{2}(\sigma_{\boldsymbol{p}} + \mathrm{i}\sigma_{\boldsymbol{q}})$ *with* $\gamma \in \mathbb{R}^+$. *Then* $\rho = \frac{1}{2}(\mathbb{1}_{2^n} + \sigma_{\boldsymbol{m}}) \in \ker(\Gamma)$ *and hence is a fixed point of the corresponding Markovian semigroup generated by* $\Gamma$.

# Appendix D

# Computing Lie Wedges II: Extensions and Generalisations

## D.1 Unital Two-Qubit Systems

In this Appendix we extend the notions introduced in Chapter 4 of calculating a systems associated Lie wedge. Here we extend the wedge construction to two qubit systems which undergo various standard noise processes. The two qubits will be denoted A and B, respectively.

### D.1.1 Controllable Channels I

A fully Hamiltonian controllable two-qubit toy-model system with switchable Ising-coupling is given by the master equation

$$\dot{\rho} = -\big(\mathrm{i}\sum_j u_j \hat{\sigma}_j + \Gamma\big)\rho \tag{D.1}$$

where $\hat{\sigma}_j \in \{\hat{\sigma}_{x1}, \hat{\sigma}_{y1}; \hat{\sigma}_{1x}, \hat{\sigma}_{1y}; \hat{\sigma}_{zz}\}$ are the Hamiltonian control terms with amplitudes $\{u_j\}_{j=1}^{5} \in \mathbb{R}$.

Since $\langle \mathrm{i}\hat{\sigma}_j \,|\, j = 1, 2, \ldots, 5\rangle_{\mathsf{Lie}} = \mathrm{ad}_{\mathfrak{su}(4)}$, the edge of the wedge is $E(\mathfrak{w}) = \mathrm{ad}_{\mathfrak{su}(4)}$. Following the algorithm for an inner approximation of the Lie wedge, step (1) thus gives

$$\mathfrak{w}_1 := \mathrm{ad}_{\mathfrak{su}(4)} \oplus (-\mathbb{R}_0^+ \Gamma) \; . \tag{D.2}$$

Conjugating the dissipative component by the exponential map of the edge and then taking the convex hull yields the convex cone

$$\mathfrak{c}_0 := \mathbb{R}_0^+ \mathrm{conv}\big\{ \mathrm{ad}_U \, \Gamma \, \mathrm{Ad}_{U^\dagger} \;\; | \; U \in SU(4) \big\} \; , \tag{D.3}$$

which is the two-qubit analogue of the cone in Eqn. (4.16). The resulting associated Lie wedge is given by

$$\mathfrak{w}_0 := \mathrm{ad}_{\mathfrak{su}(4)} \oplus (-\mathfrak{c}_0) \; . \tag{D.4}$$

### D.1.2 Controllable Channels II

By shifting the Ising coupling term from the set of switchable control Hamiltonians into the (non-switchable) drift term, $\hat{\sigma}_d = \hat{\sigma}_{zz}$, one obtains the realistic and actually widely occurring type of system

$$\dot{\rho} = -\big(\mathrm{i}\hat{\sigma}_d + \mathrm{i}\sum_j u_j \hat{\sigma}_j + \Gamma\big)\rho \tag{D.5}$$

where now one just has the local control terms $\hat{\sigma}_j \in \{\hat{\sigma}_{x1}, \hat{\sigma}_{y1}; \hat{\sigma}_{1x}, \hat{\sigma}_{1y}\}$. Since $\langle \mathrm{i}\hat{\sigma}_{x1}, \mathrm{i}\hat{\sigma}_{y1}\rangle_{\mathsf{Lie}} = \mathrm{ad}_{\mathfrak{su}_{\mathsf{A}}(2)} \otimes \mathbb{1}_{\mathsf{B}}$, whereas on the other hand $\langle \mathrm{i}\hat{\sigma}_{1x}, \mathrm{i}\hat{\sigma}_{1y}\rangle_{\mathsf{Lie}} = \mathbb{1}_{\mathsf{A}} \otimes \mathrm{ad}_{\mathfrak{su}_{\mathsf{B}}(2)}$, the edge of the wedge

$$E(\mathfrak{w}_0) = \mathrm{ad}_{\mathfrak{su}_{\mathsf{A}}(2)\widehat{\oplus}\mathfrak{su}_{\mathsf{B}}(2)} \tag{D.6}$$

is in fact brought about by the Kronecker sum of local algebras

$$\mathfrak{su}_{\mathsf{A}}(2) \otimes \mathbb{1}_{\mathsf{B}} + \mathbb{1}_{\mathsf{A}} \otimes \mathfrak{su}_{\mathsf{B}}(2) =: \mathfrak{su}_{\mathsf{A}}(2)\widehat{\oplus}\mathfrak{su}_{\mathsf{B}}(2) \tag{D.7}$$

forming the generator of the group of local unitary actions

$$\exp\big(\mathfrak{su}_{\mathsf{A}}(2)\widehat{\oplus}\mathfrak{su}_{\mathsf{B}}(2)\big) = SU_{\mathsf{A}}(2) \otimes SU_{\mathsf{B}}(2)\ . \tag{D.8}$$

Remarkably, in this important class of open quantum-dynamical systems, qubits A and B are *locally* (H) controllable, respectively, while *globally* the system satisfies but the (WH) condition.

The final Lie wedge in these systems reads

$$\mathfrak{w}_{dk}^{2\oplus 2} = \mathrm{ad}_{\mathfrak{su}_{\mathsf{A}}(2)\widehat{\oplus}\mathfrak{su}_{\mathsf{B}}(2)} \oplus -\,\mathfrak{c}_{dk}^{2\oplus 2} \tag{D.9}$$

with the convex cone

$$\mathfrak{c}_{dk}^{2\oplus 2} := \mathbb{R}_0^+ \,\mathrm{conv}\,\big\{K_d^{2\oplus 2} + P_k^{2\oplus 2}\big\} \tag{D.10}$$

being given in terms of the respective $\mathfrak{k}$ and $\mathfrak{p}$-components. Here we use the short-hand notation $\hat{U}_{2\otimes 2} := \bar{U}_{2\otimes 2} \otimes U_{2\otimes 2}$ to arrive at

$$K_d^{2\oplus 2} \;:=\; \{\hat{U}_{2\otimes 2}(i\hat{\sigma}_d)\hat{U}_{2\otimes 2}^{\dagger} \,|\, U_{2\otimes 2} \in SU(2)\otimes SU(2)\}\ ,\ \text{and} \tag{D.11}$$

$$P_k^{2\oplus 2} \;:=\; \{\hat{U}_{2\otimes 2}(\Gamma)\hat{U}_{2\otimes 2}^{\dagger} \,|\, U_{2\otimes 2} \in SU(2)\otimes SU(2)\}\,. \tag{D.12}$$

As before, this immediately results from the initial wedge approximation by step (1)

$$\mathfrak{w}_1^{2\oplus 2} := \mathrm{ad}_{\mathfrak{su}_{\mathsf{A}}(2)\widehat{\oplus}\mathfrak{su}_{\mathsf{B}}(2)} \oplus (-\mathbb{R}_0^+(\mathrm{i}\hat{\sigma}_d + \Gamma)) \tag{D.13}$$

followed by conjugation with $\mathrm{Ad}_{\exp E(\mathfrak{w})} = \mathrm{Ad}_{2\otimes 2}$ to give

$$K_d^{2\oplus 2} + P_k^{2\oplus 2} := \mathcal{O}_{SU(2)\otimes SU(2)}\big(\mathrm{i}\hat{\sigma}_d + \Gamma\big)\ . \tag{D.14}$$

Step (3) then takes the convex hull and gives the final associated Lie wedge.

### D.1.3 Controllable Channels III

In the final example of a two-qubit system, the independent local controls shall even be limited to either $x$ or $y$-controls on the two qubits according to

$$\dot{\rho} = -\big(\mathrm{i}(\hat{\sigma}_d + u_{\mathsf{A}}\hat{\sigma}_{c1} + u_{\mathsf{B}}\hat{\sigma}_{1c'}) + \Gamma\big)\rho\ , \tag{D.15}$$

where now $\hat{\sigma}_d := \mathrm{i}\big(\hat{\sigma}_{z1} + \hat{\sigma}_{1z} + \hat{\sigma}_{zz}\big)$ and $\hat{\sigma}_{c1}$ with a single $c \in \{x, y\}$ and likewise $\hat{\sigma}_{1c'}$ with a single $c' \in \{x, y\}$ and $u_{\mathsf{A}}, u_{\mathsf{B}} \in \mathbb{R}$. Furthermore, assume the system undergoes *local uncorrelated noise* in each of the two subsystems in the sense that the Lindblad operators are of local form

$$V_k \in \{\sigma_{k1} \mid k \in \{x, y, z\}\}\ , \quad \text{and} \quad V_{k'} \in \{\sigma_{1k'} \mid k' \in \{x, y, z\}\}\ , \tag{D.16}$$

where $k$ and $k'$ are chosen independently $k, k' \in \{x, y, z\}$ so that one finds

$$\Gamma\ :=\ 2\gamma\hat{\sigma}_{k1}^2\ +\ 2\gamma'\hat{\sigma}_{1k'}^2\ . \tag{D.17}$$

This system satisfies but the (WH) condition both locally and globally, the latter following from

$$\langle \mathrm{i}\hat{\sigma}_{c1}, \mathrm{i}\hat{\sigma}_{1c'}, \mathrm{i}\hat{\sigma}_d\rangle_{\mathsf{Lie}} = \mathrm{ad}_{\mathfrak{su}(4)}\ . \tag{D.18}$$

The Lie wedge is given by

$$\mathfrak{w}_{kk'}^{cc'} := \mathrm{span}_{\mathbb{R}}\,\{\mathrm{i}\hat{\sigma}_c, \mathrm{i}\hat{\sigma}_{c'}\} \oplus (-\mathfrak{c}_{kk'}^{cc'})\ , \tag{D.19}$$

where the cone part $\mathfrak{c}_{kk'}^{cc'}$ is represented as

$$\mathfrak{c}_{kk'}^{cc'} := \mathbb{R}_0^+\,\mathrm{conv}\,\big\{K^c(\theta) + K^{c'}(\theta') + K^{cc'}(\theta,\theta') + P_k^c(\theta) + P_{k'}^{c'}(\theta') \,|\, \theta, \theta' \in \mathbb{R}\,\big\} \tag{D.20}$$

which is given in terms of the $\mathfrak{k}$- and $\mathfrak{p}$" components (setting $\theta := u_{\mathsf{A}}$ and $\theta' := u_{\mathsf{B}}$ and using the relations in given by Eqn. (4.18)) as

$$K^c(\theta) + K^{c'}(\theta') + K^{cc'}(\theta,\theta') = \begin{bmatrix} \cos(\theta) \\ \sin(\theta) \\ \cos(\theta') \\ \sin(\theta') \\ \cos(\theta)\cos(\theta') \\ \cos(\theta)\sin(\theta') \\ \cos(\theta')\sin(\theta) \\ \sin(\theta)\sin(\theta') \end{bmatrix} \cdot \mathrm{i} \begin{bmatrix} \hat{\sigma}_{z1} \\ \varepsilon_{czq}\hat{\sigma}_{q1} \\ \hat{\sigma}_{1z} \\ \varepsilon_{c'zq'}\hat{\sigma}_{1q'} \\ \hat{\sigma}_{zz} \\ \varepsilon_{c'zq'}\hat{\sigma}_{zq'} \\ \varepsilon_{czq}\hat{\sigma}_{qz} \\ \hat{\sigma}_{qq'} \end{bmatrix} \tag{D.21}$$

and (as in Eqn. (4.20))

$$P_k^c(\theta) = 2\gamma \begin{bmatrix} \cos^2(\theta) \\ \sin^2(\theta) \\ \cos(\theta)\sin(\theta) \end{bmatrix} \cdot \begin{bmatrix} \hat{\sigma}_{k1}^2 \\ \hat{\sigma}_{r1}^2 \\ \varepsilon_{ckr}\{\hat{\sigma}_{k1},\hat{\sigma}_{r1}\}_+ \end{bmatrix} \tag{D.22}$$

as well as

$$P_{k'}^{c'}(\theta') = 2\gamma' \begin{bmatrix} \cos^2(\theta') \\ \sin^2(\theta') \\ \cos(\theta')\sin(\theta') \end{bmatrix} \cdot \begin{bmatrix} \hat{\sigma}_{1k'}^2 \\ \hat{\sigma}_{1r'}^2 \\ \varepsilon_{c'k'r'}\{\hat{\sigma}_{1k'}^2,\hat{\sigma}_{1r'}^2\}_+ \end{bmatrix} \tag{D.23}$$

To see this, observe that by step (1), the initial wedge approximation is given by

$$\mathfrak{w}_1 := \mathrm{span}_{\mathbb{R}}\,\{\mathrm{i}\hat{\sigma}_c, \mathrm{i}\hat{\sigma}_{c'}\} \oplus (-\mathbb{R}^+(\mathrm{i}\hat{\sigma}_d + 2\gamma\hat{\sigma}_k^2 + 2\gamma'\hat{\sigma}_{k'}^2))\,, \tag{D.24}$$

**Table D.1:** Two-Qubit System Algebras Undergoing Strong or Weak Couplings

| Noise | Lindblad Term | Control | Drift | $\dim(\mathfrak{g}_\Sigma)$ | $\dim(\mathfrak{w}–\mathfrak{w})$ |
|---|---|---|---|---|---|
| Local Unital | $\sigma_\mu^{*)}$ | $\sigma_{x1}, \sigma_{1x}$ | Weak$^{**)}$ | 225 | 13 |
| Local Dephasing | $\sigma_{z1}$ | $\sigma_{x1}$ | –"– | 22 | 6 |
| –"– | –"– | $\sigma_{1x}$ | –"– | 5 | 4 |
| Local Bit Flip | $\sigma_{x1}$ | $\sigma_{x1}$ | –"– | 16 | 4 |
| –"– | –"– | $\sigma_{1x}$ | –"– | 52 | 4 |
| Local Unital | $\sigma_\mu^{*)}$ | $\sigma_{x1}, \sigma_{1x}$ | Strong$^{***)}$ | 225 | 13 |
| Local Dephasing | $\sigma_{z1}$ | $\sigma_{x1}$ | –"– | 225 | 6 |
| –"– | –"– | $\sigma_{1x}$ | –"– | 225 | 4 |
| Local Bit Flip | $\sigma_{x1}$ | $\sigma_{x1}$ | –"– | 124 | 4 |
| –"– | –"– | $\sigma_{1x}$ | –"– | 225 | 4 |

$^{*)}$ And $\mu \neq x1$ or $1x$ $^{**)}$ $H_d = \mathrm{i}(\sigma_{z1}+\sigma_{1z}+\sigma_{zz})$. $^{***)}$ $H_d = \mathrm{i}(\sigma_{z1}+\sigma_{1z}+\sigma_{xx}+\sigma_{yy}+\sigma_{zz})$.

which has to be conjugated by $\mathrm{Ad}_{\exp(E(\mathfrak{w}))}$. As usual, the edge of the wedge is invariant under such a conjugation, so we need only determine the effects on the drift components of the system as is done in Eqns. (D.21) through (D.23).

Now, the generalisation to systems with more than two qubits satisfying the (H) or (WH) condition is obvious: assuming uncorrelated noise, the $\mathfrak{p}$-parts of the Lie wedges can be immediately extended on the grounds of the previous description, since all processes are local on each qubit. Though straightforward, calculating the $\mathfrak{k}$-components becomes a bit more tedious: but the many-body coherences have to be considered just as in Eqn. (D.21).

**Table D.2:** Analysis of Two-Qubit System Algebras

| Noise | Lindblad Terms | Control | Drift | System Algebra $\mathfrak{g}_\Sigma$ | $\dim(\mathfrak{w}–\mathfrak{w})$ |
|---|---|---|---|---|---|
| Std. Unital | $\sigma_\mu{}^{*)}$ | $\mathrm{ad}_{\mathfrak{su}(4)}$ | $\mathrm{i}(\sigma_{z1}+\sigma_{1z}+\sigma_{zz})$ | $\mathfrak{g}_0^{LK}$ | 135 |
| Std. Local Unital | $\sigma_\mu{}^{**)}$ | $\mathrm{ad}_{\mathfrak{su}(2)\oplus\mathfrak{su}(2)}$ | –"– | $\mathfrak{g}_0^{LK}$ | 21 |
| Local A&B Dephasing | $\sigma_{z1}, \sigma_{1z}$ | $\sigma_{x1}, \sigma_{1x}$ | –"– | $\mathfrak{g}_0^{LK}$ | 16 |
| –"– | $\sigma_{z1}, \sigma_{1z}$ | $\sigma_{x1}, \sigma_{1x}$ | $\mathrm{i}(\sigma_{z1}+\sigma_{1z})$ | $\mathfrak{g}_0^{LK}$ | 12 |
| Global Depolarizing | $\{\sigma_\mu\}^{***)}$ | $\mathrm{ad}_{\mathfrak{su}(2)\oplus\mathfrak{su}(2)}$ | –"– | $\mathrm{ad}_{\mathfrak{su}(2)\oplus\mathfrak{su}(2)} \oplus \mathbb{R}\Gamma$ | 7 |
| Completely Depolarizing | $\{\sigma_\mu\}^{****)}$ | $\mathrm{ad}_{\mathfrak{su}(4)}$ | –"– | $\mathrm{ad}_{\mathfrak{su}(4)} \oplus \mathbb{R}\Gamma$ | 16 |
| Amplitude Damping | $\sigma_{x1} + \mathrm{i}\sigma_{y1}$ | $\mathrm{ad}_{\mathfrak{su}(4)}$ | $\mathrm{i}(\sigma_{z1}+\sigma_{1z}+\sigma_{zz})$ | $\mathfrak{g}^{LK}$ | 240 |
| –"– | –"– | $\mathrm{ad}_{\mathfrak{su}(2)\oplus\mathfrak{su}(2)}$ | –"– | $\mathfrak{g}^{LK}$ | 24 |

$^{*)}$ Any Pauli matrix. $^{*)}$ Any local Pauli matrix. $^{***)}$ The nine non-local Pauli matrices. $^{****)}$ All Pauli matrices.

## D.2 Non-Unital Two-Qubit Systems

### D.2.1 Controllable Channels I

Consider a fully Hamiltonian controllable system with switchable Ising coupling given by the master equation

$$\dot{\rho} = -\big(\mathrm{i}\sum_{j=1}^{5} u_j \hat{\sigma}_j + \Gamma\big)\rho \tag{D.25}$$

where $\hat{\sigma}_j \in \{\hat{\sigma}_{x1}, \hat{\sigma}_{y1}, \hat{\sigma}_{1x}, \hat{\sigma}_{1y}, \hat{\sigma}_{zz}\}$ are the control Hamiltonians which pair with their control amplitudes $\{u_j\}_{j=1}^{5} \in \mathbb{R}$. Moreover, $\Gamma$ shall be given by an amplitude damping term acting locally on qubit $\mathsf{A}$ and a bit-flip term acting on qubit $\mathsf{B}$, i.e. $\Gamma = \Gamma_{\mathsf{A}} + \Gamma_{\mathsf{B}}$ where

$$\Gamma_{\mathsf{A}} := 2\gamma\left(\hat{\sigma}_{x1}^2 + \hat{\sigma}_{y1}^2 + 2\mathrm{i}\hat{\sigma}_{x1}\hat{\sigma}_{y1}^{+}\right)\;, \quad \text{and} \quad \Gamma_{\mathsf{B}} := 2\gamma'\hat{\sigma}_{1x}^2\;. \tag{D.26}$$

Conjugating $\Gamma$ by the exponential of the edge of the wedge $E(\mathfrak{w}) = \mathrm{ad}_{\mathfrak{su}(4)}$ and taking the convex hull as in the fully H controllable depolarizing single-qubit case of Eqn. (4.31) yields the associated Lie wedge

$$\mathfrak{w} = \mathrm{ad}_{\mathfrak{su}(4)} \oplus (-\mathfrak{c}_{\mathsf{AB}})\;, \tag{D.27}$$

where the cone $\mathfrak{c}_{\mathsf{AB}}$ is

$$\mathfrak{c}_{\mathsf{AB}} := \mathbb{R}_0^{+}\mathrm{conv}\big\{\,\mathrm{Ad}_U\left(2\gamma(\hat{\sigma}_{x1}^2 + \hat{\sigma}_{y1}^2 + 2\mathrm{i}\hat{\sigma}_{x1}\hat{\sigma}_{y1}^{+}) + 2\gamma'\hat{\sigma}_{1x}^2\right)\mathrm{Ad}_{U^\dagger}\;\mid U \in SU(4)\big\}\;. \tag{D.28}$$

### D.2.2 Controllable Channels II

Now we consider a modified version of the previous example by limiting the number of control Hamiltonians. The master equation is given by

$$\dot{\rho} = -\big(\mathrm{i}(\hat{\sigma}_d + u_{\mathsf{A}}\hat{\sigma}_{x1} + u_{\mathsf{B}}\hat{\sigma}_{1x}) + \Gamma_{\mathsf{A}} + \Gamma_{\mathsf{B}}\big)\rho\;, \tag{D.29}$$

where $\hat{\sigma}_d := \mathrm{i}\big(\hat{\sigma}_{z1} + \hat{\sigma}_{1z} + \hat{\sigma}_{zz}\big)$, $u_{\mathsf{A}}, u_{\mathsf{B}} \in \mathbb{R}$, and $\Gamma_{\mathsf{A}}, \Gamma_{\mathsf{B}}$ are given by Eqn. (D.26). The edge of the wedge is then

$$E(\mathfrak{w}) = \mathrm{span}_{\mathbb{R}}\,\{\mathrm{i}\hat{\sigma}_{x1}, \mathrm{i}\hat{\sigma}_{1x}\}, \tag{D.30}$$

and those $\mathfrak{k}$-components of the cone that are due to conjugation of the drift Hamiltonian can readily be calculated using Eqn. (D.21)

$$K^{x1}(\theta) + K^{1x}(\theta') + K^{xx}(\theta,\theta') = \begin{bmatrix} \cos(\theta) \\ -\sin(\theta) \\ \cos(\theta') \\ -\sin(\theta') \\ \cos(\theta)\cos(\theta') \\ -\cos(\theta)\sin(\theta') \\ -\cos(\theta')\sin(\theta) \\ \sin(\theta)\sin(\theta') \end{bmatrix} \cdot \mathrm{i} \begin{bmatrix} \hat{\sigma}_{z1} \\ \hat{\sigma}_{y1} \\ \hat{\sigma}_{1z} \\ \hat{\sigma}_{1y} \\ \hat{\sigma}_{zz} \\ \hat{\sigma}_{zy} \\ \hat{\sigma}_{yz} \\ \hat{\sigma}_{yy} \end{bmatrix}\;. \tag{D.31}$$

Since $\Gamma_{\mathsf{B}} := 2\gamma'\hat{\sigma}_{1x}^2$ clearly remains invariant under $x$ controls, we only have to consider conjugation of the amplitude-damping term $\Gamma_{\mathsf{A}}$, which reduces to the example given in Section 4.4.2 (by exchanging $y$ controls against $x$ controls) to give

$$P_{y1}^{x1}(\theta) + 2\gamma\hat{\sigma}_{x1}^2 + N_{(xy)1}^{x1}(\theta) = 2\gamma \begin{bmatrix} 2\cos(\theta) \\ 2\sin(\theta) \\ \cos(\theta)\sin(\theta) \\ 1 \\ \cos^2(\theta) \\ \sin^2(\theta) \end{bmatrix} \cdot \begin{bmatrix} \mathrm{i}\,\hat{\sigma}_{x1}\hat{\sigma}_{y1}^{+} \\ \mathrm{i}\,\hat{\sigma}_{z1}\hat{\sigma}_{x1}^{+} \\ \{\hat{\sigma}_{y1},\hat{\sigma}_{z1}\}_{+} \\ \hat{\sigma}_{x1}^2 \\ \hat{\sigma}_{y1}^2 \\ \hat{\sigma}_{z1}^2 \end{bmatrix}\;. \tag{D.32}$$

Thus in terms of the above components, in total one obtains the associated Lie wedge

$$\mathfrak{w} \quad = \quad \mathrm{span}_{\mathbb{R}} \{\mathrm{i}\hat{\sigma}_{x1}, \mathrm{i}\hat{\sigma}_{1x}\} \oplus (-\mathfrak{c}) \tag{D.33}$$

where

$$\begin{aligned} \mathfrak{c} := \mathbb{R}_0^+ \operatorname{conv} \big\{ & K^{x1}(\theta) + K^{1x}(\theta') + K^{xx}(\theta,\theta') \\ & + P_{y1}^{x1}(\theta) + 2\gamma\hat{\sigma}_{x1}^2 + N_{(xy)1}^{x1}(\theta) + 2\gamma'\hat{\sigma}_{1x}^2 \,|\, \theta,\theta' \in \mathbb{R} \big\} \ . \end{aligned} \tag{D.34}$$

### D.2.3 Controllable Channels III

The previous example can easily be modified to to case where we have amplitude damping on each of the two qubits A and B expressed as

$$\Gamma_{\mathsf{A}} := 2\gamma \left( \hat{\sigma}_{x1}^2 + \hat{\sigma}_{y1}^2 + 2\mathrm{i}\, \hat{\sigma}_{x1} \hat{\sigma}_{y1}^{+} \right) \ , \quad \text{and} \quad \Gamma_{\mathsf{B}} := 2\gamma' \left( \hat{\sigma}_{1x}^2 + \hat{\sigma}_{1y}^2 \pm 2\mathrm{i}\, \hat{\sigma}_{1x} \hat{\sigma}_{1y}^{+} \right) \ . \tag{D.35}$$

The components given by Eqns. (D.30) and (D.31) remain unaltered, the only difference to the previous example is that $\Gamma_{\mathsf{B}}$ is no longer invariant and hence rise to the new components

$$P_{1y}^{1x}(\theta') + 2\gamma'\hat{\sigma}_{1x}^2 \pm N_{1(xy)}^{1x}(\theta') = 2\gamma' \begin{bmatrix} \pm 2\cos(\theta') \\ \pm 2\sin(\theta') \\ \cos(\theta')\sin(\theta') \\ 1 \\ \cos^2(\theta') \\ \sin^2(\theta') \end{bmatrix} \cdot \begin{bmatrix} \mathrm{i}\, \hat{\sigma}_{1x}\hat{\sigma}_{1y}^{+} \\ \mathrm{i}\, \hat{\sigma}_{1z}\hat{\sigma}_{1x}^{+} \\ \{\hat{\sigma}_{1y}, \hat{\sigma}_{1z}\}_{+} \\ \hat{\sigma}_{1x}^2 \\ \hat{\sigma}_{1y}^2 \\ \hat{\sigma}_{1z}^2 \end{bmatrix} \ . \tag{D.36}$$

The final wedge follows Eqn. (D.33), where $\mathfrak{c}$ is now given by

$$\begin{aligned} \mathfrak{c} := \mathbb{R}_0^+ \operatorname{conv} \big\{ & K^{x1}(\theta) + K^{1x}(\theta') + K^{xx}(\theta,\theta') + P_{y1}^{x1}(\theta) + 2\gamma\hat{\sigma}_{x1}^2 + N_{(xy)1}^{x1}(\theta) \\ & + P_{1y}^{1x}(\theta') + 2\gamma'\hat{\sigma}_{1x}^2 \pm N_{1(xy)}^{1x}(\theta') \,|\, \theta,\theta' \in \mathbb{R} \big\} \ . \end{aligned} \tag{D.37}$$

# Appendix E

# Commutation Relations: Single-Qubit System Lie Algebra

**Table E.1**

| $\left[\mathrm{i}\,\hat{\sigma}_p\,,\hat{\sigma}_q^2\right]$ | $\hat{\sigma}_x^2$ | $\hat{\sigma}_y^2$ | $\hat{\sigma}_z^2$ |
|---|---|---|---|
| $\mathrm{i}\,\hat{\sigma}_x$ | $0$ | $-\{\hat{\sigma}_y,\hat{\sigma}_z\}_+$ | $\{\hat{\sigma}_y,\hat{\sigma}_z\}_+$ |
| $\mathrm{i}\,\hat{\sigma}_y$ | $\{\hat{\sigma}_z,\hat{\sigma}_x\}_+$ | $0$ | $-\{\hat{\sigma}_z,\hat{\sigma}_x\}_+$ |
| $\mathrm{i}\,\hat{\sigma}_z$ | $-\{\hat{\sigma}_x,\hat{\sigma}_y\}_+$ | $\{\hat{\sigma}_x,\hat{\sigma}_y\}_+$ | $0$ |

**Table E.2**

| $\left[\mathrm{i}\,\hat{\sigma}_p\,,\{\hat{\sigma}_q,\hat{\sigma}_r\}_+\right]$ | $\{\hat{\sigma}_y,\hat{\sigma}_z\}_+$ | $\{\hat{\sigma}_z,\hat{\sigma}_x\}_+$ | $\{\hat{\sigma}_x,\hat{\sigma}_y\}_+$ |
|---|---|---|---|
| $\mathrm{i}\,\hat{\sigma}_x$ | $2(\hat{\sigma}_y^2-\hat{\sigma}_z^2)$ | $\{\hat{\sigma}_x,\hat{\sigma}_y\}_+$ | $-\{\hat{\sigma}_z,\hat{\sigma}_x\}_+$ |
| $\mathrm{i}\,\hat{\sigma}_y$ | $-\{\hat{\sigma}_x,\hat{\sigma}_y\}_+$ | $2(\hat{\sigma}_z^2-\hat{\sigma}_x^2)$ | $\{\hat{\sigma}_y,\hat{\sigma}_z\}_+$ |
| $\mathrm{i}\,\hat{\sigma}_z$ | $\{\hat{\sigma}_z,\hat{\sigma}_x\}_+$ | $-\{\hat{\sigma}_y,\hat{\sigma}_z\}_+$ | $2(\hat{\sigma}_x^2-\hat{\sigma}_y^2)$ |

**Table E.3**

| $\left[\{\hat{\sigma}_p, \hat{\sigma}_q\}_+, \{\hat{\sigma}_r, \hat{\sigma}_s\}_+\right]$ | $\{\hat{\sigma}_y, \hat{\sigma}_z\}_+$ | $\{\hat{\sigma}_z, \hat{\sigma}_x\}_+$ | $\{\hat{\sigma}_x, \hat{\sigma}_y\}_+$ |
|---|---|---|---|
| $\{\hat{\sigma}_y, \hat{\sigma}_z\}_+$ | $0$ | $-\mathrm{i}\hat{\sigma}_z$ | $\mathrm{i}\hat{\sigma}_y$ |
| $\{\hat{\sigma}_z, \hat{\sigma}_x\}_+$ | $\mathrm{i}\hat{\sigma}_z$ | $0$ | $-\mathrm{i}\hat{\sigma}_x$ |
| $\{\hat{\sigma}_x, \hat{\sigma}_y\}_+$ | $-\mathrm{i}\hat{\sigma}_y$ | $\mathrm{i}\hat{\sigma}_x$ | $0$ |

**Table E.4**

| $\left[\hat{\sigma}_p^2, \{\hat{\sigma}_q, \hat{\sigma}_r\}_+\right]$ | $\{\hat{\sigma}_y, \hat{\sigma}_z\}_+$ | $\{\hat{\sigma}_z, \hat{\sigma}_x\}_+$ | $\{\hat{\sigma}_x, \hat{\sigma}_y\}_+$ |
|---|---|---|---|
| $\hat{\sigma}_x^2$ | $0$ | $-\mathrm{i}\hat{\sigma}_y$ | $\mathrm{i}\hat{\sigma}_z$ |
| $\hat{\sigma}_y^2$ | $\mathrm{i}\hat{\sigma}_x$ | $0$ | $-\mathrm{i}\hat{\sigma}_z$ |
| $\hat{\sigma}_z^2$ | $-\mathrm{i}\hat{\sigma}_x$ | $\mathrm{i}\hat{\sigma}_y$ | $0$ |

**Table E.5**

| $\left[\mathrm{i}\,\hat{\sigma}_p\hat{\sigma}_q^+, \mathrm{i}\,\hat{\sigma}_r\right]$ | $\mathrm{i}\,\hat{\sigma}_x$ | $\mathrm{i}\,\hat{\sigma}_y$ | $\mathrm{i}\,\hat{\sigma}_z$ |
|---|---|---|---|
| $\mathrm{i}\,\hat{\sigma}_y\hat{\sigma}_z^+$ | $\mathbf{0}$ | $-\mathrm{i}\,\hat{\sigma}_x\hat{\sigma}_y^+$ | $\mathrm{i}\,\hat{\sigma}_z\hat{\sigma}_x^+$ |
| $\mathrm{i}\,\hat{\sigma}_z\hat{\sigma}_x^+$ | $\mathrm{i}\,\hat{\sigma}_x\hat{\sigma}_y^+$ | $\mathbf{0}$ | $-\mathrm{i}\,\hat{\sigma}_y\hat{\sigma}_z^+$ |
| $\mathrm{i}\,\hat{\sigma}_x\hat{\sigma}_y^+$ | $-\mathrm{i}\,\hat{\sigma}_z\hat{\sigma}_x^+$ | $\mathrm{i}\,\hat{\sigma}_y\hat{\sigma}_z^+$ | $\mathbf{0}$ |

**Table E.6**

| $[\mathrm{i}\,\hat{\sigma}_p\hat{\sigma}_q^+, \mathrm{i}\,\hat{\sigma}_r^2]$ | $\mathrm{i}\,\hat{\sigma}_x^2$ | $\mathrm{i}\,\hat{\sigma}_y^2$ | $\mathrm{i}\,\hat{\sigma}_z^2$ |
|---|---|---|---|
| $\mathrm{i}\,\hat{\sigma}_y\hat{\sigma}_z^+$ | $\mathbf{0}$ | $\mathrm{i}\,\hat{\sigma}_y\hat{\sigma}_z^+$ | $\mathrm{i}\,\hat{\sigma}_y\hat{\sigma}_z^+$ |
| $\mathrm{i}\,\hat{\sigma}_z\hat{\sigma}_x^+$ | $\mathrm{i}\,\hat{\sigma}_z\hat{\sigma}_x^+$ | $\mathbf{0}$ | $\mathrm{i}\,\hat{\sigma}_z\hat{\sigma}_x^+$ |
| $\mathrm{i}\,\hat{\sigma}_x\hat{\sigma}_y^+$ | $\mathrm{i}\,\hat{\sigma}_x\hat{\sigma}_y^+$ | $\mathrm{i}\,\hat{\sigma}_x\hat{\sigma}_y^+$ | $\mathbf{0}$ |

**Table E.7**

| $\left[\mathrm{i}\,\hat{\sigma}_p\hat{\sigma}_q^+, \{\hat{\sigma}_r, \hat{\sigma}_s\}_+\right]$ | $\{\hat{\sigma}_x, \hat{\sigma}_y\}_+$ | $\{\hat{\sigma}_y, \hat{\sigma}_z\}_+$ | $\{\hat{\sigma}_z, \hat{\sigma}_x\}_+$ |
|---|---|---|---|
| $\mathrm{i}\,\hat{\sigma}_x\hat{\sigma}_y^+$ | $\mathbf{0}$ | $\mathrm{i}\,\hat{\sigma}_z\hat{\sigma}_x^+$ | $\mathrm{i}\,\hat{\sigma}_y\hat{\sigma}_z^+$ |
| $\mathrm{i}\,\hat{\sigma}_y\hat{\sigma}_z^+$ | $\mathrm{i}\,\hat{\sigma}_z\hat{\sigma}_x^+$ | $\mathbf{0}$ | $\mathrm{i}\,\hat{\sigma}_x\hat{\sigma}_y^+$ |
| $\mathrm{i}\,\hat{\sigma}_z\hat{\sigma}_x^+$ | $\mathrm{i}\,\hat{\sigma}_y\hat{\sigma}_z^+$ | $\mathrm{i}\,\hat{\sigma}_x\hat{\sigma}_y^+$ | $\mathbf{0}$ |